%% file: main.tex
\documentclass[10pt,twocolumn,letterpaper]{article}
\usepackage[pagenumbers]{cvpr}

\input{preamble}

\definecolor{cvprblue}{rgb}{0.21,0.49,0.74}
\usepackage[pagebackref,breaklinks,colorlinks,allcolors=cvprblue]{hyperref}

\begin{document}

\title{Tight Inversion: Image-Conditioned Inversion for Real Image Editing}

\author{
  Edo Kadosh\thanks{Equal contribution.}\textsuperscript{\enspace 1} \quad
  Nir Goren\footnotemark[1]\textsuperscript{\enspace 1} \quad
  Or Patashnik\textsuperscript{1,2} \quad
  Daniel Garibi\textsuperscript{1} \quad
  Daniel Cohen-Or\textsuperscript{1,2} \\[5pt]
  \textsuperscript{1}Tel Aviv University \quad \textsuperscript{2}Snap Research \\
}

\maketitle
\begin{abstract}
    \input{sec/0_abstract}

\end{abstract}
\input{sec/1_intro}

\input{sec/2_related_work}

\input{sec/3_method}

\input{sec/4_experiments}

\input{sec/5_conclusion}

{
    \small
    \bibliographystyle{ieeenat_fullname}
    \bibliography{main}
}

\clearpage
\input{sec/6_figures_only}

\end{document}

%% file: preamble.tex
\usepackage{multirow}
\usepackage{pgfplots}
\usepackage{amssymb}
\usepgfplotslibrary{groupplots}
\usepackage{pgfplotstable}
\usepackage{soul}
\usepackage{subcaption}
\pgfplotsset{compat=newest}
\usepackage{siunitx}

%% file: sec/0_abstract.tex
Text-to-image diffusion models offer powerful image editing capabilities.
To edit real images, many methods rely on the inversion of the image into Gaussian noise.
A common approach to invert an image is to gradually add noise to the image, where the noise is determined by reversing the sampling equation. 
This process has an inherent tradeoff between reconstruction and editability, limiting the editing of challenging images such as highly-detailed ones.
Recognizing the reliance of text-to-image models inversion on a text condition, this work explores the importance of the condition choice.
We show that a condition that precisely aligns with the input image significantly improves the inversion quality.
Based on our findings, we introduce Tight Inversion, an inversion method that utilizes the most possible precise condition -- the input image itself.
This tight condition narrows the distribution of the model's output and enhances both reconstruction and editability.
We demonstrate the effectiveness of our approach when combined with existing inversion methods through extensive experiments, evaluating the reconstruction accuracy as well as the integration with various editing methods.

%% file: sec/1_intro.tex
\section{Introduction}
\label{sec:intro}

\input{figures/teaser}

Text-to-image diffusion models have seen remarkable advancements in recent years \cite{ho2020denoisingdiffusionprobabilisticmodels, ramesh2022hierarchicaltextconditionalimagegeneration}. These models generate images through an iterative denoising process, where each step is conditioned on the input text prompt. 
This condition text prompt dictates the conditional distribution from which the generated image is sampled, and guides each step towards this distribution.

The ability of these models to produce high-quality and diverse images has sparked significant interest in their potential for editing \textit{real} images, which often fall outside the model's native distribution. 
To edit real images, inversion techniques are often employed to derive an initial noise that faithfully reconstructs the real image through the model's denoising process~\cite{dhariwal2021diffusionmodelsbeatgans, Mokady_2023_CVPR, hubermanspiegelglas2024editfriendlyddpmnoise}. Having the initial noise that reconstructs the image allows to steer the denoising process towards the target edit~\cite{hertz2022prompt, tumanyan2022plugandplaydiffusionfeaturestextdriven, hubermanspiegelglas2024editfriendlyddpmnoise, Parmar_2023, cao2023masactrltuningfreemutualselfattention, patashnik2023localizingobjectlevelshapevariations}.

Inverting a real image presents a significant challenge, as it requires balancing the tradeoff between accurately reconstructing the image and ensuring the editability of the resulting initial noise~\cite{tov2021designing}. DDIM inversion~\cite{song2022denoisingdiffusionimplicitmodels, dhariwal2021diffusionmodelsbeatgans} is a widely used approach and serves as the basis for many other inversion techniques~\cite{Mokady_2023_CVPR, garibi2024renoise, miyake2023negative, samuel2024lightningfastimageinversionediting}. This method reverses the sampling process by performing forward diffusion according to the reversed algorithm, utilizing the diffusion model at each step. When applying DDIM inversion in text-to-image diffusion models, the process also relies on setting an appropriate text prompt, which conditions the model’s prediction at every forward step.

In this work, we investigate the role of the specific condition used during the inversion process. Our findings reveal that conditioning the inversion on a text prompt that accurately describes the input image improves both the reconstruction quality and the editability of the inversion results. 
These findings are illustrated in Figure~\ref{fig:motivation-figure}. We present there reconstruction results of DDIM inversion using three levels of text prompt specificity.
As shown, closely aligning the condition with the source image effectively narrows and tightens the model's target distribution, shifting it from a broad range of images to those closely resembling the source image. 

Building on these insights, we propose an inversion approach that employs the ultimate condition: the source image itself. We call this method \textit{Tight Inversion}, as image conditions are inherently more precise than text conditions. This tight conditioning significantly improves inversion quality and enhances editing performance. We show that Tight Inversion integrates seamlessly with various inversion methods beyond DDIM inversion, consistently enhancing their performance.

To evaluate the effectiveness of our approach, we conduct extensive experiments, emphasizing both reconstruction accuracy and editability. While reconstruction ensures that the inverted noise reproduces the given image, it is not a meaningful goal on its own. The true purpose of inversion is to enable meaningful edits to the reconstructed image. Thus, our evaluation emphasizes how well the inversion facilitates edits while preserving fidelity to the original content. Our experiments primarily target the inversion of complex and challenging images, as this is where the strength of our method truly stands out. We demonstrate the effectiveness of our method using three types of models: a standard diffusion model, a few-step diffusion model, and a flow model.

\input{figures/motivation_figure}

%% file: figures/teaser.tex
\begin{figure}
    \centering
    \setlength{\tabcolsep}{1pt}
    \includegraphics[width=\linewidth]{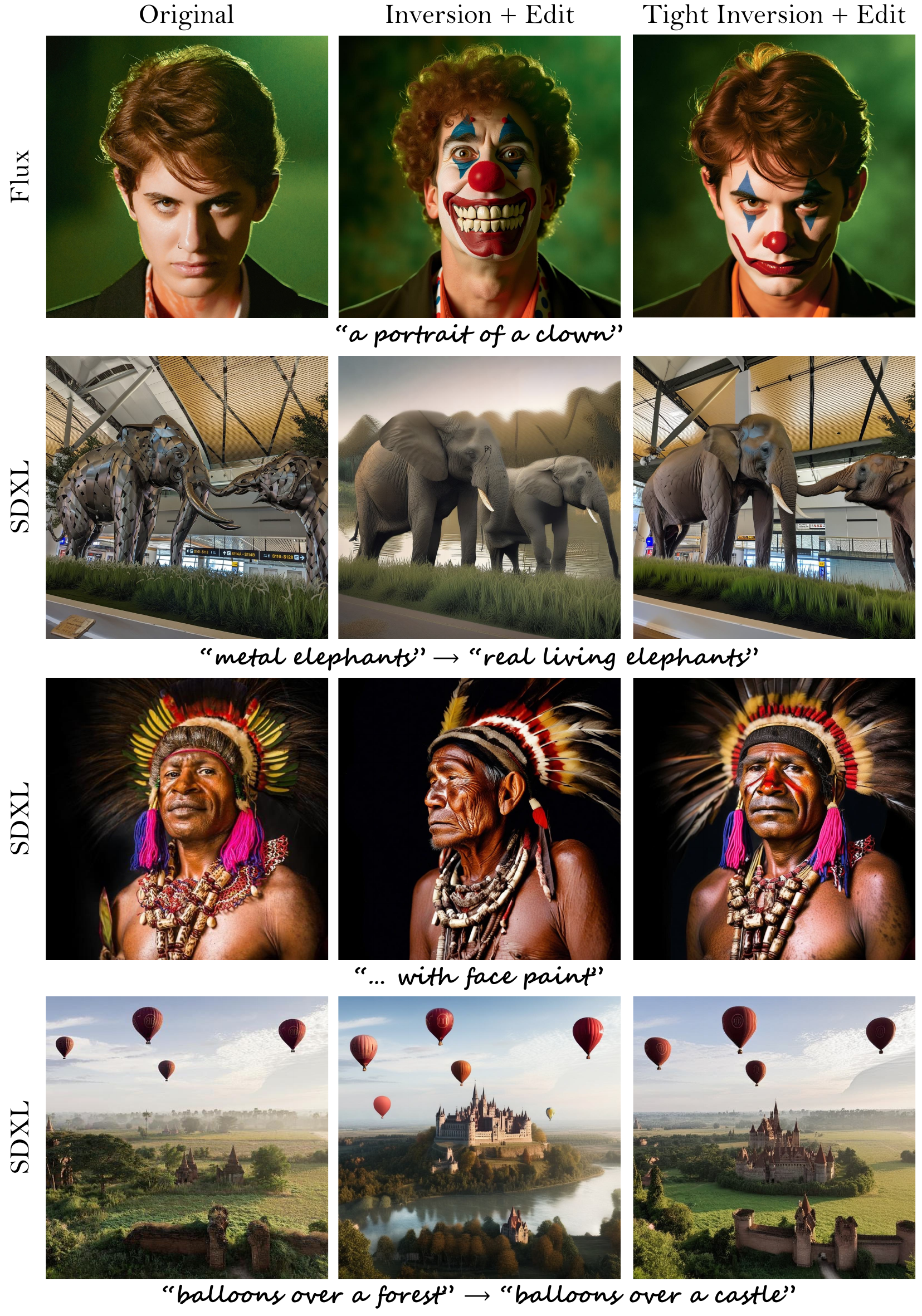}
    \vspace{-16pt}
    \caption{Our Tight Inversion method facilitates the editing of highly-detailed challenging real images across different models. 
    }
    \vspace{-16pt}
    \label{fig:teaser}
\end{figure}

%% file: figures/motivation_figure.tex
\begin{figure}
    \centering
    \setlength{\tabcolsep}{1pt}
    \begin{tabular}{ccccc}
        \includegraphics[width=0.19\linewidth]{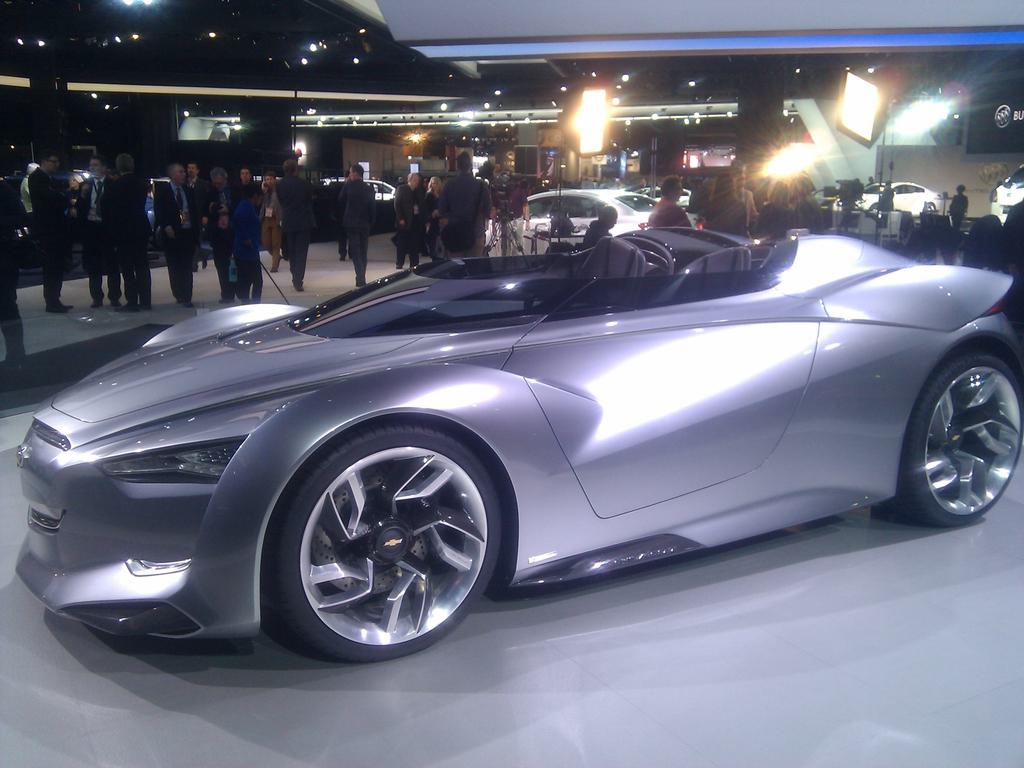} &
        \includegraphics[width=0.19\linewidth]{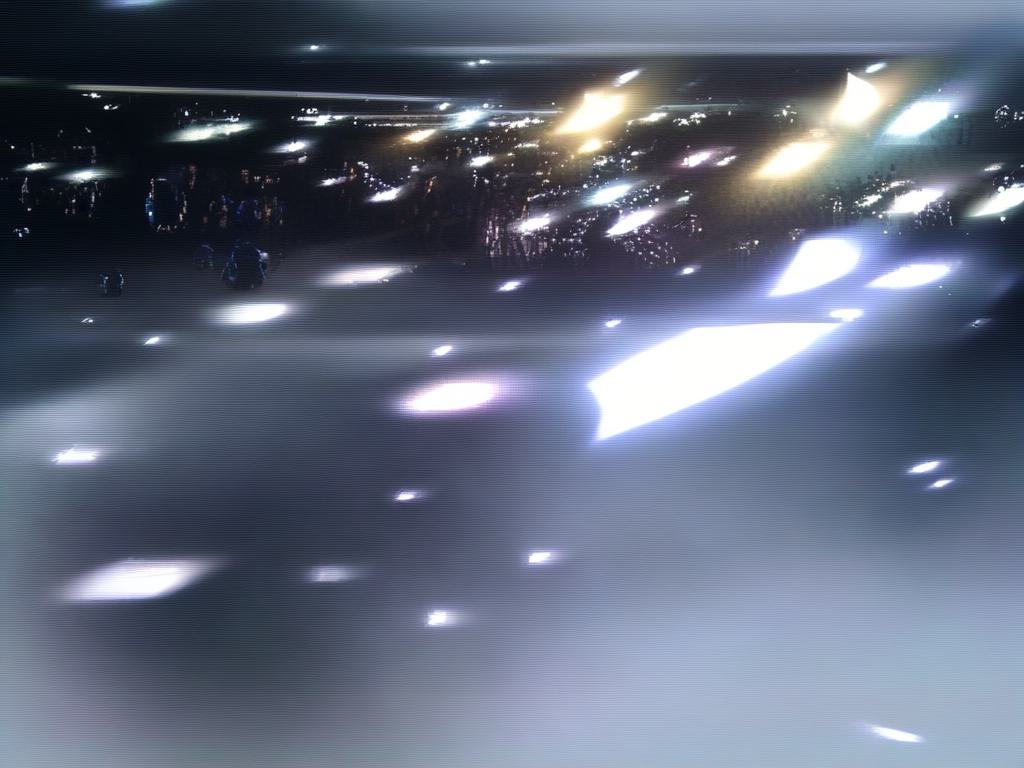} &
        \includegraphics[width=0.19\linewidth]{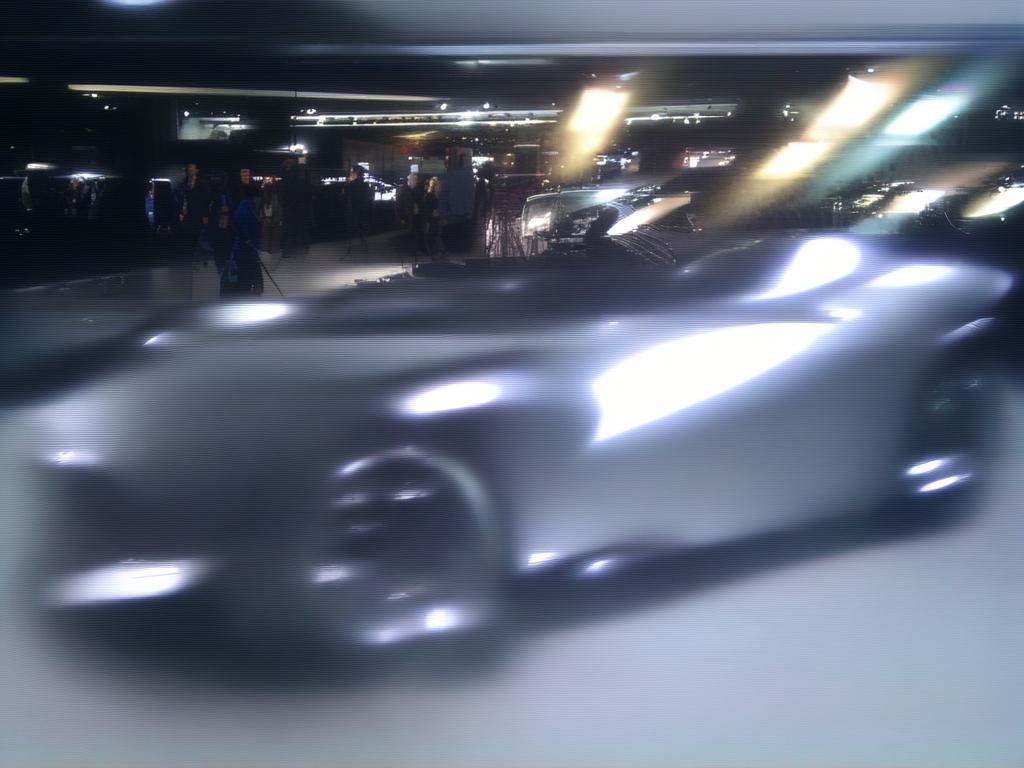} &
        \includegraphics[width=0.19\linewidth]{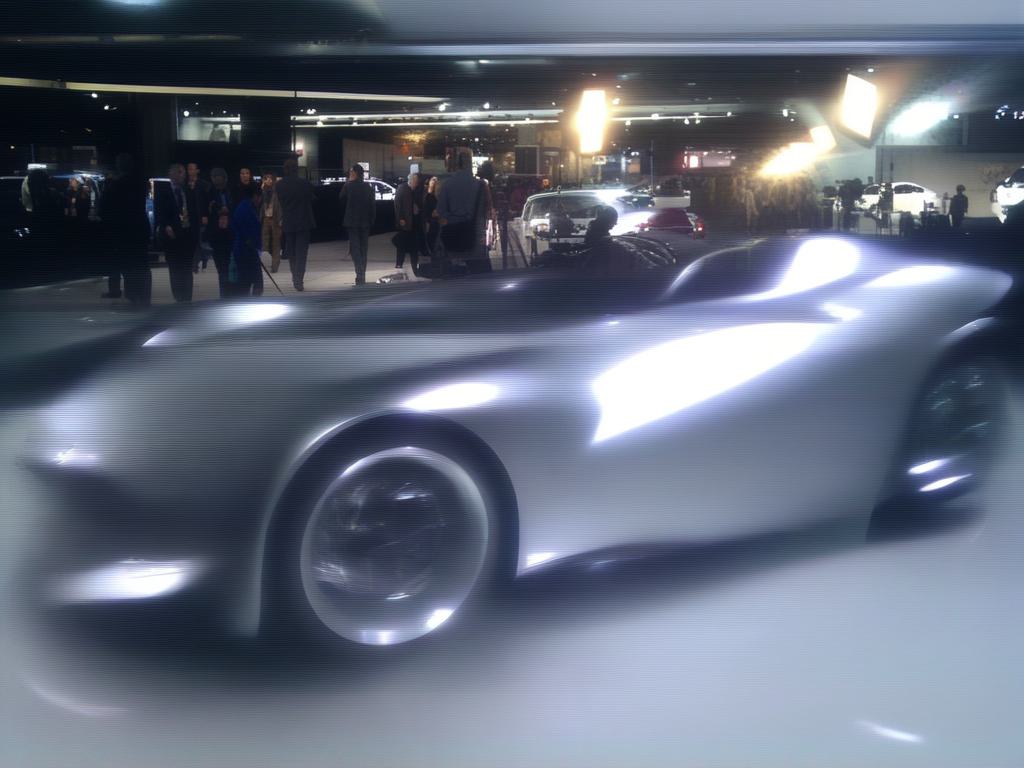} &
        \includegraphics[width=0.19\linewidth]{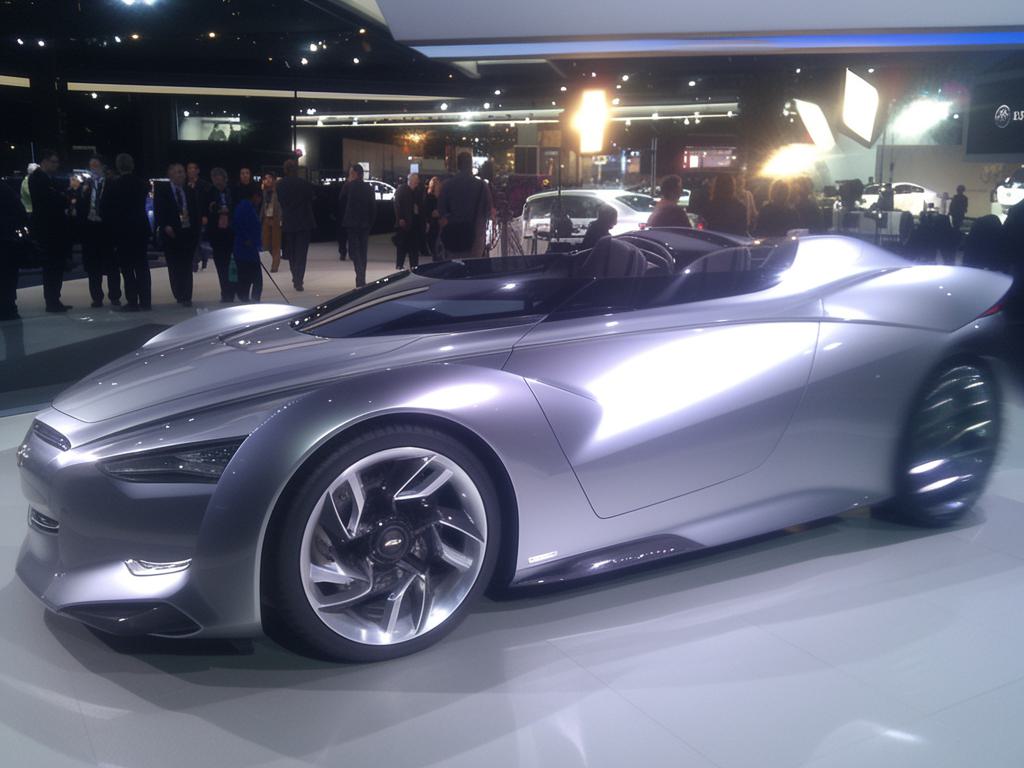} \\
        \includegraphics[width=0.19\linewidth]{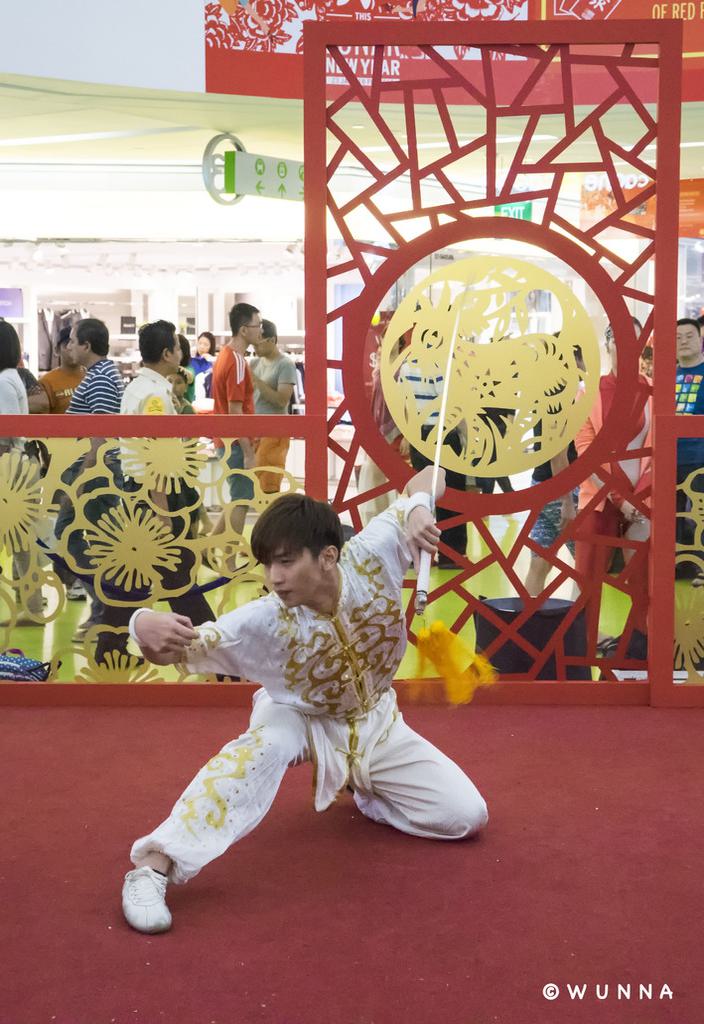} &
        \includegraphics[width=0.19\linewidth]{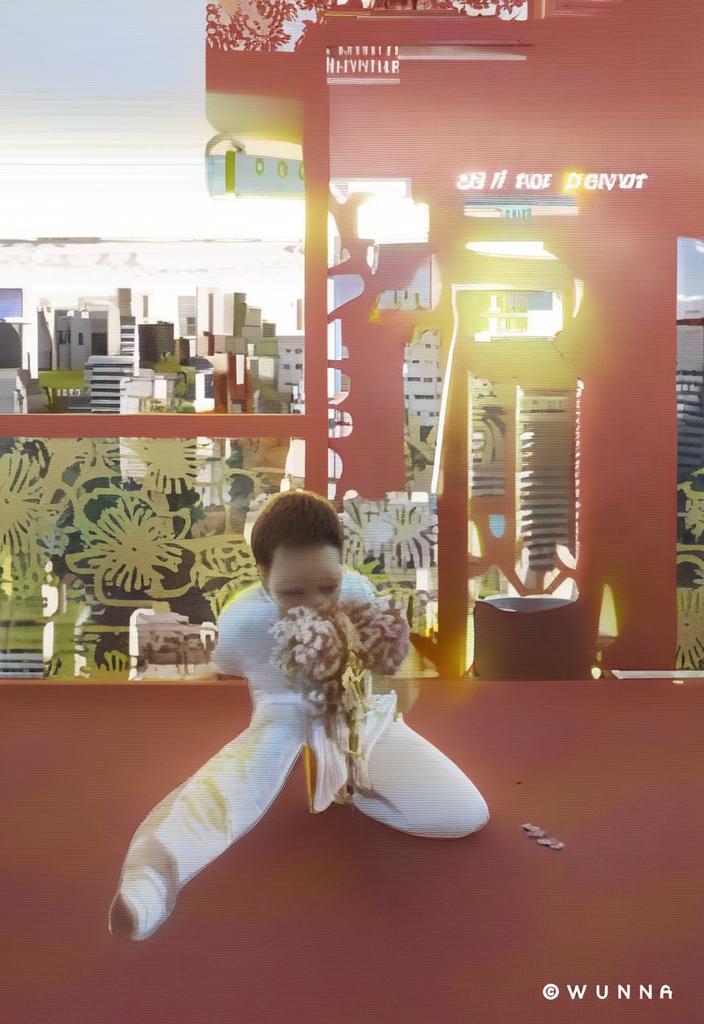} &
        \includegraphics[width=0.19\linewidth]{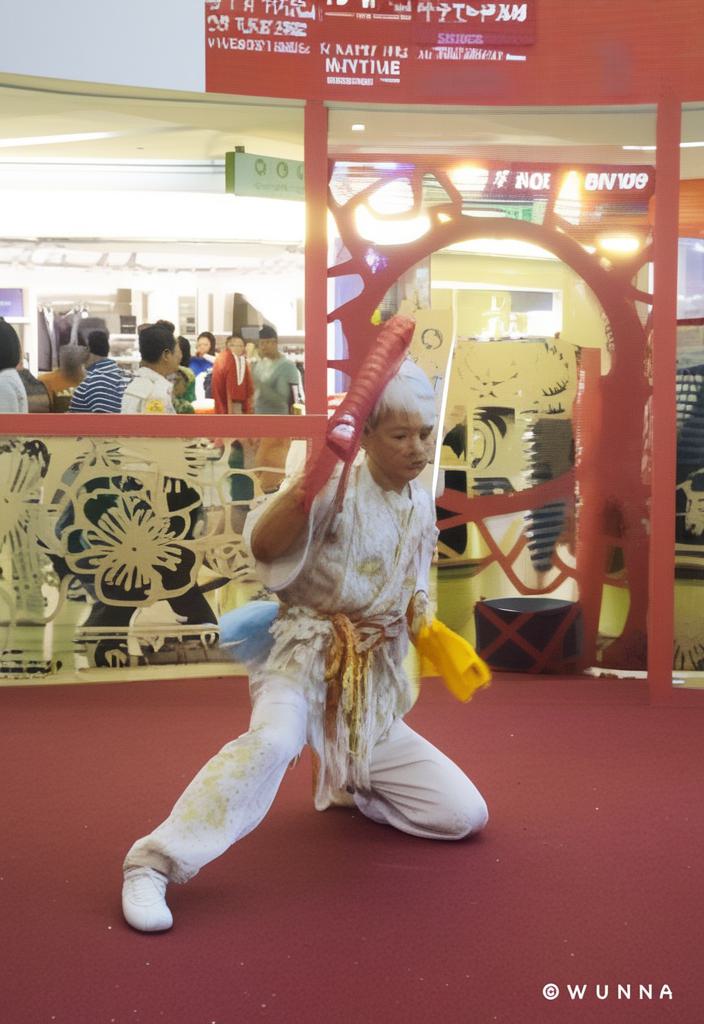} &
        \includegraphics[width=0.19\linewidth]{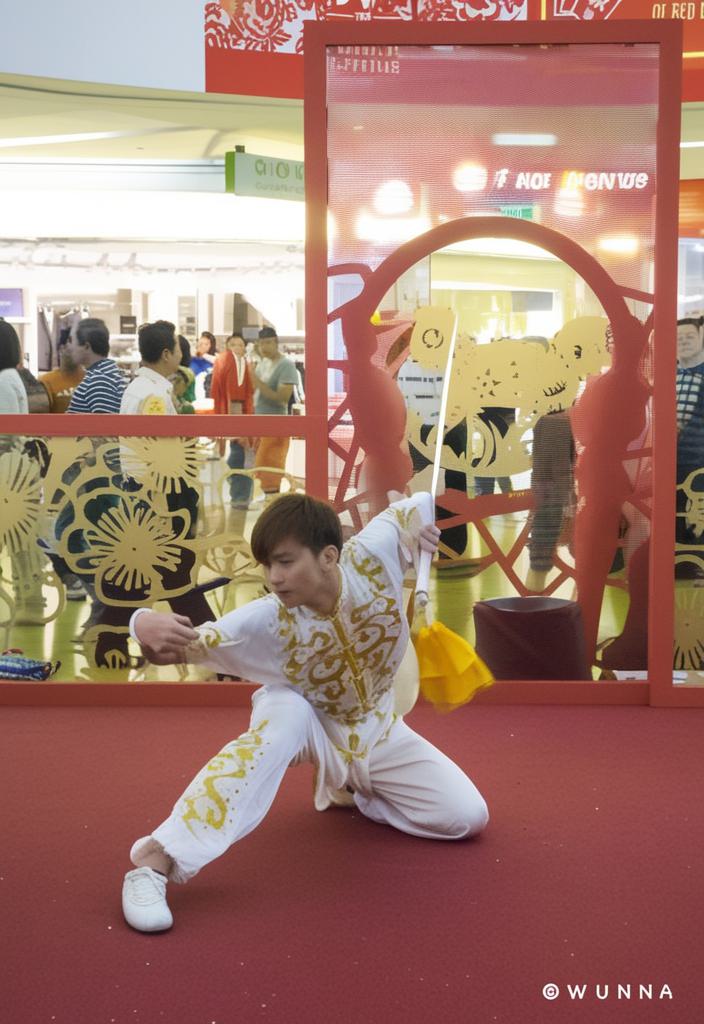} &
        \includegraphics[width=0.19\linewidth]{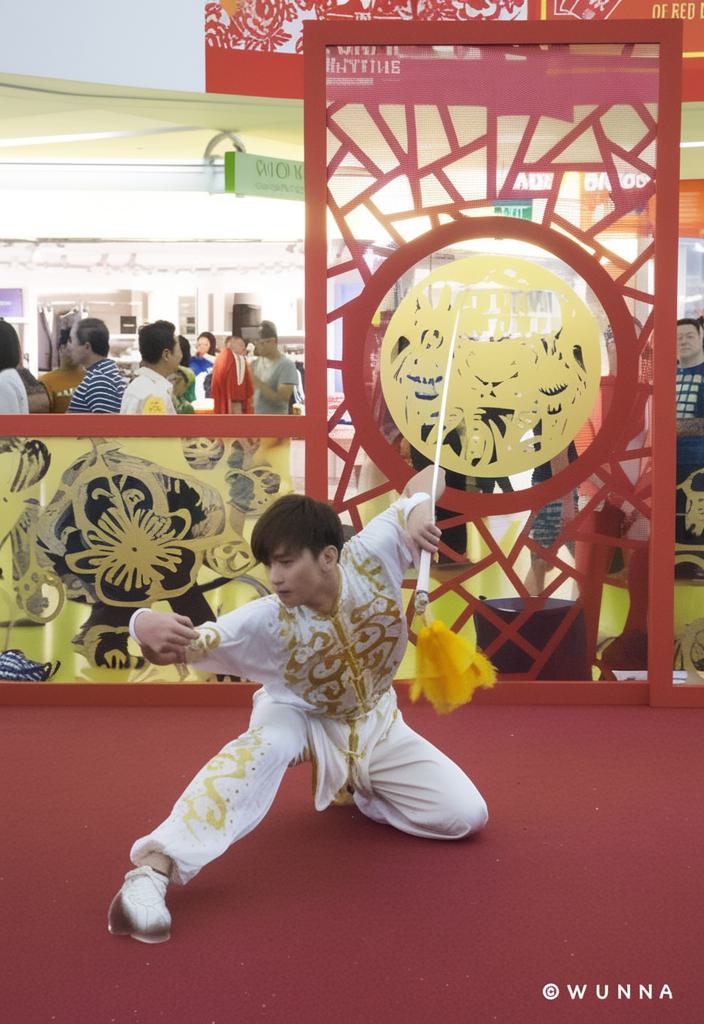} \\
        Input & Empty & Short & Long & Image \\
    \end{tabular}
    \caption{
    Each row presents a real, highly detailed image followed by reconstruction results using progressively more precise conditions during inversion and denoising. As shown, increasing the precision of the condition enhances reconstruction accuracy. In the rightmost column, we use the ultimate condition -- the input image itself -- resulting in the highest reconstruction fidelity. In all presented results, no CFG was applied during either the inversion or denoising processes.}
    \label{fig:motivation-figure}
\end{figure}

%% file: sec/2_related_work.tex
\section{Related Work}
\label{sec:related_work}

\paragraph{\textbf{Image Editing with Diffusion Models}}
In recent years, diffusion models \cite{ho2020denoisingdiffusionprobabilisticmodels, nichol2022glide, song2022denoisingdiffusionimplicitmodels, ramesh2022hierarchicaltextconditionalimagegeneration, saharia2022photorealistictexttoimagediffusionmodels, songscore} have shown rapid improvements in generating high-quality images from text prompts.
However, editing real images using textual prompts remains a challenge, as these models are not inherently designed to modify existing images. Image editing requires a careful balance between preserving key attributes of the original image (e.g., structure, semantics) and introducing controlled changes (e.g., style, pose, or specific objects).
To address this task, various approaches have been proposed. A notable line of work builds on the observation that images generated from the same initial noise tend to share semantic and structural similarities when conditioned on different signals. To further preserve original attributes, these methods manipulate the denoising process by injecting features from the source image into the edited output~\cite{hertz2022prompt, Parmar_2023, tumanyan2022plugandplaydiffusionfeaturestextdriven, cao2023masactrltuningfreemutualselfattention, alaluf2023crossimageattentionzeroshotappearance, patashnik2023localizingobjectlevelshapevariations, Mokady_2023_CVPR, ge2023expressive, lu2023tf, tokenflow2023, avrahami2024diffuhaul}. To apply these methods for real-image editing, an inversion technique is needed to predict the initial noise $z_T$ that reconstructs the image.

Other approaches for diffusion-based image editing include partially noising an input image followed by denoising with a different text condition~\cite{meng2022sdeditguidedimagesynthesis, hubermanspiegelglas2024editfriendlyddpmnoise, tsaban2023leditsrealimageediting, brack2024ledits}, fine-tuning the base model to accept an input image as a condition~\cite{Rombach_2022_CVPR, brooks2023instructpix2pixlearningfollowimage, Zhang2023MagicBrush, Avrahami_2023_CVPR, zhang2023addingconditionalcontroltexttoimage}, and utilizing masks to enable localized edits~\cite{Avrahami_2022, avrahami2023blendedlatent, couairondiffedit, mirzaei2025watch}.

\paragraph{\textbf{Diffusion Models Inversion}}
To edit an image $I$ using a diffusion model, many methods require obtaining an initial noise $z_T$ such that denoising $z_T$ reconstructs $I$. A common approach for this is DDIM inversion~\cite{song2022denoisingdiffusionimplicitmodels, dhariwal2021diffusionmodelsbeatgans}, which reverses the denoising process to approximate the initial noise. This inversion relies on solving an implicit equation by assuming that consecutive points in the denoising trajectory are close to each other. However, this assumption often does not hold during typical use with a practical number of denoising steps and introduces inaccuracies.
To address these inaccuracies, some methods~\cite{garibi2024renoise, samuel2024lightningfastimageinversionediting, Pan_2023} employ different algorithms to solve the implicit equation. Another limitation of DDIM inversion arises from the use of classifier-free guidance~\cite{ho2021classifierfree} during denoising~\cite{Mokady_2023_CVPR}. To address this, some methods optimize the null-text embedding~\cite{Mokady_2023_CVPR}, use empty prompts during inversion~\cite{cao2023masactrltuningfreemutualselfattention}, or use negative prompts~\cite{miyake2023negative, proxedit}.
As we demonstrate in this work, DDIM inversion is sensitive to the prompts used during the inversion process. Therefore, integrating DDIM inversion based methods with our approach can significantly improve both reconstruction and editability, particularly for challenging images.

Another line of work focuses on the non-deterministic DDPM denoising process \cite{ho2020denoisingdiffusionprobabilisticmodels}, inverting the image into the intermediate noise maps introduced throughout the stochastic process~\cite{hubermanspiegelglas2024editfriendlyddpmnoise, deutch2024turboedittextbasedimageediting, tsaban2023leditsrealimageediting, cyclediffusion}. While these methods ensure perfect reconstruction of the input image, they often struggle to preserve fidelity to the original image during editing, particularly for challenging cases. Our approach enhances the editability of these methods, achieving better preservation of the original image.

\paragraph{\textbf{Image Conditioned Diffusion Models}}
Some methods train encoders (or adapters) that take an image as input and produce a latent representation, which is then injected into a pretrained text-to-image model~\cite{ye2023ipadaptertextcompatibleimage, gal2024lcmlookahead, gal20232e4t, Wei_2023, parmar2025viscomposer, patashnik2025nested, guo2024pulid, zeng2024jedi, arar2023agnostic}. These approaches typically aim to personalize the text-to-image model, enabling it to generate a subject in new contexts and styles.
In our work, we utilize IP-Adapter~\cite{ye2023ipadaptertextcompatibleimage, xlabs-flux-ip-adapter} and PuLID~\cite{guo2024pulid} to condition the model on an image.
IP-Adapter was trained on a broad domain with the objective of reconstructing the input image. While it does not fully reconstruct the image in practice and instead produces semantic variations, it serves as an effective tool to transform text-conditioned models into models conditioned on both text and images. PuLID is trained on images containing faces with the goal of preserving identity in the generated image with minimal disruption to the original model's behavior.

\input{figures/method}

%% file: figures/method.tex
\begin{figure*}
    \centering
    \setlength{\tabcolsep}{1pt}
    \begin{subfigure}{0.24\linewidth}
        \centering
        \includegraphics[width=\linewidth]{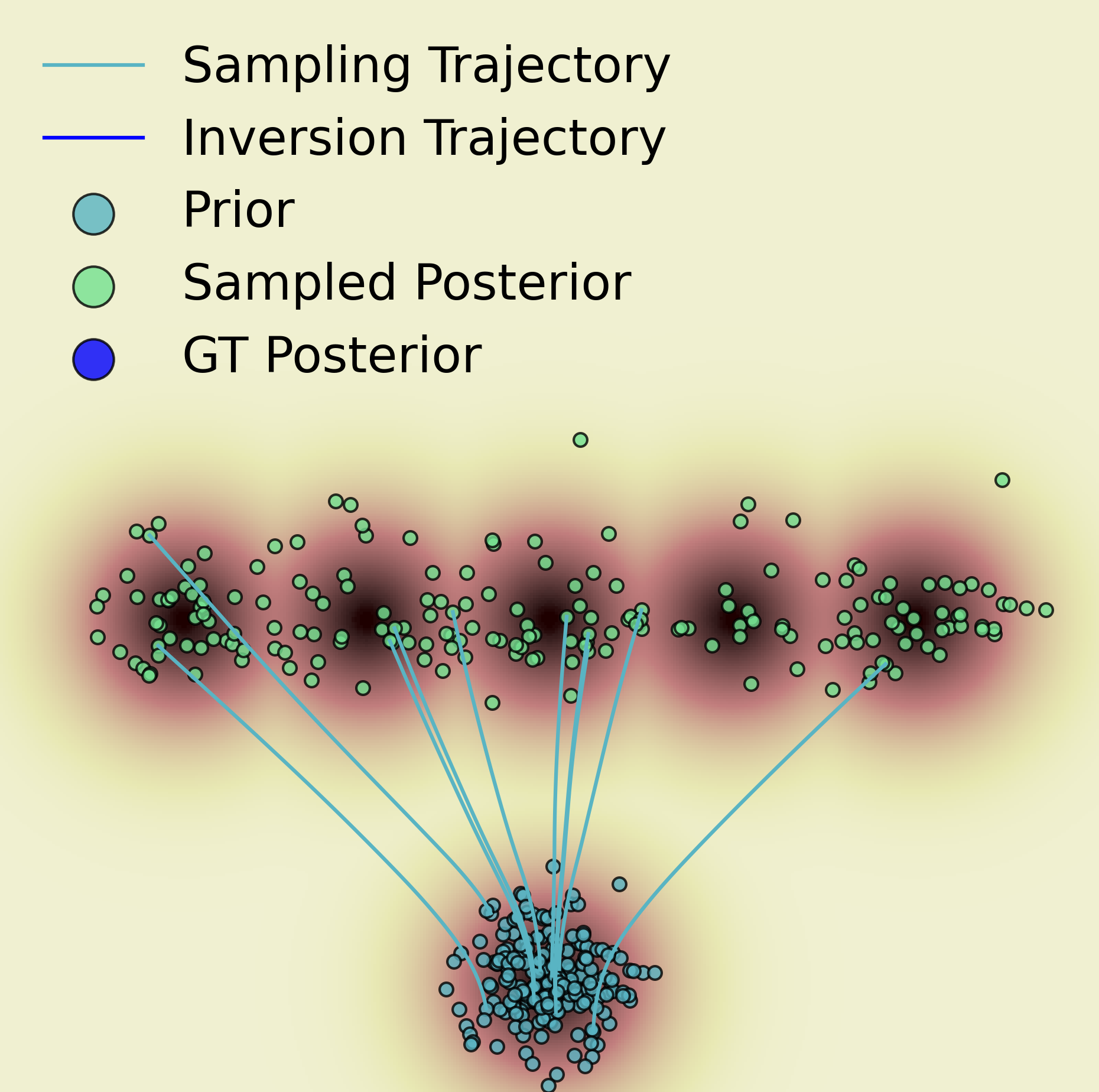}
        \caption{CNF}
        \label{fig:toy-flow}
    \end{subfigure}
    \begin{subfigure}{0.24\linewidth}
        \centering
        \includegraphics[width=\linewidth]{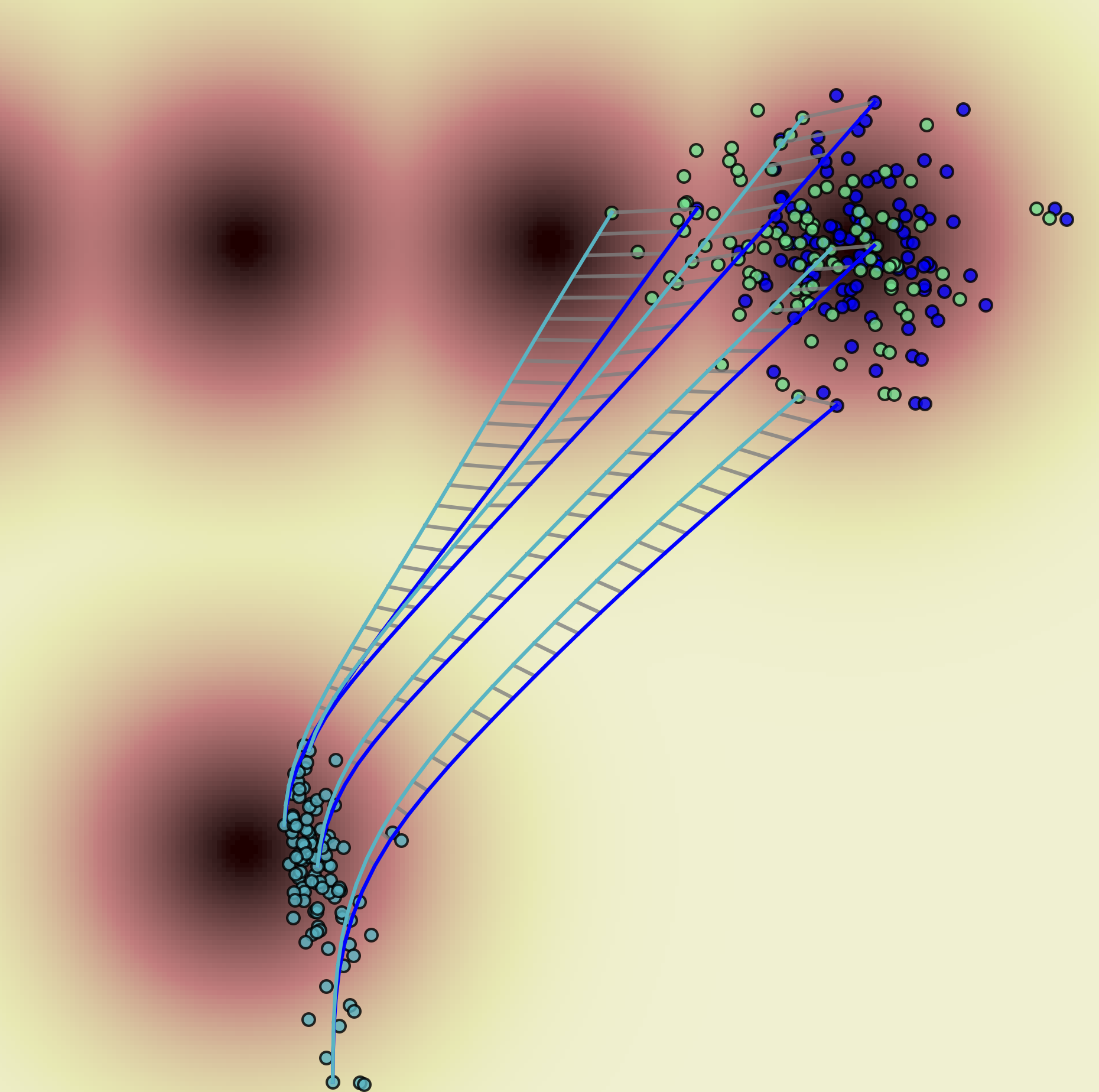}
        \caption{Inversion w/ null condition}
        \label{fig:toy-uncond}
    \end{subfigure}
    \begin{subfigure}{0.24\linewidth}
        \centering
        \includegraphics[width=\linewidth]{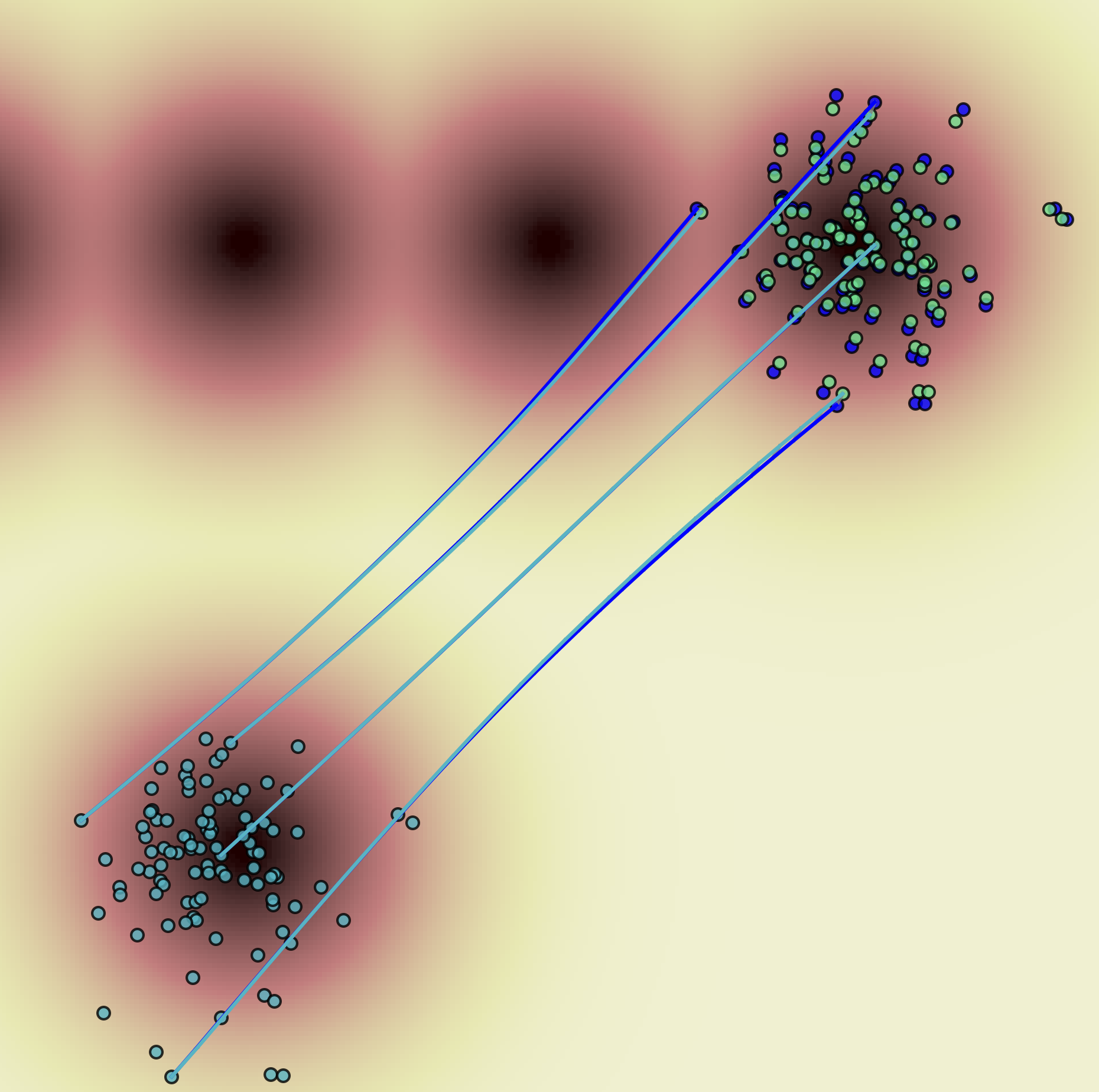}
        \caption{Inversion w/ accurate condition}
        \label{fig:toy-cond}
    \end{subfigure}
    \begin{subfigure}{0.24\linewidth}
        \centering
        \includegraphics[width=\linewidth]{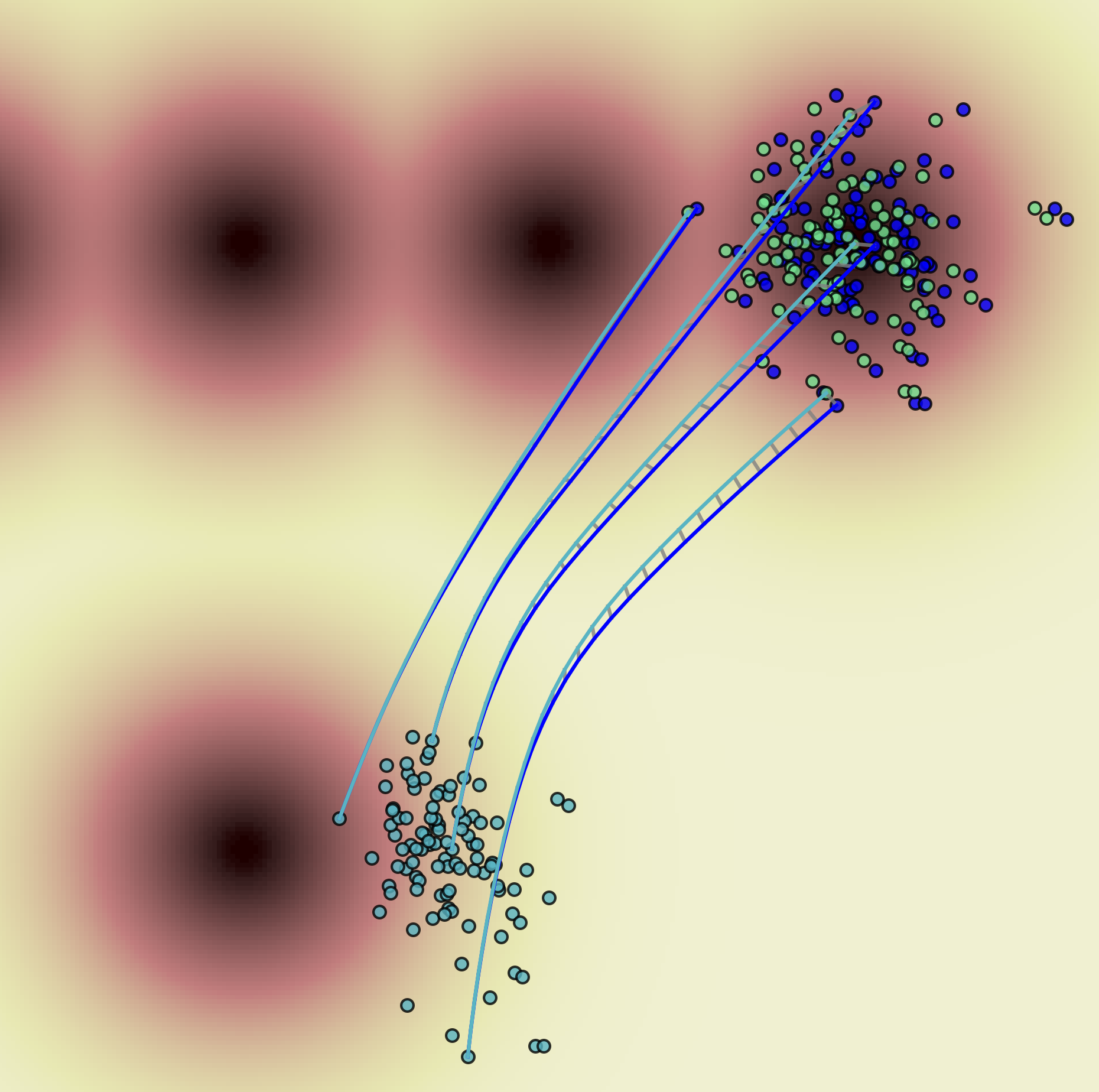}
        \caption{Inversion w/ inaccurate condition}
        \label{fig:toy-incorrect-cond}
    \end{subfigure}
    \vspace{-8pt}
    \caption{We train a toy conditional CNF model to analyze the importance of the condition used during inversion. The prior distribution is a single Gaussian, and the posterior consists of five Gaussians. (a) shows denoising trajectories from the prior, and (b)-(d) show inversion and denoising trajectories for points from the posterior. In (b), a null condition is used for both processes, in (c), the condition matches the Gaussian from which the point was sampled, and in (d), the condition corresponds to the adjacent Gaussian. Lines connect points on the inversion and denoising trajectories to illustrate offsets between these processes.}
    \vspace{-8pt}
    \label{fig:method}
\end{figure*}

%% file: sec/3_method.tex
\section{Tight Inversion}

Given a real image $I$, the goal of our method is to predict a noise image $z_T$ such that denoising $z_T$ yields $I$ back. Importantly, it should be possible to edit $I$ when using a target text prompt during the denoising process. Our work builds on DDIM inversion~\cite{dhariwal2021diffusionmodelsbeatgans, song2022denoisingdiffusionimplicitmodels} and begins by analyzing it.

\paragraph{\textbf{Background and Motivation}}
To invert a real image $I$, DDIM inversion iteratively adds noise to the image, forming a trajectory from the real data distribution to the Gaussian distribution. 
Each point $z_t$ in the inversion trajectory is defined as:
\begin{equation} \label{eq:inversion}
    z_{t} = A_t z_{t-1} - B_t \epsilon_{\theta}(z_{t-1}, t, c),
\end{equation}
where $\epsilon_{\theta}$ is the pretrained diffusion model, $c$ is the text condition fed into the model, $A_t, B_t$ are constants defined by DDIM~\cite{song2022denoisingdiffusionimplicitmodels}, and $z_0=I$.
To reconstruct the image, the same condition $c$ as the one used during the inversion is used in the denoising process.

Previous work~\cite{songscore} has shown that a pretrained diffusion model can be viewed as a score function, resulting in
\begin{equation}
    \epsilon_{\theta}(z_t, t, c) \propto \nabla_{z_t} \log{p_{\theta}(z_t | c)}.
\end{equation} 
Therefore, a more detailed and precise condition $c$ should lead to a narrower conditional distribution $p_{\theta}(z_t | c)$, which in turn should improve the accuracy of $\epsilon_{\theta}(z_t, t, c)$. Since Equation~\ref{eq:inversion} relies on $\epsilon_{\theta}(z_t, t, c)$, we expect that its increased accuracy will result in a more accurate inversion process.
We verify this intuition through the following experiment. 

First, we generate a set of elaborated text prompts using an LLM, and sample a single image for each prompt. Then, we apply DDIM inversion~\cite{dhariwal2021diffusionmodelsbeatgans} on each image with three different text conditions: (i) the text prompt used to generate the image (full), (ii) a shortened version of this prompt (short), and (iii) an empty prompt. 
We re-generate the images from the inverted noises with the same condition used in the inversion, and do not use classifier-free guidance (CFG).
We measure $L_2$, PSNR, SSIM and LPIPS~\cite{zhang2018perceptual} between the sampled image and the reconstructed one and display the results in Table~\ref{tab:prompt_levels_experiment}. 
As observed from the results, across all the metrics using a short prompt results in a better reconstruction than using an empty prompt, and using a detailed prompt results in a better reconstruction than using a short prompt.

\input{tables/prompts_levels.tex}

\paragraph{\textbf{Toy Example}}
To further illustrate the motivation behind our method, we explore the role of the condition used during inversion through a toy example depicted in Figure~\ref{fig:method}. In this setup, we train a CNF (Flow Matching) model $\phi: \mathbb{R}^2 \rightarrow \mathbb{R}^2$ \cite{lipmanflow, chen2018neural}. The prior distribution, $\mathcal{N}(\textbf{0}, 1)$, is represented by the bottom Gaussian, while the posterior (target) distribution consists of five Gaussians $\{\mathcal{N}((5 \cdot c_i, 10), 1)\}_{i=1}^5$ with $c_i \in \{-2, -1, 0, 1, 2\}$, corresponds to the five Gaussians on the top (see Figure~\ref{fig:toy-flow}). The model $\phi$ is trained as a conditional model, where each sample from the posterior distribution is assigned a condition corresponding to the index of the Gaussian from which it was drawn. Additionally, in $50\%$ of the training iterations, a null condition is used.
Figure~\ref{fig:toy-flow} shows the denoising trajectories of (light blue) points sampled from the prior distribution when using the null condition.
In Figures~\ref{fig:toy-uncond}, \ref{fig:toy-cond}, \ref{fig:toy-incorrect-cond}, we sample points (shown in blue) from the posterior distribution of $c_5$ that were not seen during $\phi$'s training, then invert and reconstruct them. The inverted points are depicted in light blue, while the reconstructed points are shown in green. We show the inversion and reconstruction trajectories for a subset of the points to provide further insight. The inversion trajectory is depicted in blue while the reconstruction trajectory is depicted in light blue. For each timestep $t$, we connect the corresponding points on the inversion and reconstruction trajectories (see Figure~\ref{fig:toy-uncond}).

In Figure~\ref{fig:toy-uncond}, we inverted and reconstructed the points using the null condition. As shown, blue points located outside the dense regions of the posterior distribution tend to exhibit higher reconstruction errors. Additionally, the inverted points cluster within a small region of the prior distribution. 
Moreover, the inversion and reconstruction trajectories do not overlap, as illustrated by the lines connecting corresponding points on the inversion and reconstruction trajectories.
In Figure~\ref{fig:toy-cond}, we performed inversion using the correct condition for the blue points. This results in accurate reconstruction, with the inverted points distributed in better alignment with the prior distribution. Here, the inversion and reconstruction trajectories coincide, and therefore the lines connecting corresponding points on them are not seen. Finally, in Figure~\ref{fig:toy-incorrect-cond}, we inverted points sampled from the Gaussian matching $c_5$ but used the condition $c_4$ during inversion and reconstruction. Using an incorrect condition again leads to higher reconstruction errors. Furthermore, the inverted points are mapped to low-probability regions of the prior distribution, which suggests a reduction in the editability of these points.

\input{figures/recon_qualitative}

\paragraph{\textbf{Image-conditioned Inversion}}
Given a real image $I$, we opt to find a condition $c$ that best aligns with it. Unlike the synthetic samples from the previous experiments for which we know the conditions that were used to generate them, for real images we do not have such condition prompts.
A common approach is to use a VLM to generate such prompts. 
The key idea of our method is that the most descriptive condition for an image is the image itself. That is, the conditional distribution $p_{\theta}(z_t | c)$ where $c$ is set as $I$ is more narrow than any other condition we can potentially use. 
However, the conditioning mechanism of the text-to-image model was trained to take textual tokens as input rather than images. 

Notably, many recent methods~\cite{ye2023ipadaptertextcompatibleimage, xlabs-flux-ip-adapter} train image encoders or adapters, to condition the generation process on images. Specifically, we utilize IP-Adapter (Image Prompt Adapter)~\cite{ye2023ipadaptertextcompatibleimage} which adds cross-attention layers that operate in parallel to the existing cross-attention layers of the model, and take as input image tokens rather than textual tokens. Instead of using the original text-to-image model, $\epsilon_\theta$, in Tight Inversion we use the one that integrates with IP-Adapter, $\bar{\epsilon}_\theta$, during both inversion and denoising processes. In the last row of Table~\ref{tab:prompt_levels_experiment}, we show the reconstruction results obtained by utilizing the input image as a condition through IP-Adapter~\cite{ye2023ipadaptertextcompatibleimage}, where the condition text is set as an empty prompt. As observed by the table, using the input image as the model's condition results in superior inversion results.

We note that Tight Inversion can be easily integrated with previous inversion methods (e.g., Edit Friendly DDPM, ReNoise) by employing $\bar{\epsilon}_\theta$ instead of $\epsilon_\theta$. As we demonstrate in the next section, Tight Inversion consistently improves such methods in terms of both reconstruction and editability.

%% file: tables/prompts_levels.tex
\begin{table}
    \centering
    \caption{DDIM inversion with various level of details prompts.}
    \vspace{-8pt}
    \begin{tabular}{l c c c c}
      \toprule
      Prompt& $L_2$ $\downarrow$ & PSNR $\uparrow$ & SSIM $\uparrow$ & LPIPS $\downarrow$\\
      \midrule
      Empty                                   & 58.972 & 25.107 & 0.756 & 0.264 \\
      Short                                  & 38.937 & 28.807 & 0.858 & 0.126 \\
      Full                                 & 21.944 & 32.526 & 0.929 & 0.040 \\
      \midrule
      Image prompt                             & 20.497 & 32.903 & 0.932 & 0.035 \\
      \bottomrule
      \end{tabular}
    \label{tab:prompt_levels_experiment}
  \end{table}

%% file: figures/recon_qualitative.tex
\begin{figure}
    \centering
    \setlength{\tabcolsep}{1pt}
    \begin{tabular}{ccccc}
        \includegraphics[width=0.19\linewidth]{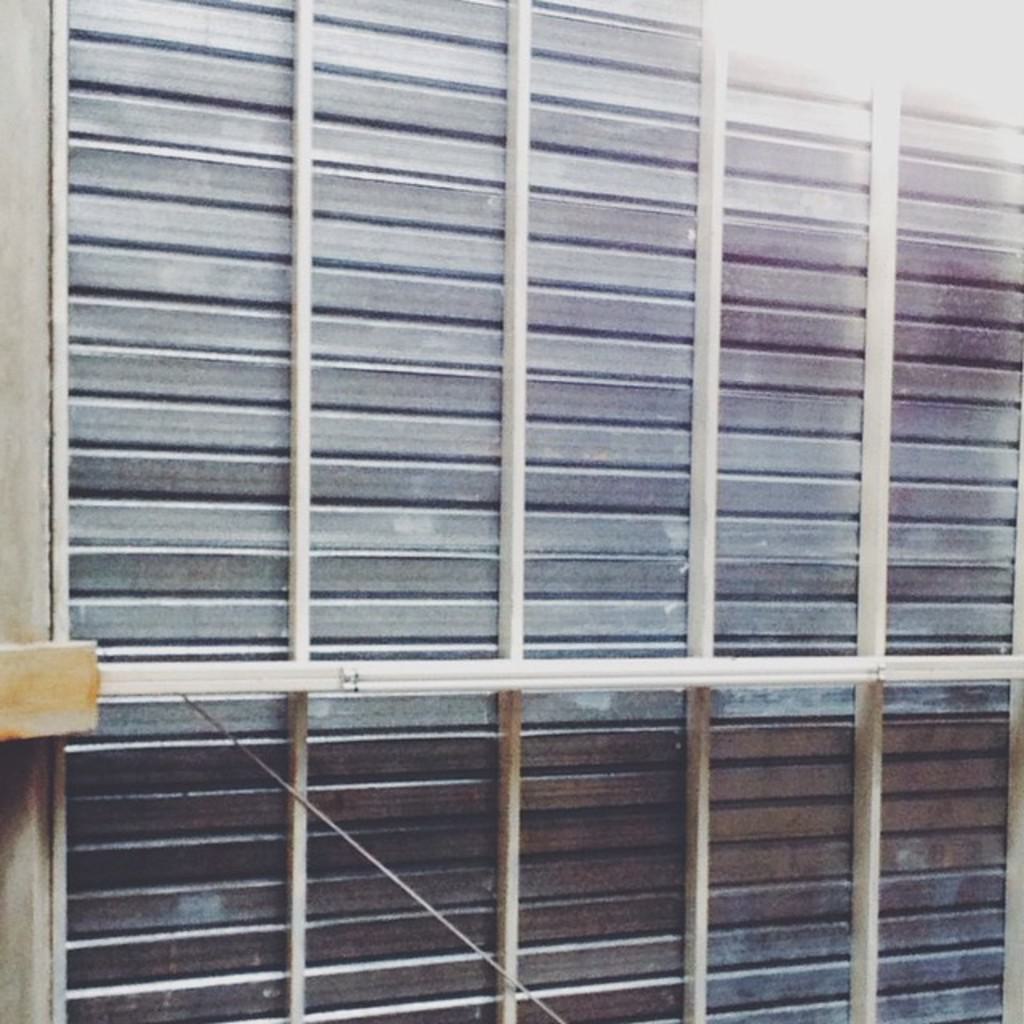} &
        \includegraphics[width=0.19\linewidth]{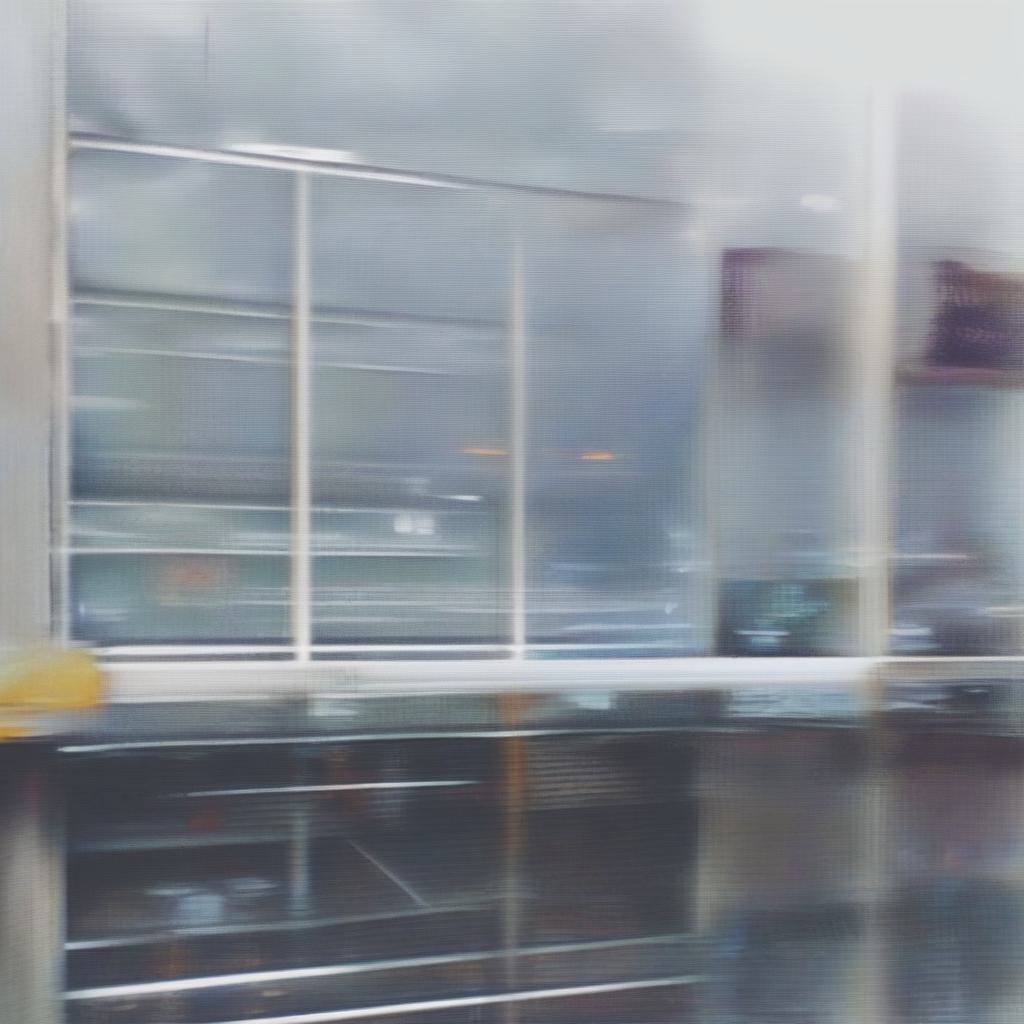} &
        \includegraphics[width=0.19\linewidth]{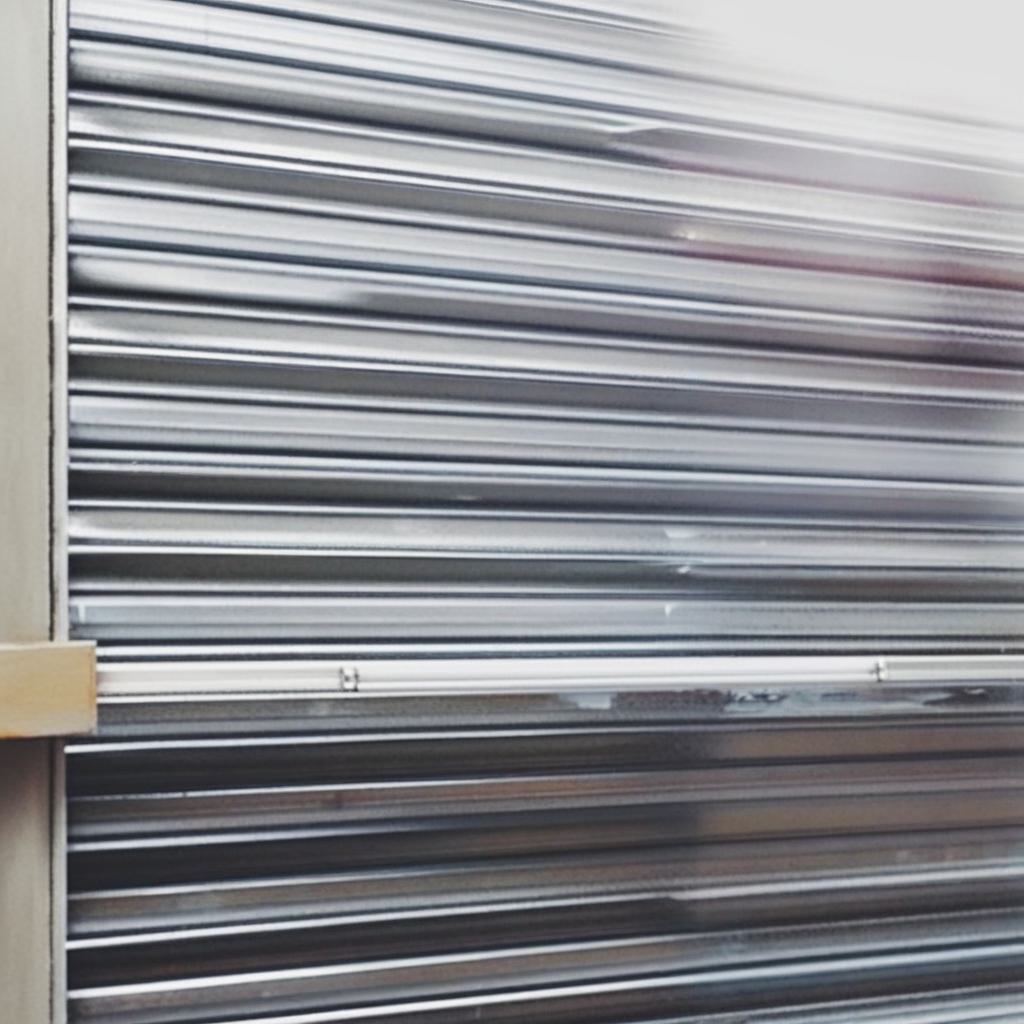} &
        \includegraphics[width=0.19\linewidth]{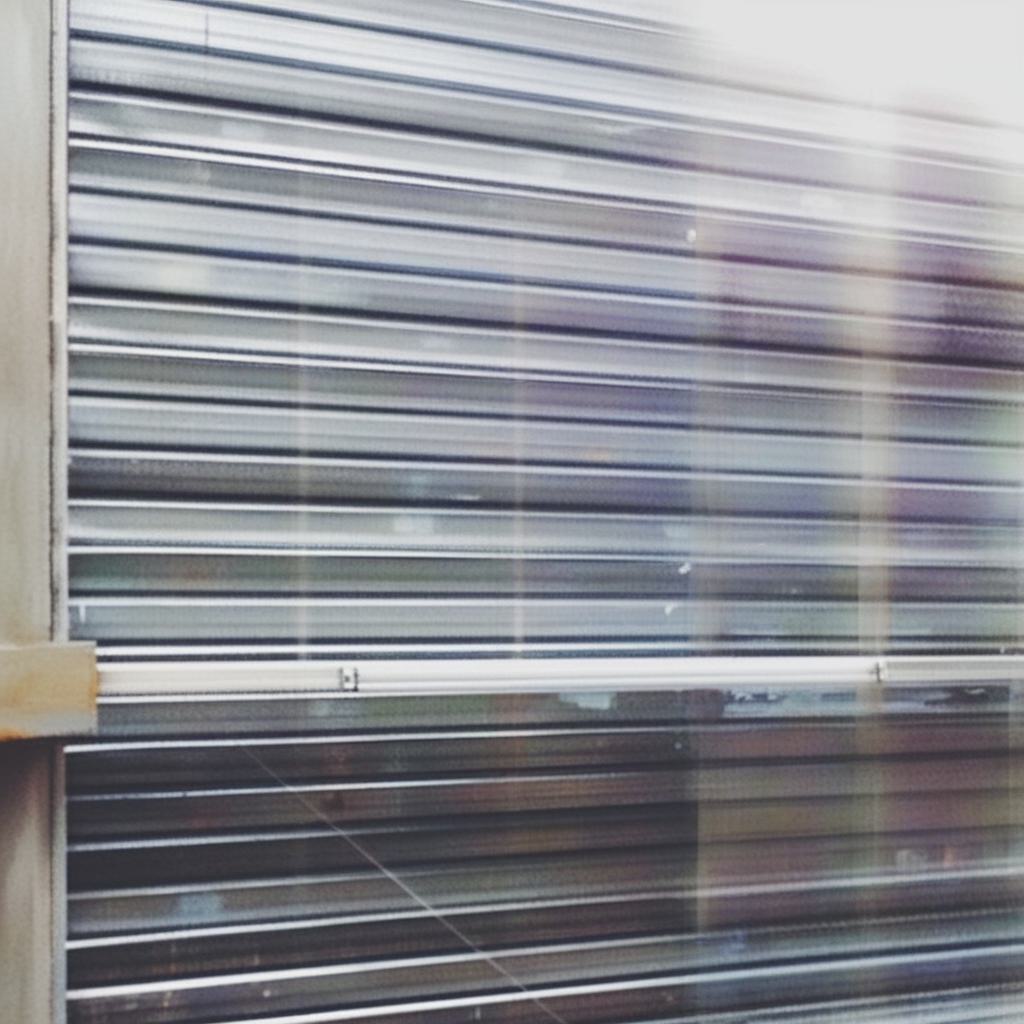} &
        \includegraphics[width=0.19\linewidth]{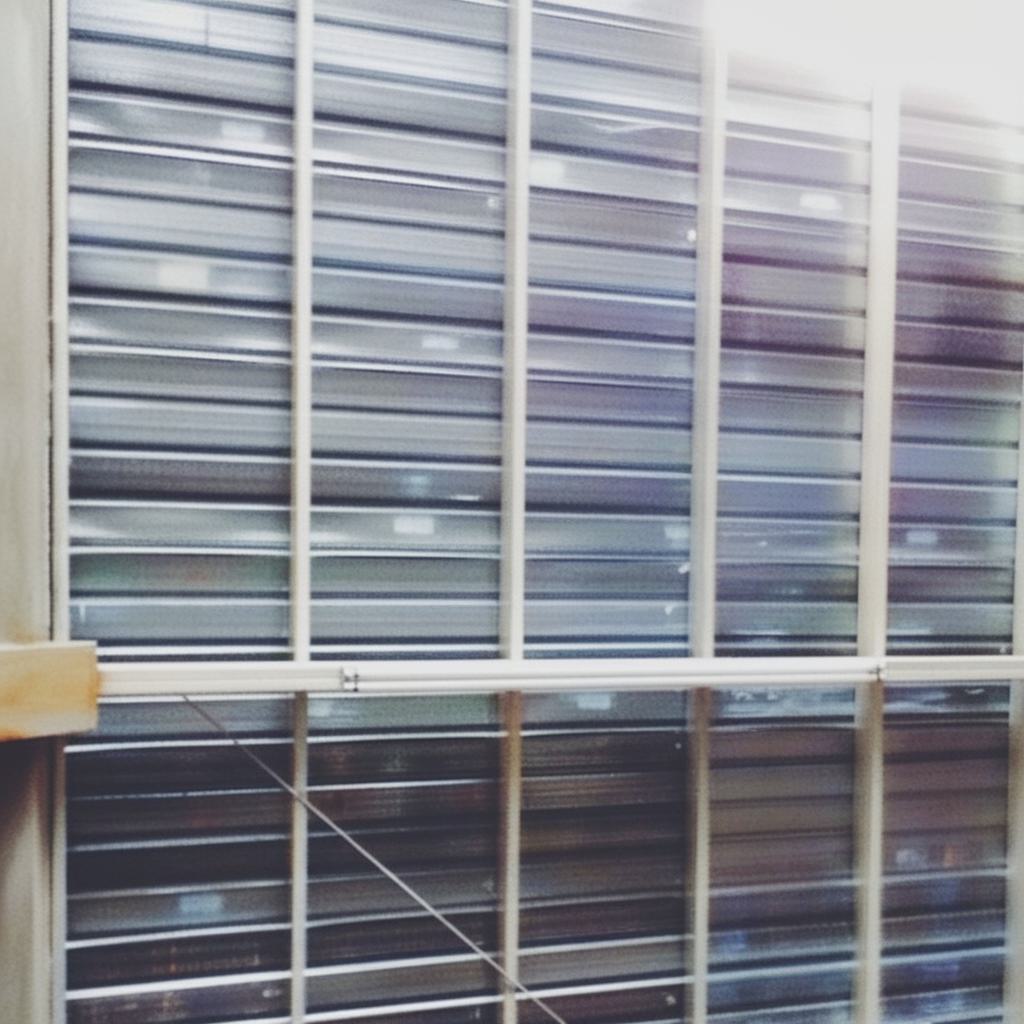} \\
        \includegraphics[width=0.19\linewidth]{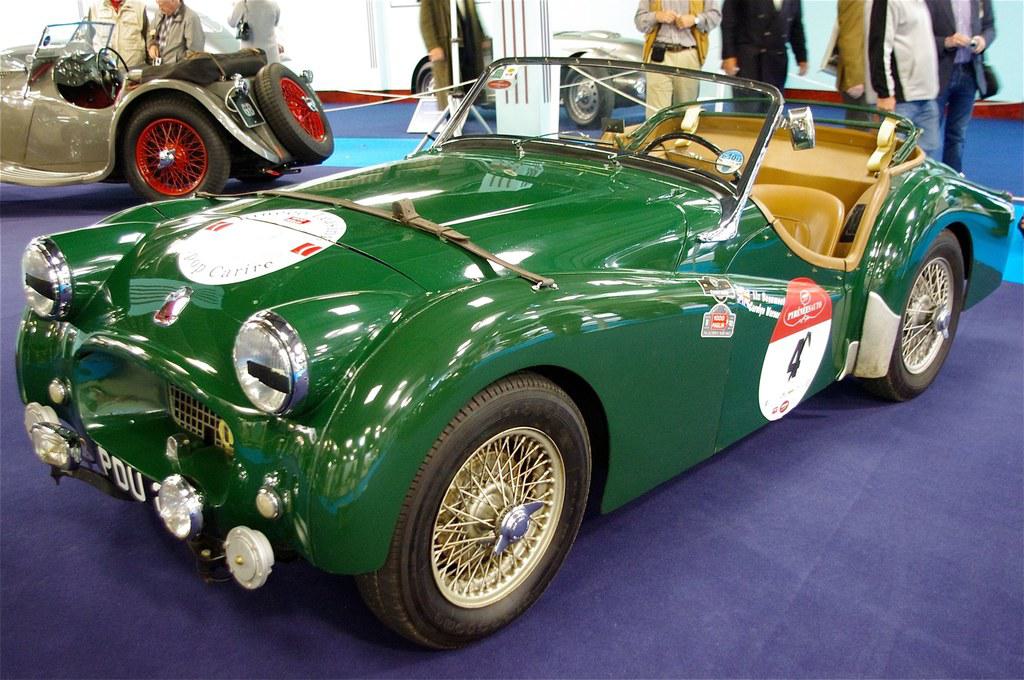} &
        \includegraphics[width=0.19\linewidth]{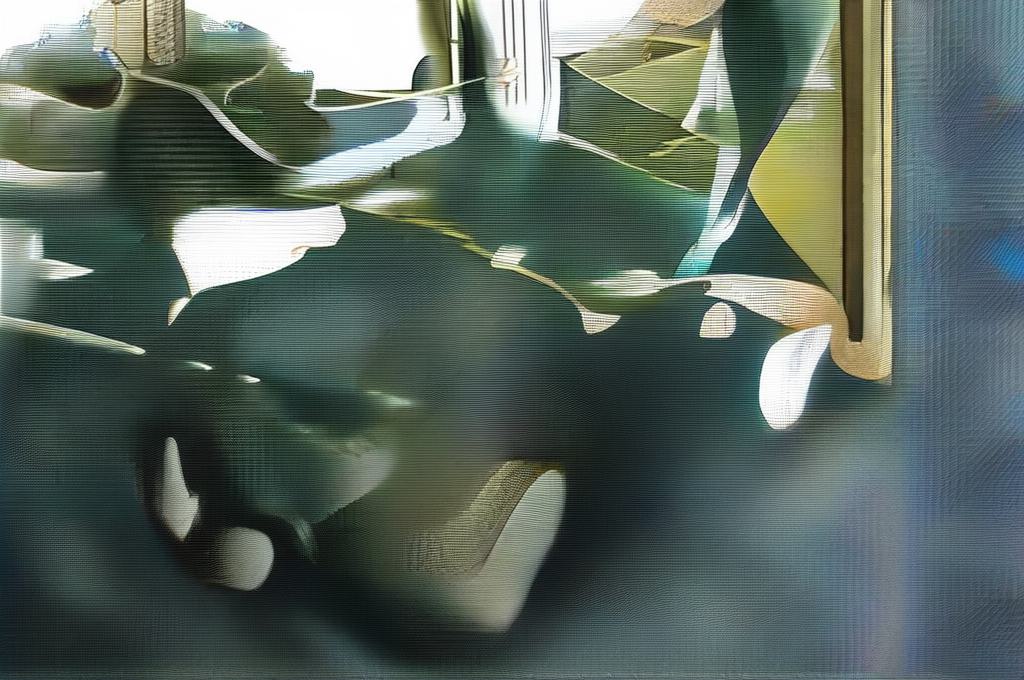} &
        \includegraphics[width=0.19\linewidth]{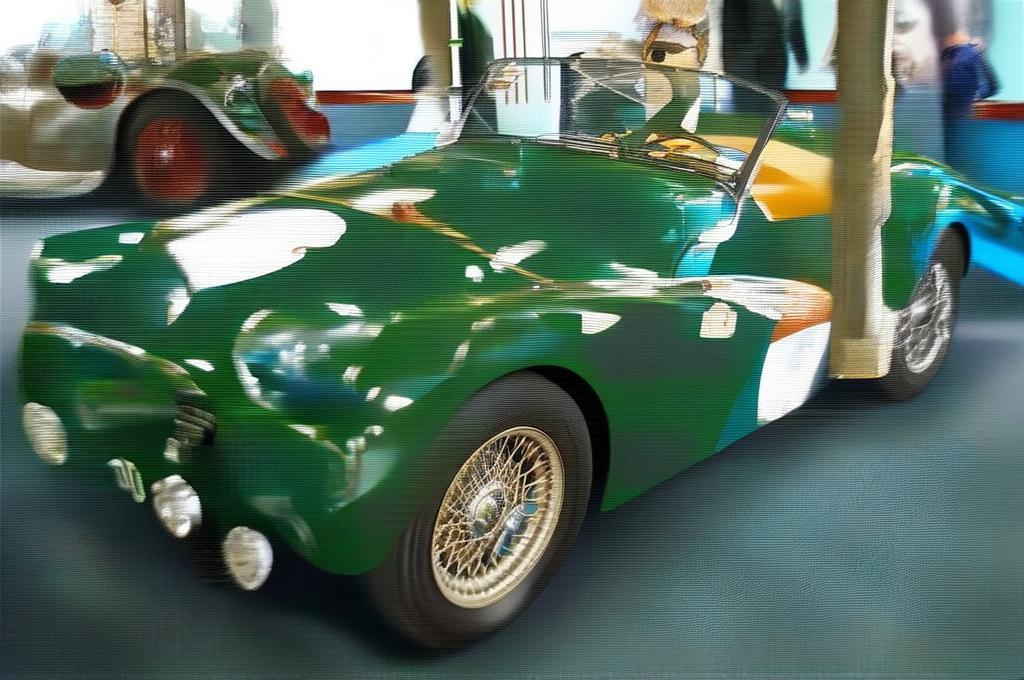} &
        \includegraphics[width=0.19\linewidth]{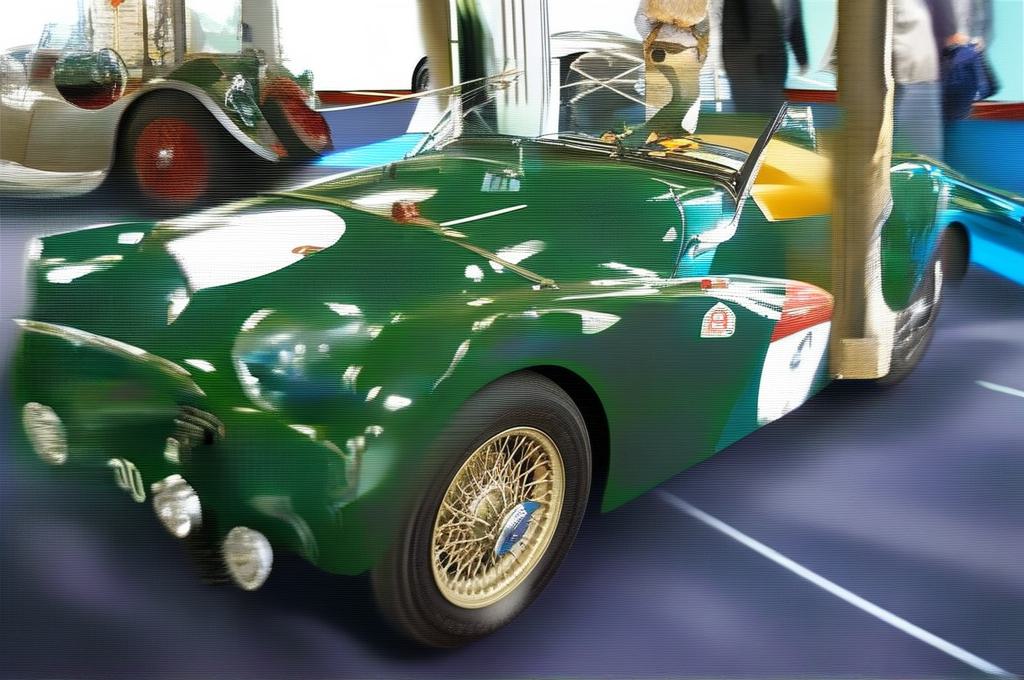} &
        \includegraphics[width=0.19\linewidth]{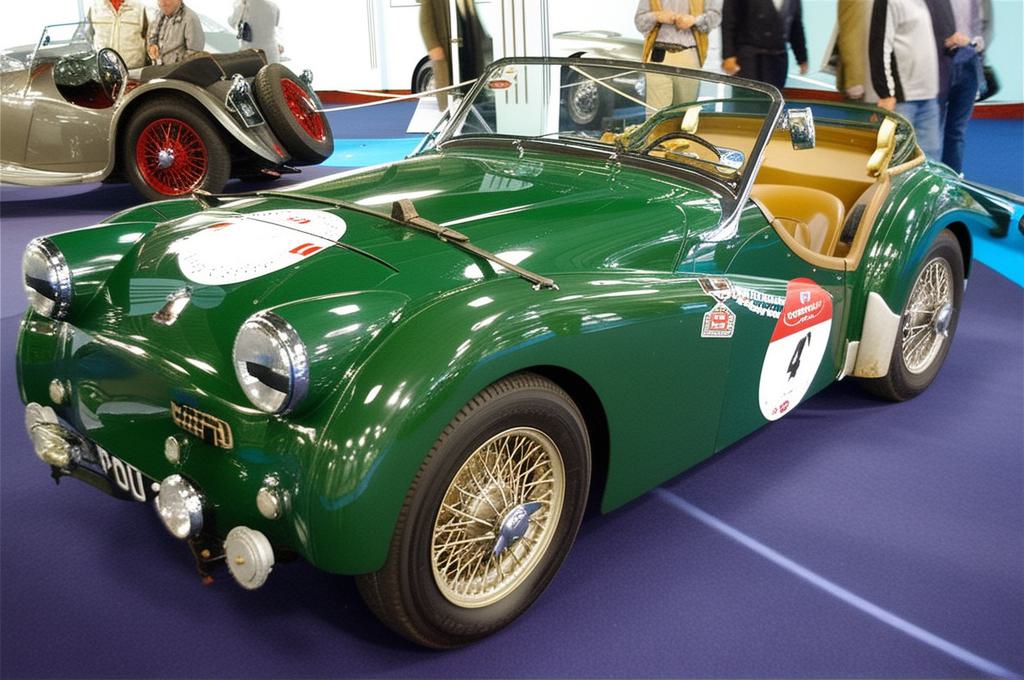} \\
        \includegraphics[width=0.19\linewidth]{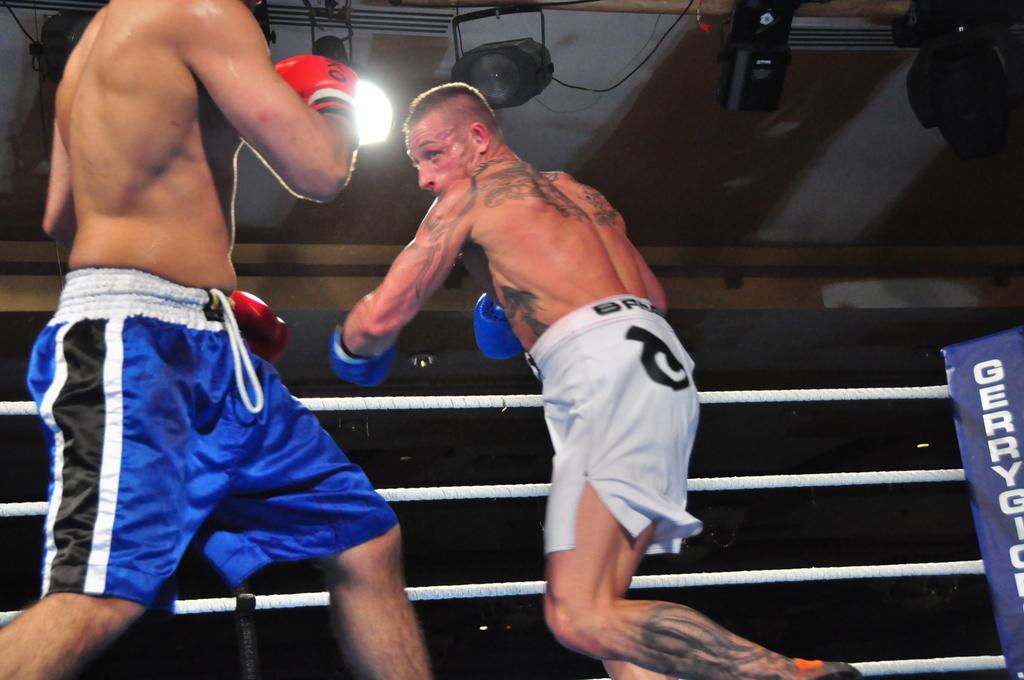} &
        \includegraphics[width=0.19\linewidth]{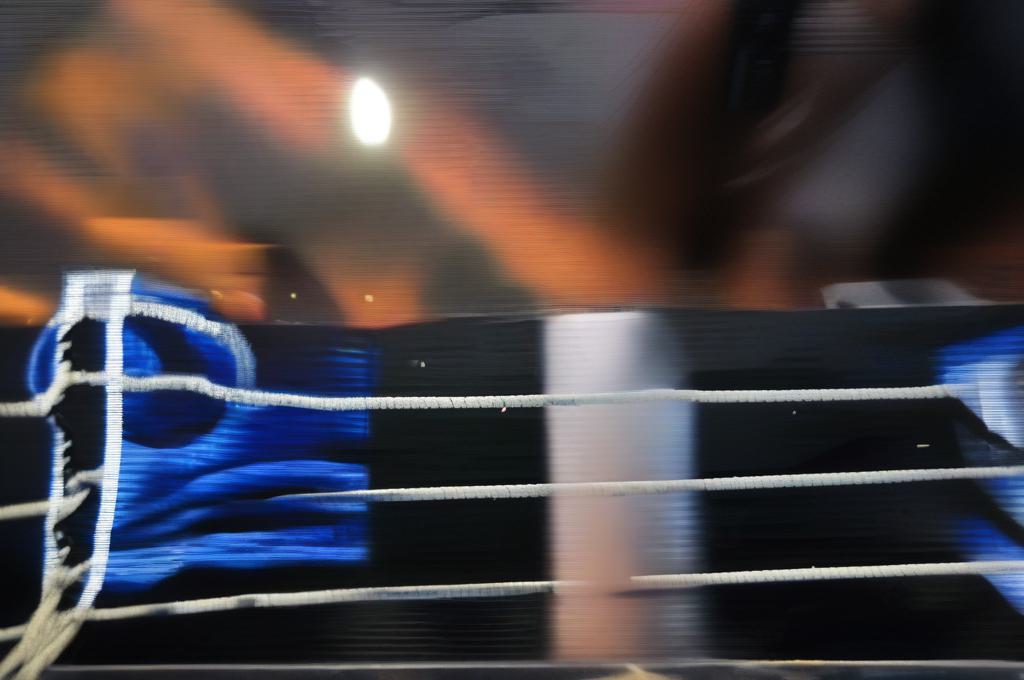} &
        \includegraphics[width=0.19\linewidth]{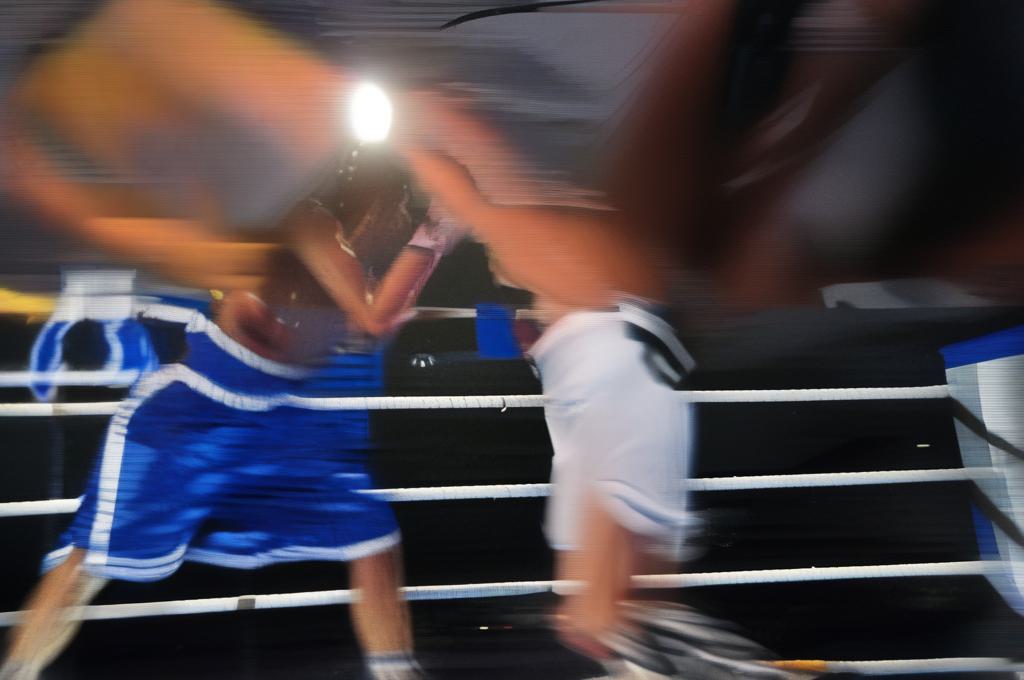} &
        \includegraphics[width=0.19\linewidth]{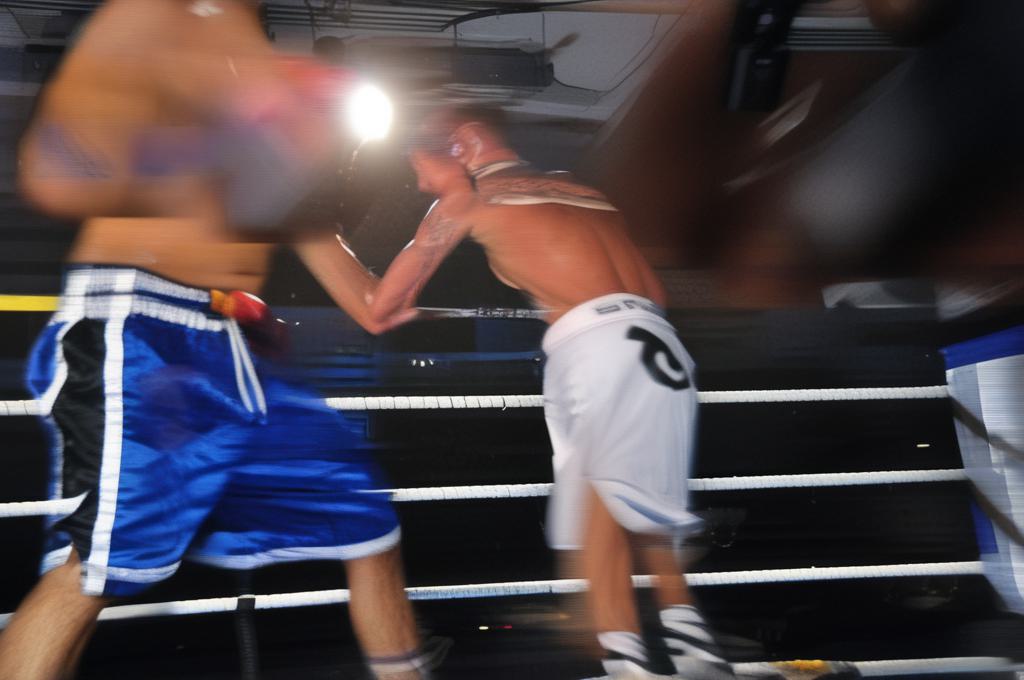} &
        \includegraphics[width=0.19\linewidth]{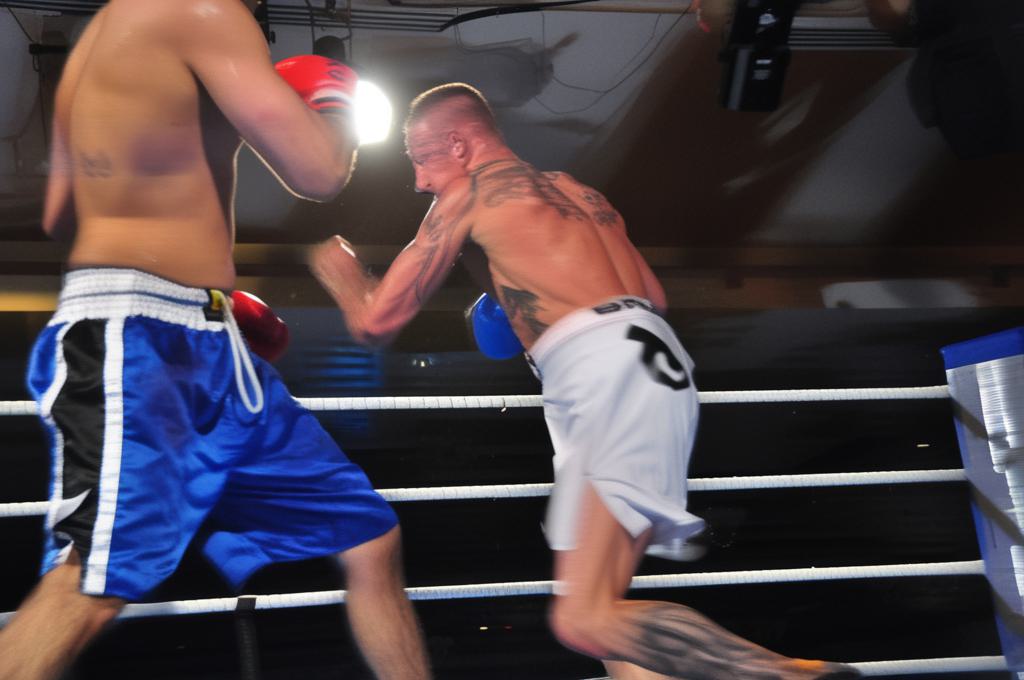} \\
        \includegraphics[width=0.19\linewidth]{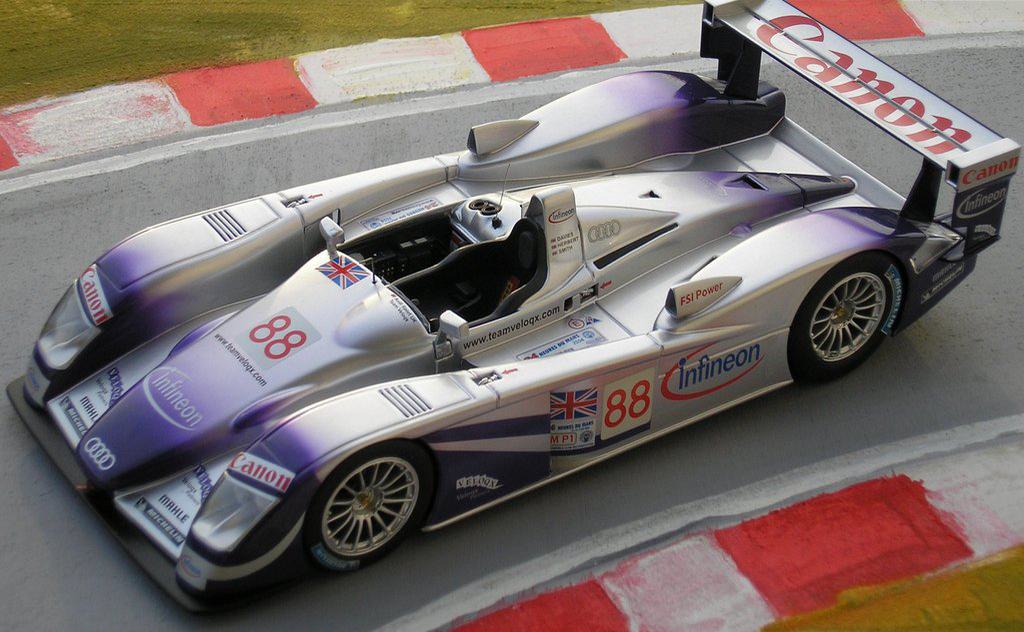} &
        \includegraphics[width=0.19\linewidth]{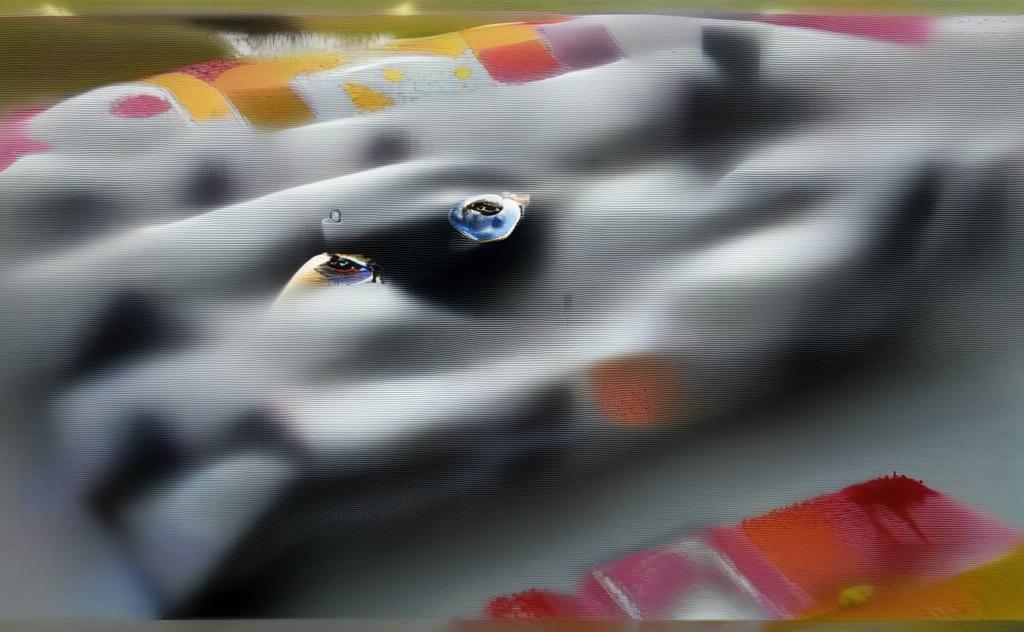} &
        \includegraphics[width=0.19\linewidth]{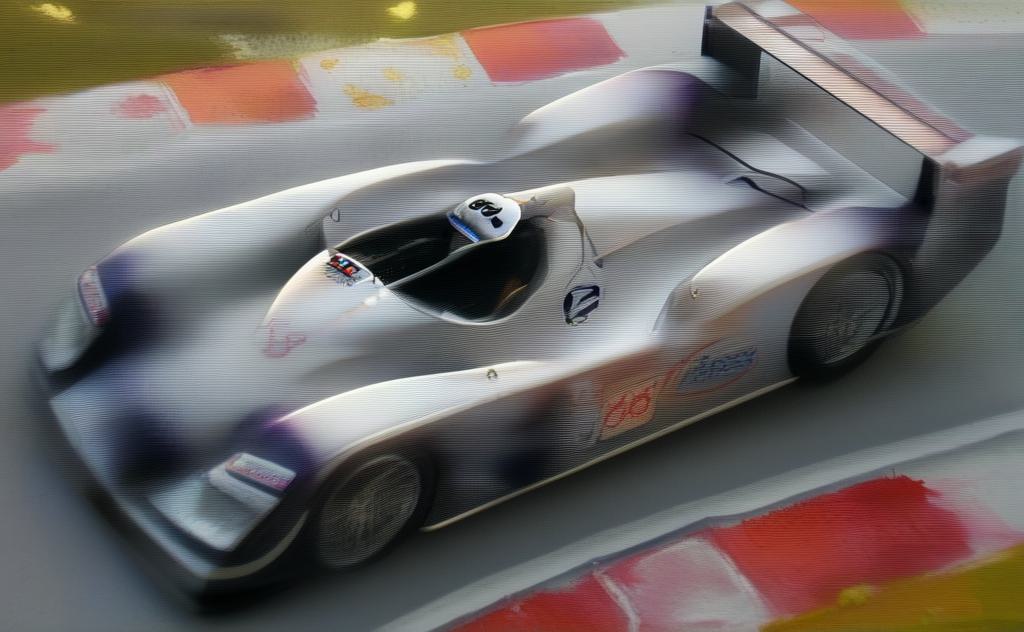} &
        \includegraphics[width=0.19\linewidth]{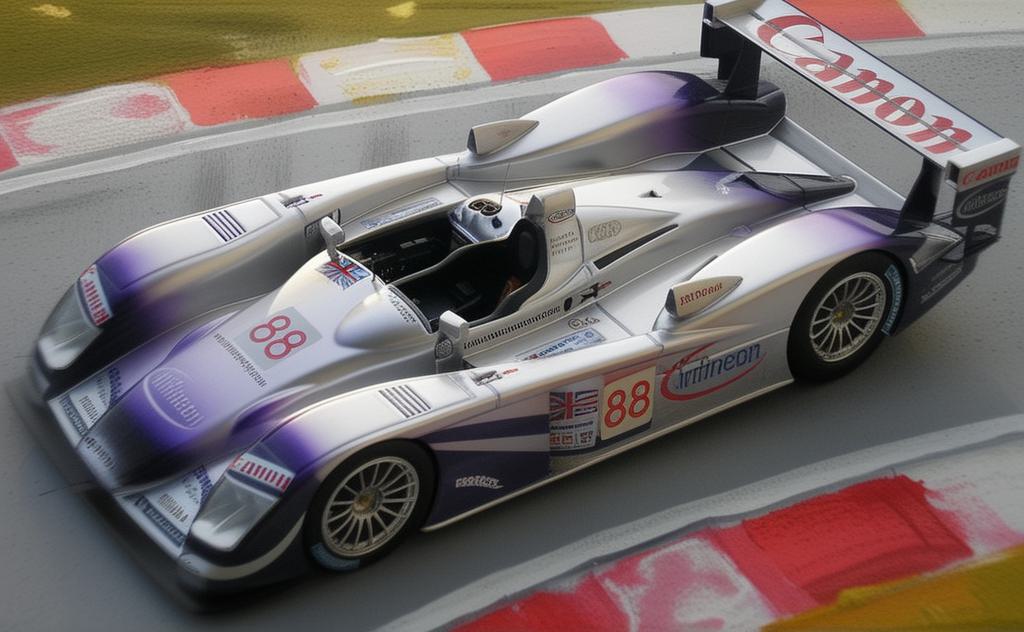} &
        \includegraphics[width=0.19\linewidth]{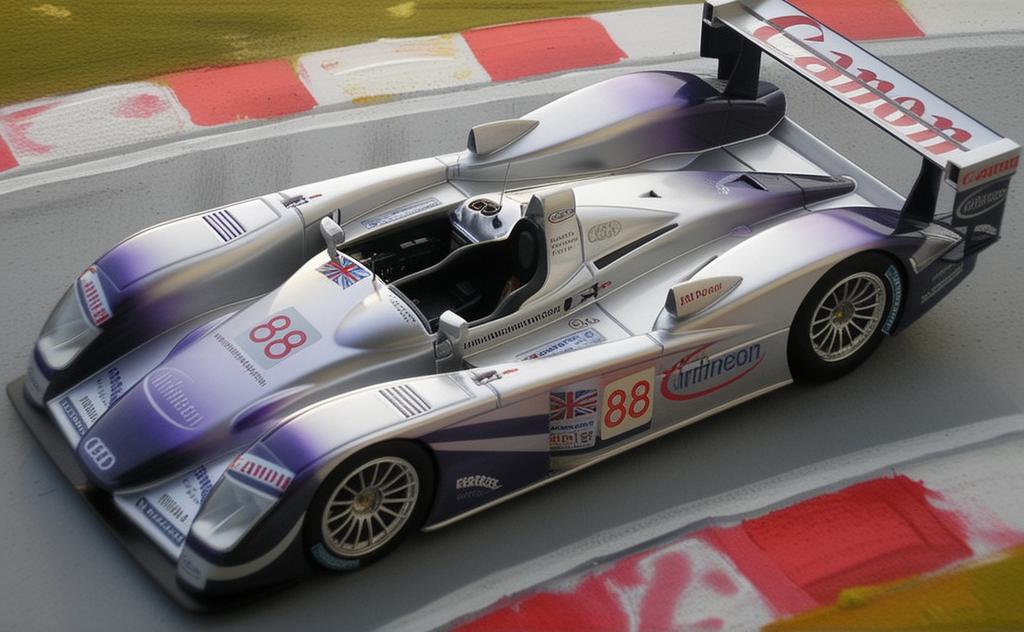} \\
        Input & Empty & Short & Long & Image (ours)
    \end{tabular}
    \caption{Using a descriptive condition in DDIM inversion results in improved reconstruction. As shown, image conditioning outperforms text conditioning. The benefit of our method is particularly evident in challenging images with intricate details.}
    \vspace{-10pt}
    \label{fig:recon-qualitative}
\end{figure}

%% file: sec/4_experiments.tex
\input{figures/recon_comparison_qual}

\section{Experiments}
\label{sec:experiments}
\graphicspath{../images/}

We evaluate our inversion method based on both reconstruction accuracy and editability. To demonstrate editability, we utilize a variety of existing image editing techniques, each excelling in different types of edits, and apply them to the inverted images.

\input{tables/reconstruction_quant}
\input{figures/recon_comparison_qual_flux}

Unless stated otherwise, our experiments use SDXL~\cite{podell2024sdxl} with DDIM scheduler~\cite{song2022denoisingdiffusionimplicitmodels}. All experiments utilize $50$ denoising steps with a default guidance scale of $7.5$. For image conditioning, we employ IP-Adapter-plus\_sdxl\_vit-h~\cite{ye2023ipadaptertextcompatibleimage}.
In few-step diffusion experiments, we use SDXL-Turbo~\cite{sauer2023adversarialdiffusiondistillation} with an Euler scheduler and perform $4$ denoising steps. We also explore Flux~\cite{flux} using FLUX.1-dev where we condition the model with PulID-Flux~\cite{guo2024pulid} and use RF-Inversion~\cite{rout2024rfinversion} with 28 steps. As PulID was trained only on human faces, we focus on this domain for evaluating our method with Flux.

\subsection{Reconstruction}

We evaluate reconstruction both qualitatively and quantitatively. For quantitative evaluation, we measure $L_2$ distance, PSNR, SSIM and LPIPS~\cite{zhang2018perceptual}.
Figures~\ref{fig:motivation-figure} and \ref{fig:recon-qualitative} present qualitative results of DDIM inversion~\cite{dhariwal2021diffusionmodelsbeatgans} under increasingly descriptive conditions. These examples highlight that conditioning the inversion process on an image significantly improves reconstruction in highly detailed regions. Notably, in the third example of Figure~\ref{fig:recon-qualitative}, our method successfully reconstructs the tattoo on the back of the right boxer. Furthermore, the boxer's leg pose is more accurately preserved, and the tattoo on the leg becomes visible.

\vspace{-6pt}
\paragraph{\textbf{Comparisons}}
We integrate Tight Inversion with several existing inversion methods and demonstrate that it enhances their reconstruction performance. Specifically, we combine our method with DDIM inversion~\cite{dhariwal2021diffusionmodelsbeatgans}, ReNoise~\cite{garibi2024renoise}, and RF-Inversion~\cite{rout2024rfinversion}. Note that DDPM-based inversion methods typically guarantee perfect reconstruction, so we compare with these methods only in terms of editability.
Qualitative results are shown in Figures~\ref{fig:recon-qualitative-comp-sdxl} and \ref{fig:recon-qualitative-comp-flux}. As illustrated, integrating Tight Inversion with existing methods consistently improves reconstruction. For example, in Figure~\ref{fig:recon-qualitative-comp-sdxl}, our method accurately reconstructs the handrail in the leftmost example and the man with the blue shirt in the rightmost example.

We further validate the improvement quantitatively. Following previous works~\cite{Mokady_2023_CVPR, garibi2024renoise}, we utilize the test set of MS-COCO~\cite{lin2015microsoftcococommonobjects} and present the results in Table~\ref{tab:reconstruction_comparison_quant}. As observed from the table, our method improves reconstruction of existing inversion methods across all metrics.

\paragraph{\textbf{Ablation Studies}}
We conduct ablation studies to evaluate the importance of combining image conditioning with an inversion method. Since IP-Adapter is trained to reconstruct images from image conditions, it is reasonable to explore whether accurate reconstruction can be achieved solely by conditioning on the image, without requiring a carefully selected noise initialization.
Figure~\ref{fig:ablation1} explores this possibility. In the first row, a random noise is sampled, and the denoising process is conditioned on the input image. While the semantics and colors are captured, the reconstructed image poorly matches the original one. This demonstrates that precise reconstruction still requires a specific initial noise.
In the second row, DDIM inversion is performed using only a text prompt, while denoising is conditioned on the input image. The results show slight over-saturation and the disappearance of the phone in the man's hand.
In the third row, our Tight Inversion method is applied, conditioning both inversion and denoising on an input image. Our method significantly outperforms the alternatives, faithfully reconstructing both colors and fine details, including the phone.

\input{figures/ablation1}

We further explore the impact of image conditioning strength. Specifically, IP-Adapter provides a guidance scale, $s$, which controls the influence of the input image on the generated output. Setting $s$ to zero is equivalent to using the text-to-image model without IP-Adapter. Figure~\ref{fig:ablation2} presents the results for different values of $s$.
As expected, we observe that reconstruction quality (second row) improves with higher IP-Adapter scales, emphasizing the importance of precise conditioning.

\input{figures/ablation2}

\subsection{Editing}

Next, we evaluate Tight Inversion in the context of image editing. Specifically, we analyze the impact of integrating our technique with various image editing methods (prompt2prompt~\cite{hertz2022prompt}, Edit Friendly DDPM~\cite{hubermanspiegelglas2024editfriendlyddpmnoise}, LEDITS++ \cite{brack2024ledits}, RF-Inversion \cite{rout2024rfinversion}). We demonstrate that, in addition to providing accurate reconstruction, our method significantly enhances editability. Specifically, we perform different types of edit and show that our approach consistently improves editing results, both qualitatively and quantitatively.

\begin{figure}[h!]
    \centering
    \input{figures/scale_clip_similarity}
    \caption{CLIP Similarity of the edited text prompt and the edited image vs. CLIP Similarity of the source image and edited imaged for IPA scales in the range of 0 (without tight inversion, marked with a cross) to 0.7 (strong conditioning on the source image). For both axes, higher is better.}
    \label{fig:clip_similarity}
\end{figure}

\input{figures/editing_qualitative}

\paragraph{\textbf{Qualitative Comparison}}
We present qualitative results obtained with SDXL \cite{podell2024sdxl} and Flux~\cite{flux} in Figure~\ref{fig:edit-qualitative-comp} (more results are in Figures~\ref{fig:more-results1} and \ref{fig:more-results2}). In the first and second rows, we perform a na\"ive edit by changing the prompt during the denoising process. 
In the third row, we apply DDIM inversion and denoise the inverted noise using prompt2prompt \cite{hertz2022prompt}. 
The next two rows utilize the inversion and denoising methods from Edit Friendly DDPM \cite{hubermanspiegelglas2024editfriendlyddpmnoise} and LEDITS++ \cite{brack2024ledits}, respectievly.
In the last three rows, we use RF-Inversion \cite{rout2024rfinversion} with Flux, and we use PulID \cite{guo2024pulid} as the conditioning mechanism for our Tight Inversion method.
In each row, we show the input image, followed by the reconstruction results (with and without Tight Inversion), and then the edited images obtained from the inverted noises (with and without Tight Inversion).

Note that both Edit Friendly DDPM Inversion and LEDITS++ guarantee perfect reconstruction. For the other methods, we select examples where the reconstruction, even without Tight Inversion, is accurate. This choice emphasizes that, even when competing methods produce plausible reconstructions, our method outperforms them in terms of editability.

As shown in the results, our method better preserves the original image, maintaining the structure of the diner in the first row, the patterns on the snow and the animal's expression in the third row, and the horse's pose in the fifth row. In the results obtained with Flux, our method preserves the identity of the individual significantly better in the edited image, even when the reconstruction is comparable (e.g., the shape of LeCun's head).

In Figure~\ref{fig:turbo_results}, we present results with SDXL-Turbo~\cite{sauer2023adversarialdiffusiondistillation}. Here, we use ReNoise inversion~\cite{garibi2024renoise} combined with Tight Inversion. To edit the inverted noise, we denoise it with a target text prompt. As shown, Tight Inversion results in better preservation of the cups in the top example and the background in the bottom example.

\input{figures/challenging_edits_turbo}

\paragraph{\textbf{Quantitative Comparisons}}
Next, we evaluate our editing results quantitatively. We use the MagicBrush benchmark~\cite{Zhang2023MagicBrush} for the evaluation, as it contains diverse and challenging images and edits. Following previous work~\cite{brooks2023instructpix2pixlearningfollowimage} we evaluate the edit quality in terms of the preservation of the input image, and the adherence to the target prompt, and we use CLIP~\cite{radford2021learning} to measure both.
We present the results with DDIM Inversion and LEDITS++ in Figure~\ref{fig:clip_similarity}. In both graphs the tradeoff between image preservation and adherence to the target edit is clearly observed~\cite{tov2021designing}. Tight Inversion provides better control on this tradeoff, and better preserves the input image while still aligning with the edit prompt as also evident in Figure~\ref{fig:edit-qualitative-comp}. Note, that a CLIP similarity of above 0.3 between an image and a text prompt indicates plausible alignment between the image and the prompt.

\paragraph{\textbf{Ablation Studies}}
In Figure~\ref{fig:ablation1}, we edit the image by denoising using a modified prompt. In the first row, where we use a random noise, the resulting image significantly differs from the input. In the second row, where the inversion is not conditioned on the input image, the red hat is not added, which may result from the initial noise being slightly out of distribution. This makes it more difficult to edit, particularly when an image condition is used. In the third row, a red hat is added to the man while the input image is successfully preserved.

We explore the IP-Adapter guidance scale effect on the edit in Figure~\ref{fig:ablation2}. In the third row, we add a cowboy hat to the deer, where various guidance scales are used for the inversion and denoising. 
We observe a clear reconstruction-editability tradeoff associated with the IP-Adapter scale. While increasing the scale improves reconstruction quality, it progressively limits editing capabilities, eventually preserving the original image intact. In practice, we found that an IP-Adapter scale of 0.4 strikes an effective balance for most cases.

%% file: figures/recon_comparison_qual.tex
\begin{figure}
    \centering
    \setlength{\tabcolsep}{1pt}
    \scriptsize{
    \begin{tabular}{cc cccc}
        \multicolumn{2}{c}{\raisebox{22pt}{\rotatebox[origin=t]{90}{Input}}} &
        \includegraphics[width=0.228\linewidth]{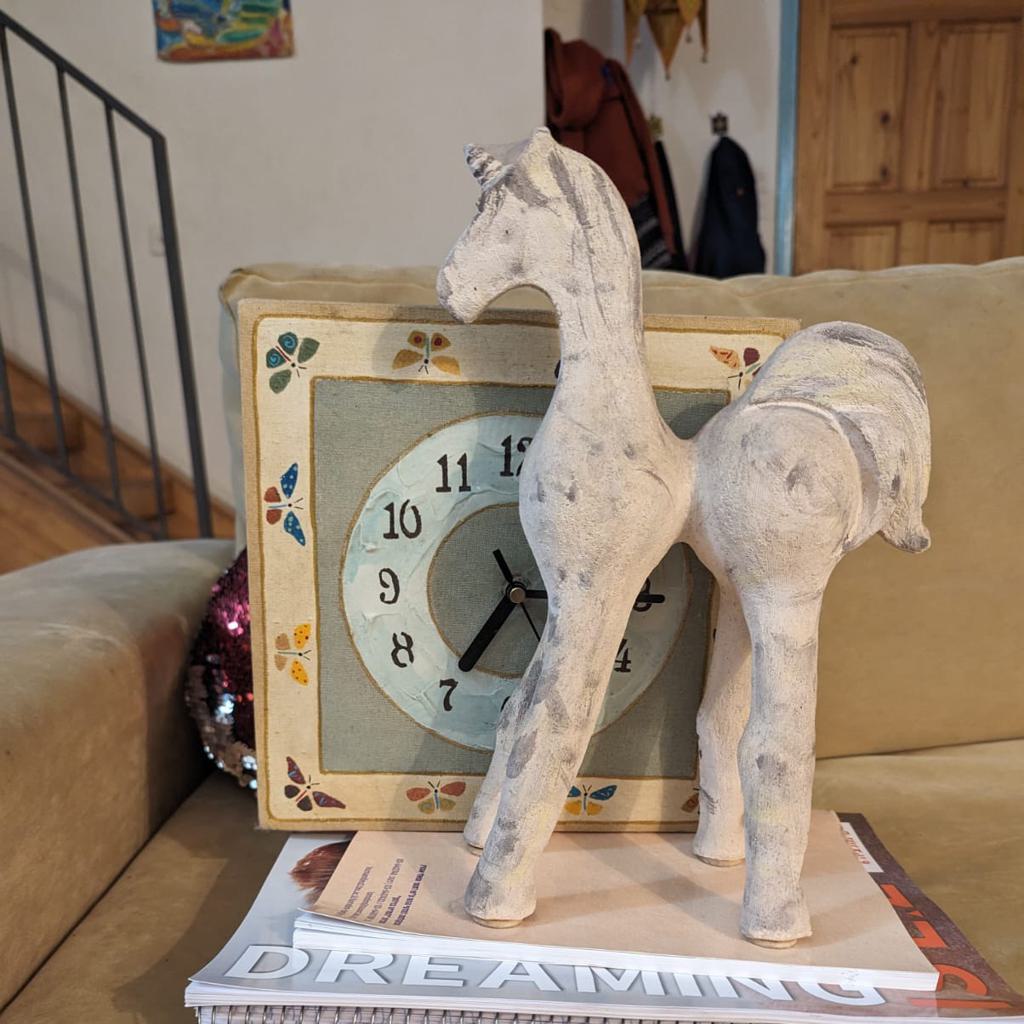} &
        \includegraphics[width=0.228\linewidth]{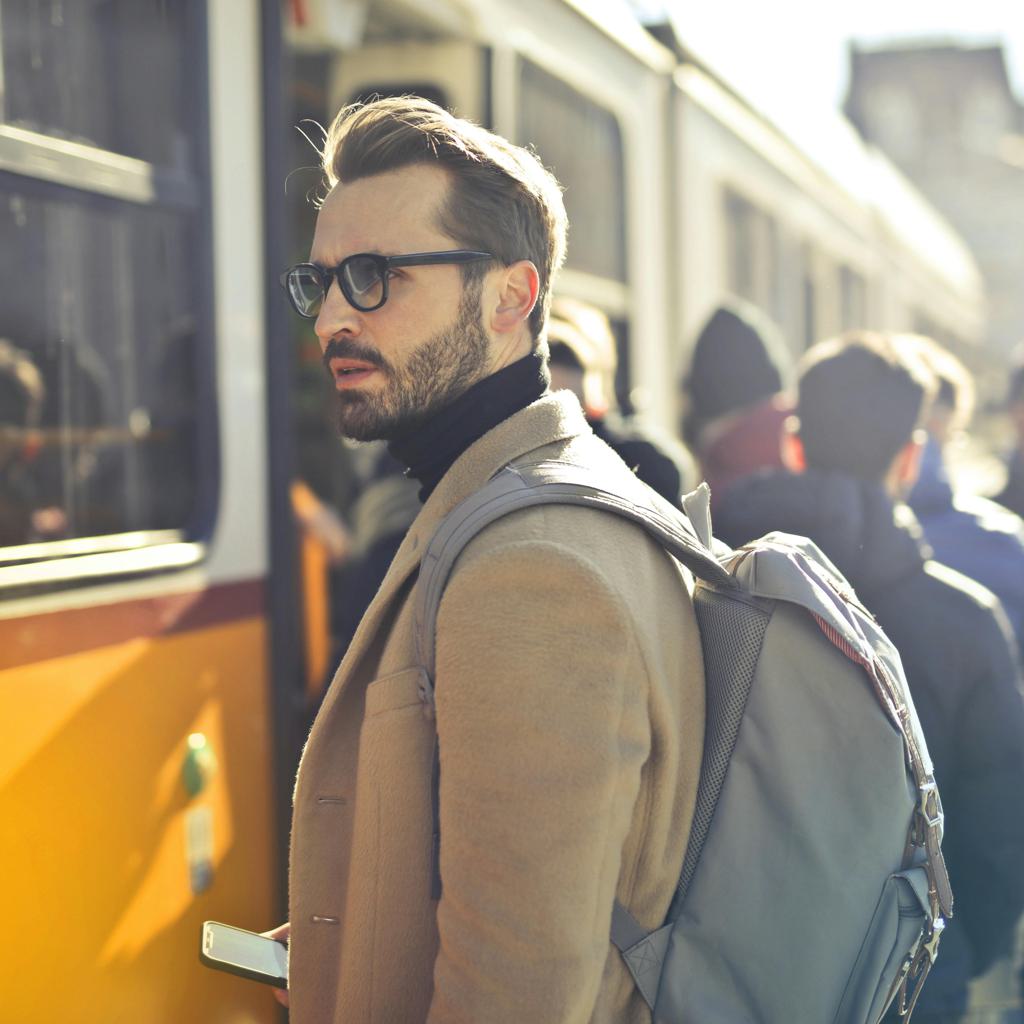} &
        \includegraphics[width=0.228\linewidth]{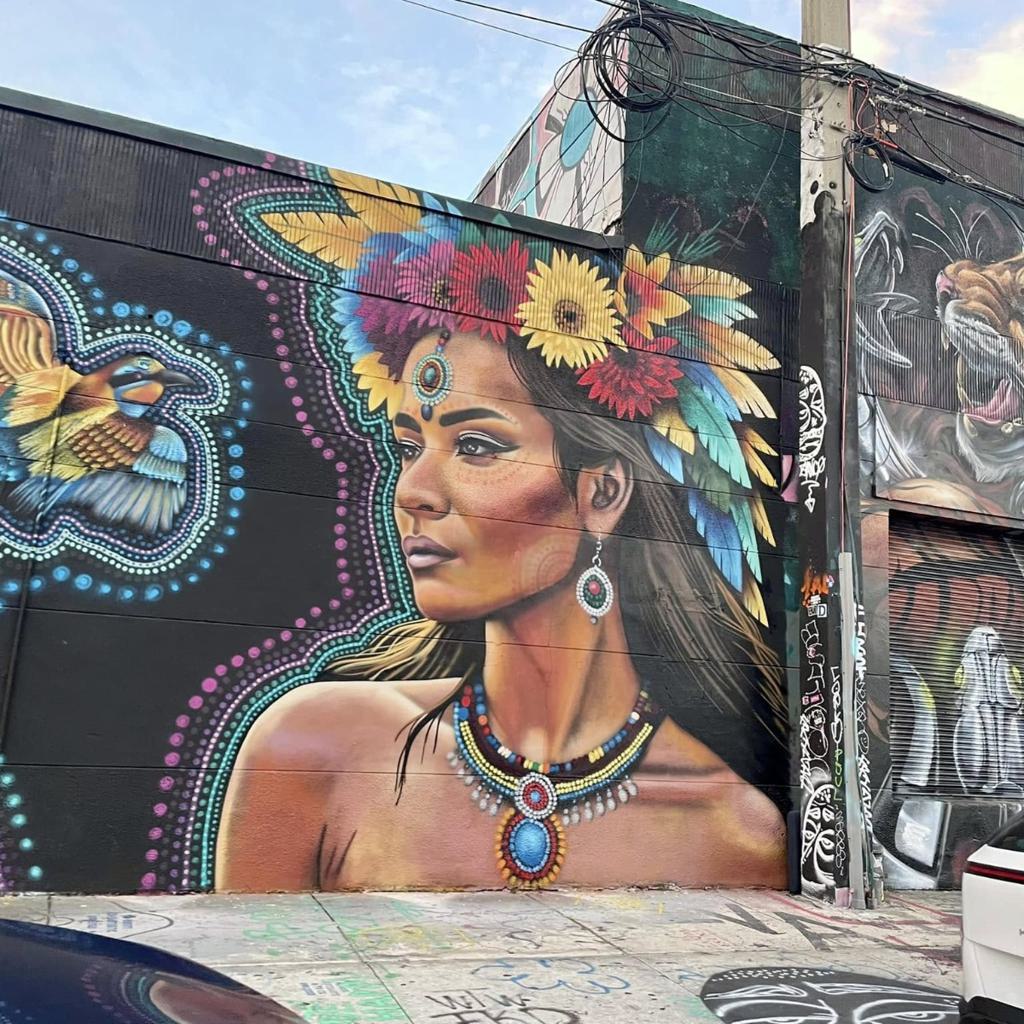} &
        \includegraphics[width=0.228\linewidth]{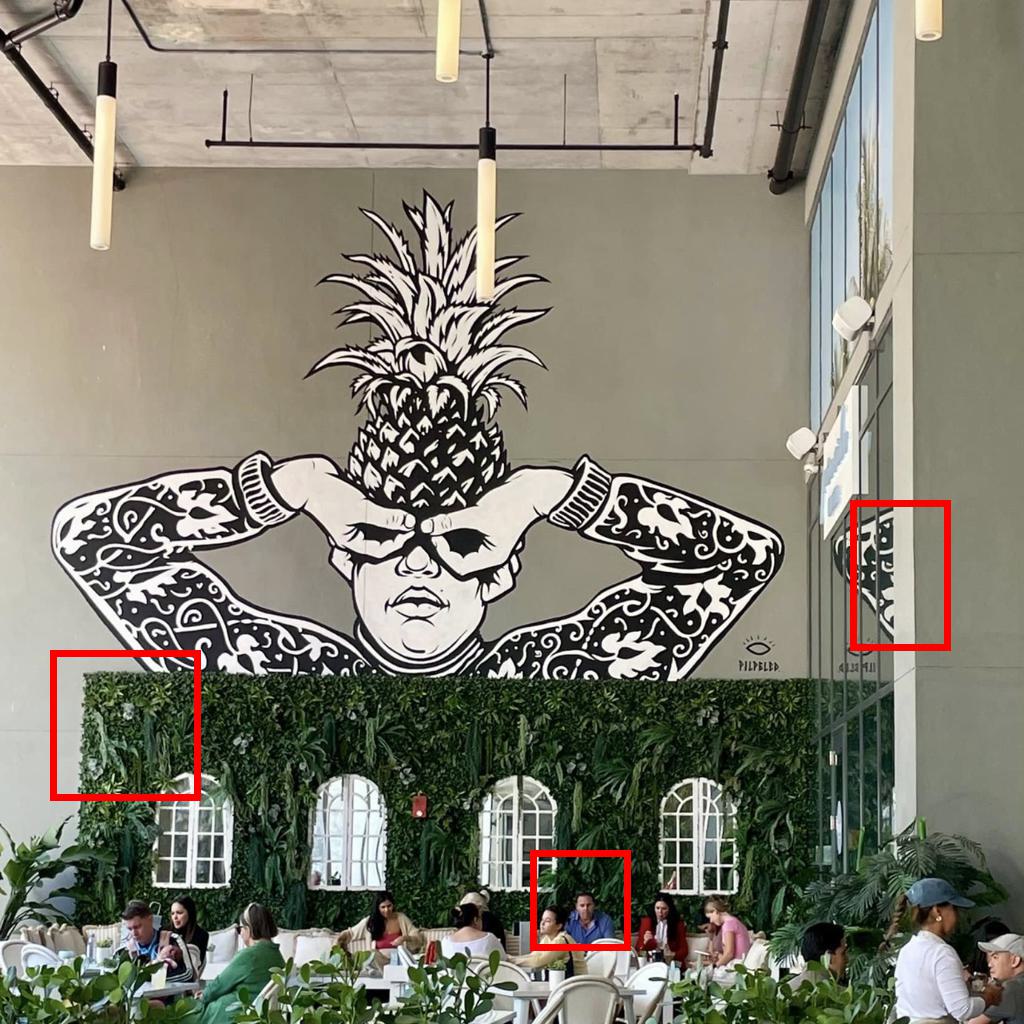} \\
        
        \multicolumn{2}{c}{\raisebox{22pt}{\rotatebox[origin=t]{90}{DDIM Inv.}}} &
        \includegraphics[width=0.228\linewidth]{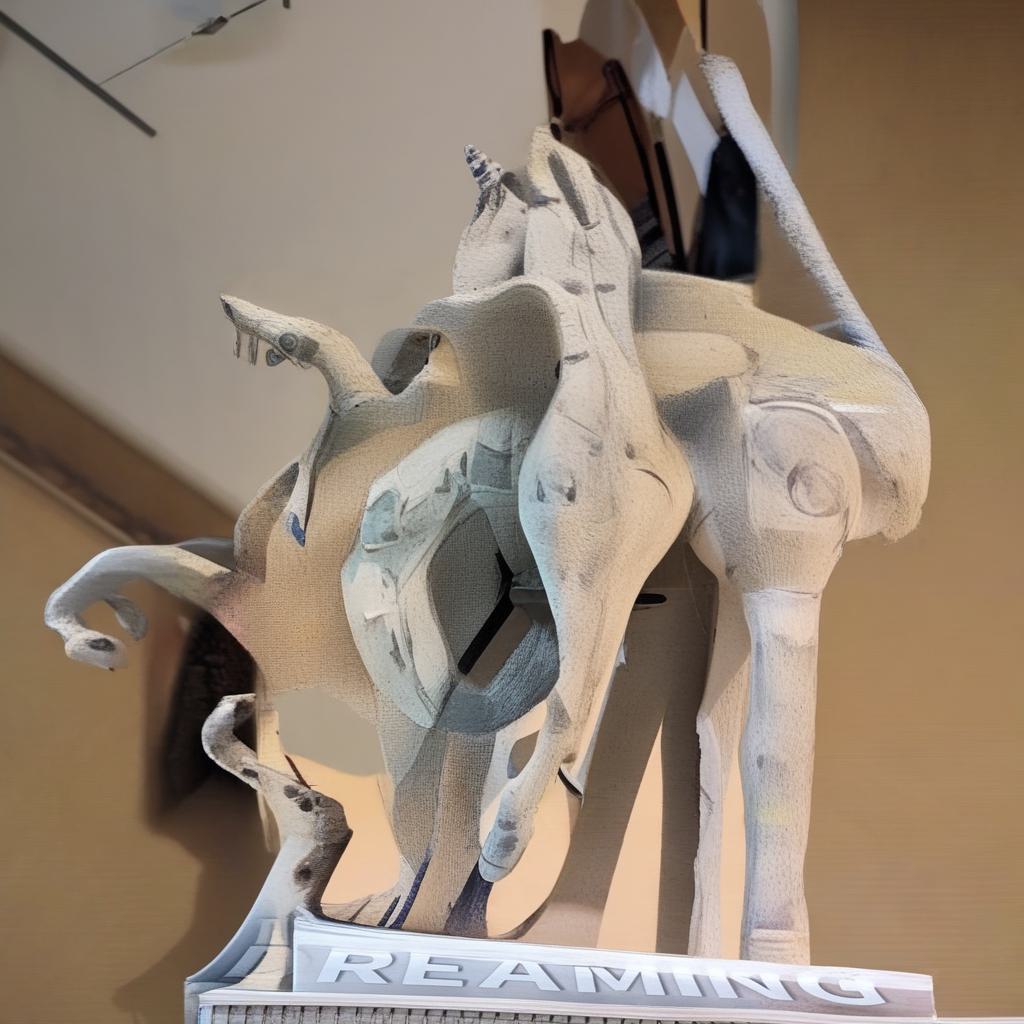} &
        \includegraphics[width=0.228\linewidth]{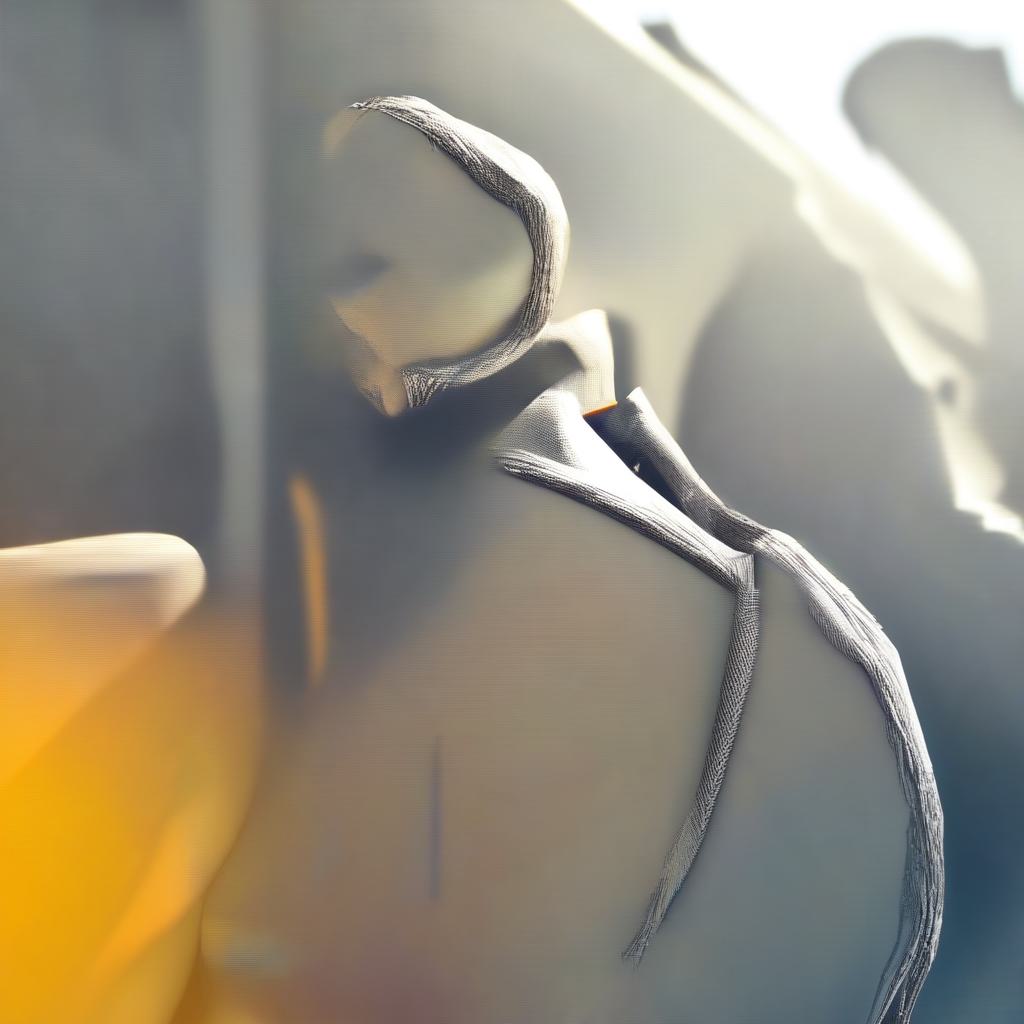} &
        \includegraphics[width=0.228\linewidth]{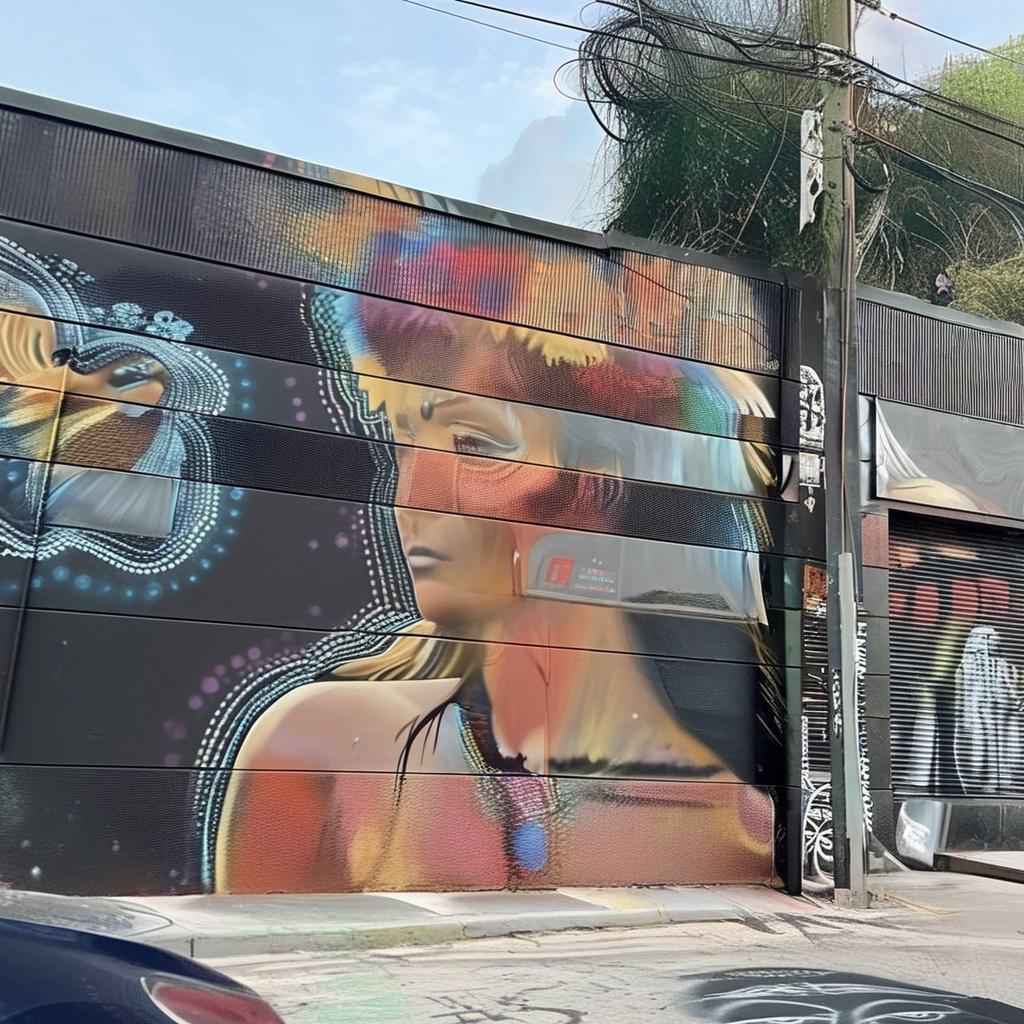} &
        \includegraphics[width=0.228\linewidth]{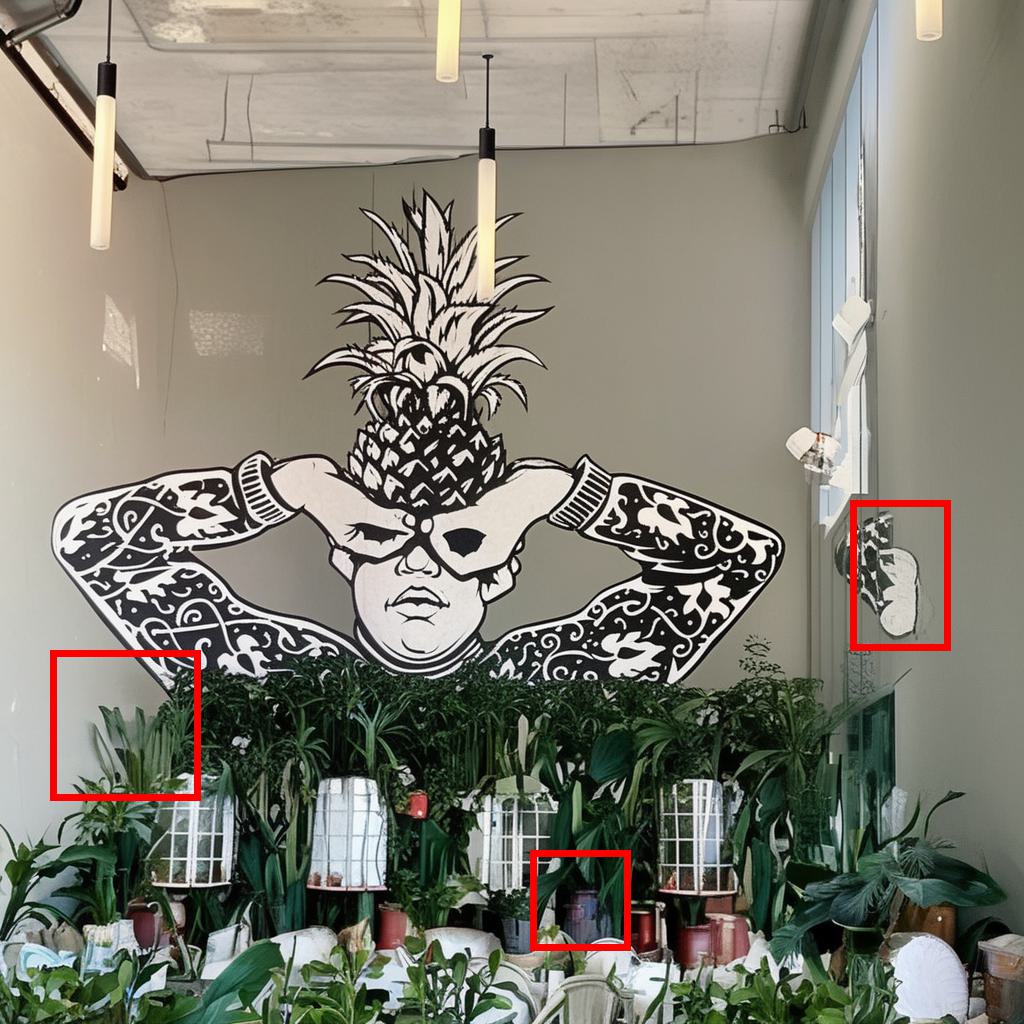} \\
        
        \raisebox{22pt}{\rotatebox[origin=t]{90}{DDIM Inv.}} &
        \raisebox{22pt}{\rotatebox[origin=t]{90}{+ Tight Inversion}} &
        \includegraphics[width=0.228\linewidth]{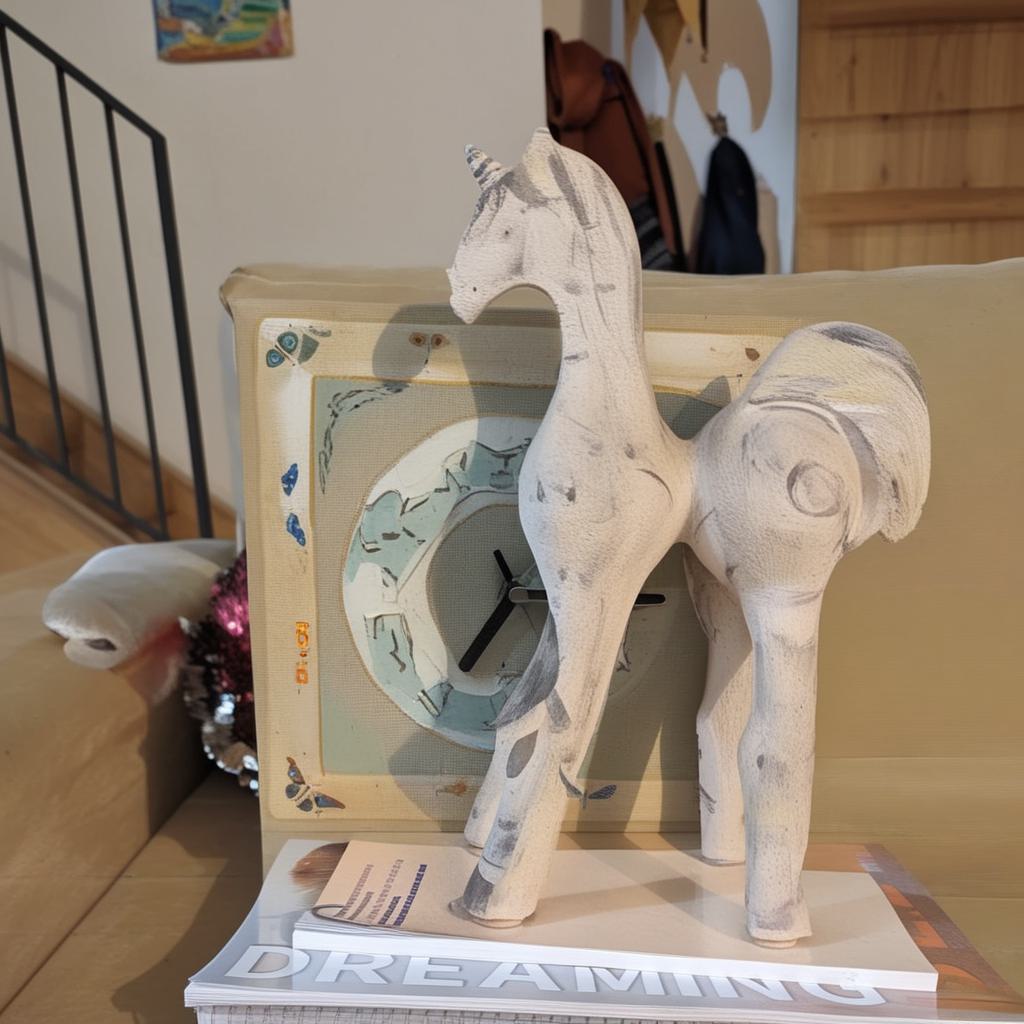} &
        \includegraphics[width=0.228\linewidth]{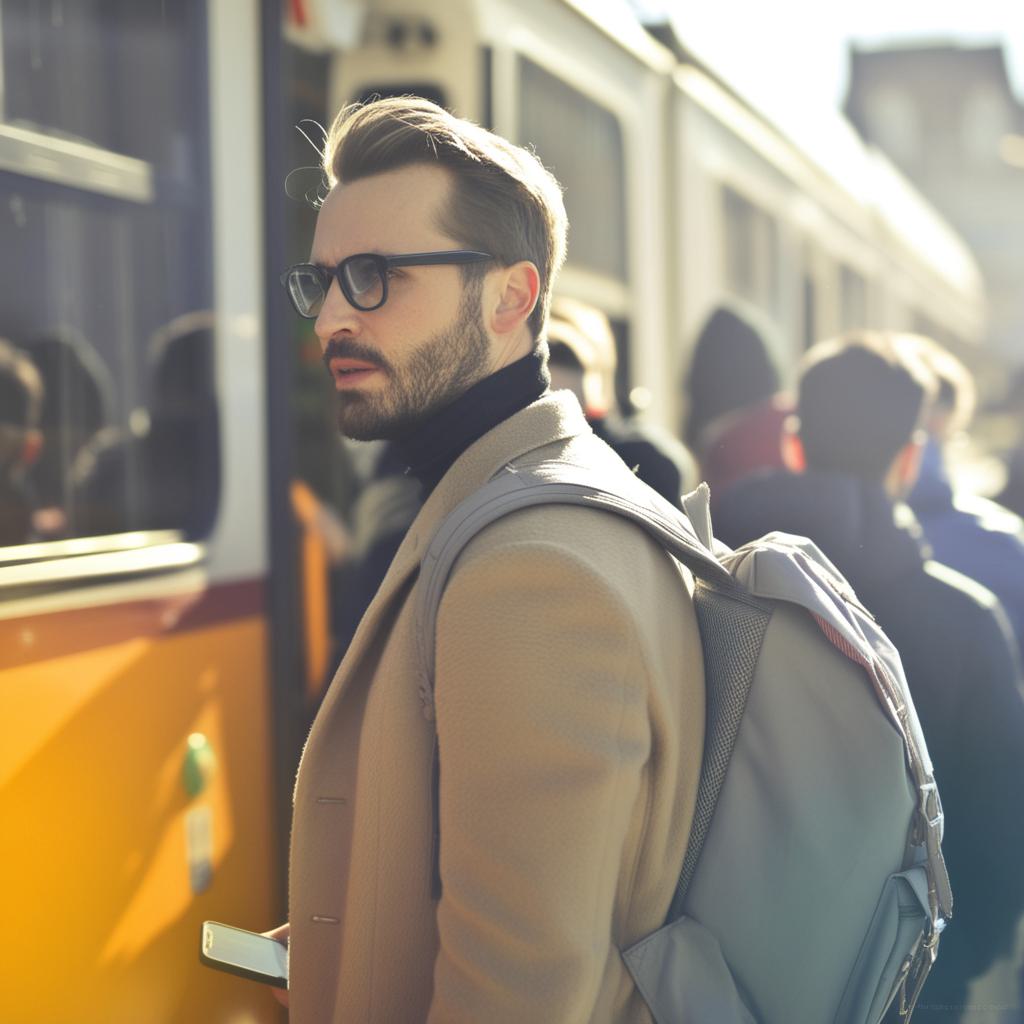} &
        \includegraphics[width=0.228\linewidth]{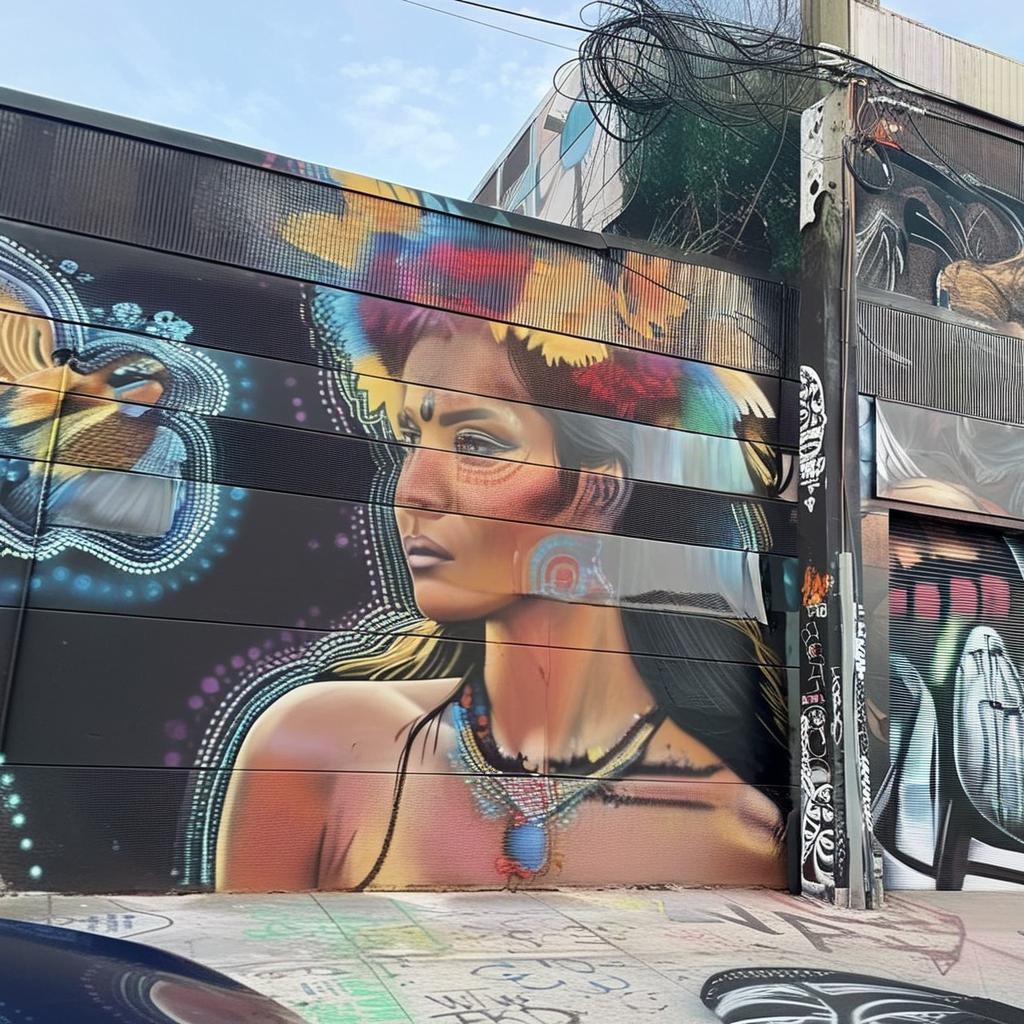} &
        \includegraphics[width=0.228\linewidth]{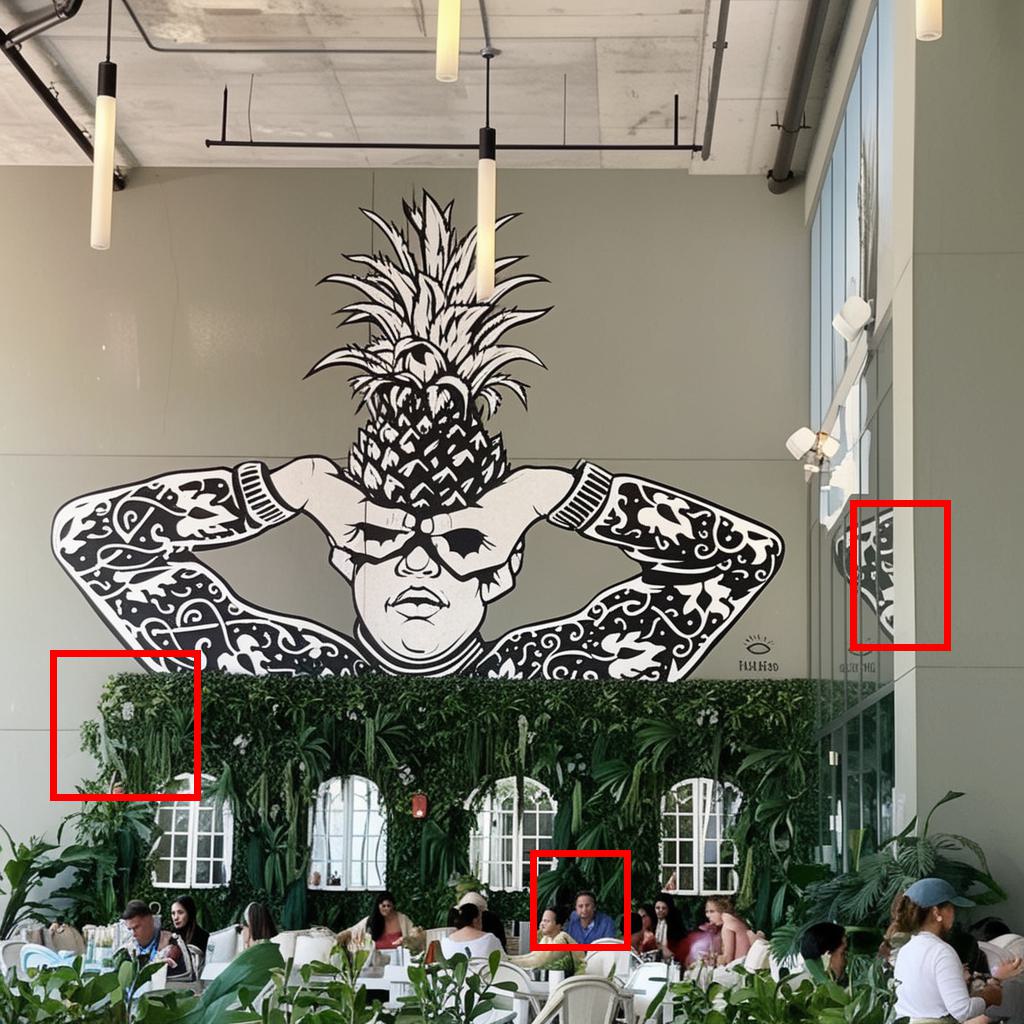} \\
        
        \multicolumn{2}{c}{\raisebox{22pt}{\rotatebox[origin=t]{90}{ReNoise}}} &
        \includegraphics[width=0.228\linewidth]{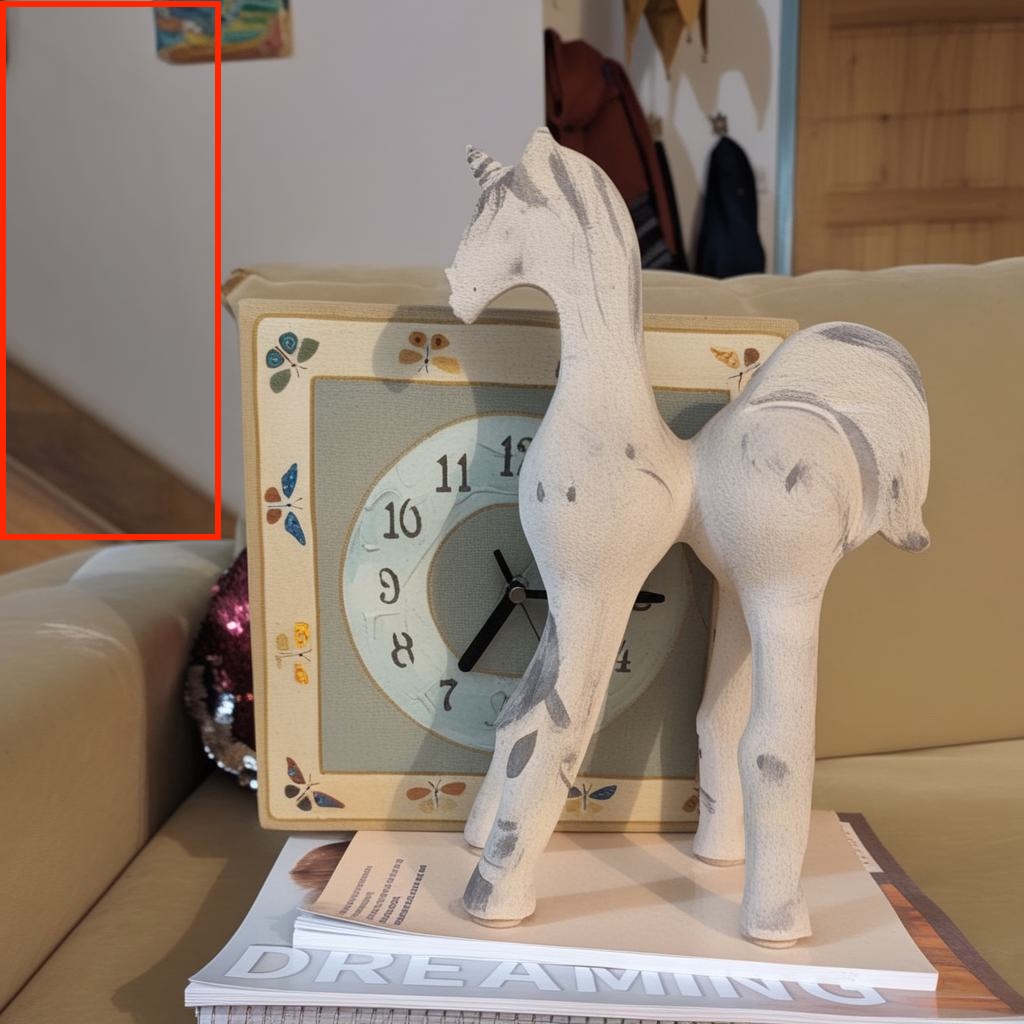} &
        \includegraphics[width=0.228\linewidth]{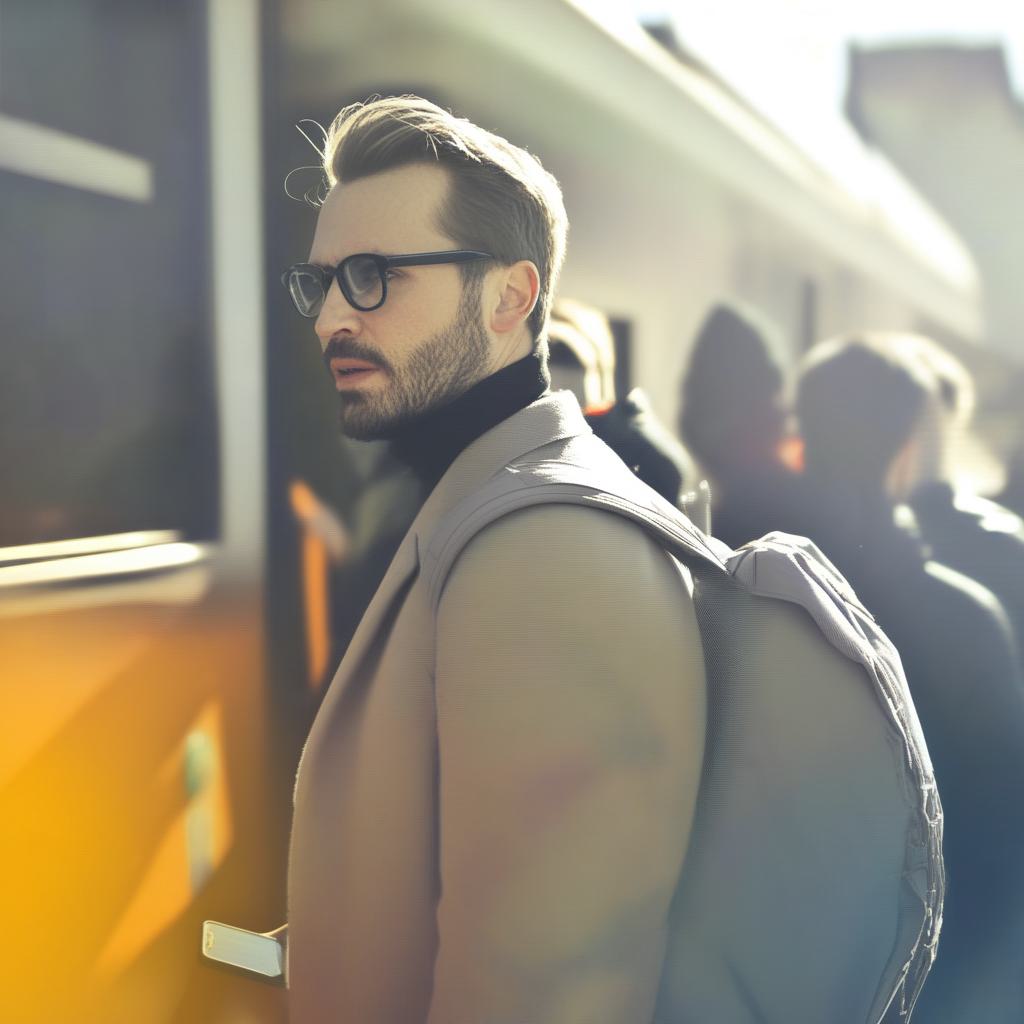} &
        \includegraphics[width=0.228\linewidth]{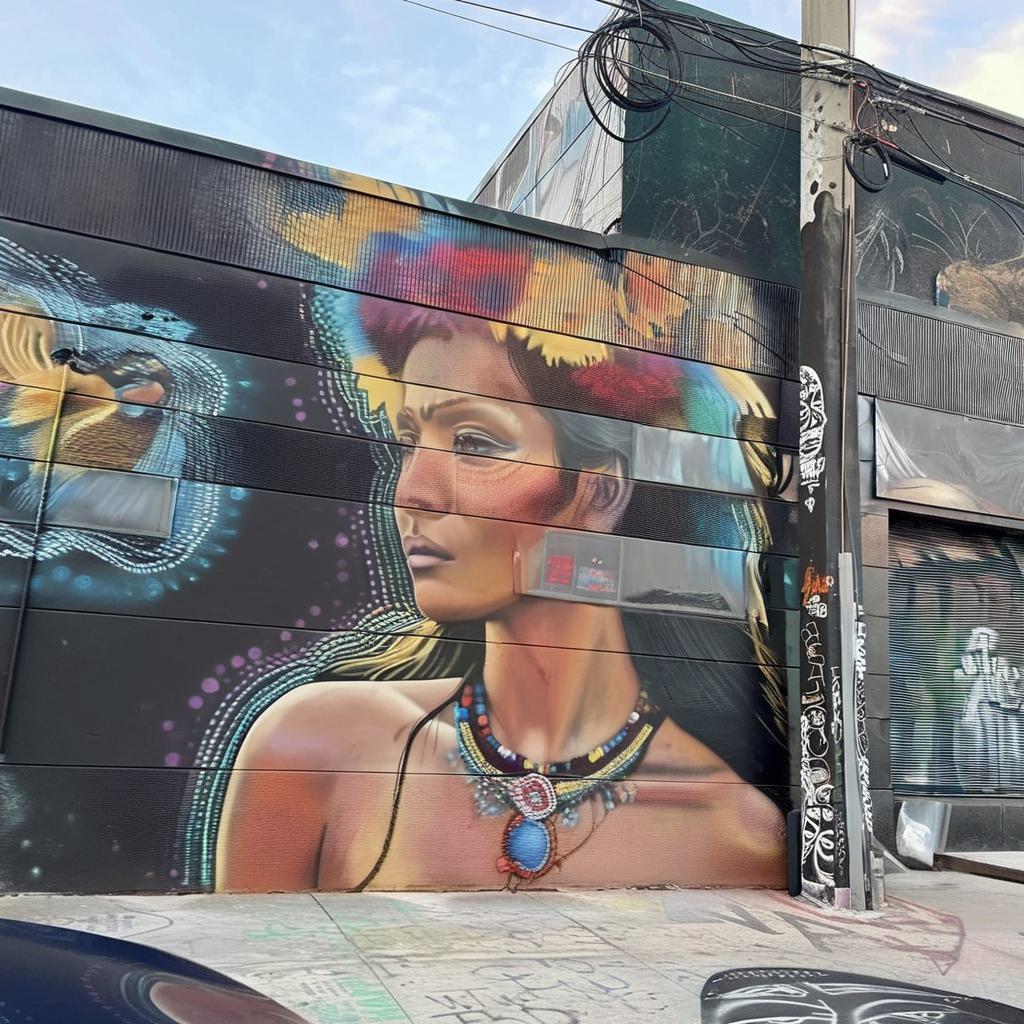} &
        \includegraphics[width=0.228\linewidth]{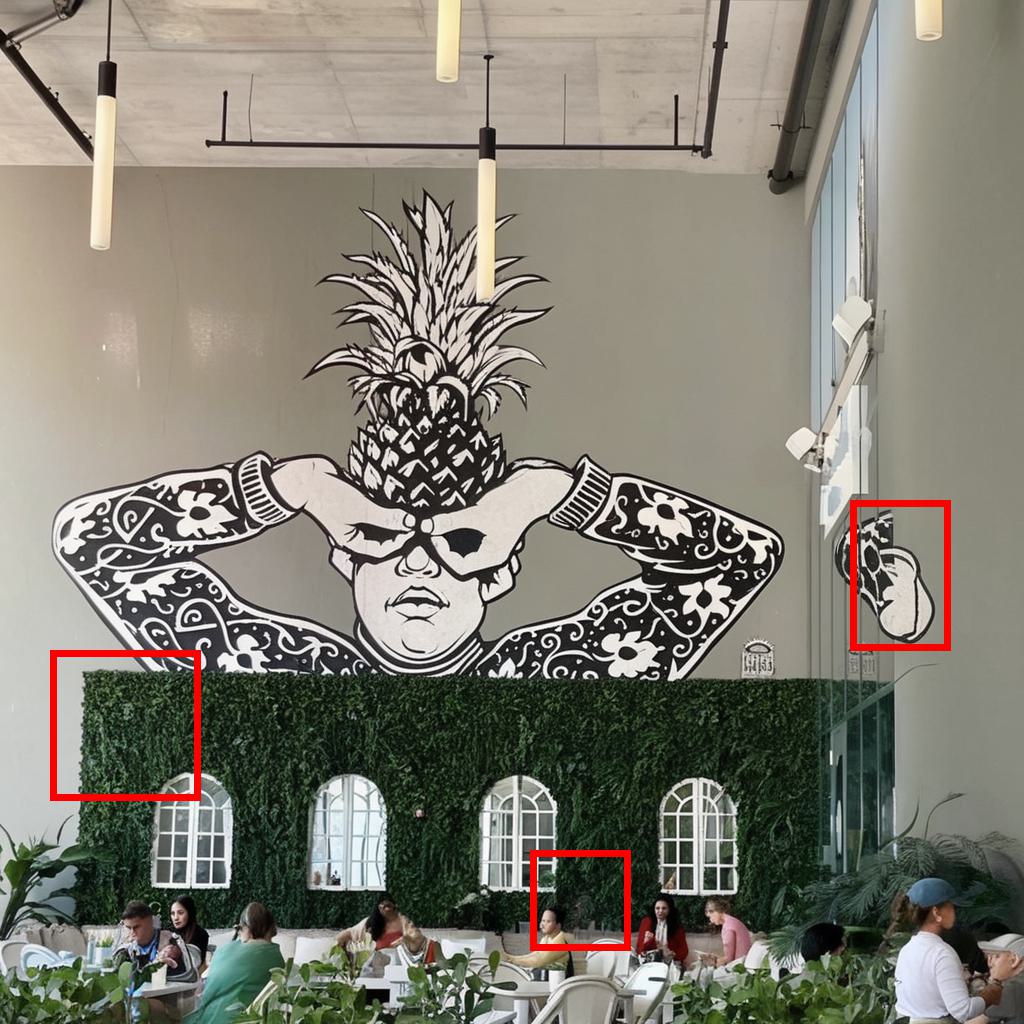} \\
        
        \raisebox{22pt}{\rotatebox[origin=t]{90}{ReNoise}} &
        \raisebox{22pt}{\rotatebox[origin=t]{90}{+ Tight Inversion}} &
        \includegraphics[width=0.228\linewidth]{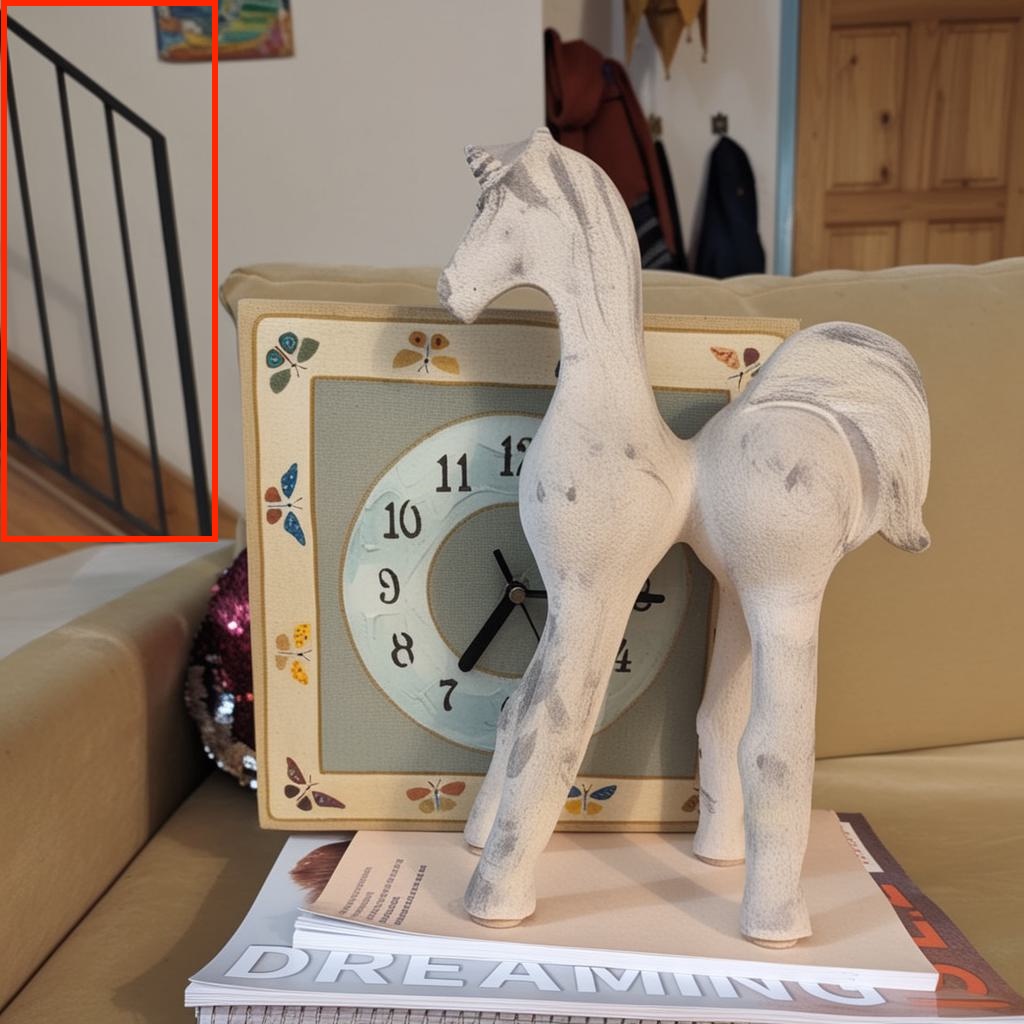} &
        \includegraphics[width=0.228\linewidth]{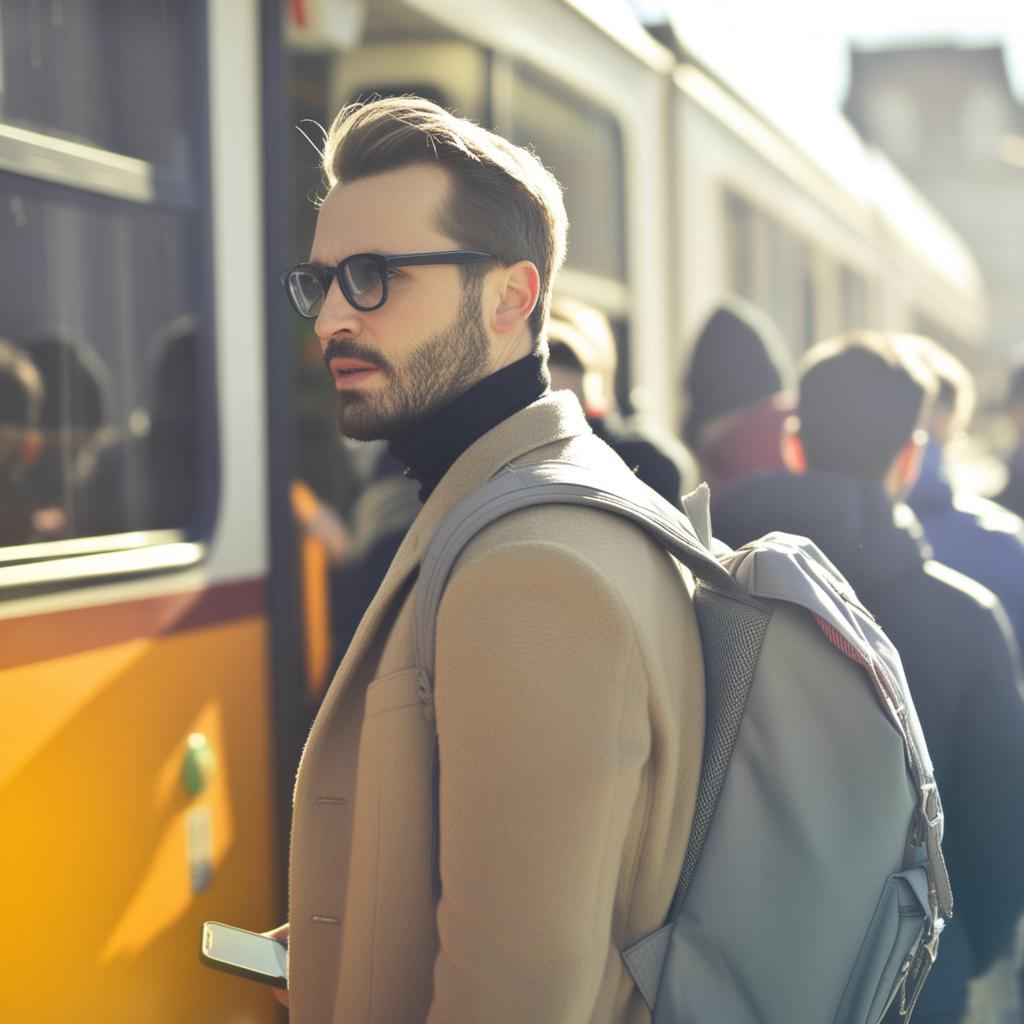} &
        \includegraphics[width=0.228\linewidth]{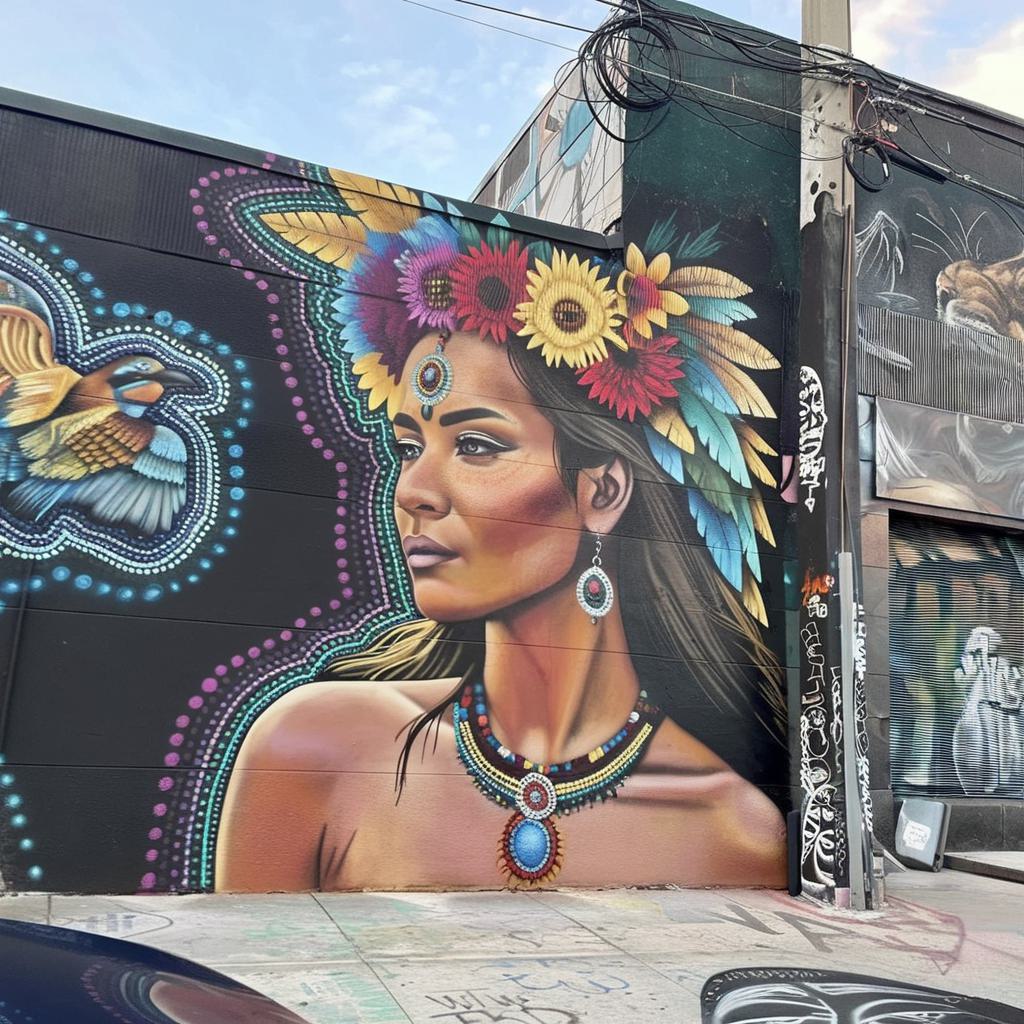} &
        \includegraphics[width=0.228\linewidth]{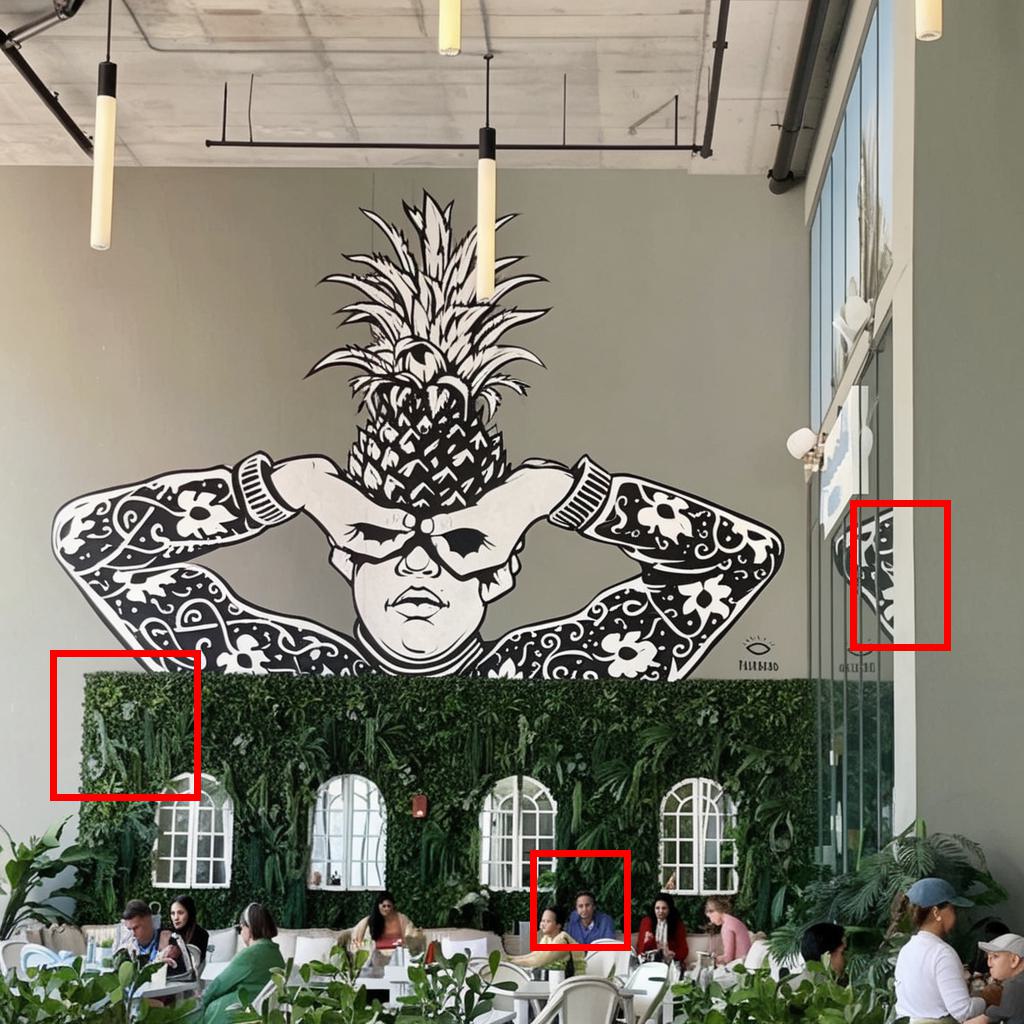} \\
        
        && \multicolumn{4}{c}{\textbf{SDXL}}       
    \end{tabular}
    }
    \caption{Qualitative reconstruction results with SDXL. Integrating Tight Inversion with various inversion methods enhances reconstruction. Observe the reflection on the window in the second column.}
    \label{fig:recon-qualitative-comp-sdxl}
\end{figure}

%% file: tables/reconstruction_quant.tex
\begin{table}
{\small
    \centering
    \caption{Quantitative comparison of various existing inversion methods with and without Tight Inversion.}
    \vspace{-8pt}
    \setlength{\tabcolsep}{3.5pt}
    \begin{tabular}{l c c c c}
      \toprule
      Method & $L_2$ $\downarrow$ & PSNR $\uparrow$ & SSIM $\uparrow$ & LPIPS $\downarrow$ \\
      \midrule
      DDIM Inversion & 50.5897 & 25.3404 & 0.7699 & 0.1485 \\
      DDIM Inversion + Ours & 42.8394 & 26.9030 & 0.7981 & 0.1055 \\
      \midrule
      ReNoise & 42.9509 & 27.1584 & 0.7928 & 0.1179 \\
      ReNoise + Ours & 37.8595 & 28.0413 & 0.8134 & 0.0877 \\
      \bottomrule
      \end{tabular}
    \label{tab:reconstruction_comparison_quant}
    }
  \end{table}

%% file: figures/recon_comparison_qual_flux.tex
\begin{figure}
    \centering
    \setlength{\tabcolsep}{1pt}
    \scriptsize{
    \begin{tabular}{cc cccc}
        \multicolumn{2}{c}{\raisebox{22pt}{\rotatebox[origin=t]{90}{Input}}} &
        \includegraphics[width=0.228\linewidth]{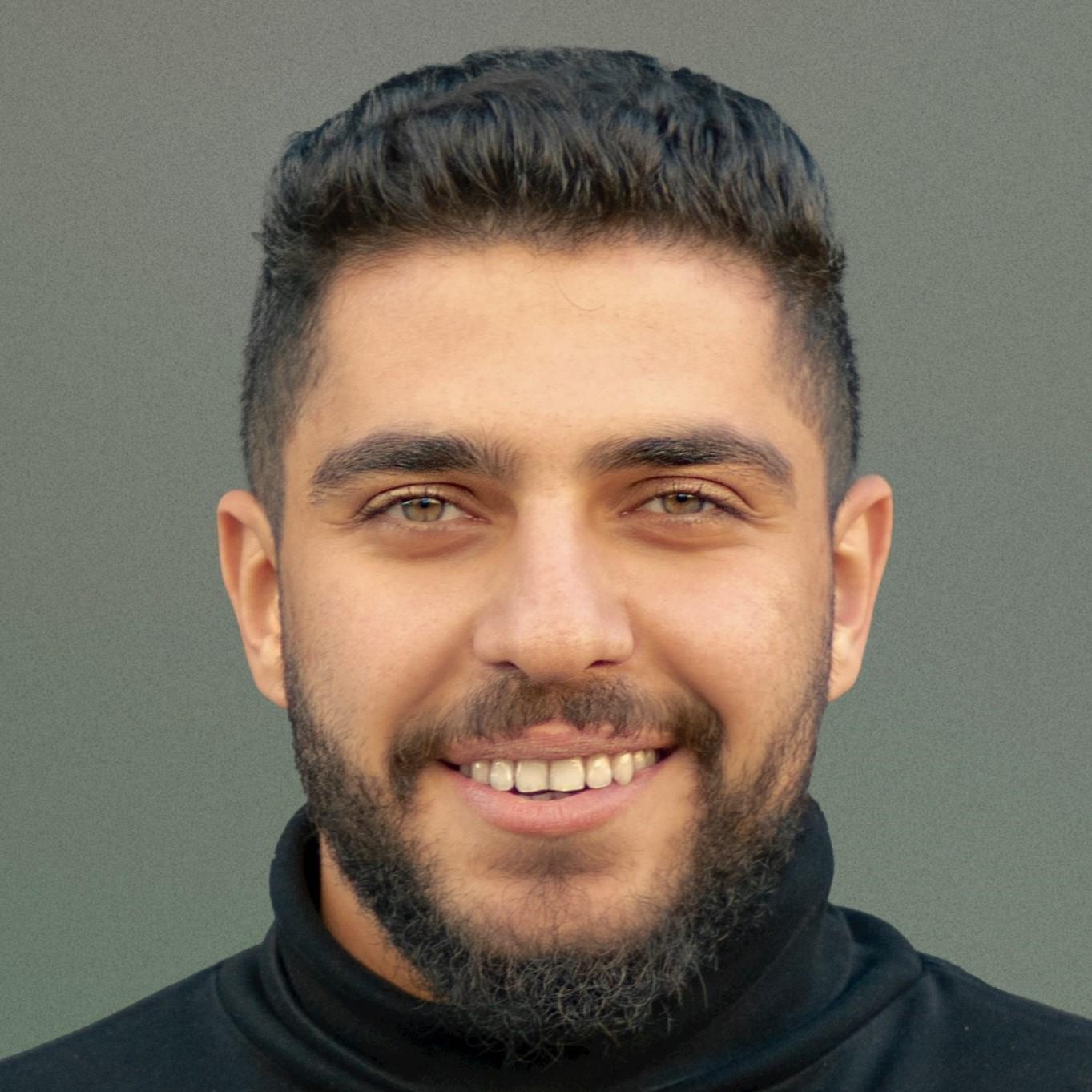} &
        \includegraphics[width=0.228\linewidth]{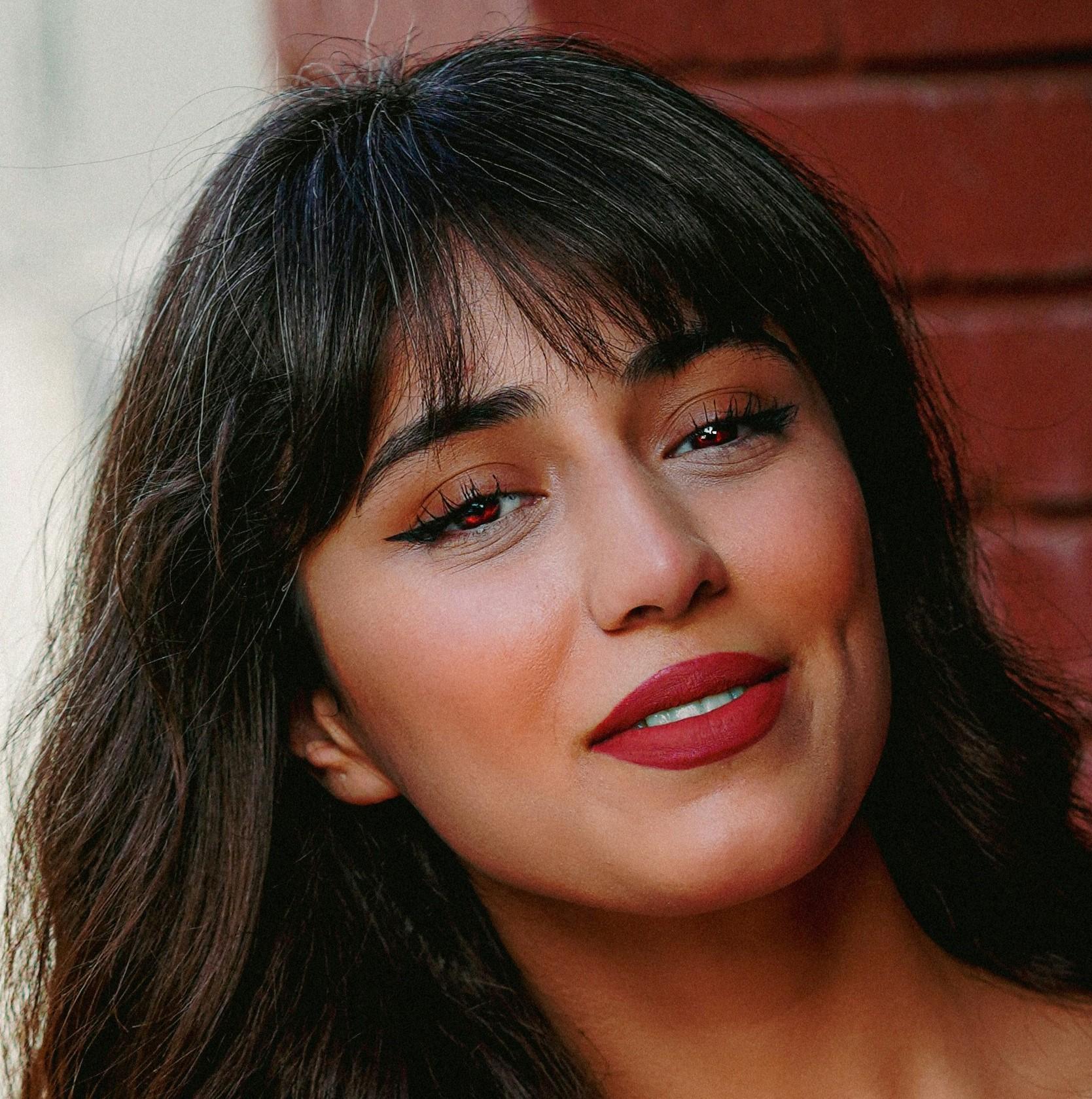} &
        \includegraphics[width=0.228\linewidth]{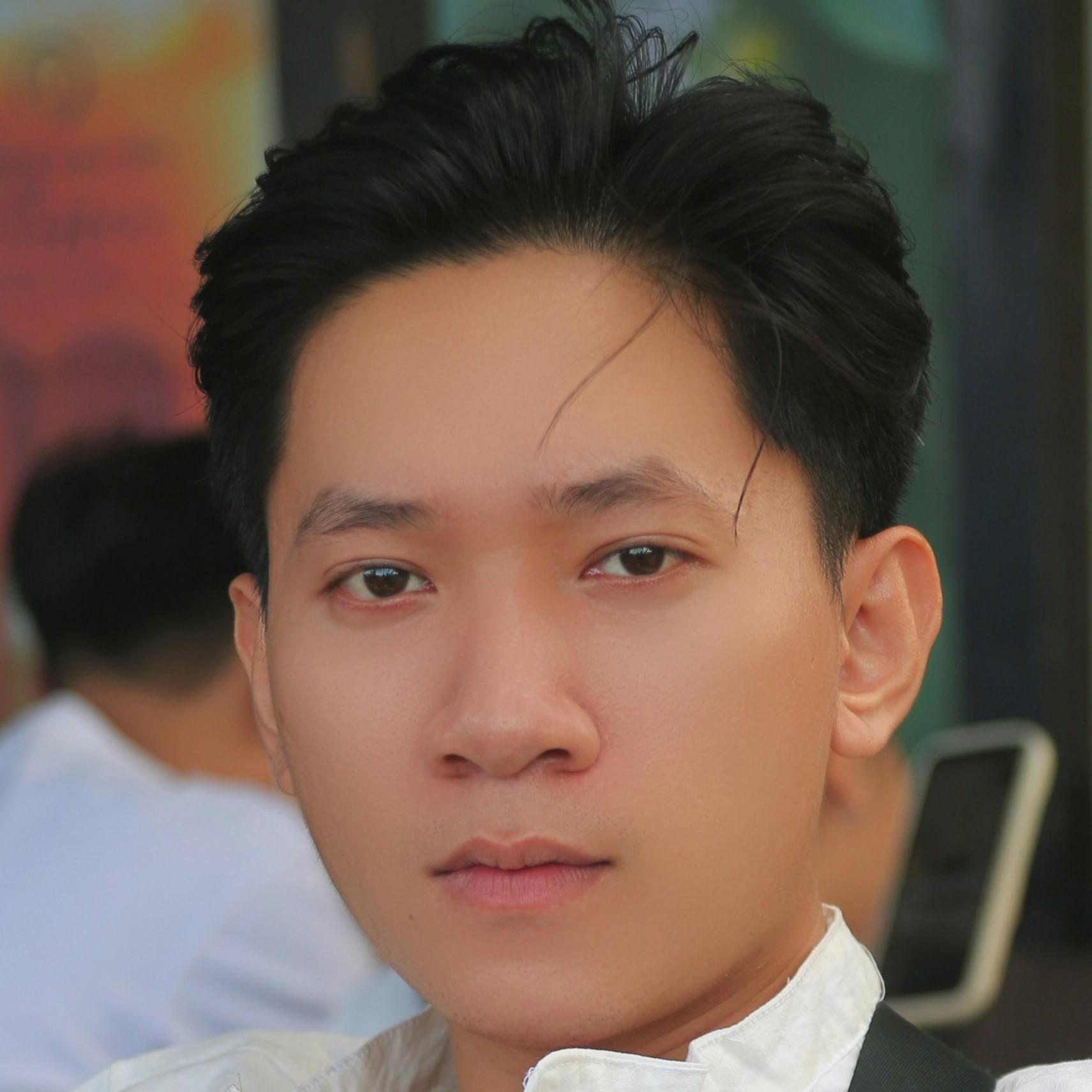} &
        \includegraphics[width=0.228\linewidth]{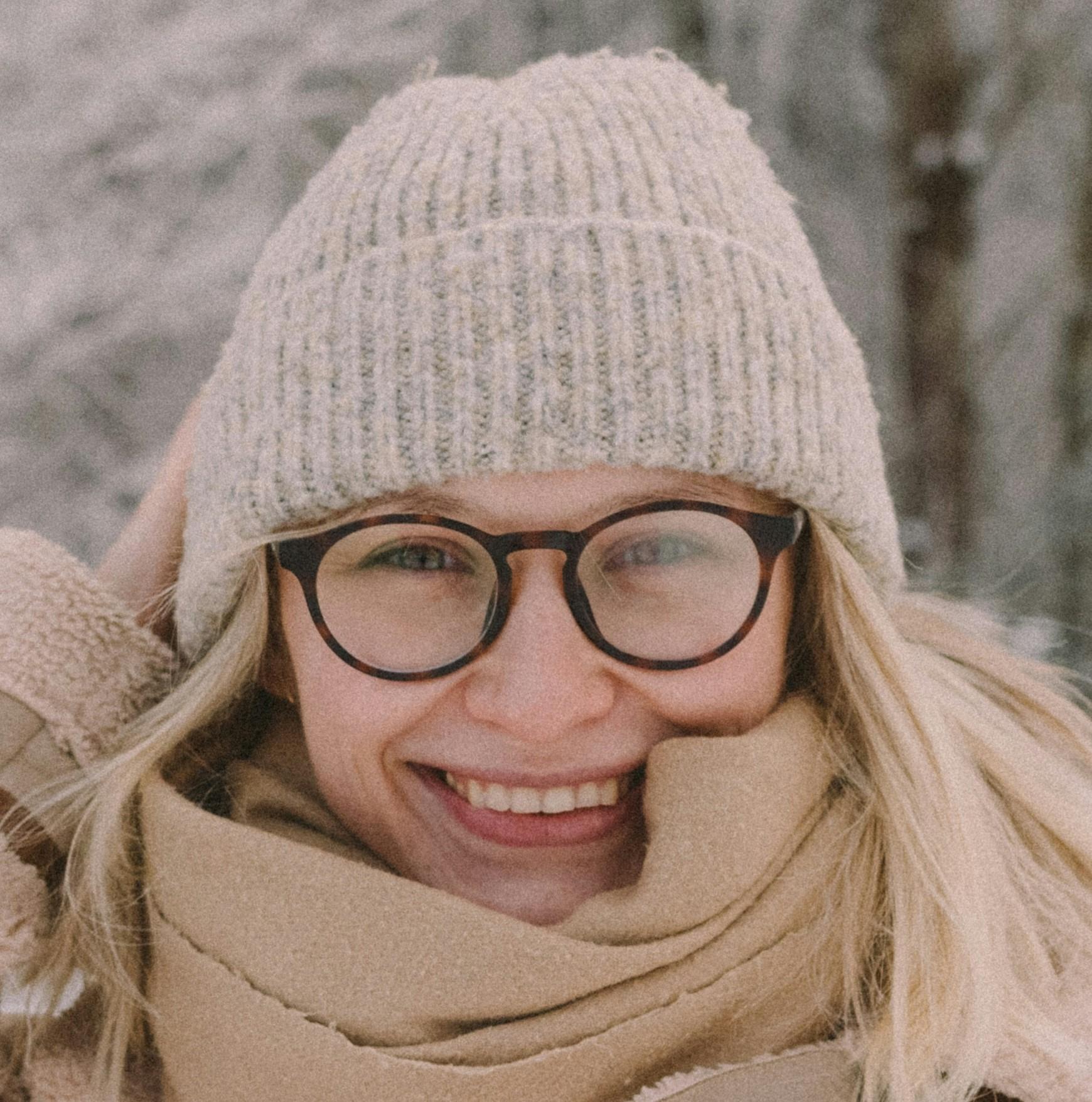} \\
        
        \multicolumn{2}{c}{\raisebox{22pt}{\rotatebox[origin=t]{90}{RF Inv.}}} &
        \includegraphics[width=0.228\linewidth]{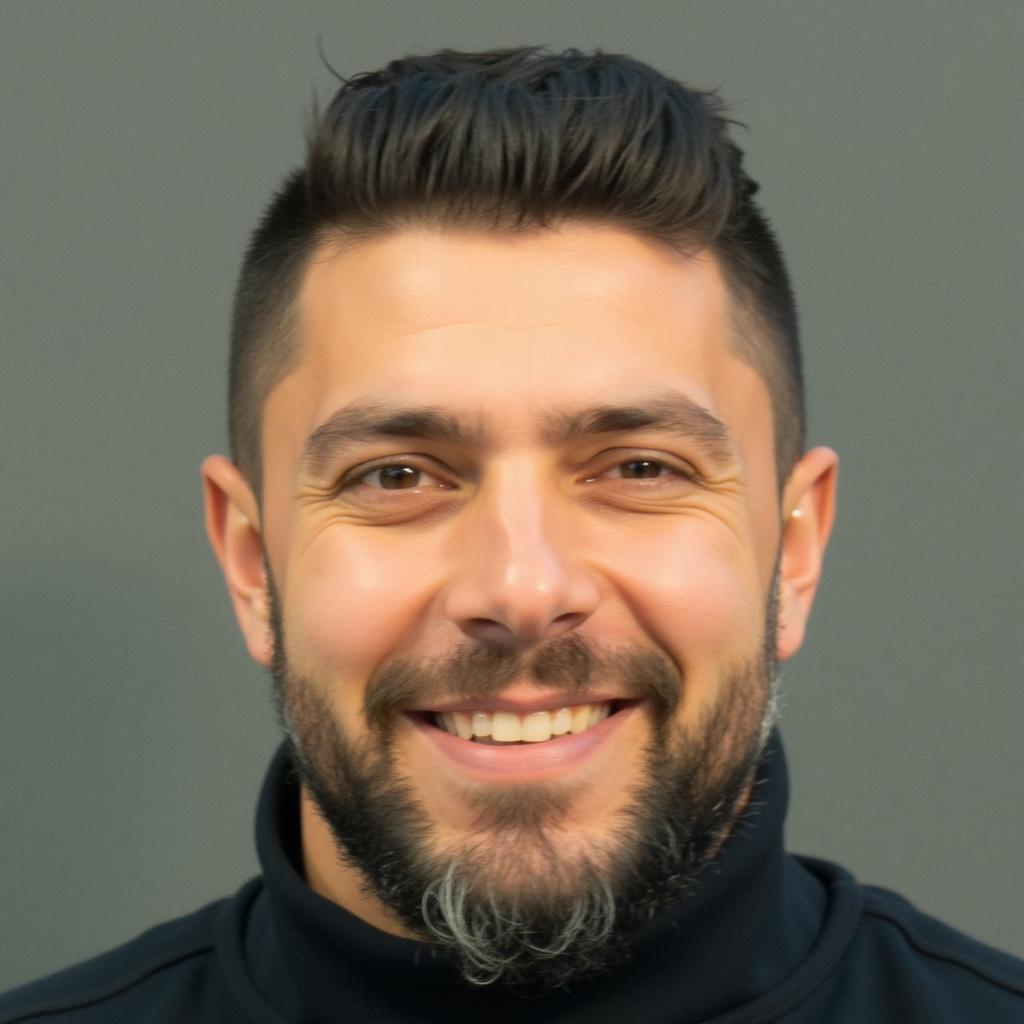} &
        \includegraphics[width=0.228\linewidth]{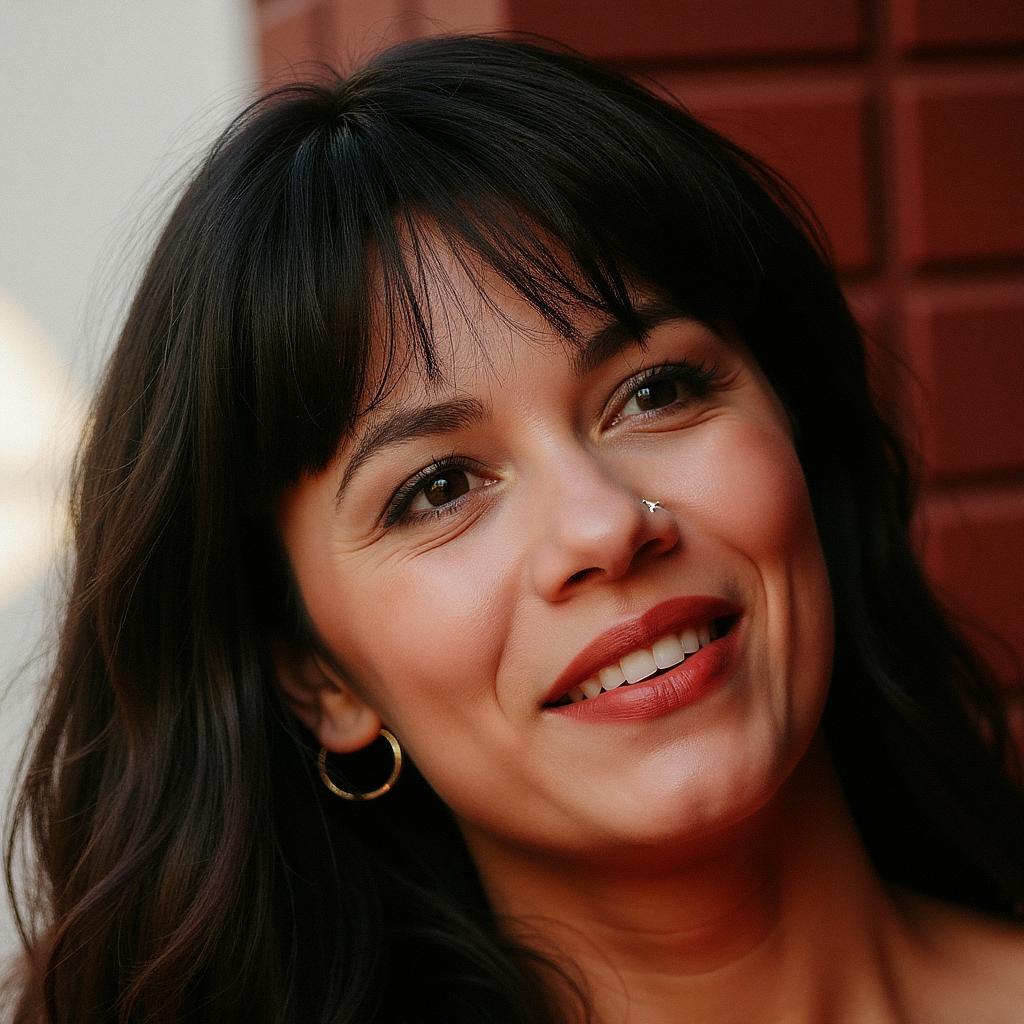} &
        \includegraphics[width=0.228\linewidth]{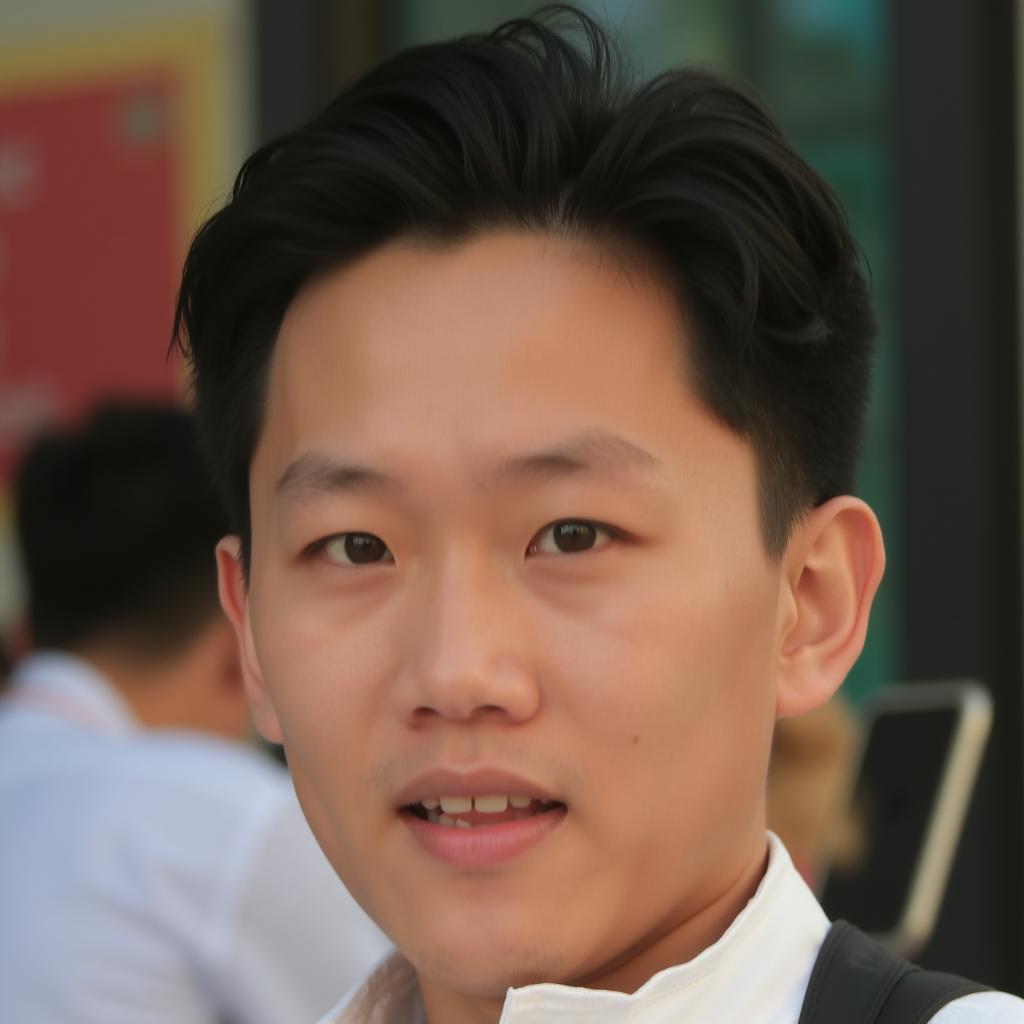} &
        \includegraphics[width=0.228\linewidth]{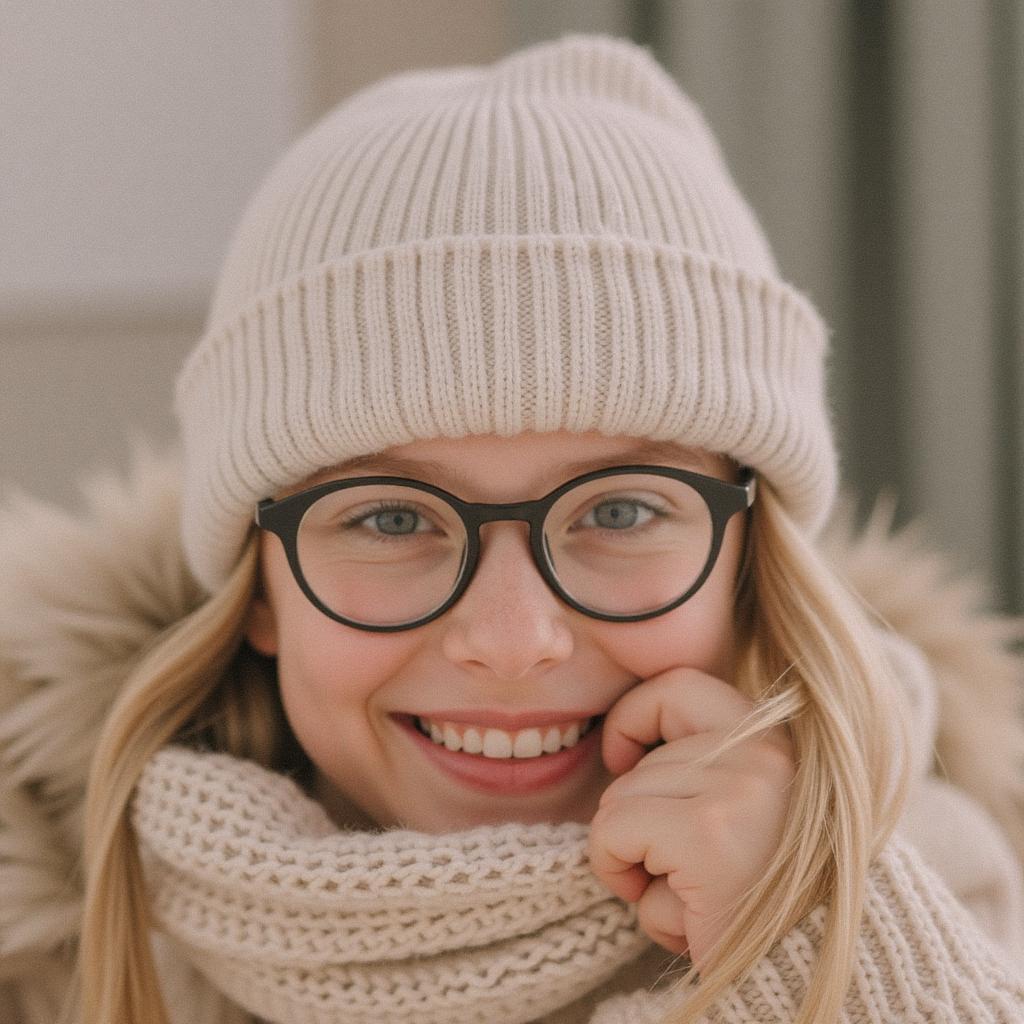} \\
        
        \raisebox{22pt}{\rotatebox[origin=t]{90}{RF Inv.}} &
        \raisebox{22pt}{\rotatebox[origin=t]{90}{+ Tight Inversion}} &
        \includegraphics[width=0.228\linewidth]{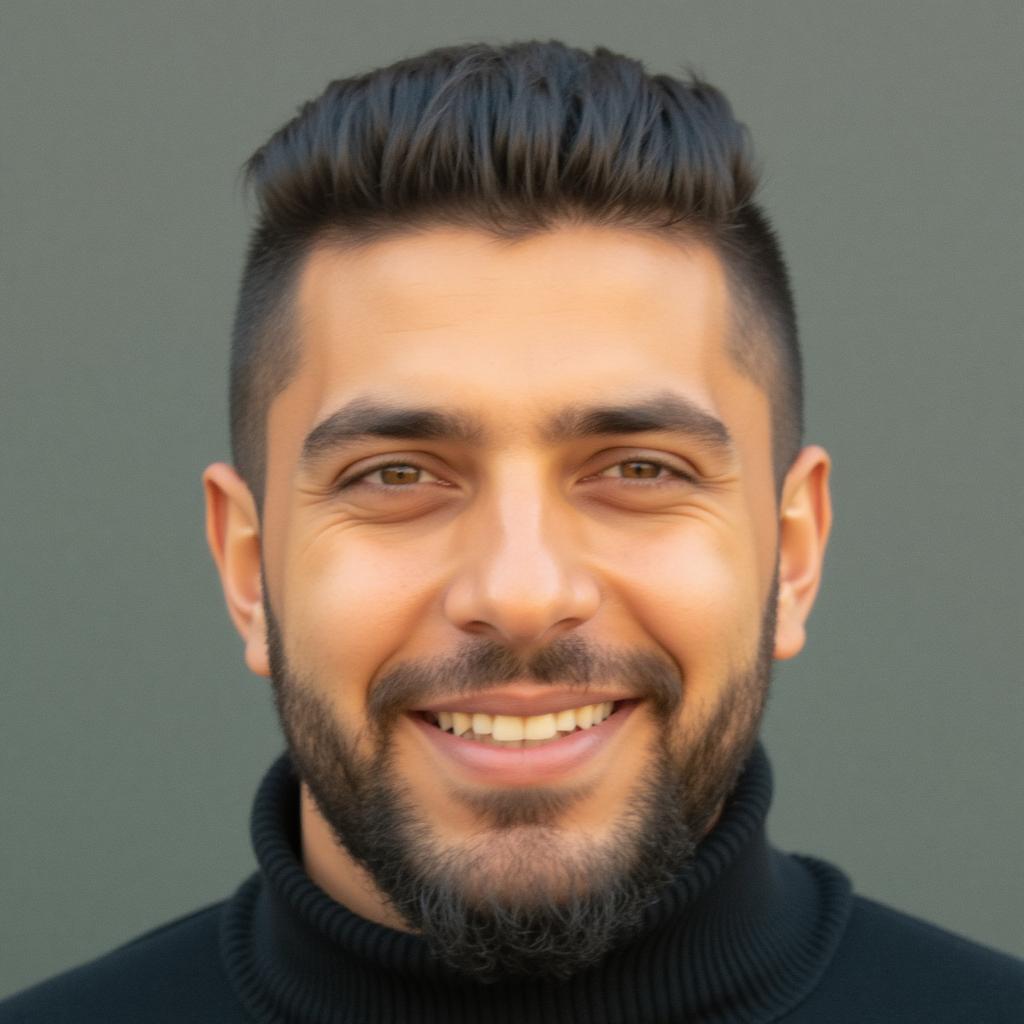} &
        \includegraphics[width=0.228\linewidth]{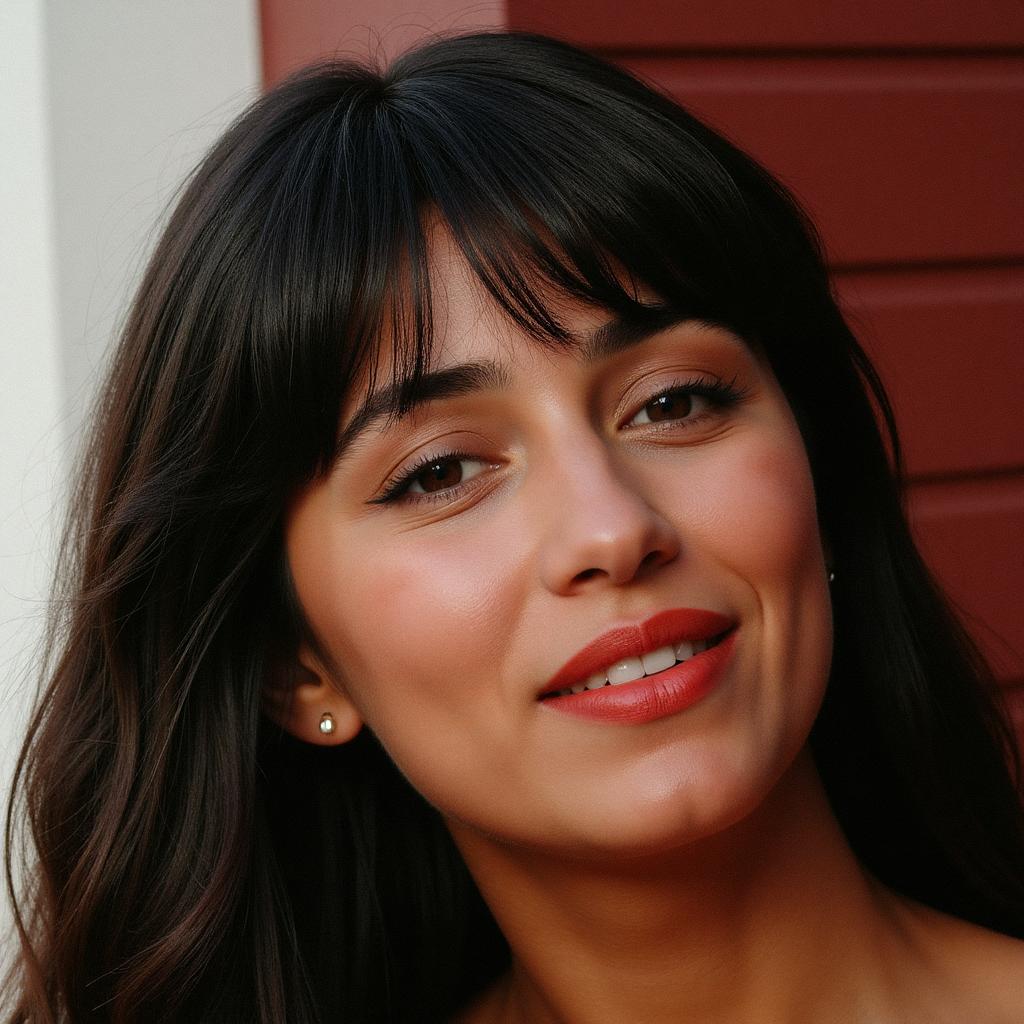} &
        \includegraphics[width=0.228\linewidth]{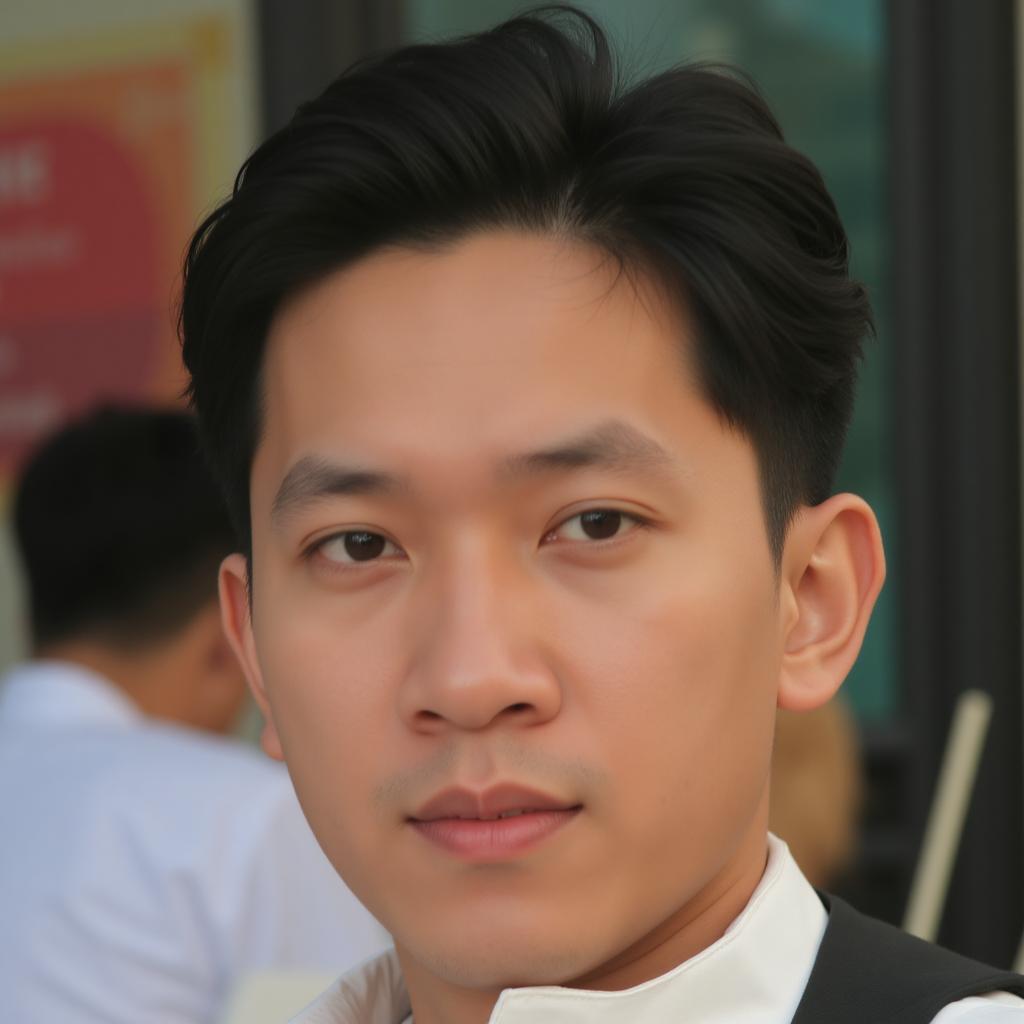} &
        \includegraphics[width=0.228\linewidth]{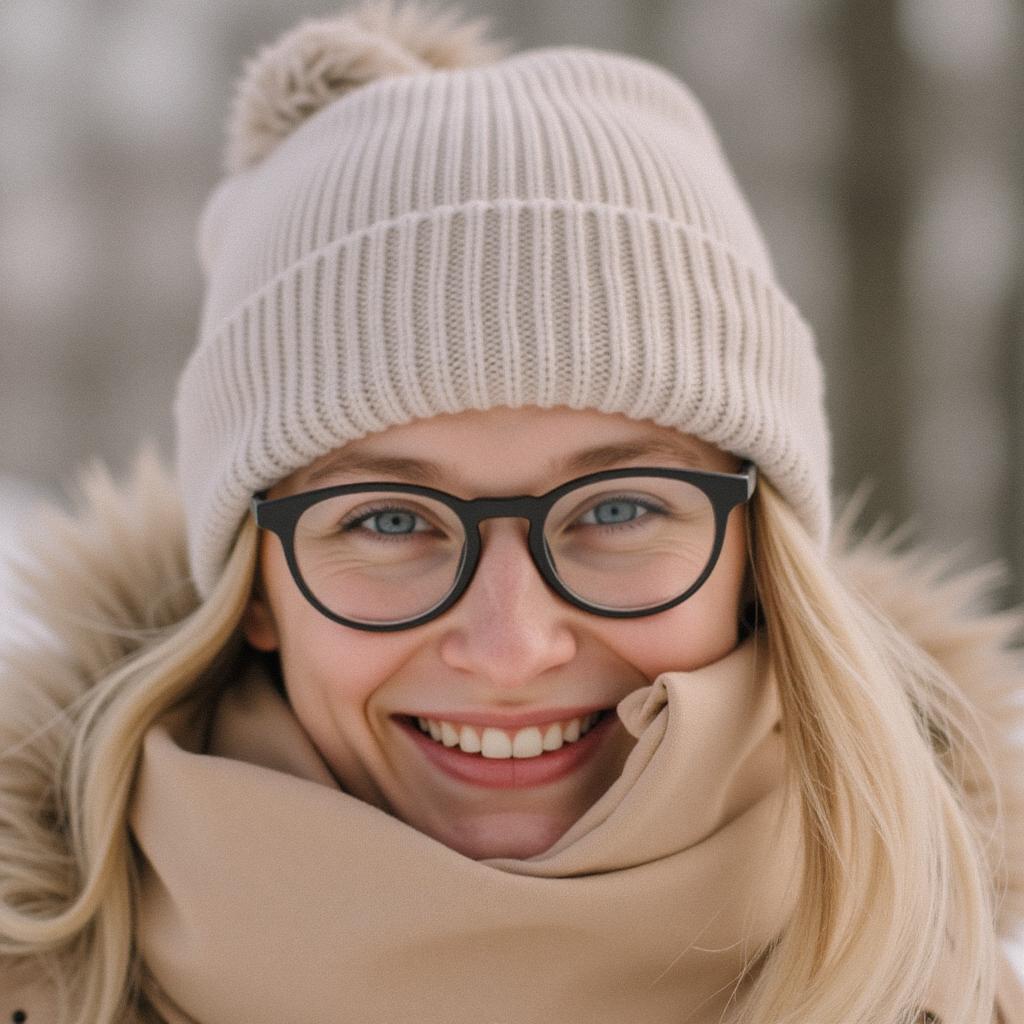} \\
        
        && \multicolumn{4}{c}{\textbf{Flux}}       
    \end{tabular}
    }
    \caption{Qualitative reconstruction results with Flux. Integrating Tight Inversion with RF-Inversion enhances the identity preservation of the reconstruction.}
    \label{fig:recon-qualitative-comp-flux}
\end{figure}

%% file: figures/ablation1.tex
\begin{figure}
    \setlength{\tabcolsep}{1pt}
  \centering
  \begin{tabular}{cccc}
   \raisebox{30pt}{\rotatebox[origin=t]{90}{Random noise}} &
    \includegraphics[width=0.32\linewidth]{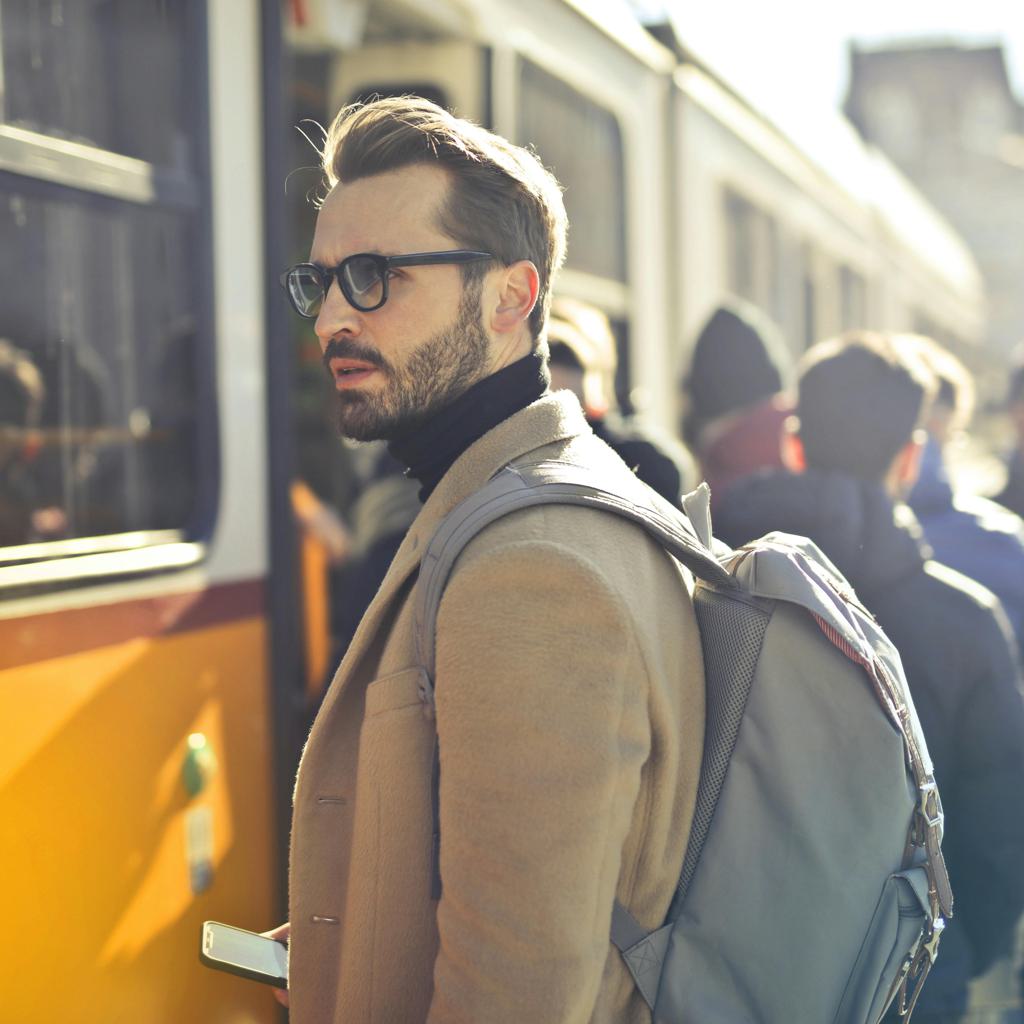} &
    \includegraphics[width=0.32\linewidth]{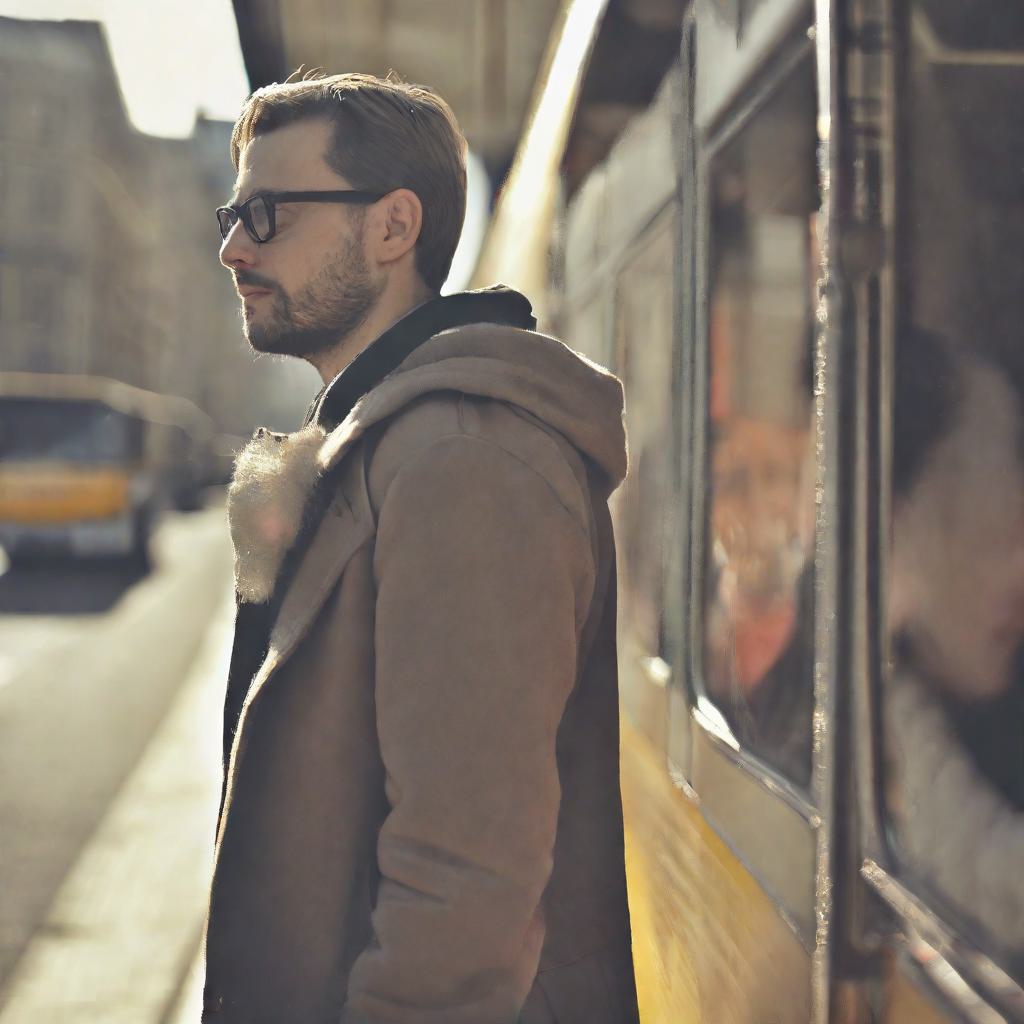} &
    \includegraphics[width=0.32\linewidth]{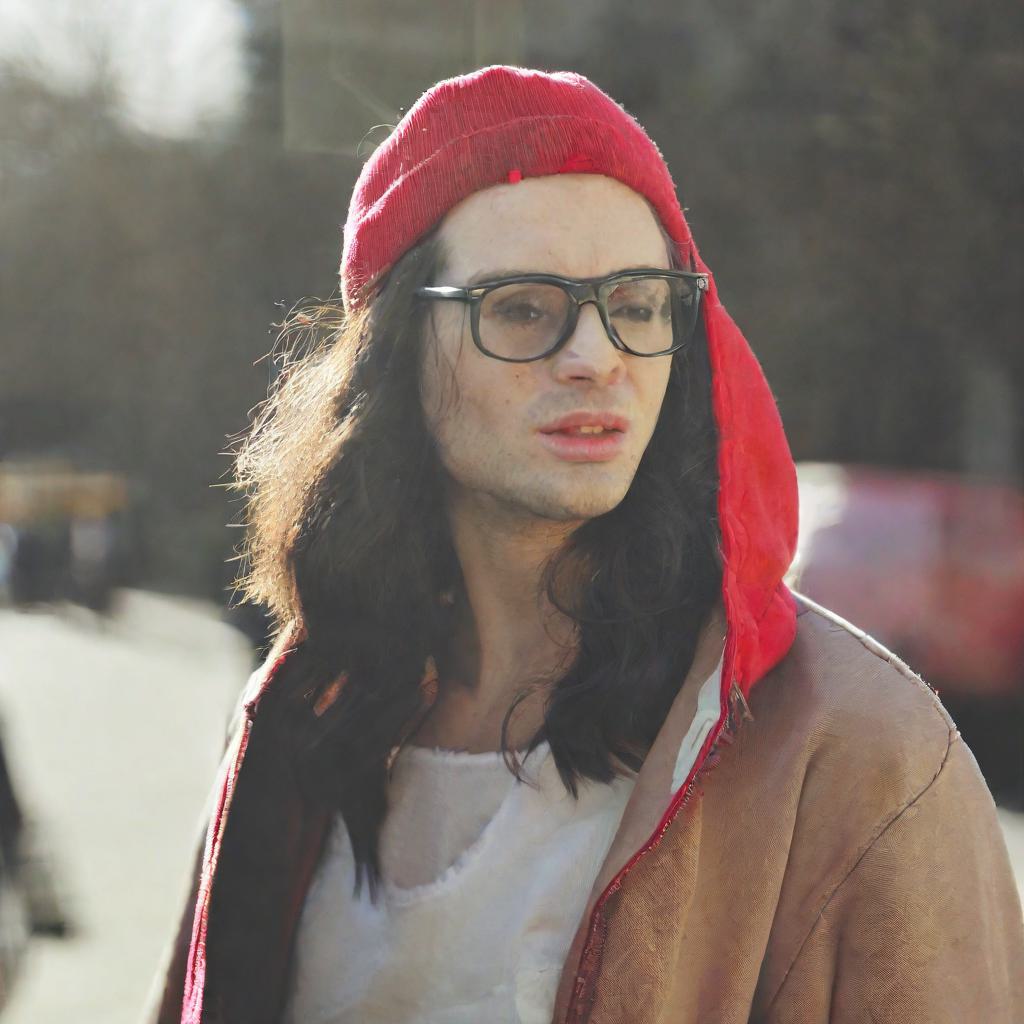} \\
    \raisebox{30pt}{\rotatebox[origin=t]{90}{DDIM Inversion}} &
    \includegraphics[width=0.32\linewidth]{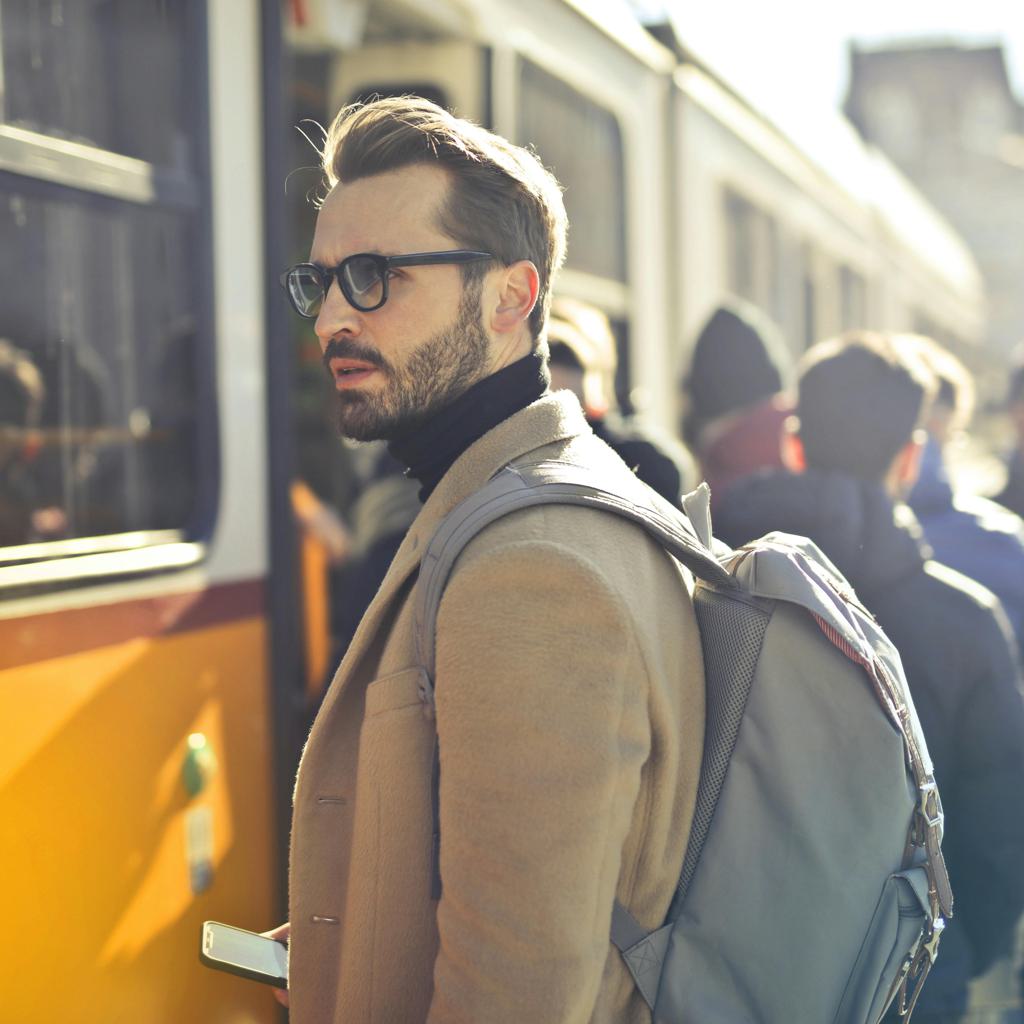} &
    \includegraphics[width=0.32\linewidth]{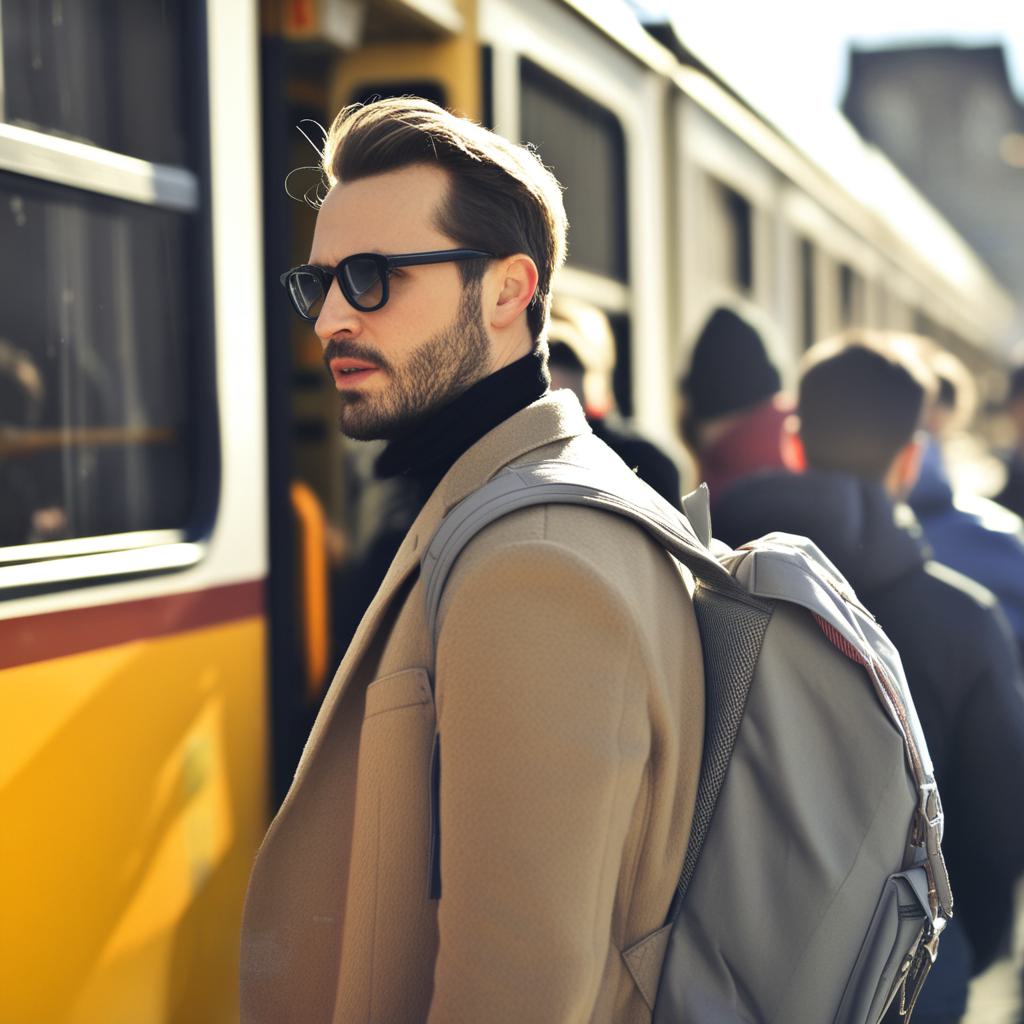} &
    \includegraphics[width=0.32\linewidth]{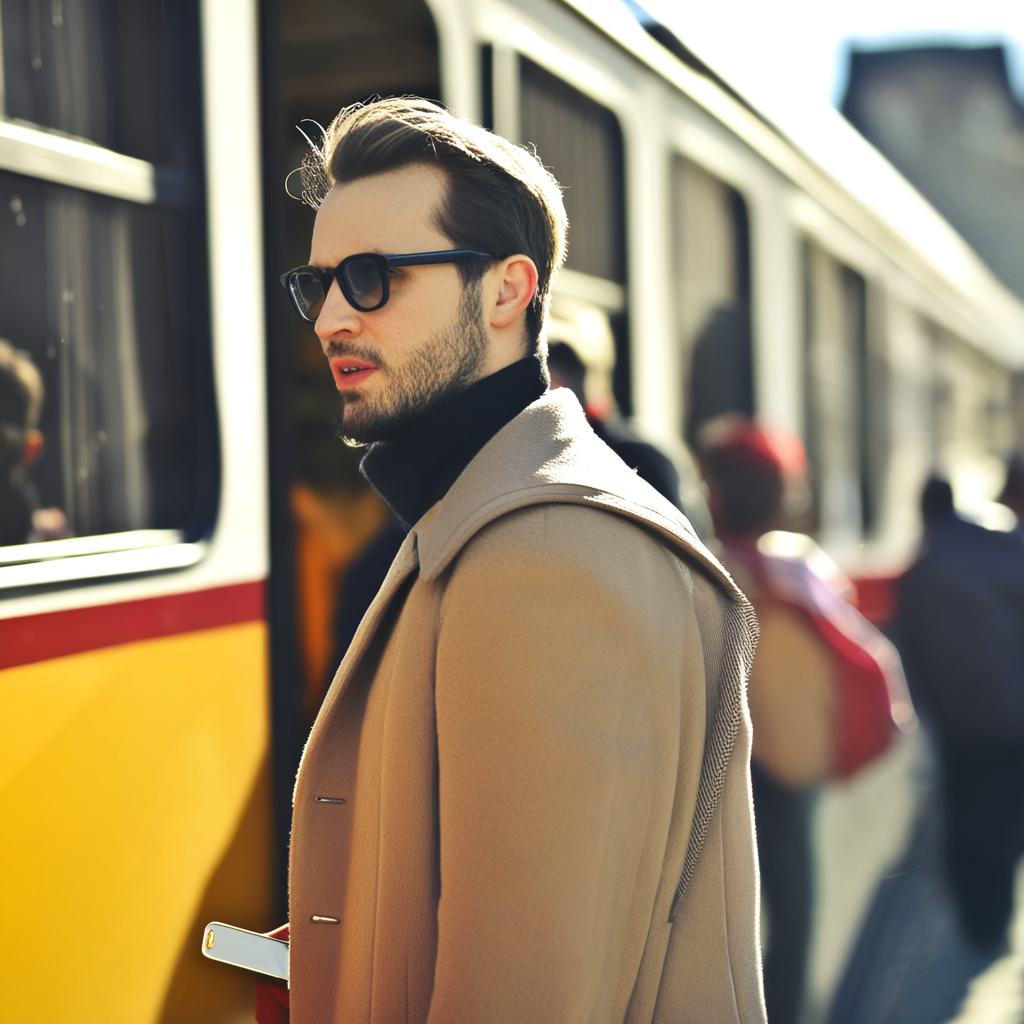} \\
    \raisebox{30pt}{\rotatebox[origin=t]{90}{Tight Inversion}} &
    \includegraphics[width=0.32\linewidth]{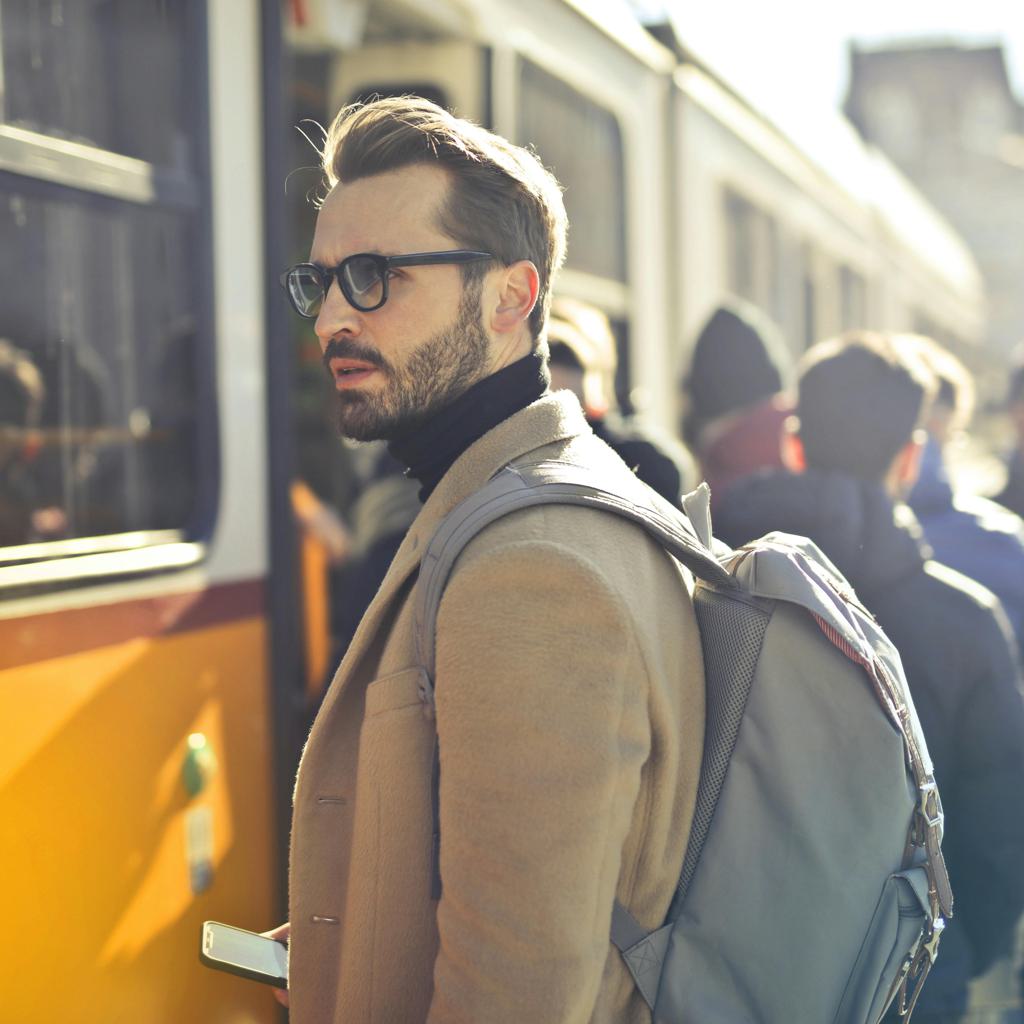} &
    \includegraphics[width=0.32\linewidth]{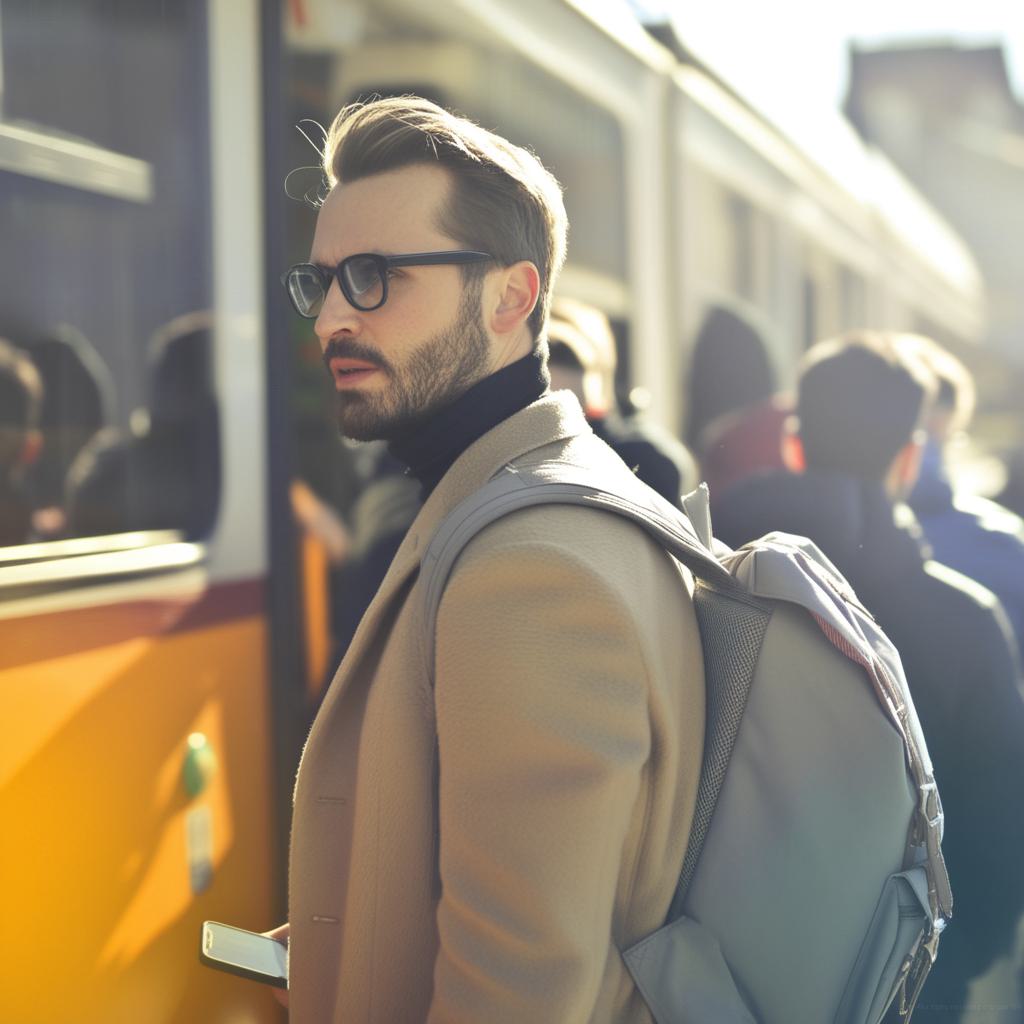} &
    \includegraphics[width=0.32\linewidth]{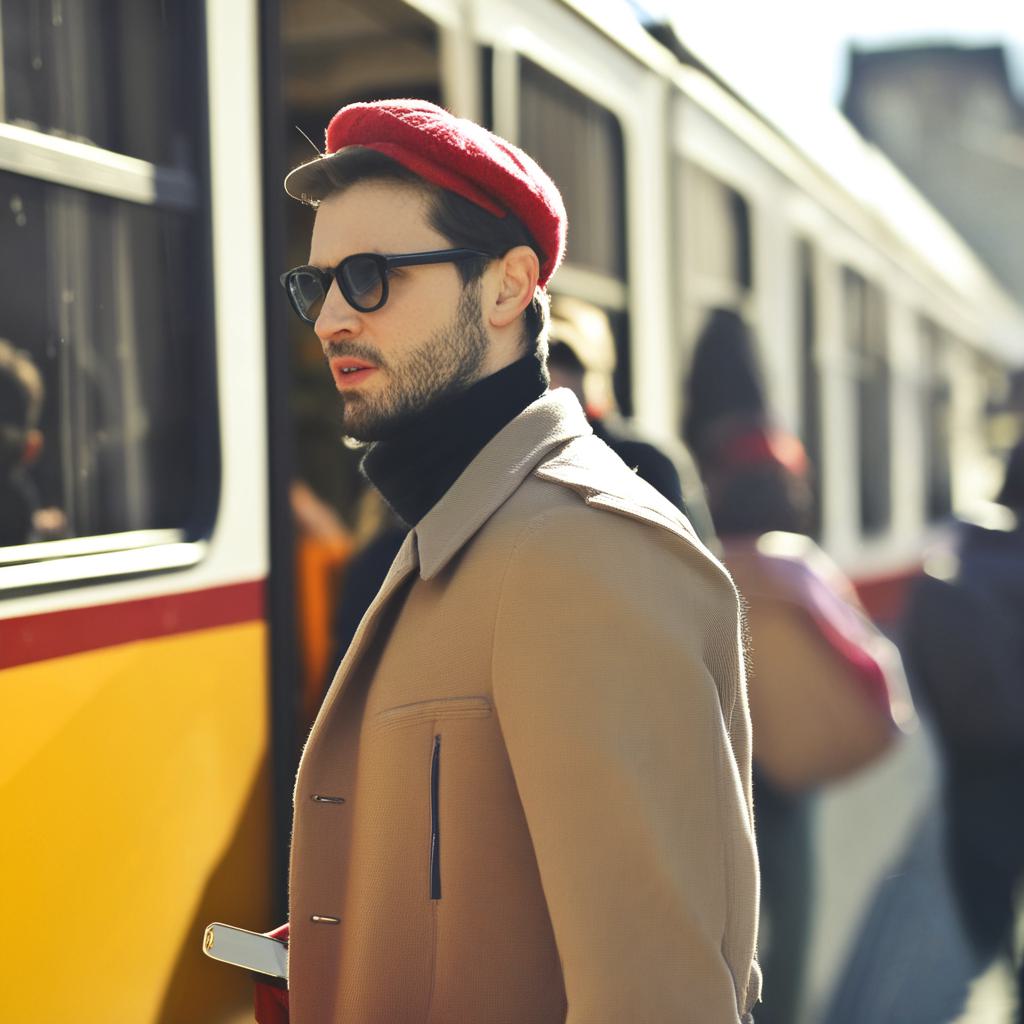} \\
    & Input & Reconstruction & $+$ ``a red hat'' \\
  \end{tabular}
  \caption{In all three rows, the denoising process is conditioned on the input image. In the first row, a random noise is sampled instead of inverting the image. In the second row, we apply vanilla DDIM inversion conditioned on a text prompt only. In the third row, we apply Tight Inversion, conditioning both the inversion and the denoising process on the input image. }
  \label{fig:ablation1}
\end{figure}

%% file: figures/ablation2.tex
\begin{figure}
    \setlength{\tabcolsep}{1pt}
    \centering
    {\scriptsize
    \begin{tabular}{cccccc}
        & 0.1 & 0.3 & 0.4 & 0.5 & 0.7 \\
        \raisebox{19pt}{\rotatebox[origin=t]{90}{Input}} &
        \includegraphics[width=0.19\linewidth]{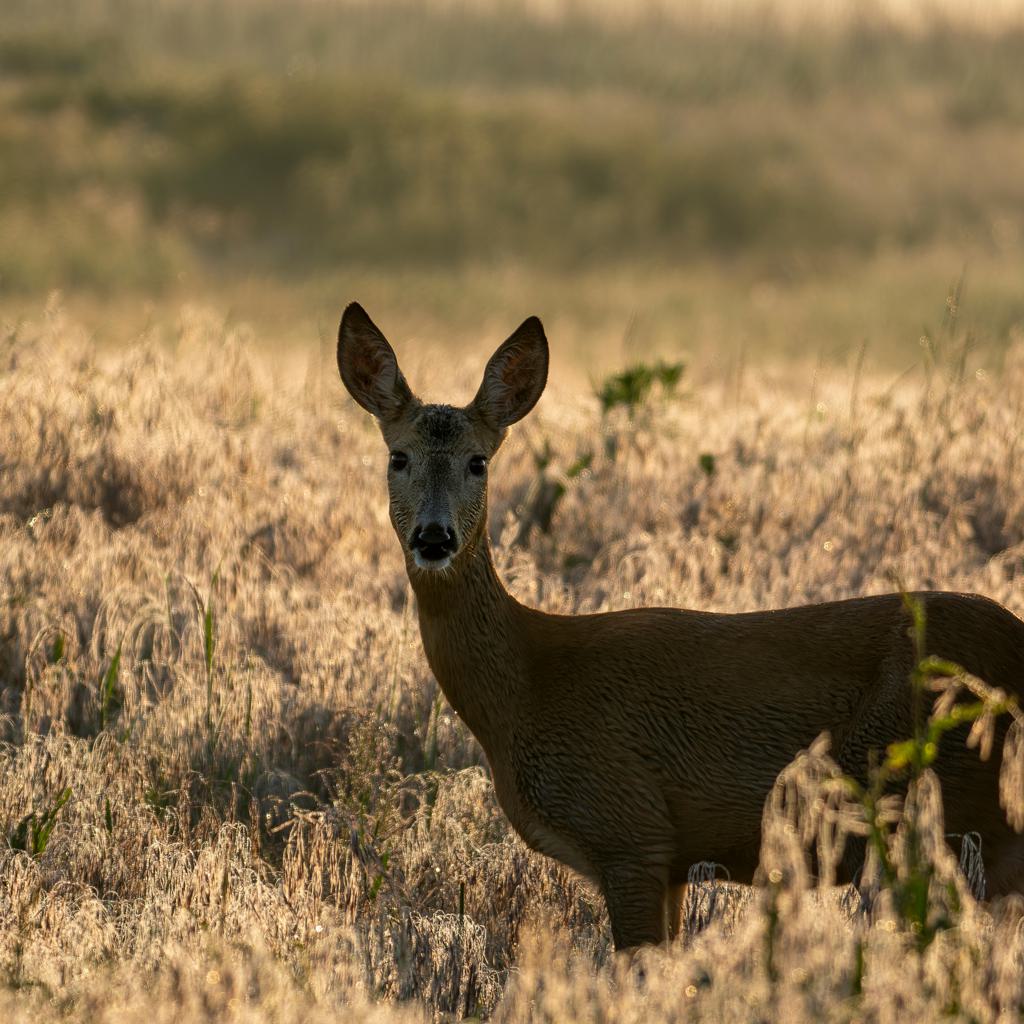} &
        \includegraphics[width=0.19\linewidth]{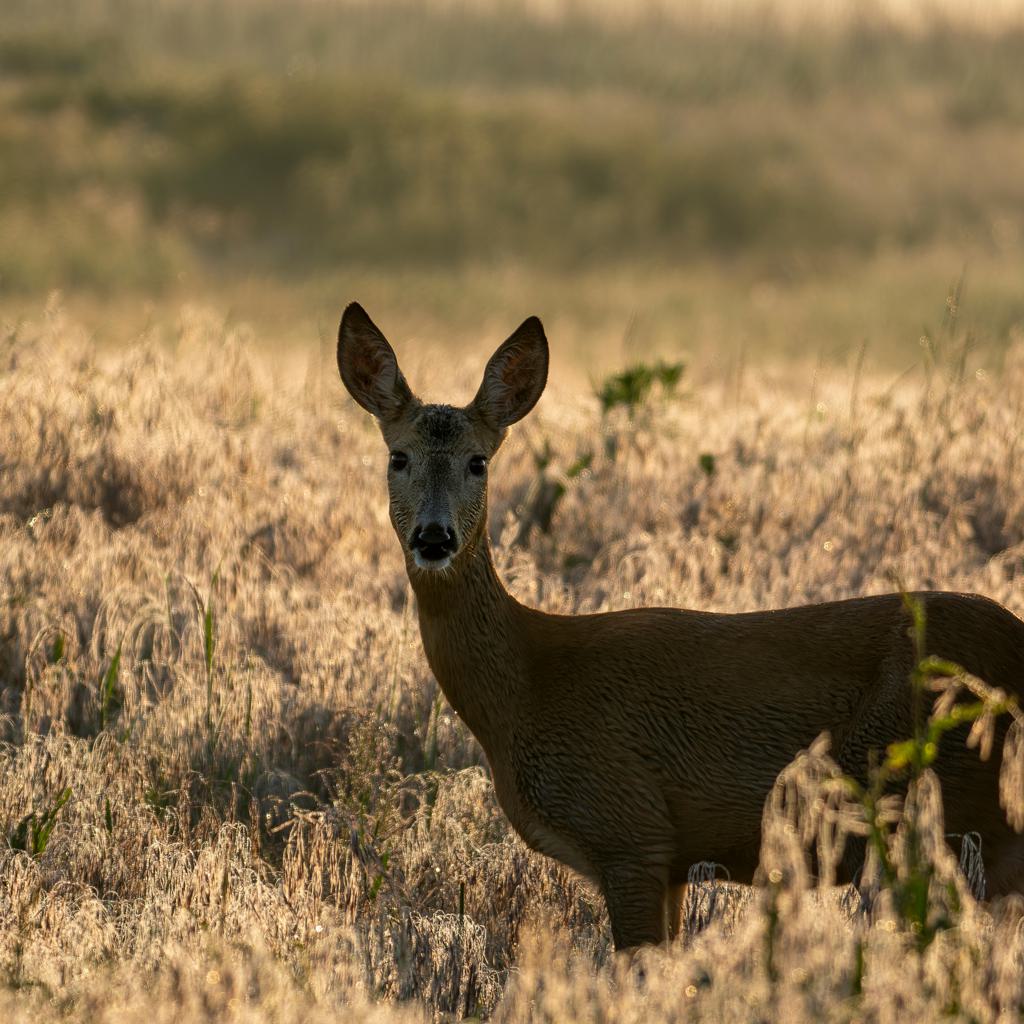} &
        \includegraphics[width=0.19\linewidth]{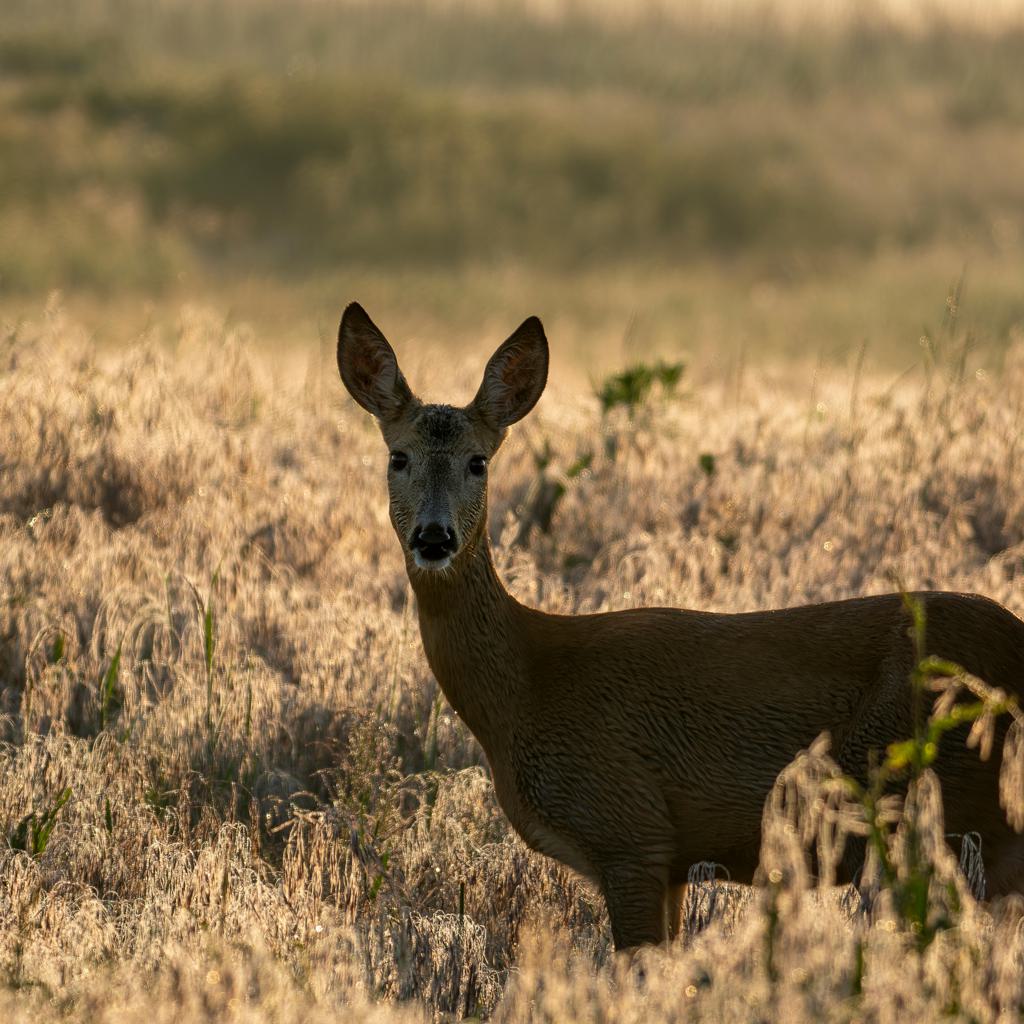} &
        \includegraphics[width=0.19\linewidth]{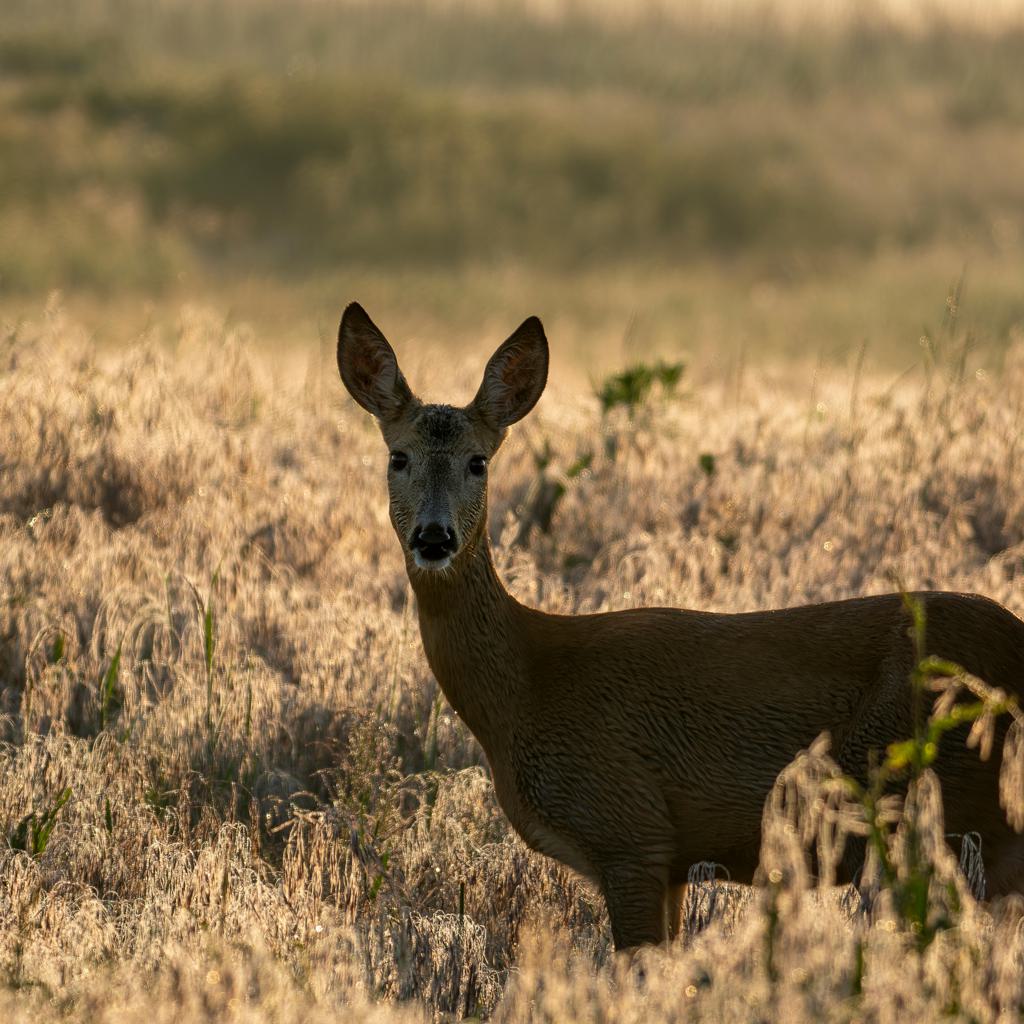} &
        \includegraphics[width=0.19\linewidth]{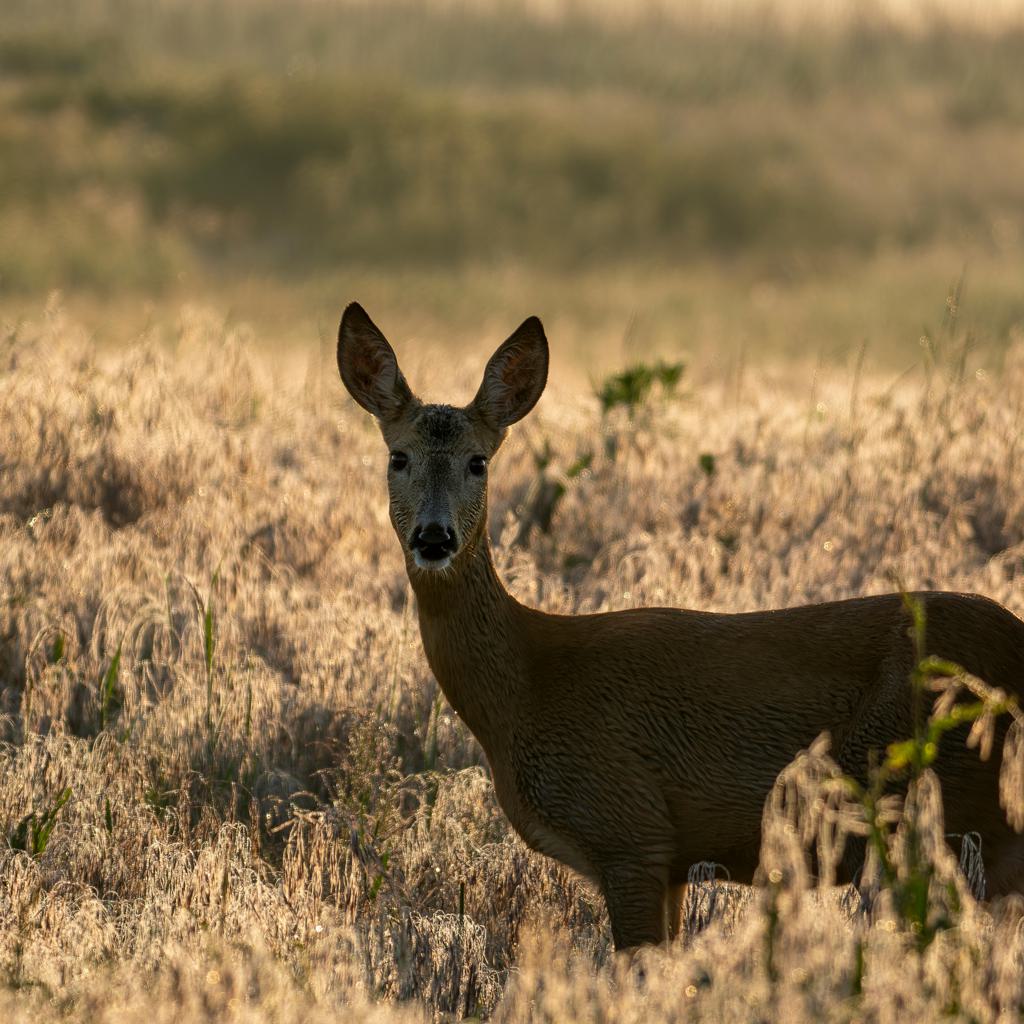} \\
        \raisebox{19pt}{\rotatebox[origin=t]{90}{Recon.}} &
        \includegraphics[width=0.19\linewidth]{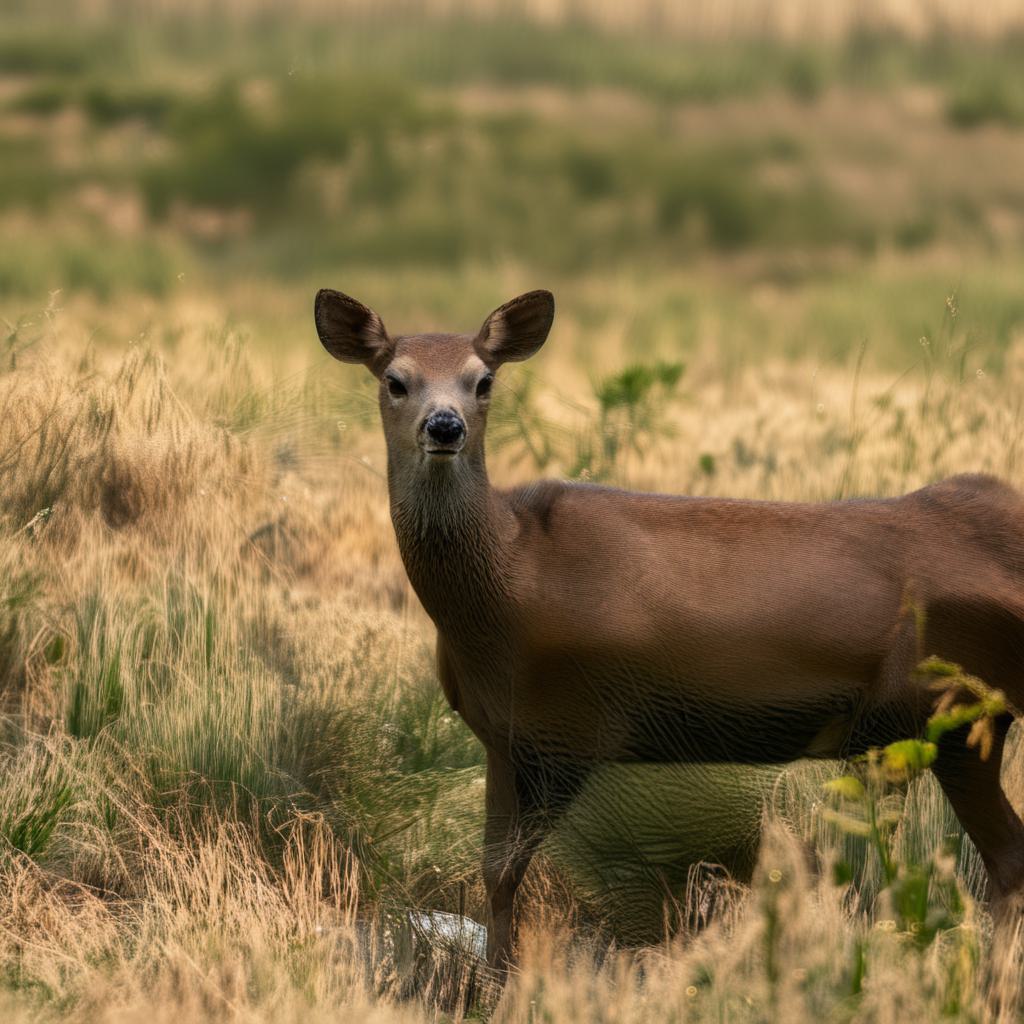} &
        \includegraphics[width=0.19\linewidth]{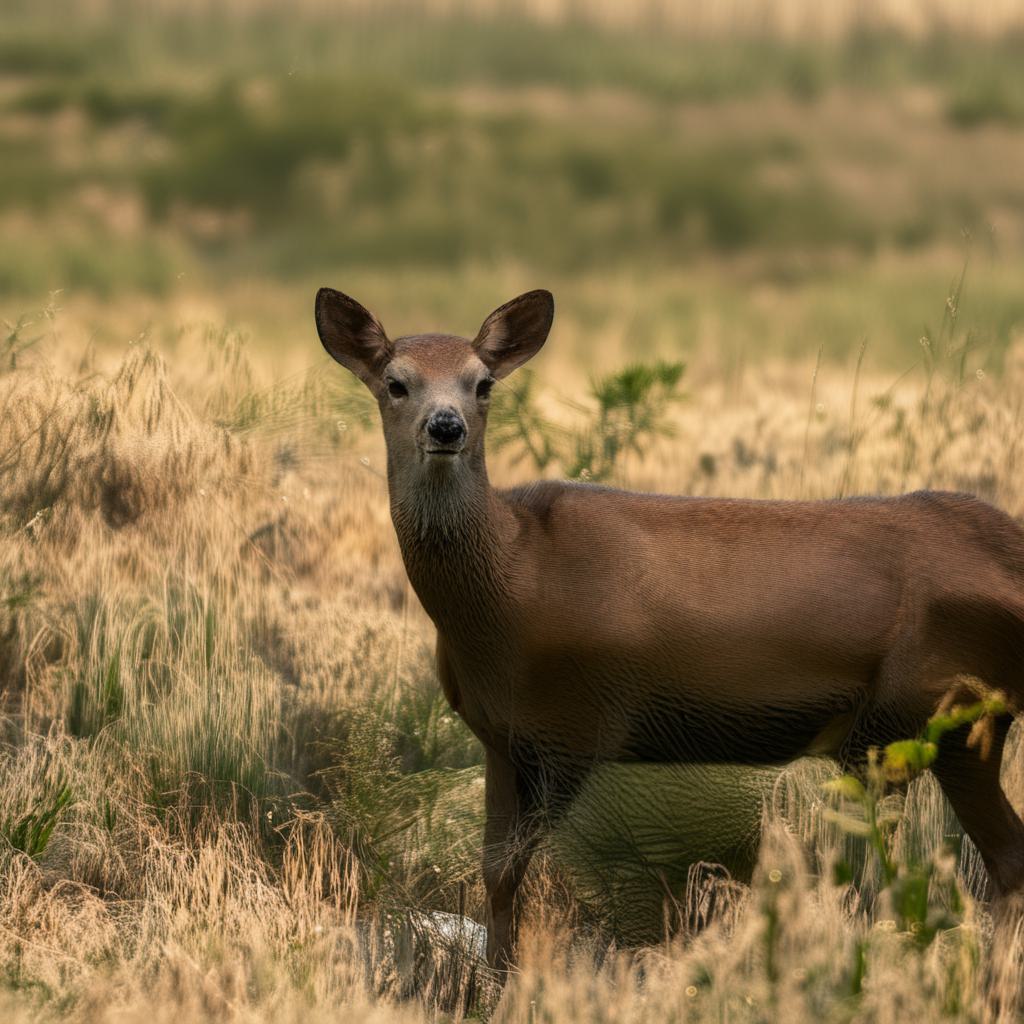} &
        \includegraphics[width=0.19\linewidth]{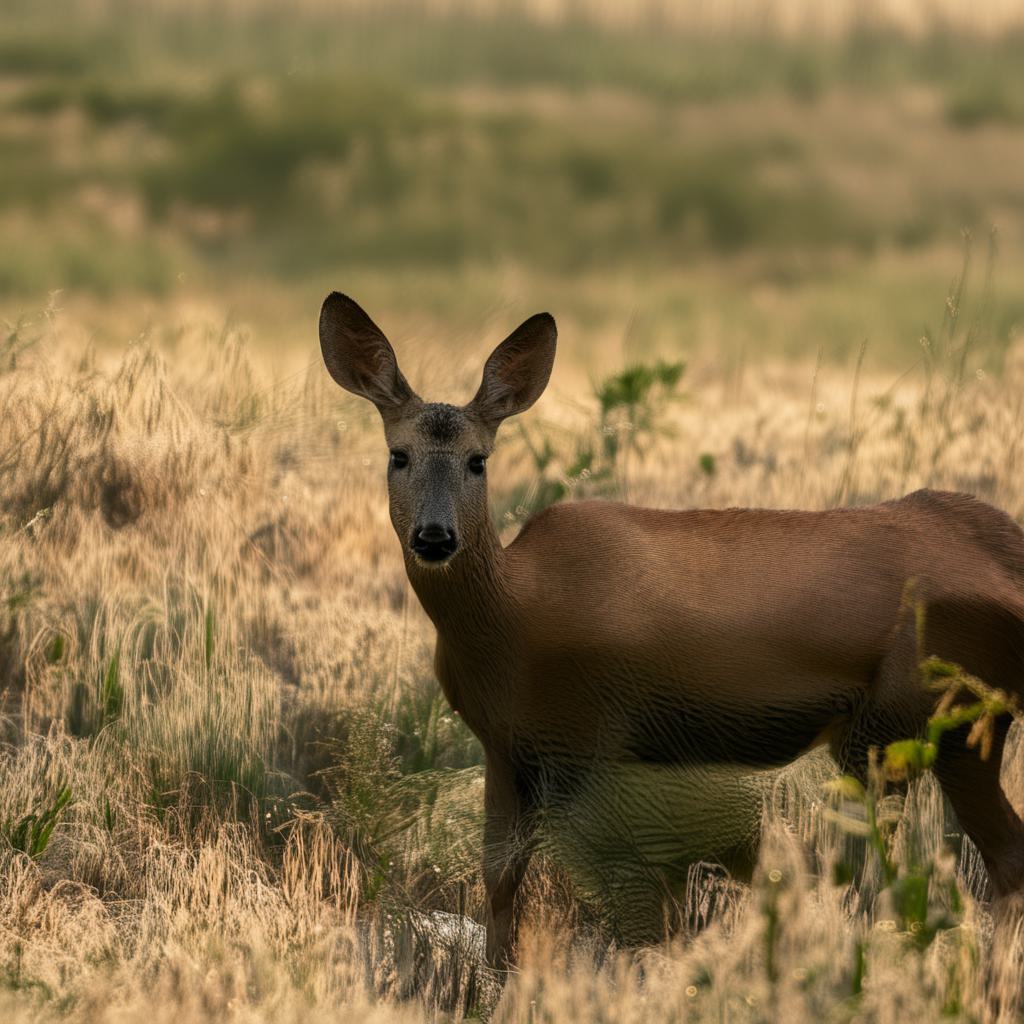} &
        \includegraphics[width=0.19\linewidth]{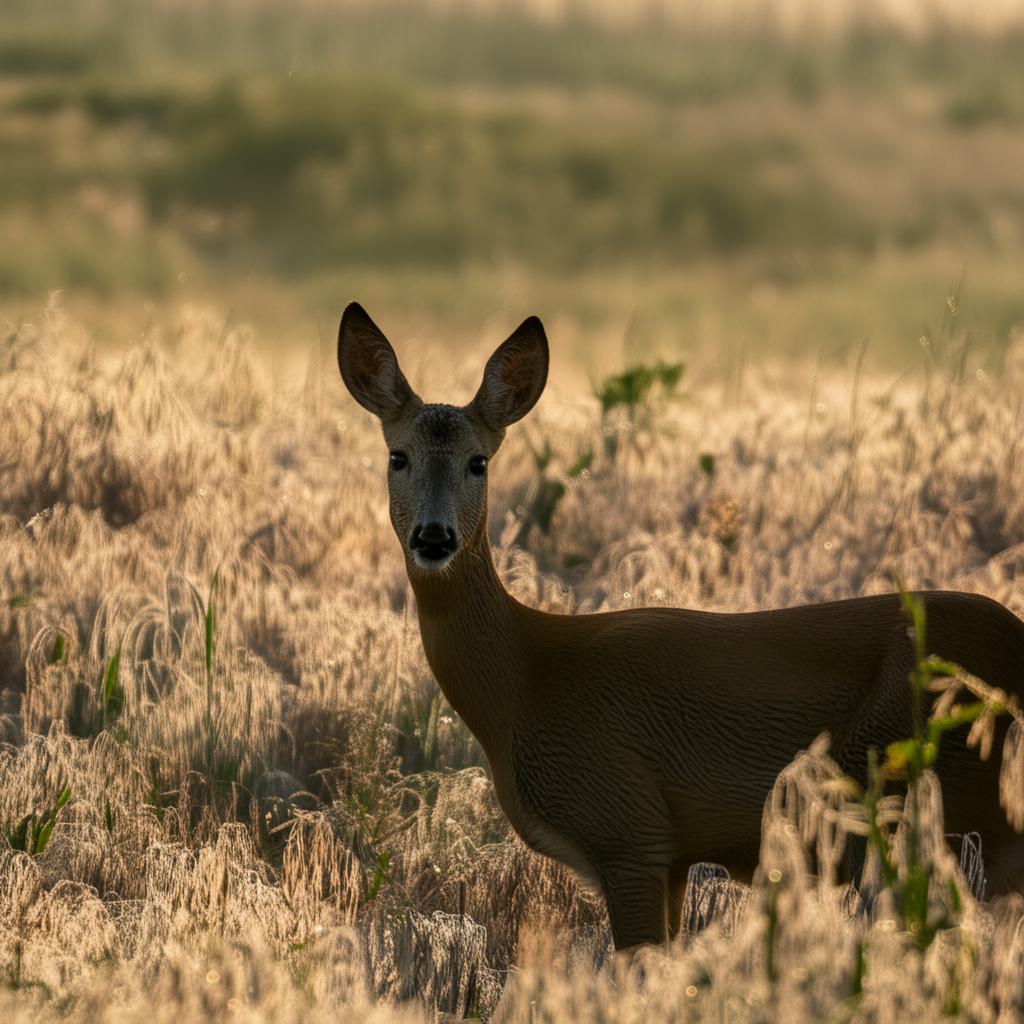} &
        \includegraphics[width=0.19\linewidth]{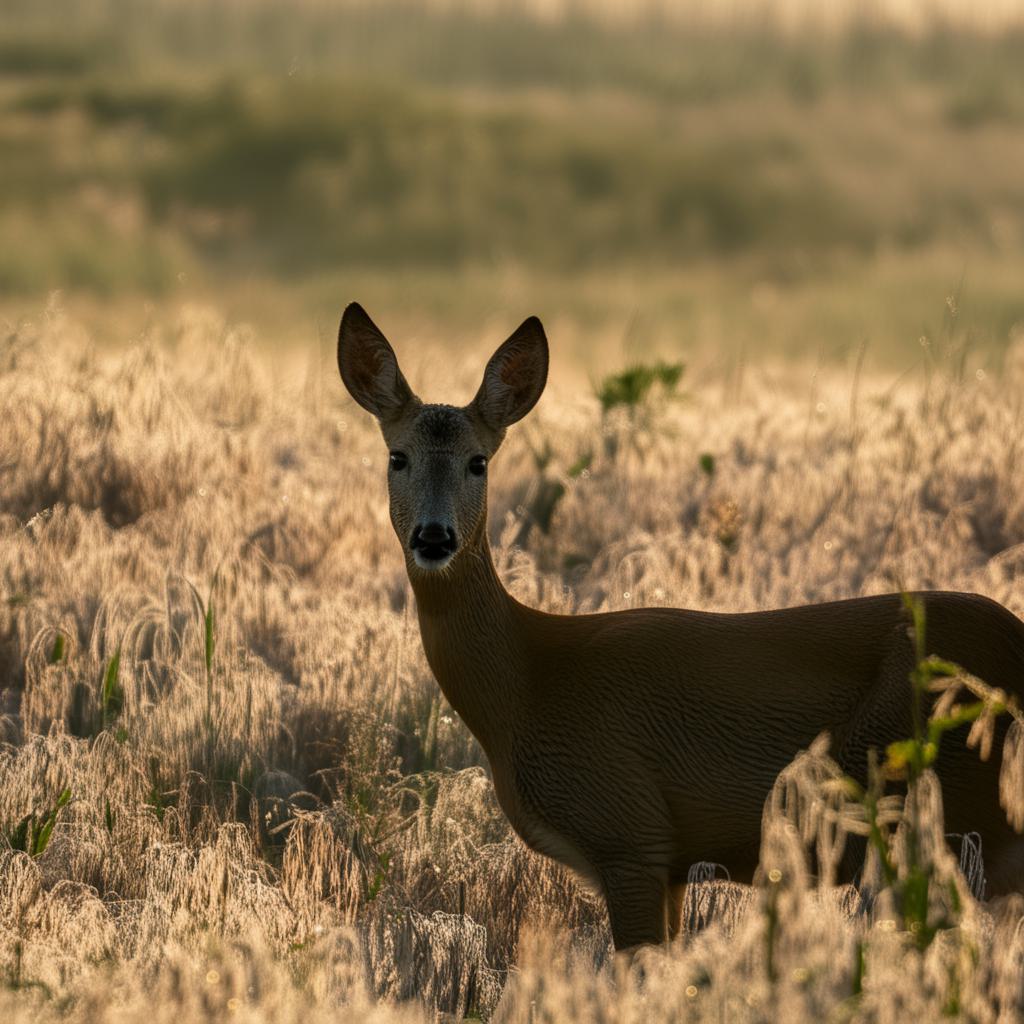} \\
        \raisebox{19pt}{\rotatebox[origin=t]{90}{``cowboy hat''}} &
        \includegraphics[width=0.19\linewidth]{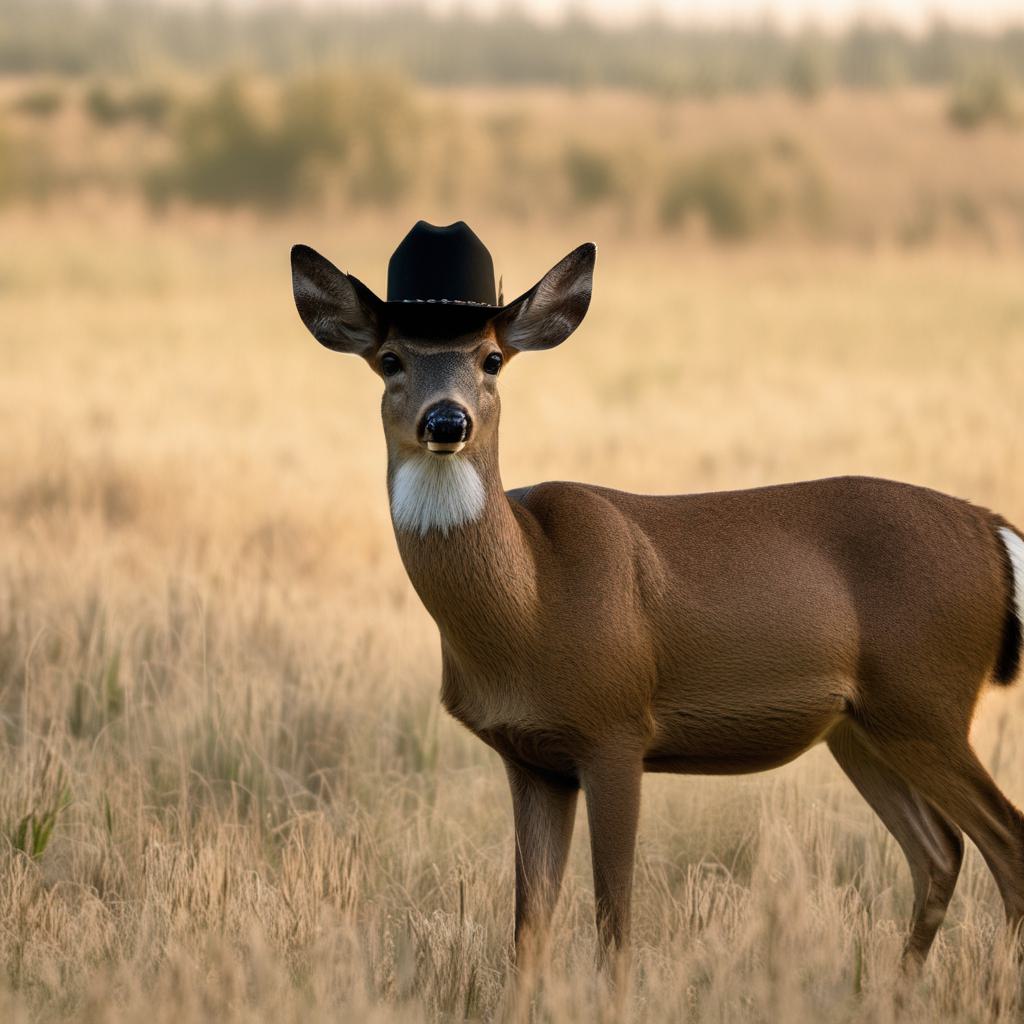} &
        \includegraphics[width=0.19\linewidth]{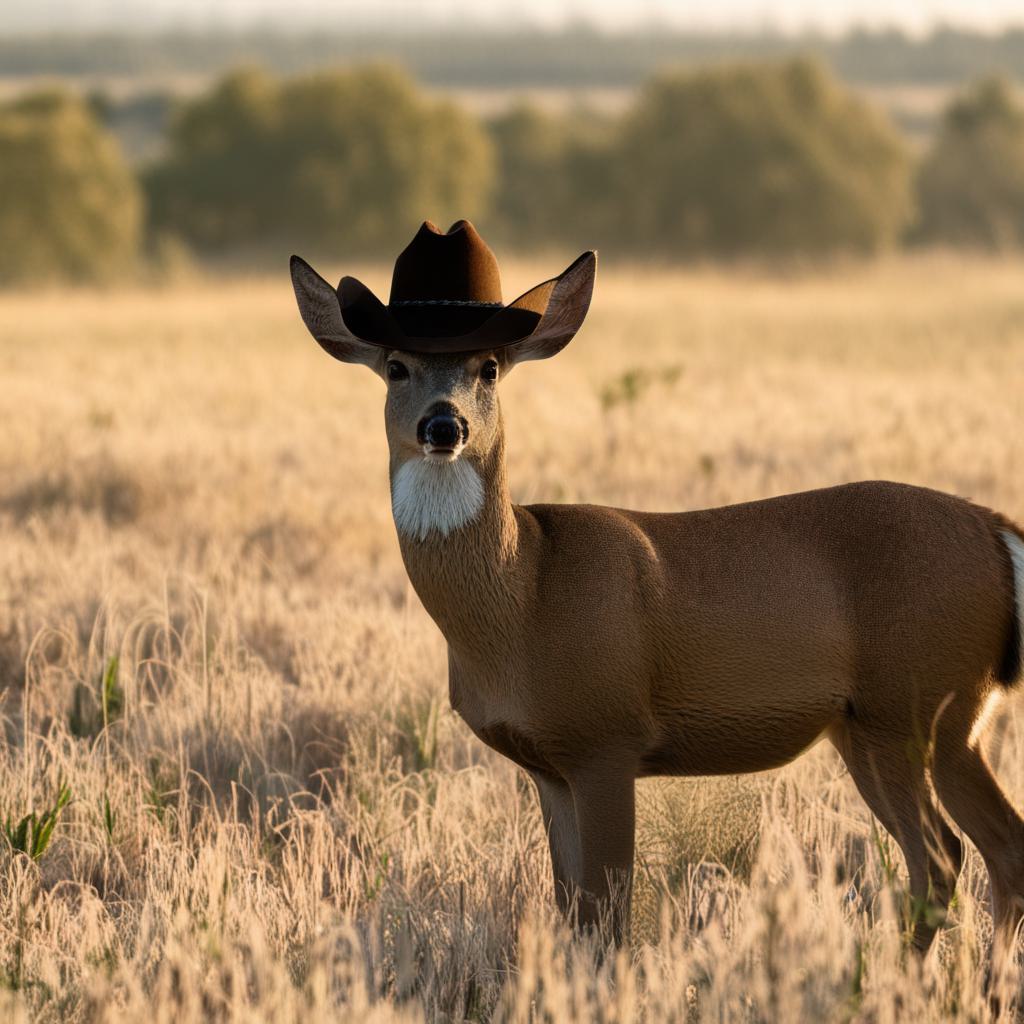} &
        \includegraphics[width=0.19\linewidth]{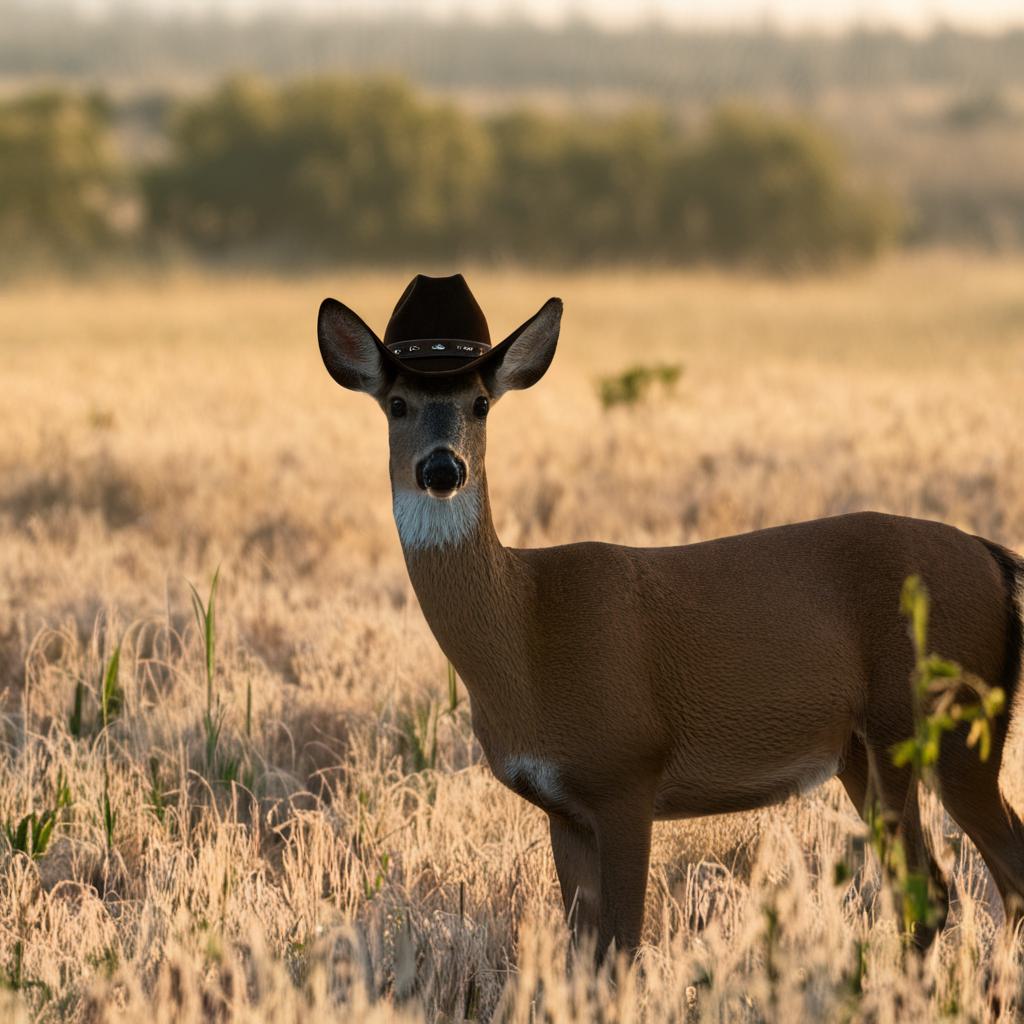} &
        \includegraphics[width=0.19\linewidth]{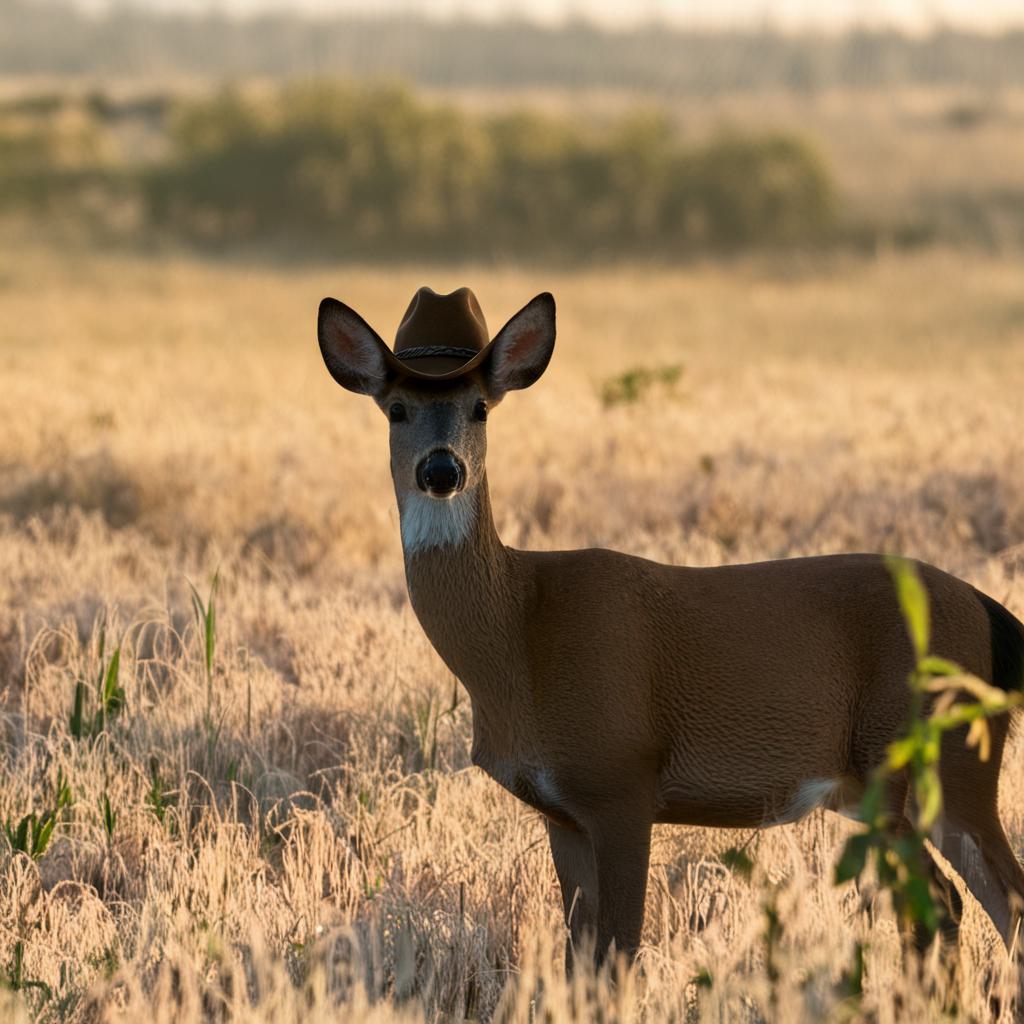} &
        \includegraphics[width=0.19\linewidth]{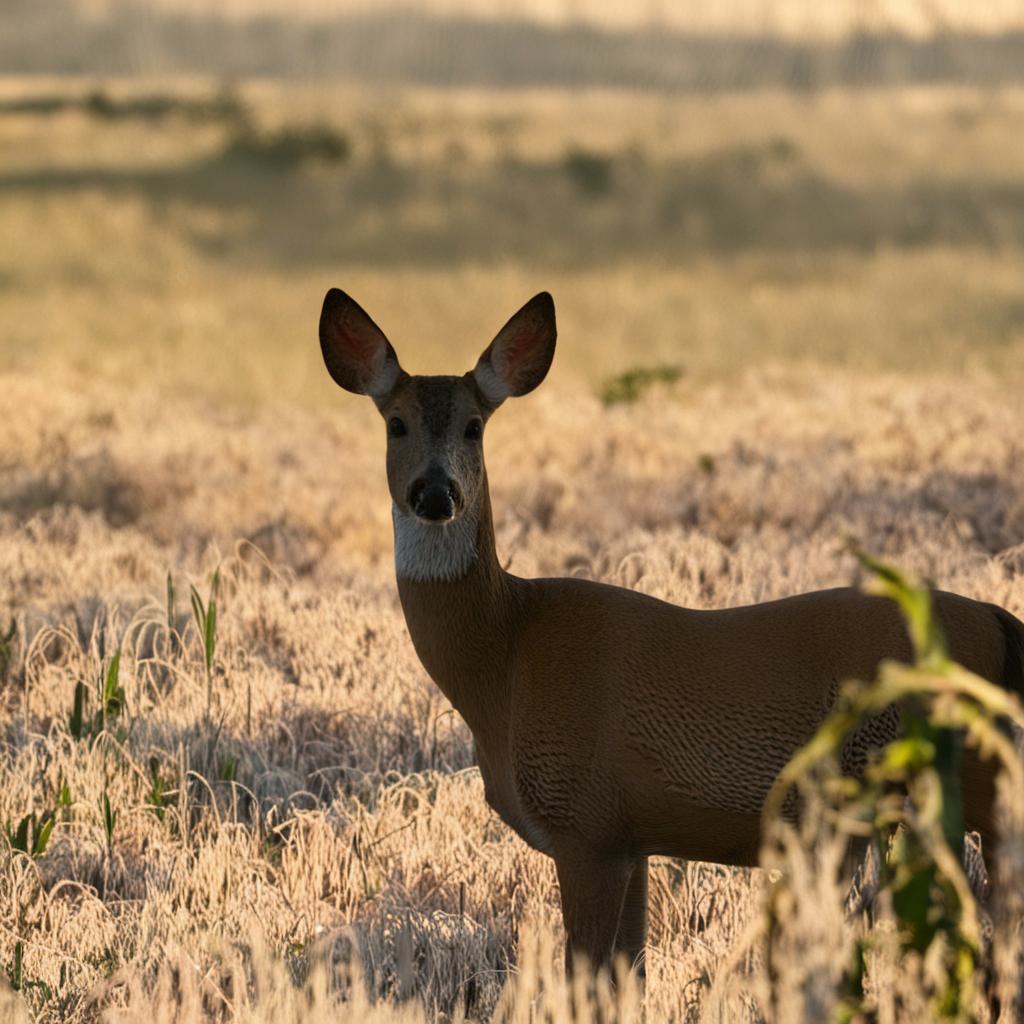} \\
    \end{tabular}
    }
    \caption{Ablating the guidance scales used in IP-Adapter. Increasing the scale results in a better reconstruction and better preservation of the original image in the edited one. However, using an overly strong scale limits the capability to edit the image.}
    \label{fig:ablation2}
\end{figure}

%% file: figures/scale_clip_similarity.tex
\begin{tikzpicture}
    \small{
    \begin{axis}[
        hide axis,
        scale only axis,
        height=0pt,
        colorbar horizontal,
        colormap/viridis, %
        colorbar style={
            width=0.8\linewidth, %
            height=0.3cm,
            yshift=-1cm,
            xlabel={IP-Adapter Scale}, %
            xticklabel style={/pgf/number format/.cd, fixed, precision=2},
        },
        point meta min=0.0, %
        point meta max=0.7, %
    ]
    \end{axis}

    \begin{axis}[
        name=plot1,
        width=4.2cm, height=4.0cm,
        xlabel={\footnotesize{Text-Img CLIP Sim.}},
        ylabel={\footnotesize{Img-Img CLIP Sim.}},
        title={DDIM Inversion},
        xmin=0.29, xmax=0.35,
        ymin=0.7, ymax=0.92,
        grid=both,
        legend pos=north west,
        colormap/viridis,
        colorbar=false, %
        point meta min=0.0, %
        point meta max=0.7, %
    ]
        \addplot[
            scatter,
            only marks,
            mark=*,
            mark size=2pt,
            scatter src=explicit, %
        ] table [meta=colormap] { %
            x       y       colormap
            0.3413  0.7676  0.1
            0.3352  0.8137  0.2
            0.3294  0.8442  0.3
            0.3198  0.8725  0.4
            0.3114  0.8851  0.5
            0.3021  0.8950  0.6
            0.2956  0.8944  0.7
        };

        \addplot[
            scatter,
            only marks,
            mark=x, %
            mark size=3pt, %
            scatter src=explicit,
        ] table [meta=colormap] {
            x       y       colormap
            0.3417  0.7143  0.0
        };
        
        \addplot[
            smooth,
            thick,
            blue,
        ] table {
            x       y
            0.3413  0.7676
            0.3352  0.8137
            0.3294  0.8442
            0.3198  0.8725
            0.3114  0.8851
            0.3021  0.8950
            0.2956  0.8944
        };
    \end{axis}

    \hspace{8pt}
    \begin{axis}[
        name=plot2,
        at=(plot1.right of south east), anchor=left of south west,
        width=4.2cm, height=4.0cm,
        xlabel={\footnotesize{Text-Img CLIP Sim.}},
        ylabel={\footnotesize{Img-Img CLIP Sim.}},
        title={LEDITS++},
        xmin=0.292, xmax=0.318,
        ymin=0.86, ymax=0.935,
        grid=both,
        legend pos=north west,
        colormap/viridis,
        colorbar=false, %
        point meta min=0.0, %
        point meta max=0.7, %
    ]
        \addplot[
            scatter,
            only marks,
            mark=*,
            mark size=2pt,
            scatter src=explicit, %
        ] table [meta=colormap] { %
            x       y       colormap
            0.3105  0.8780  0.1
            0.3056  0.8969  0.2
            0.3015  0.9105  0.3
            0.2979  0.9191  0.4
            0.2946  0.9248  0.5
        };

        \addplot[
            scatter,
            only marks,
            mark=x, %
            mark size=3pt, %
            scatter src=explicit,
        ] table [meta=colormap] {
            x       y       colormap
            0.3138  0.8644  0.0
        };
        
        \addplot[
            smooth,
            thick,
            blue,
        ] table {
            x       y
            0.3105  0.8780
            0.3056  0.8969
            0.3015  0.9105
            0.2979  0.9191
            0.2946  0.9248
        };
    \end{axis}
    }
\end{tikzpicture}

%% file: figures/editing_qualitative.tex
\begin{figure}
    \centering
    \setlength{\tabcolsep}{1pt}
    \scriptsize{
    \begin{tabular}{cccccc}
        \multicolumn{5}{c}{``people in a diner'' $\longrightarrow$ ``robots in a diner'', DDIM Inversion} \\
        \includegraphics[width=0.195\linewidth]{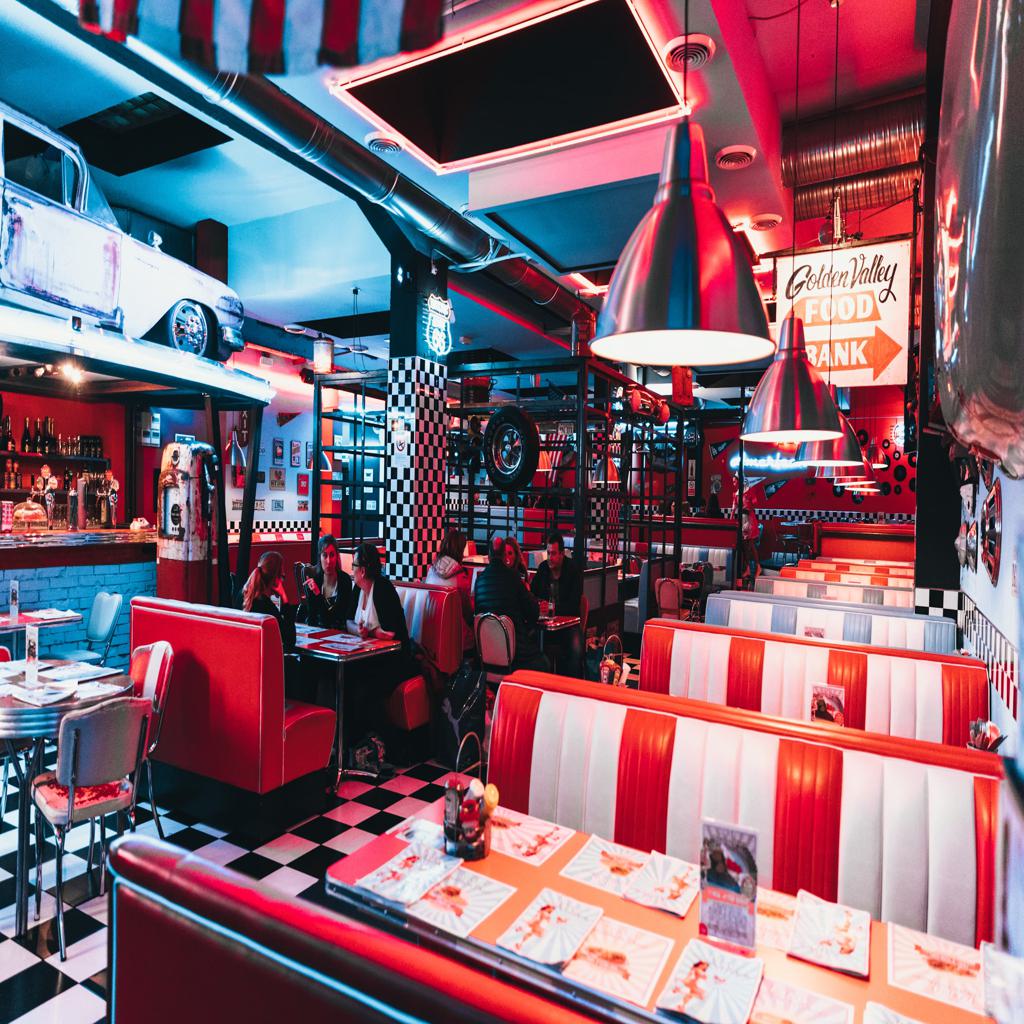} &
        \includegraphics[width=0.195\linewidth]{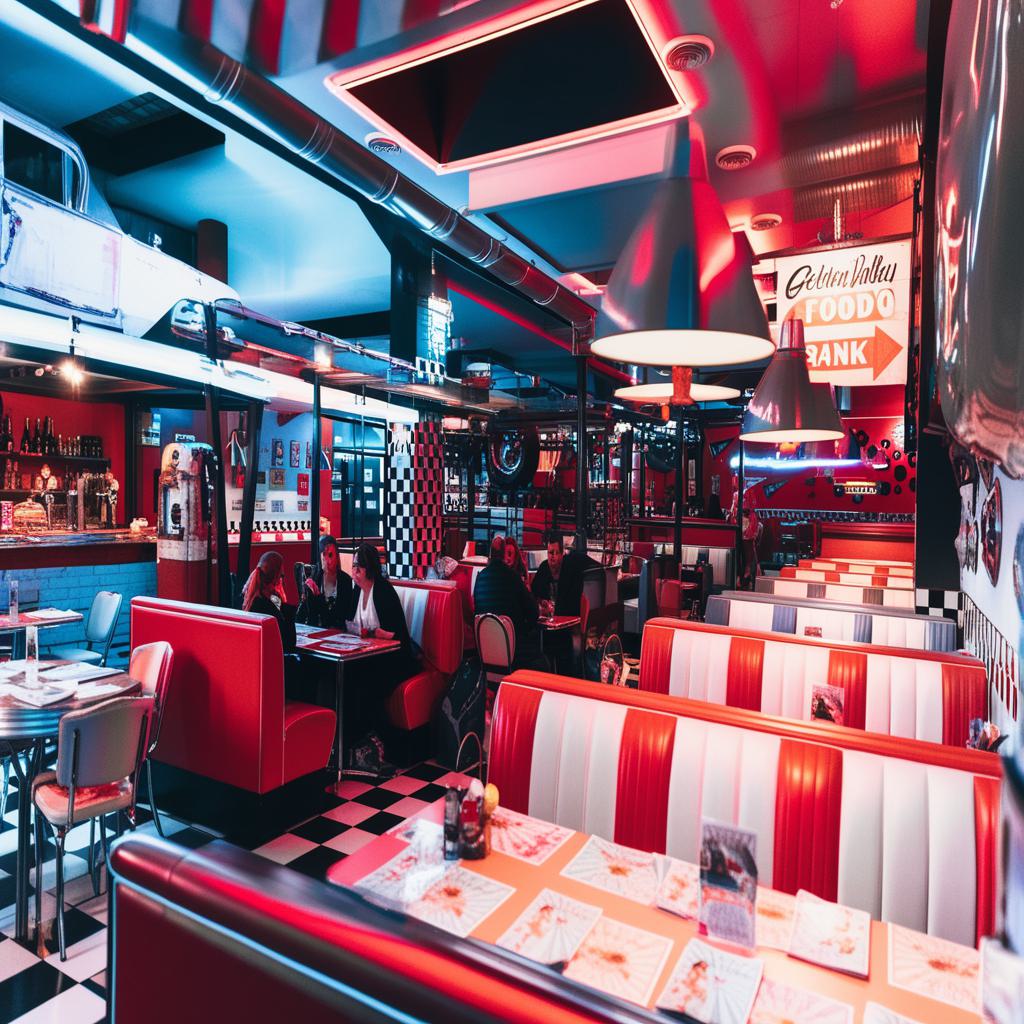} &
        \includegraphics[width=0.195\linewidth]{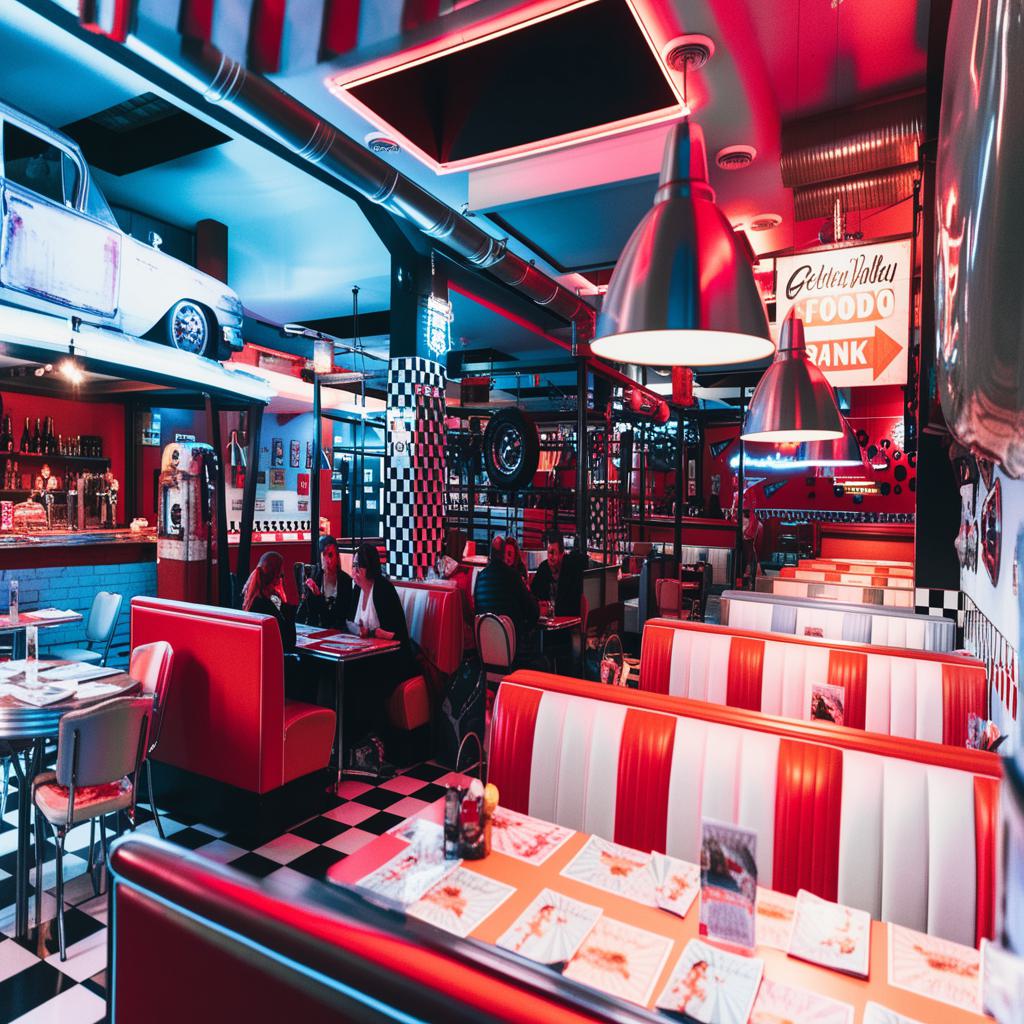} &
        \includegraphics[width=0.195\linewidth]{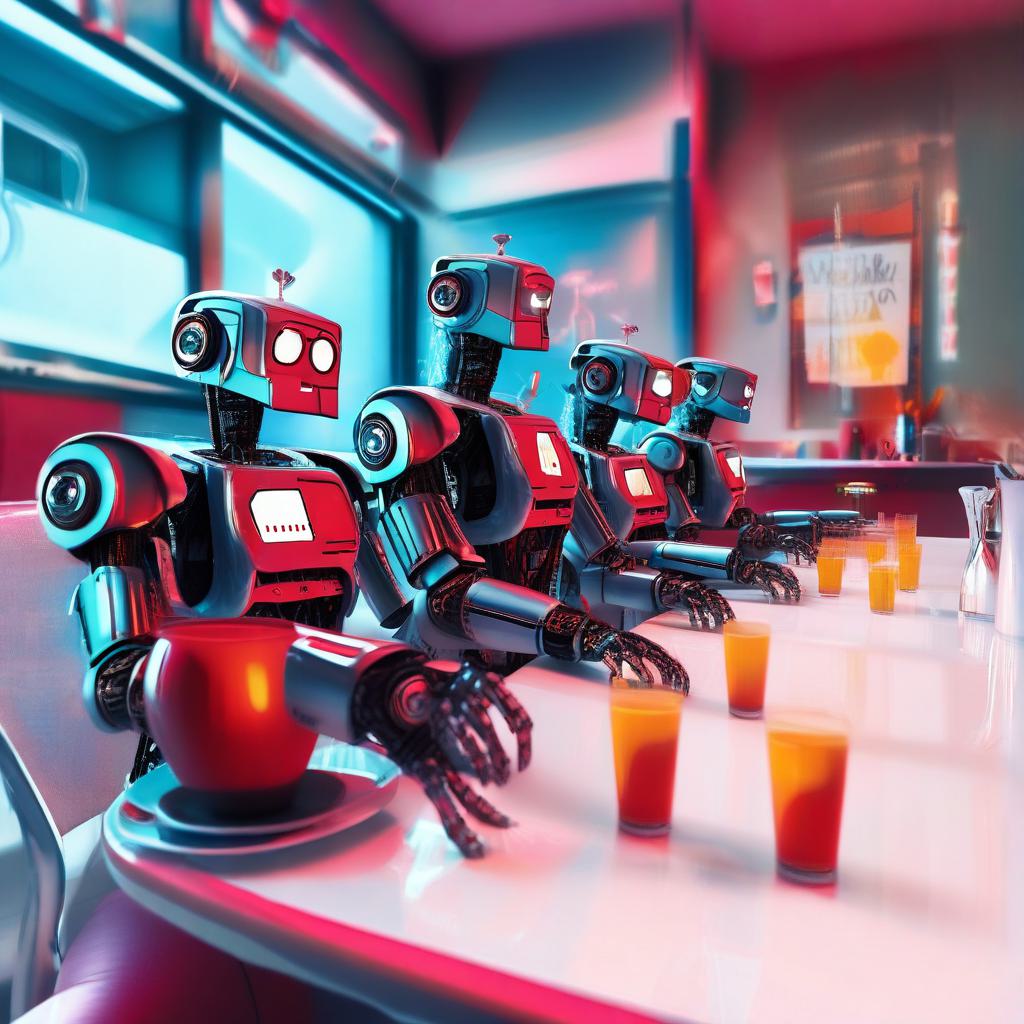} &
        \includegraphics[width=0.195\linewidth]{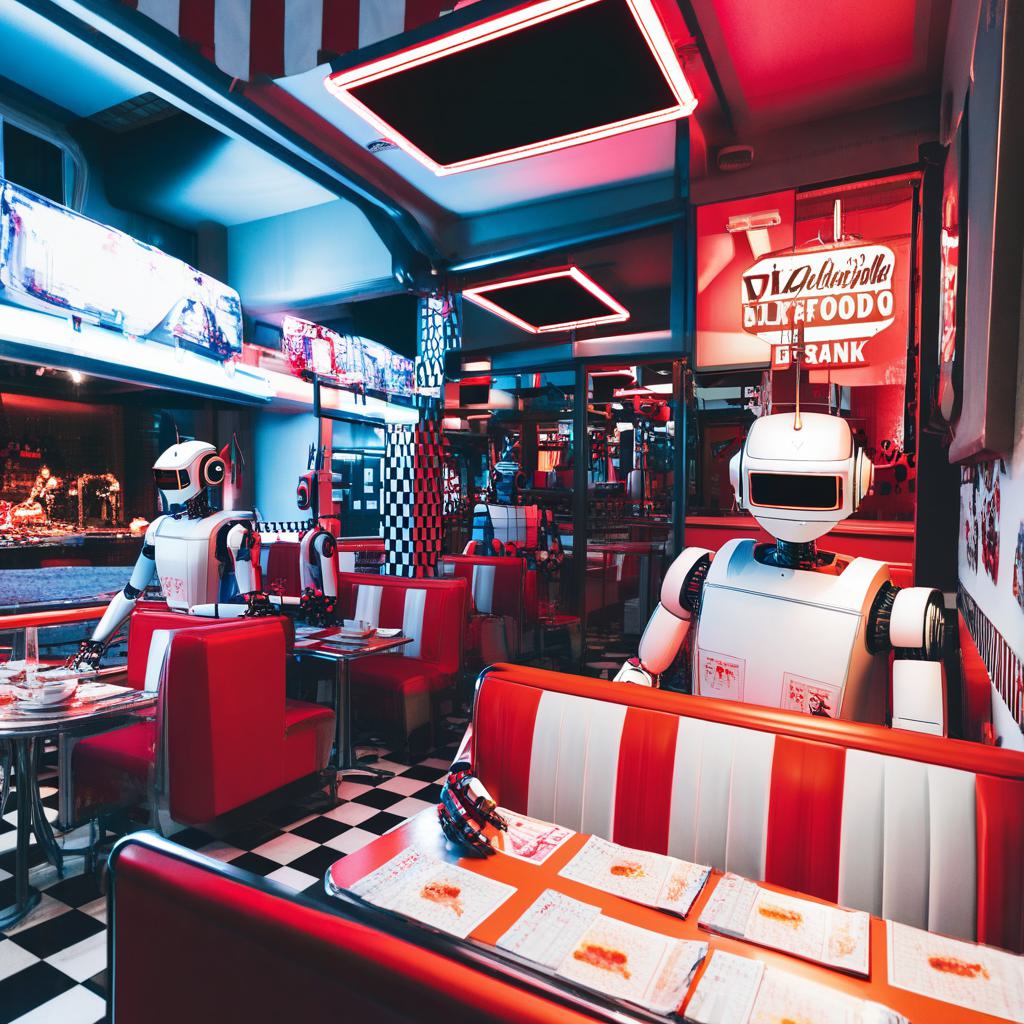} \\
        \multicolumn{5}{c}{``an antelope in the field'' $\longrightarrow$ ``... jumping in the field'', DDIM Inversion} \\
        \includegraphics[width=0.195\linewidth]{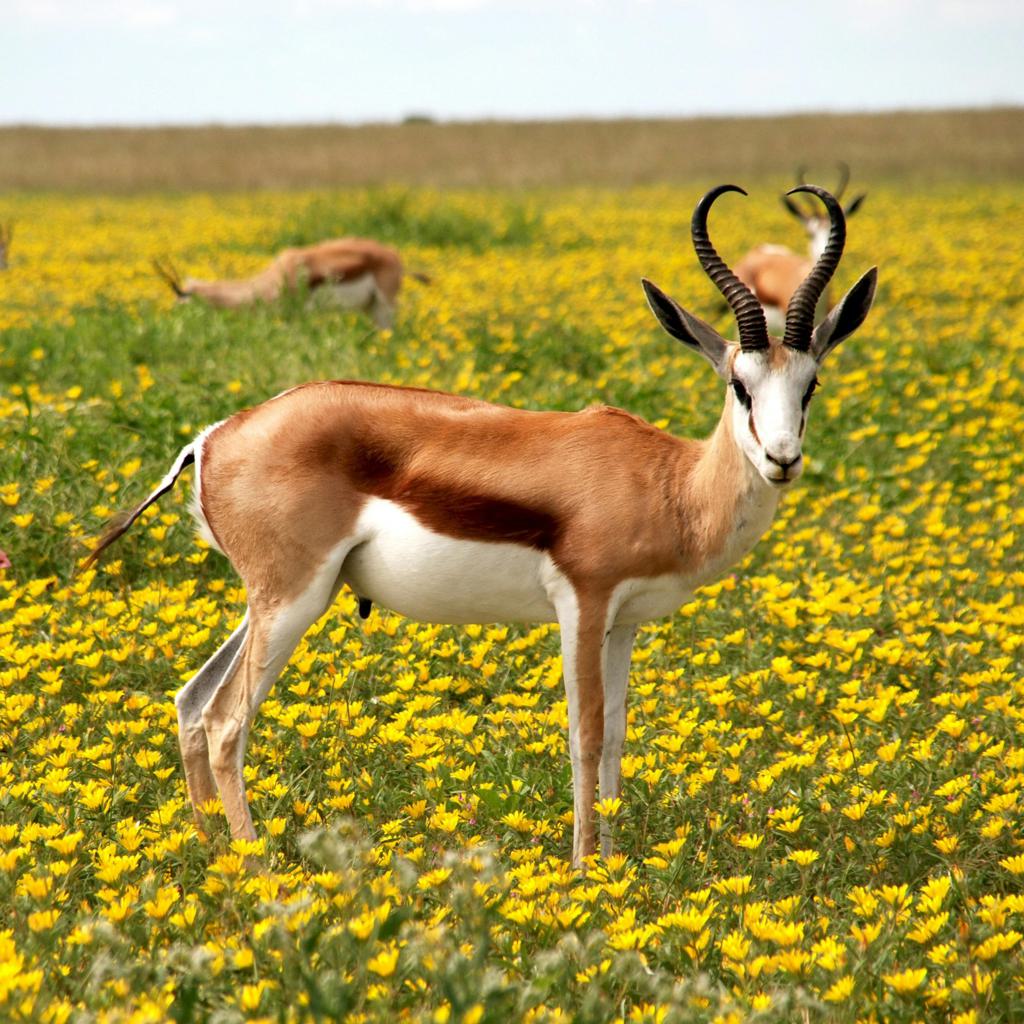} &
        \includegraphics[width=0.195\linewidth]{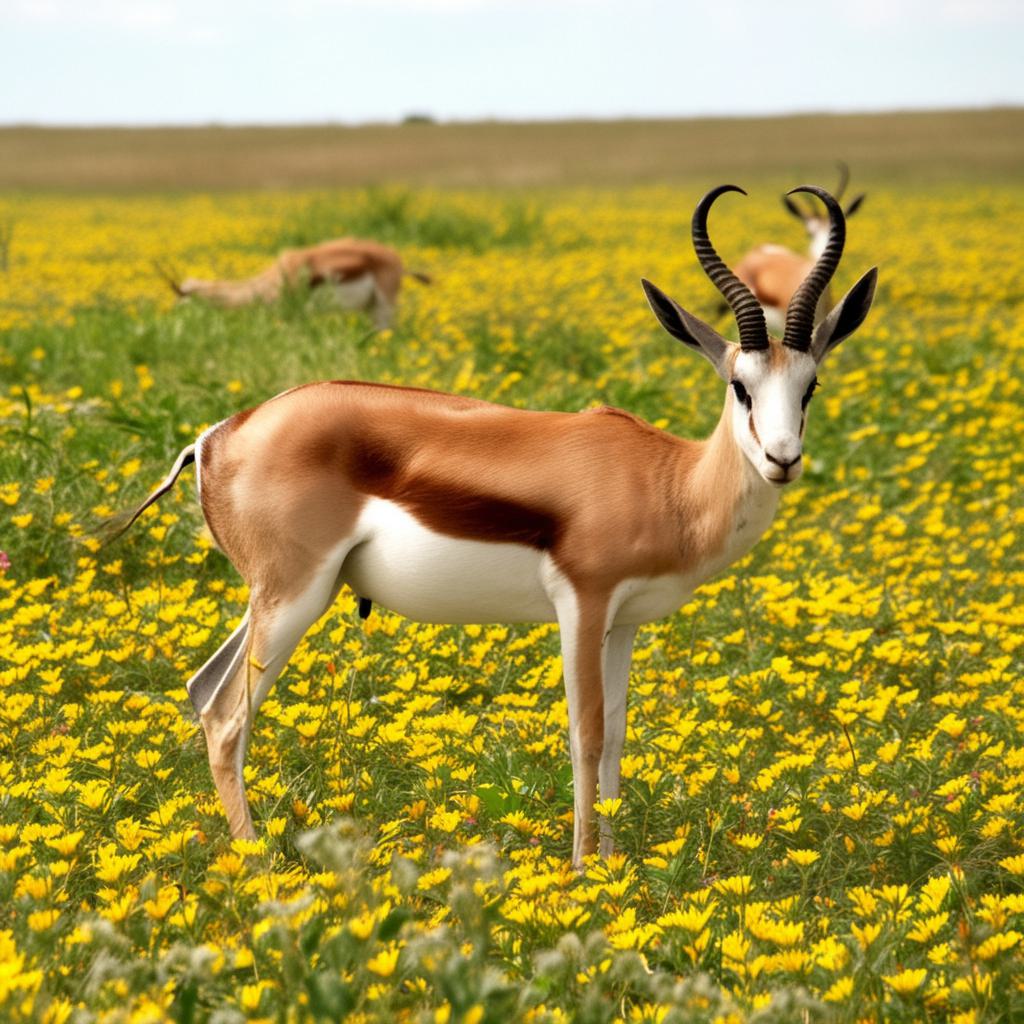} &
        \includegraphics[width=0.195\linewidth]{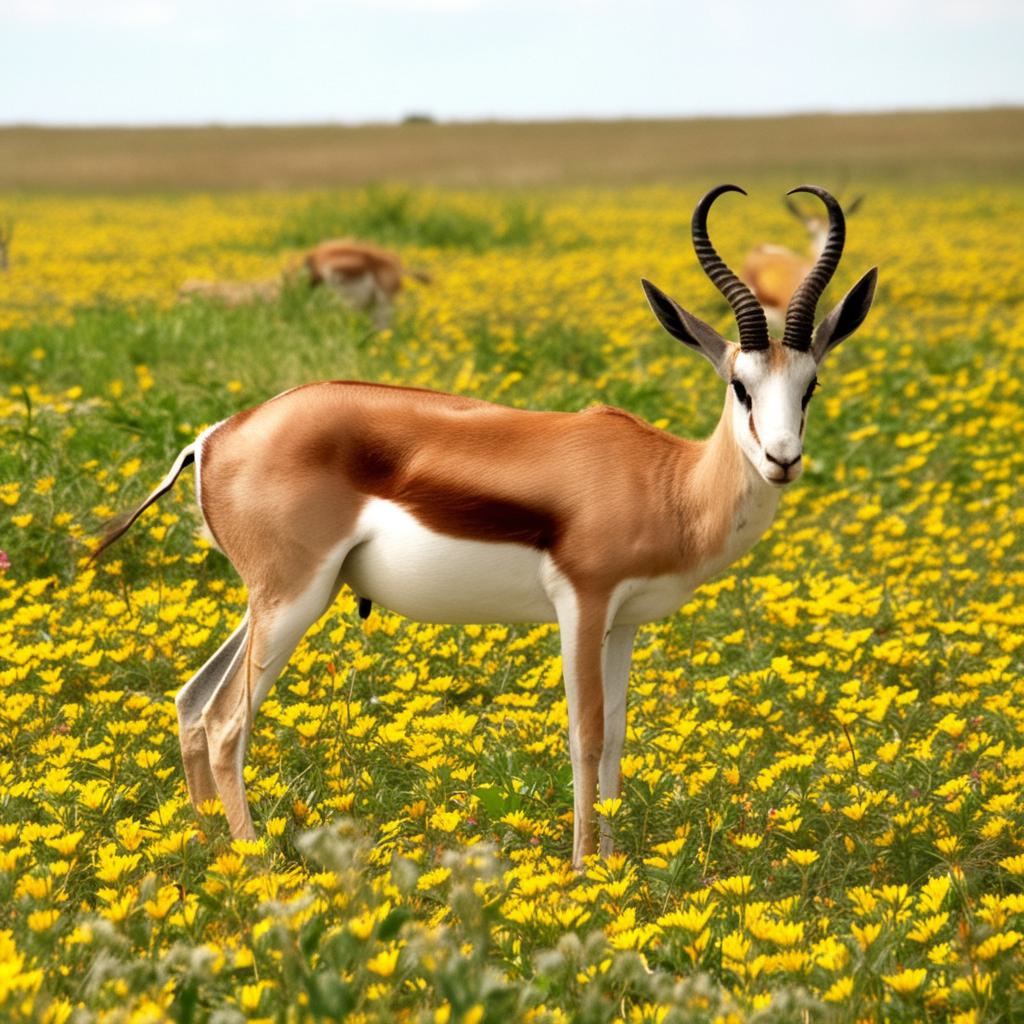} &
        \includegraphics[width=0.195\linewidth]{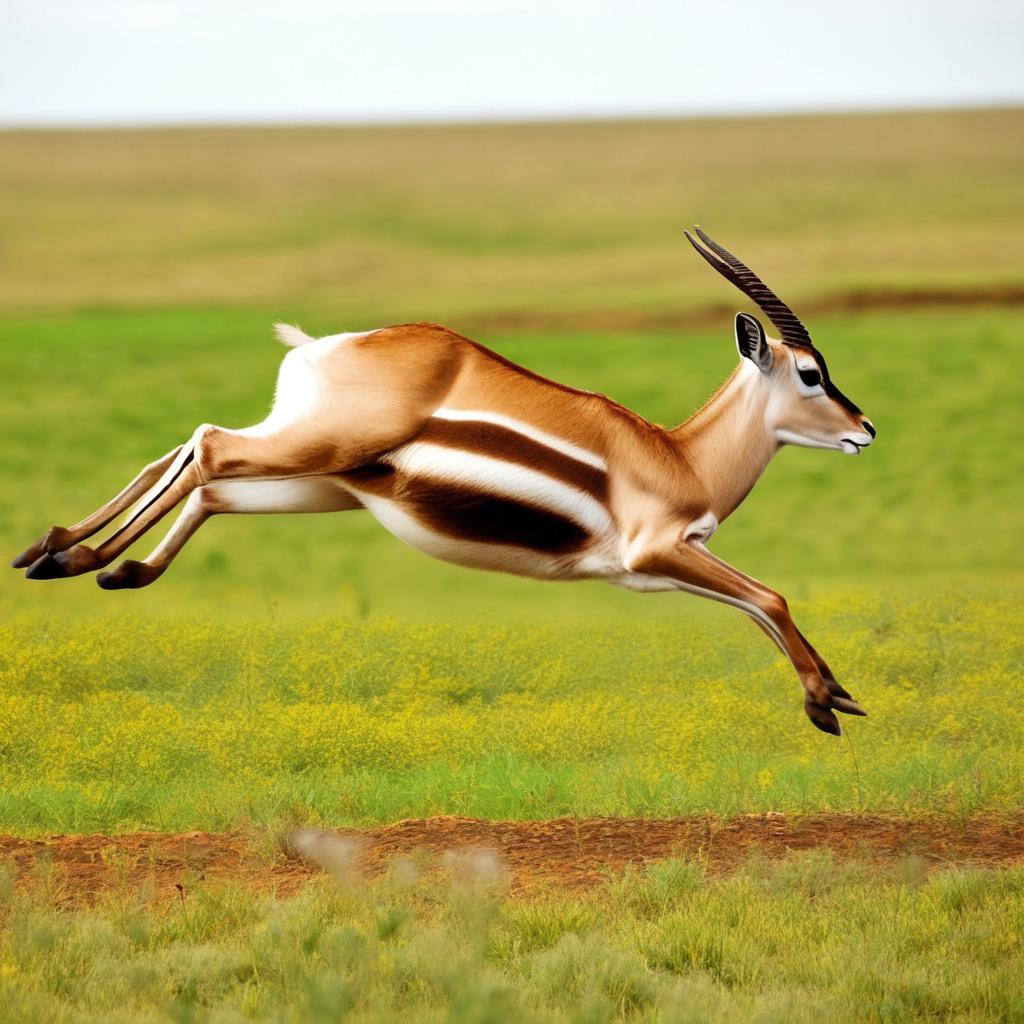} &
        \includegraphics[width=0.195\linewidth]{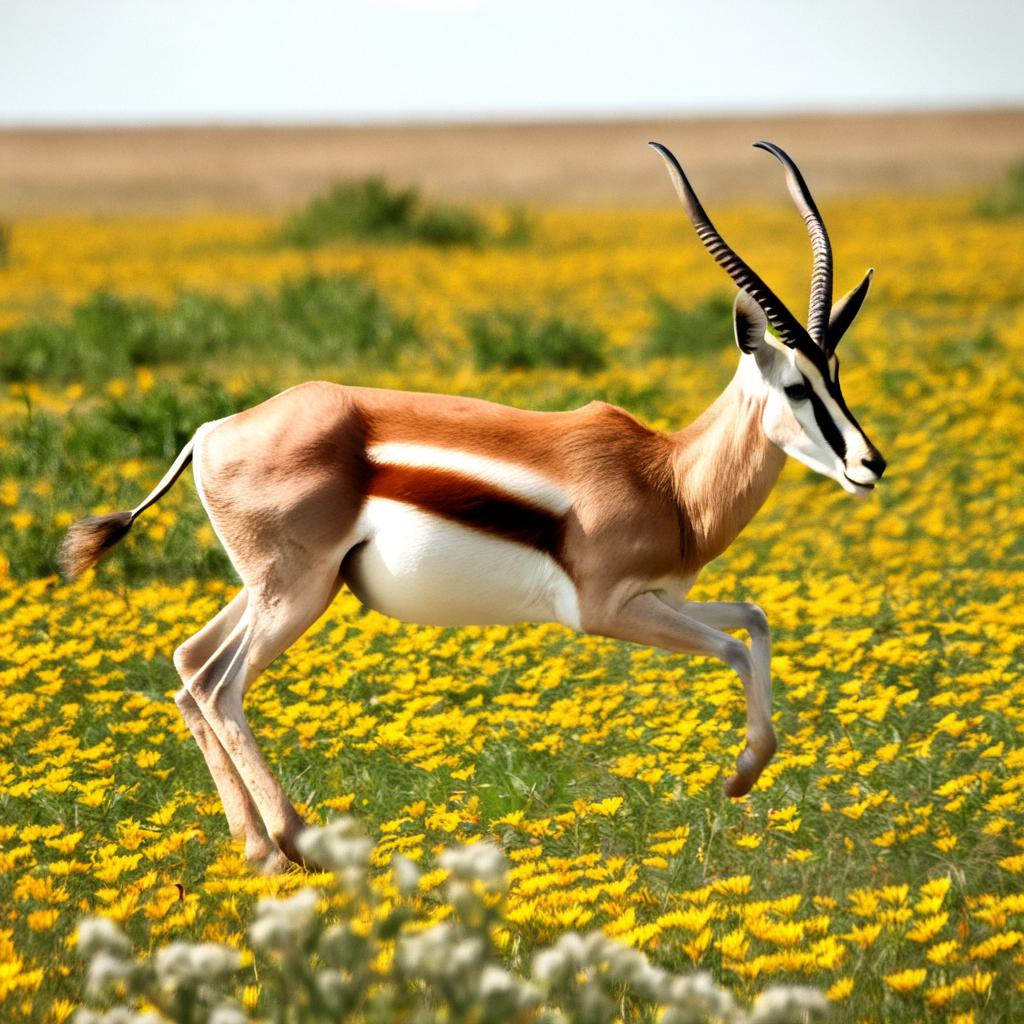} \\
        \multicolumn{5}{c}{``a dog in the snow'' $\longrightarrow$ ``a cat in the snow'', prompt2prompt} \\
        \includegraphics[width=0.195\linewidth]{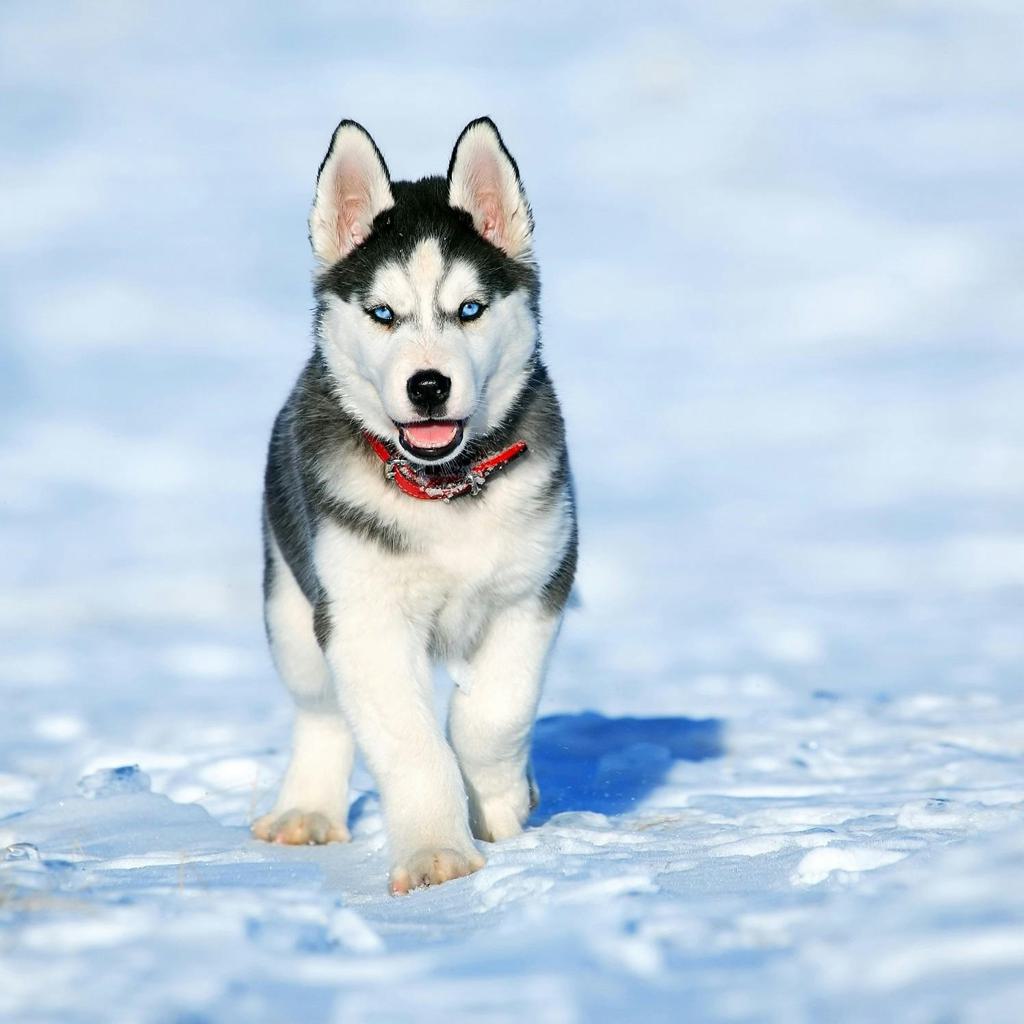} &
        \includegraphics[width=0.195\linewidth]{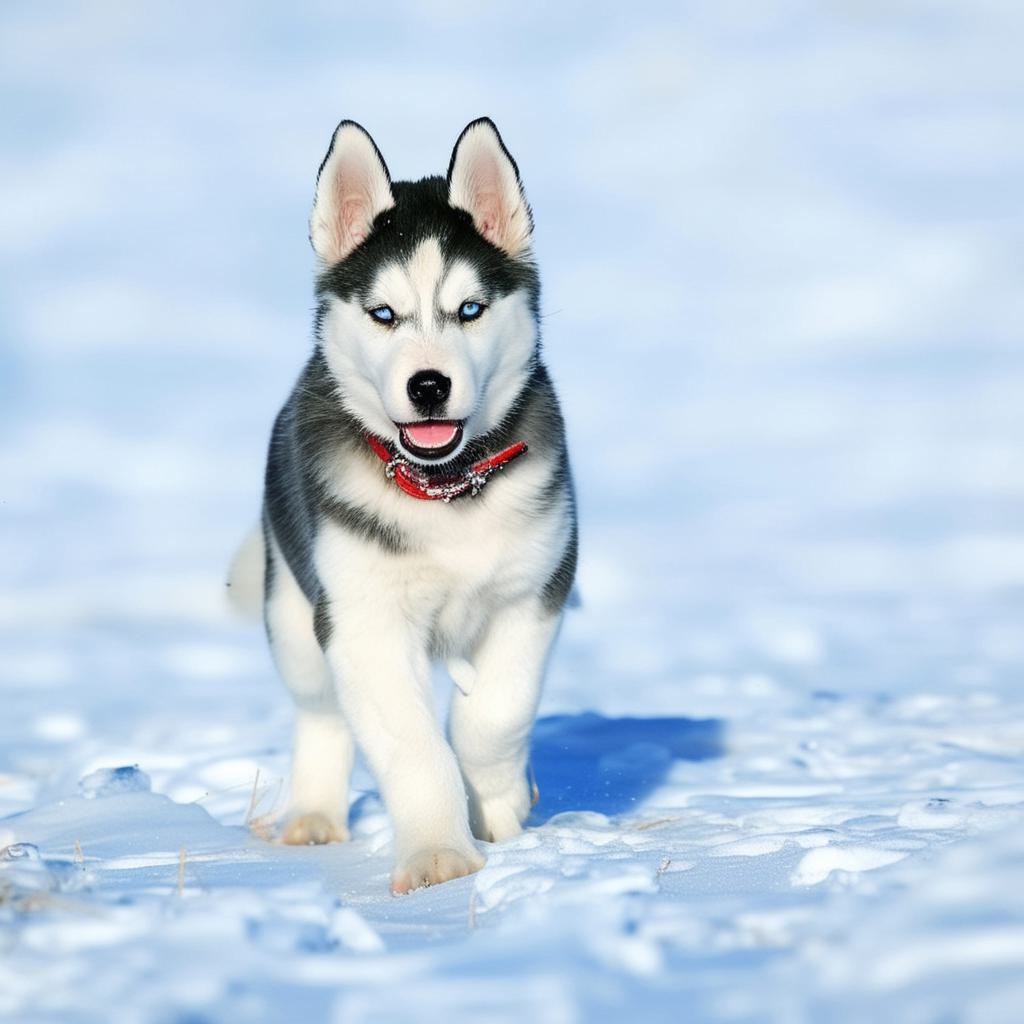} &
        \includegraphics[width=0.195\linewidth]{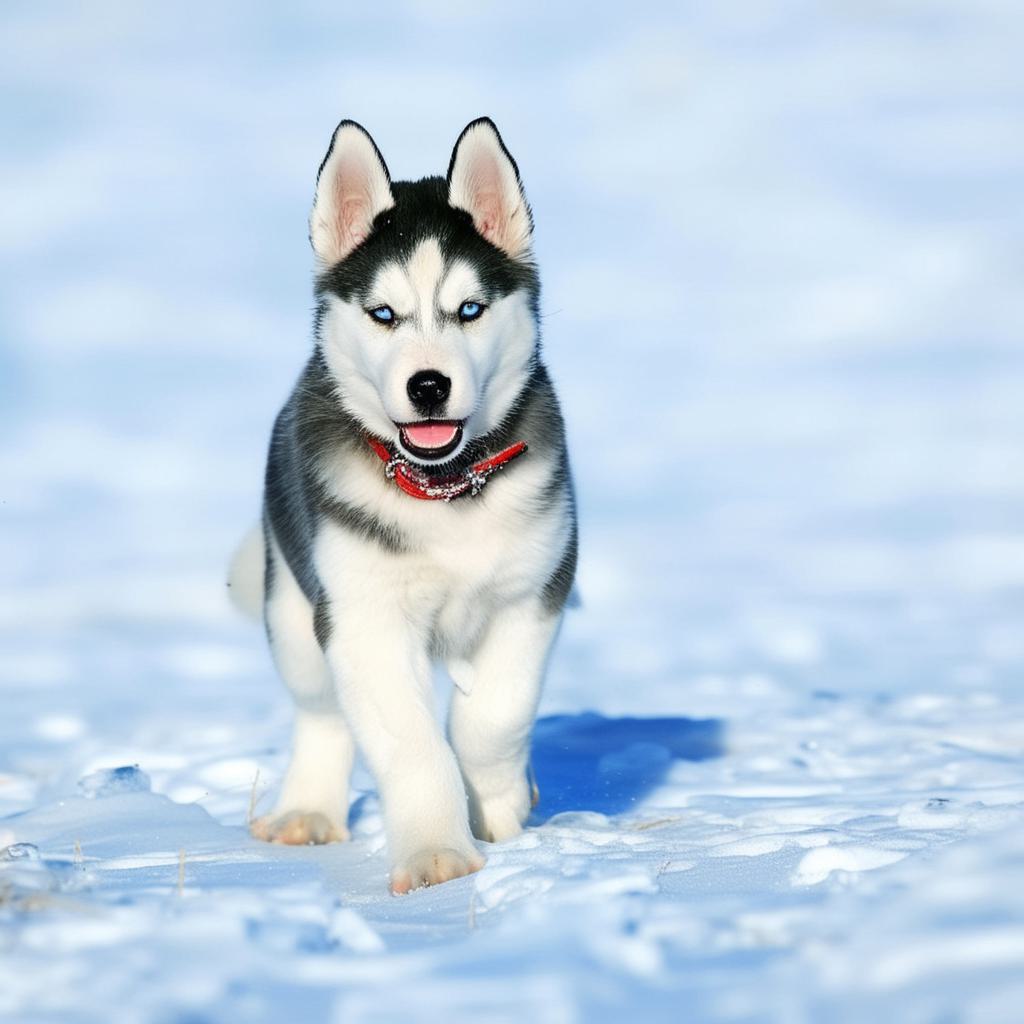} &
        \includegraphics[width=0.195\linewidth]{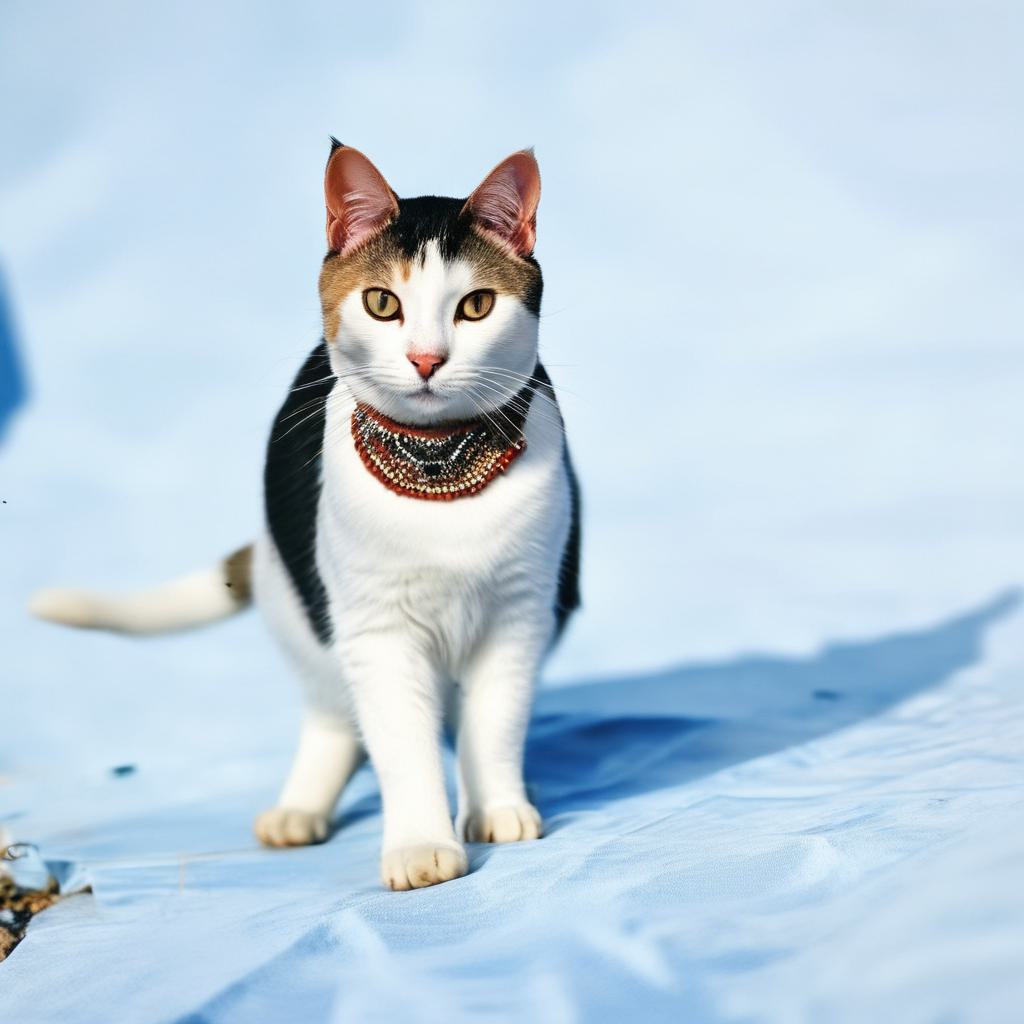} &
        \includegraphics[width=0.195\linewidth]{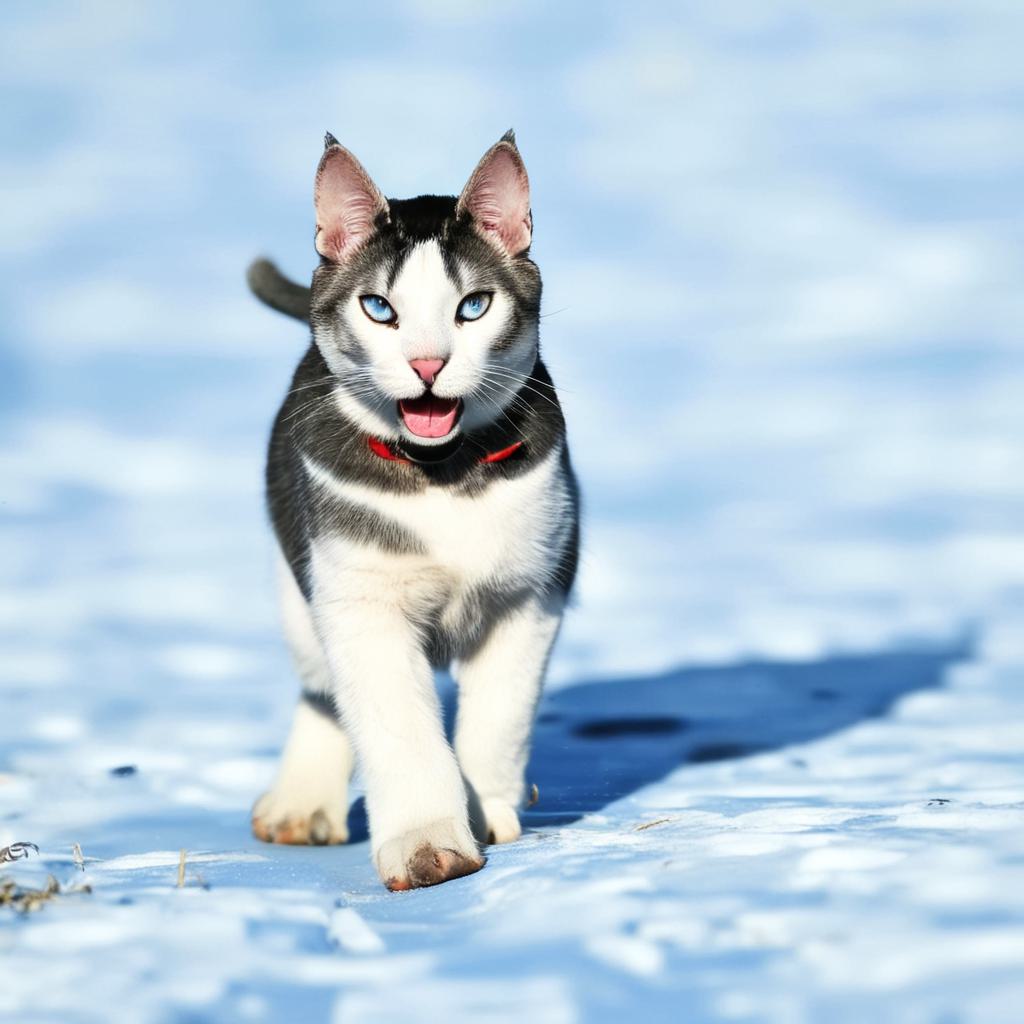} \\
        \multicolumn{5}{c}{``a person'' $\longrightarrow$ ``a person with a thick beard'', Edit Friendly DDPM Inversion} \\
        \includegraphics[width=0.195\linewidth]{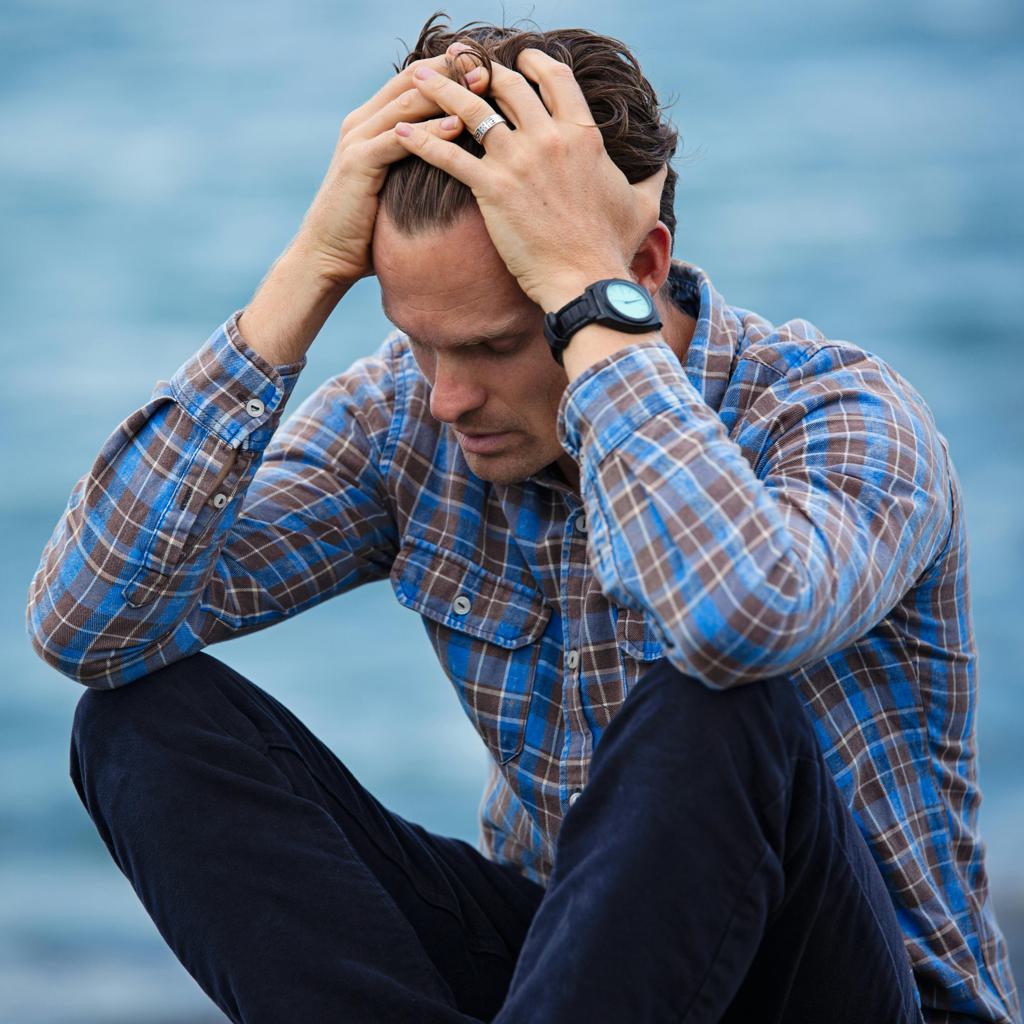} &
        \includegraphics[width=0.195\linewidth]{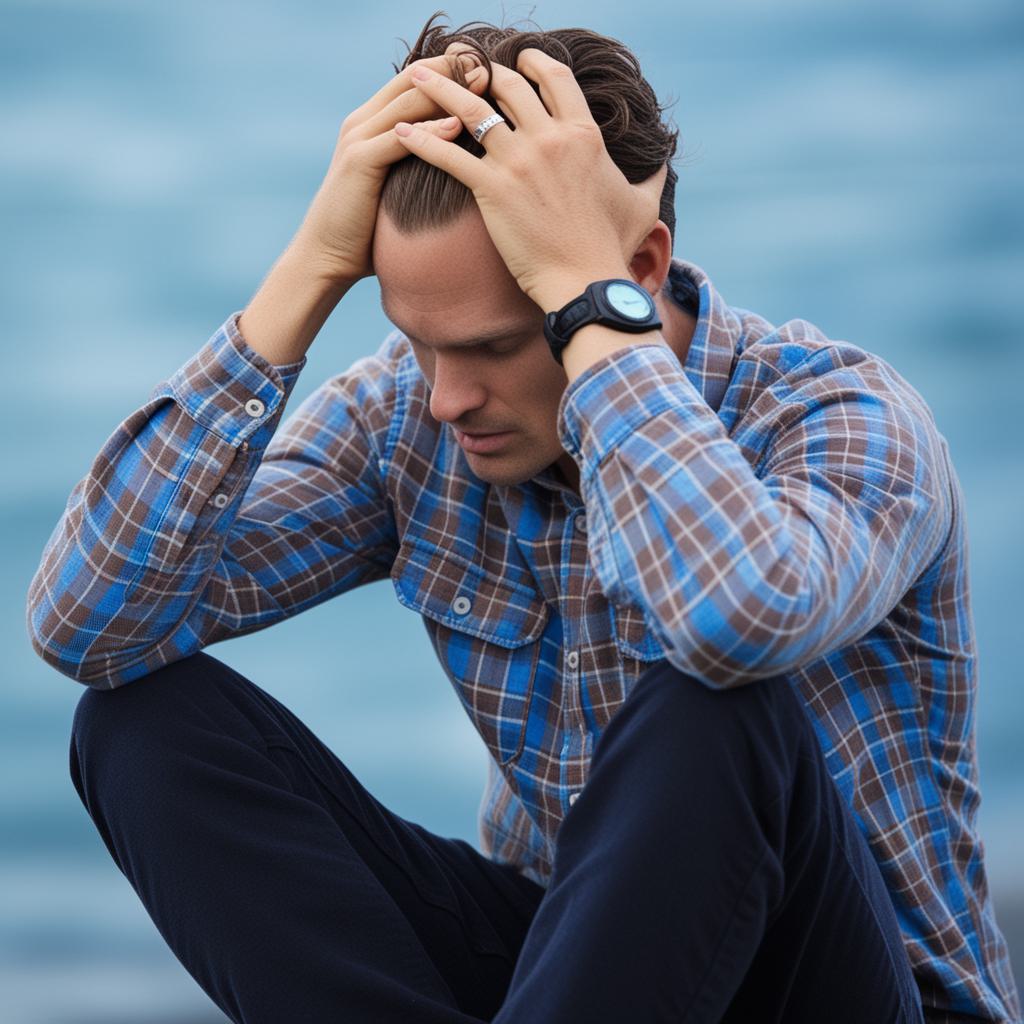} &
        \includegraphics[width=0.195\linewidth]{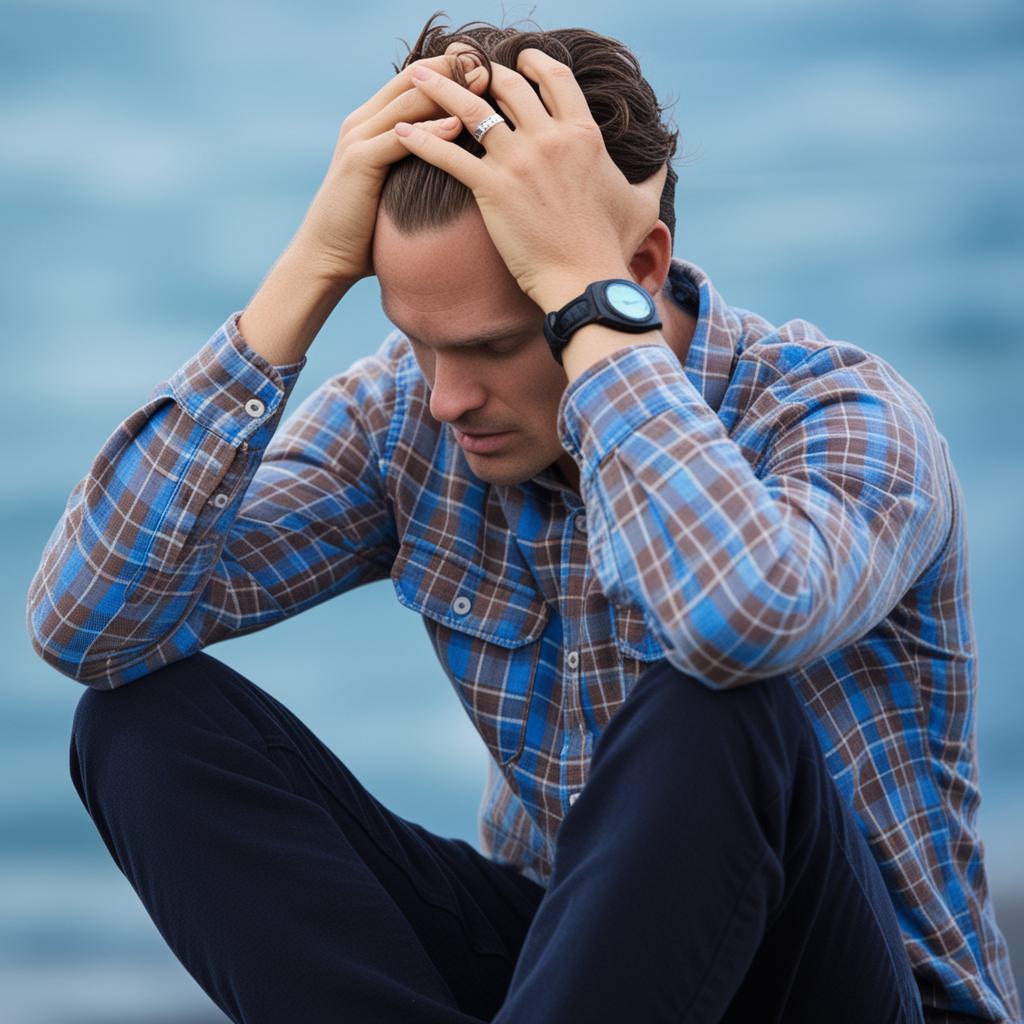} &
        \includegraphics[width=0.195\linewidth]{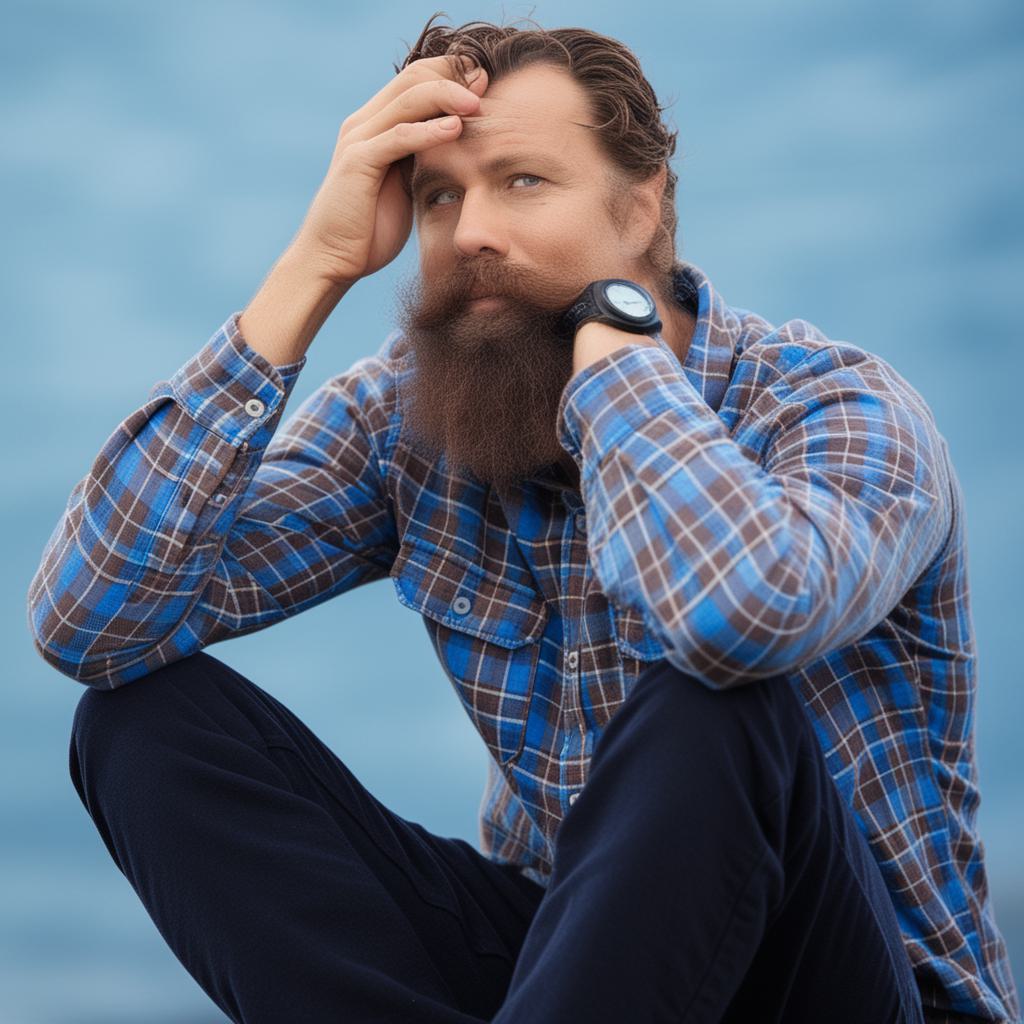} &
        \includegraphics[width=0.195\linewidth]{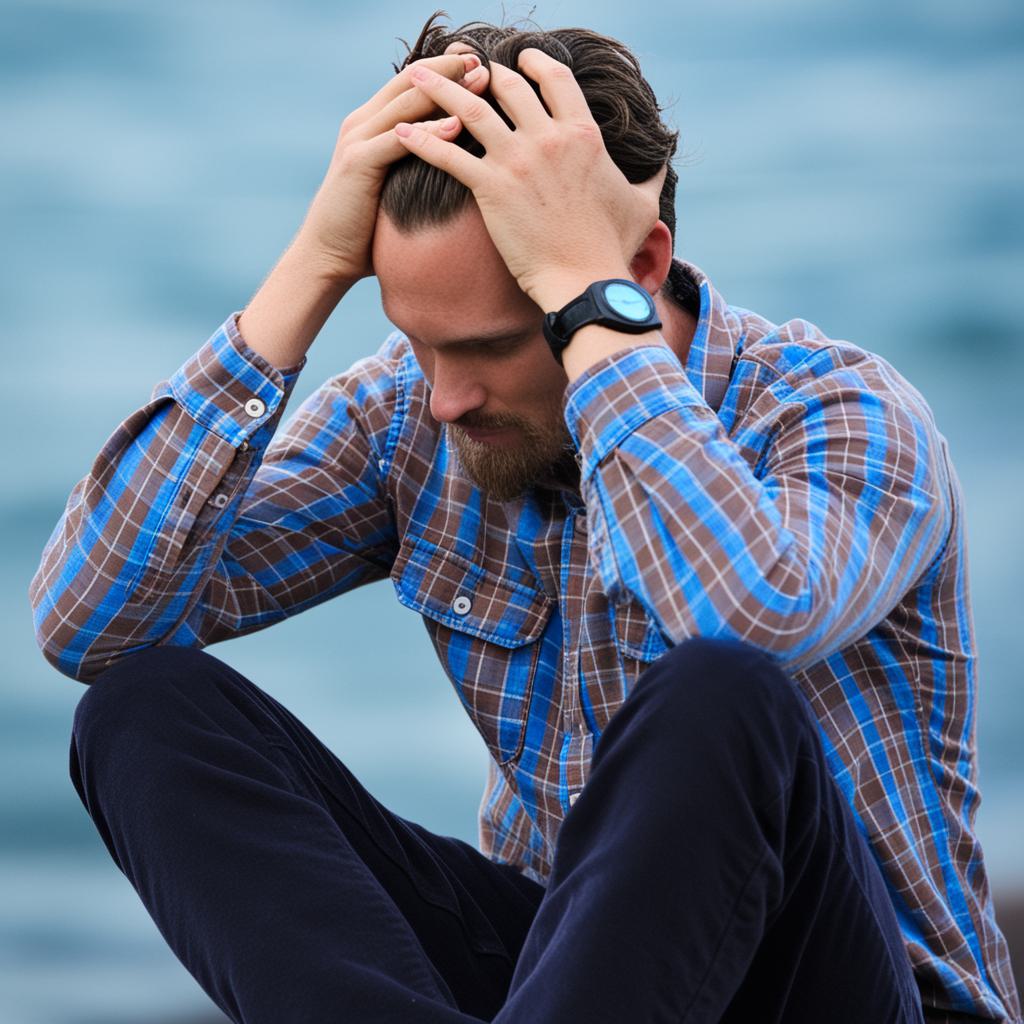} \\
        \multicolumn{5}{c}{``a wooden horse in the room'' $\longrightarrow$ ``a horse made of lego ...'', LEDITS++} \\
        \includegraphics[width=0.195\linewidth]{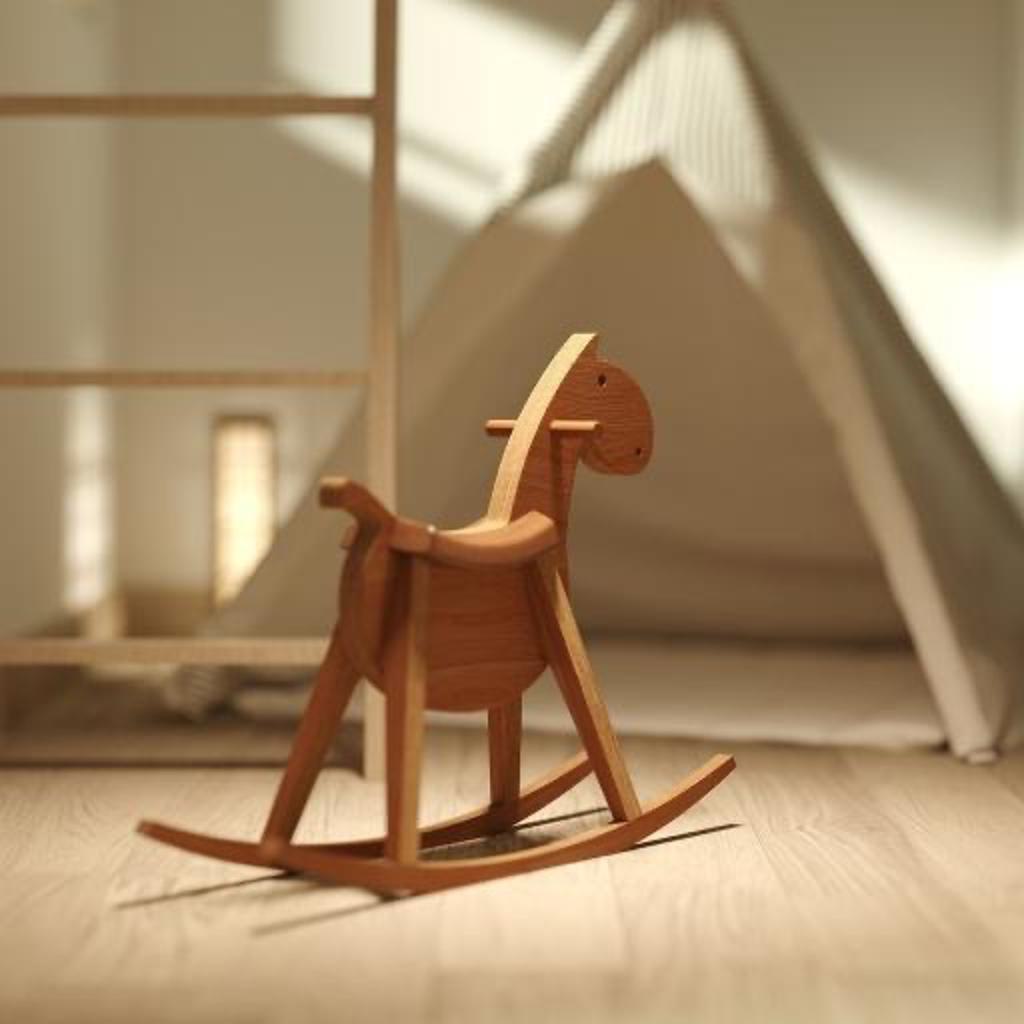} &
        \includegraphics[width=0.195\linewidth]{images/ledits/wooden_horse/wooden_horse_original.jpg} &
        \includegraphics[width=0.195\linewidth]{images/ledits/wooden_horse/wooden_horse_original.jpg} &
        \includegraphics[width=0.195\linewidth]{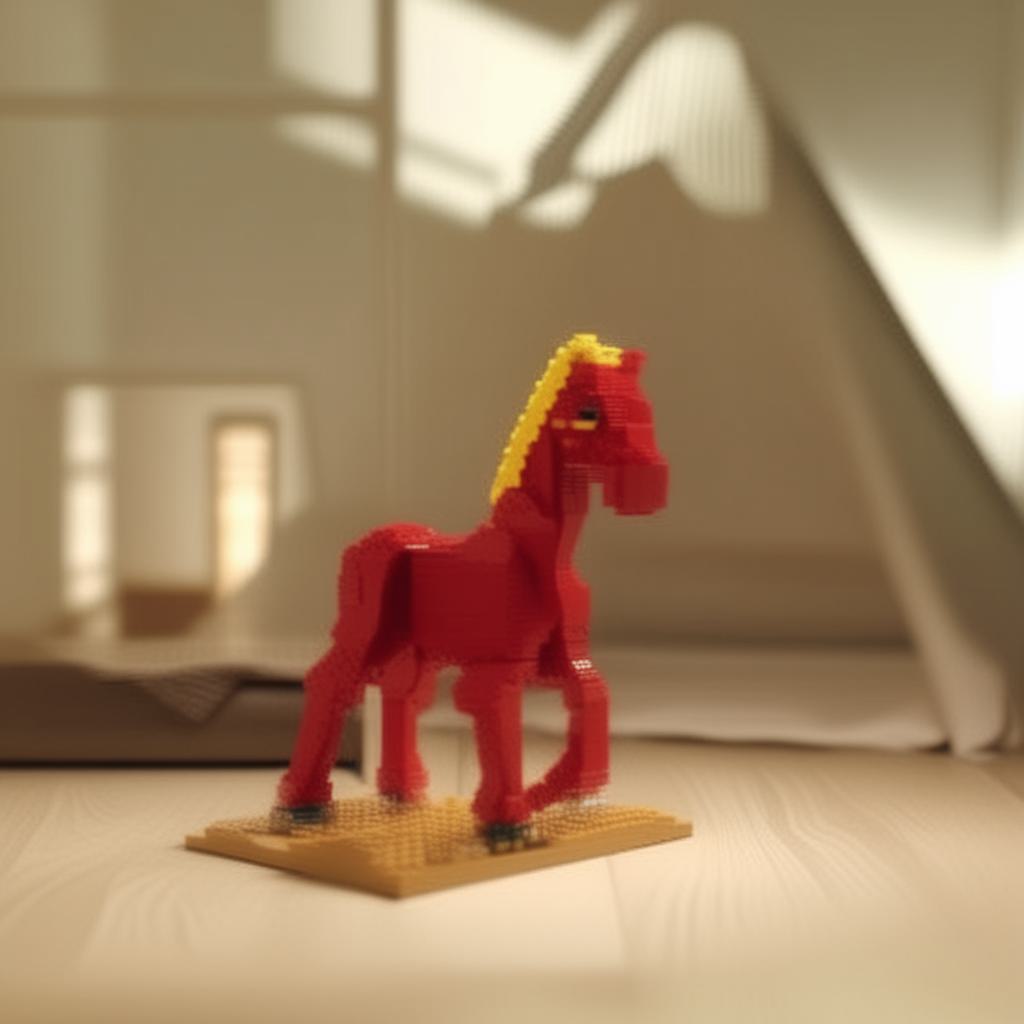} &
        \includegraphics[width=0.195\linewidth]{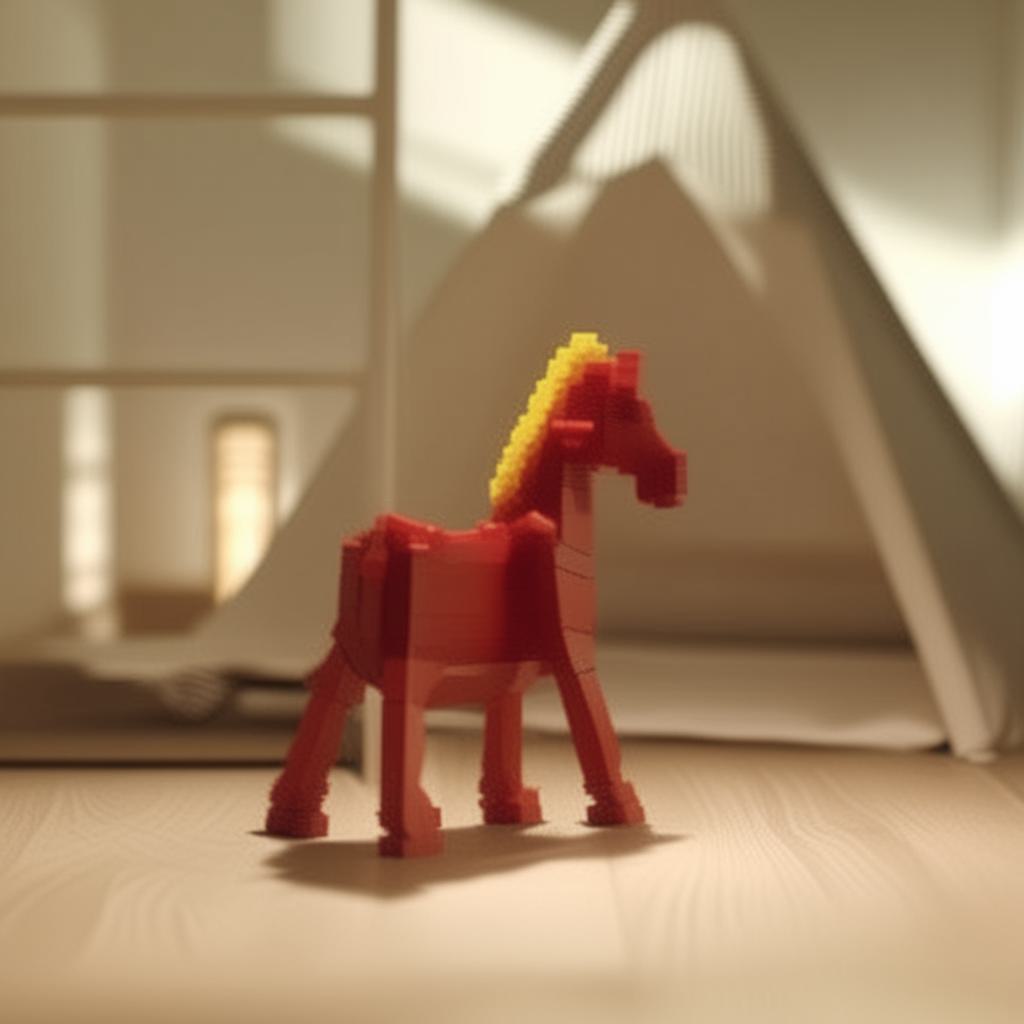} \\
        \multicolumn{5}{c}{``'' $\longrightarrow$ ``A portrait of an elf'', RF-Inversion (Flux)} \\
        \includegraphics[width=0.195\linewidth]{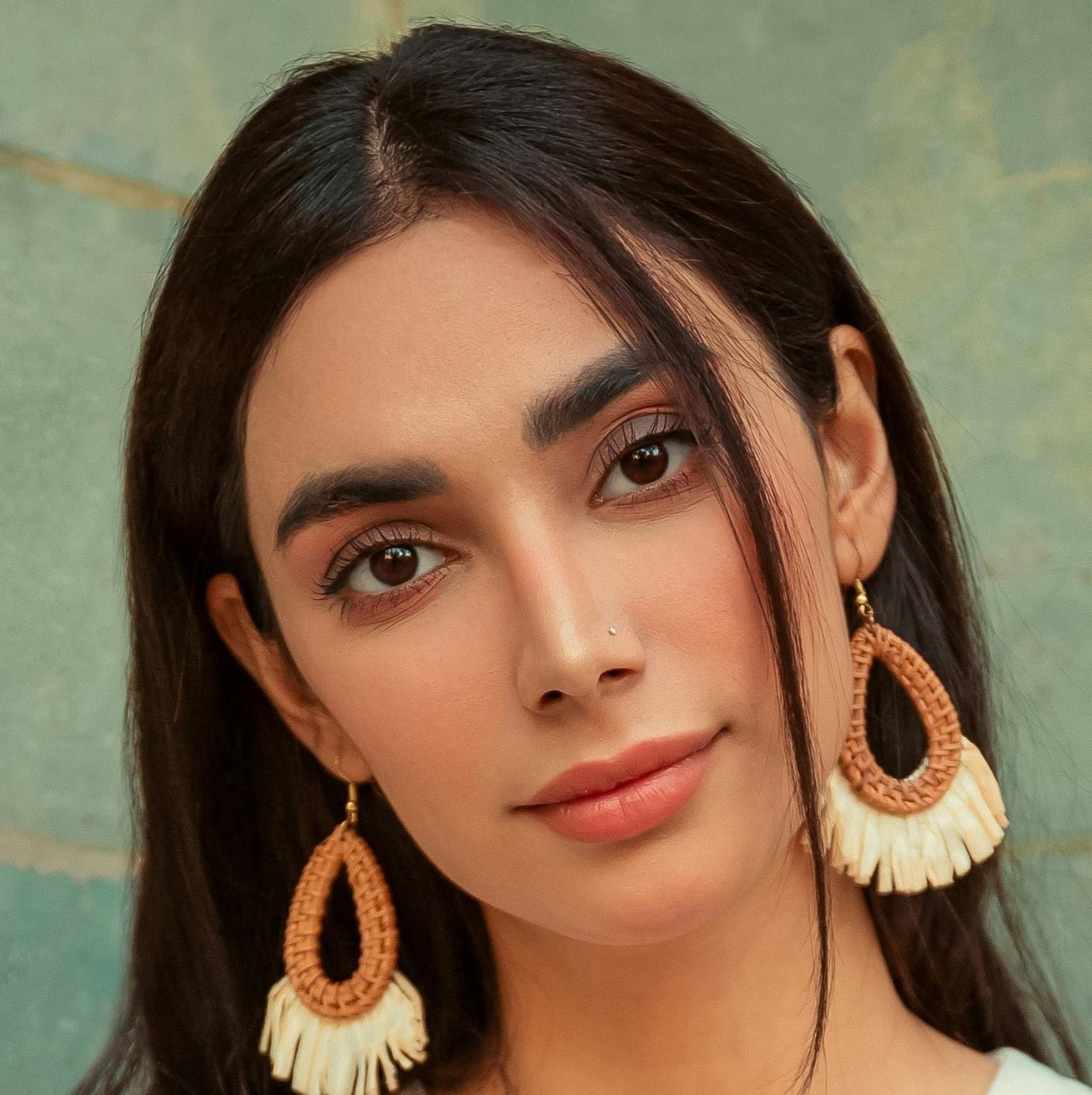}
        &
        \includegraphics[width=0.195\linewidth]{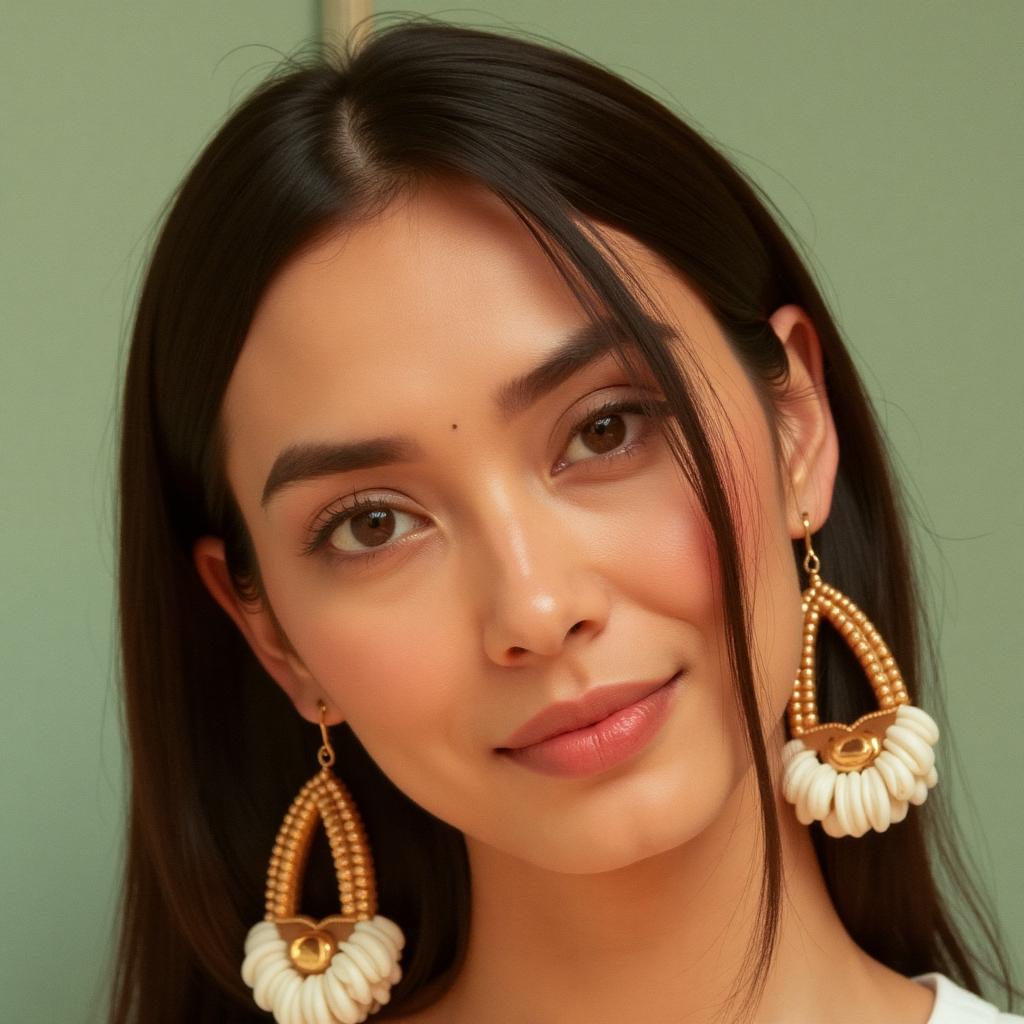} &
        \includegraphics[width=0.195\linewidth]{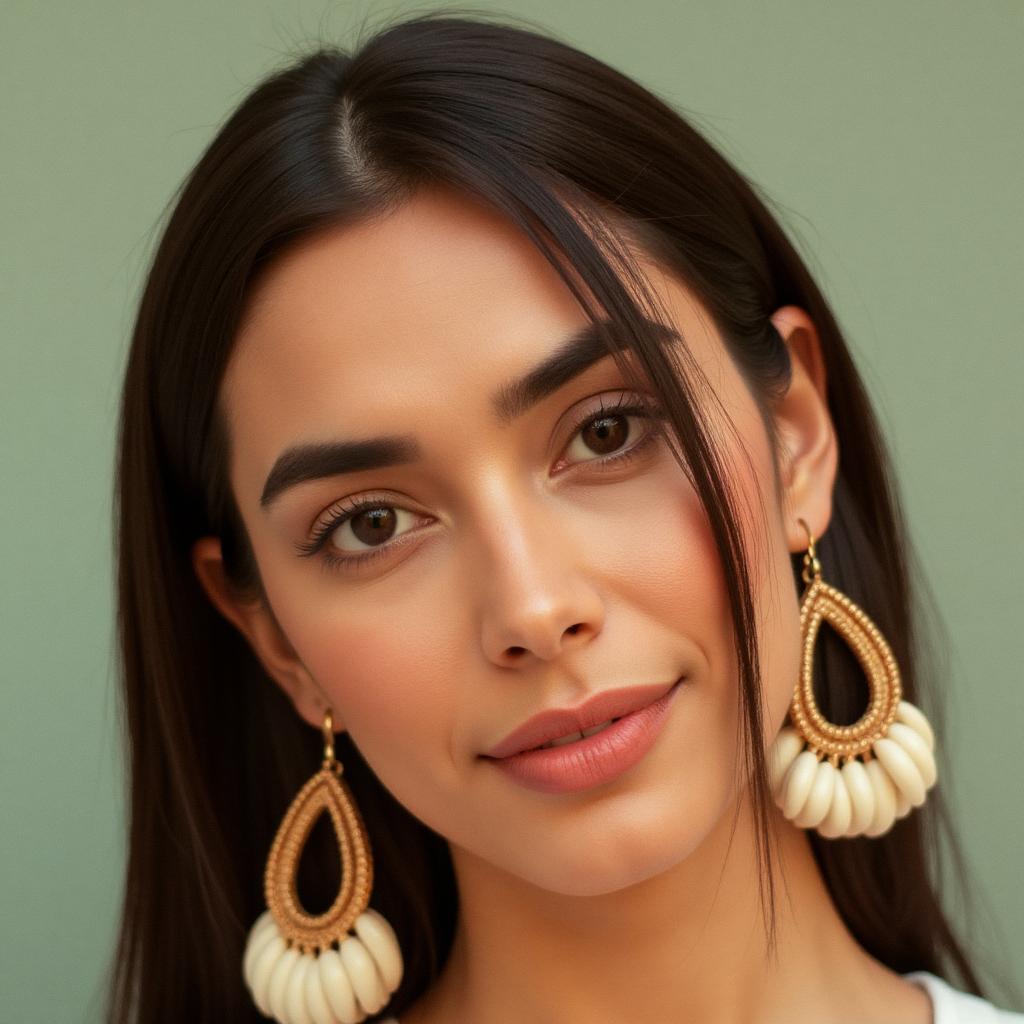} &
        \includegraphics[width=0.195\linewidth]{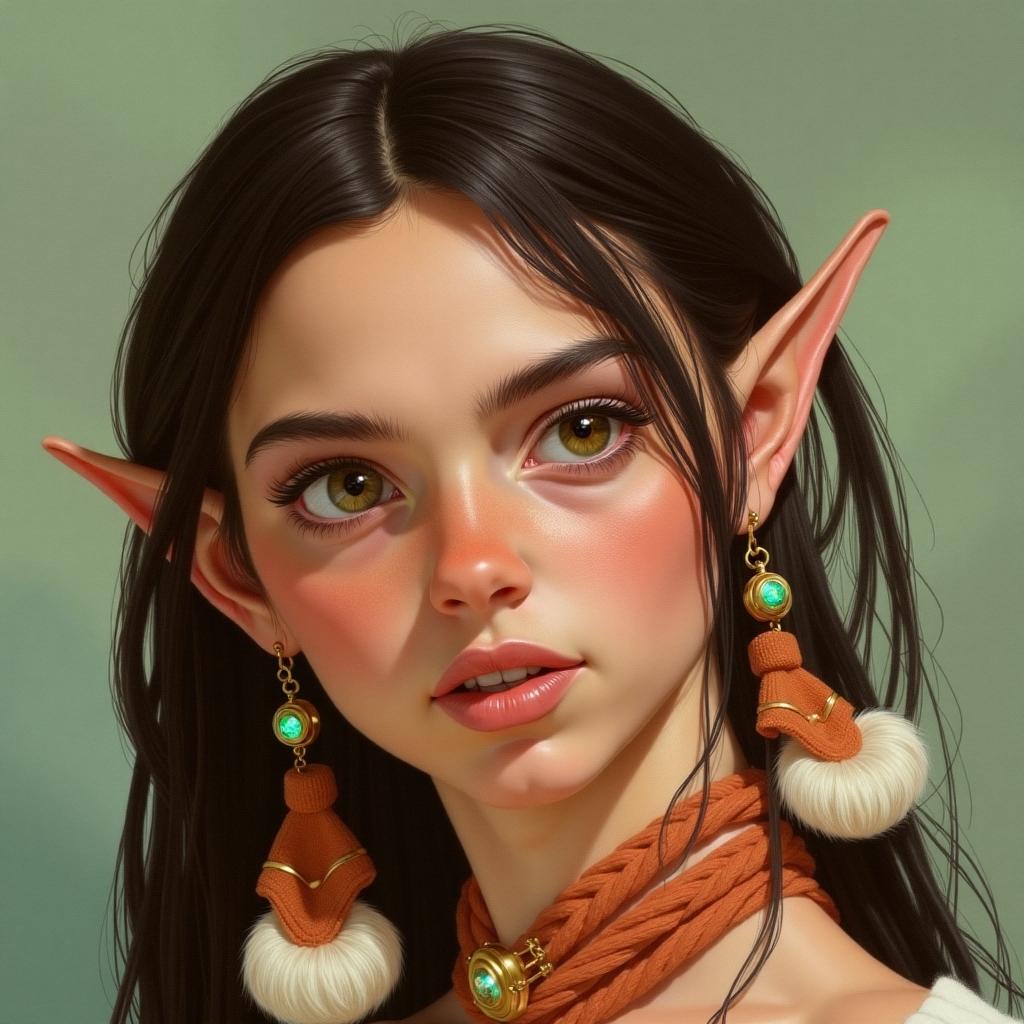} &
        \includegraphics[width=0.195\linewidth]{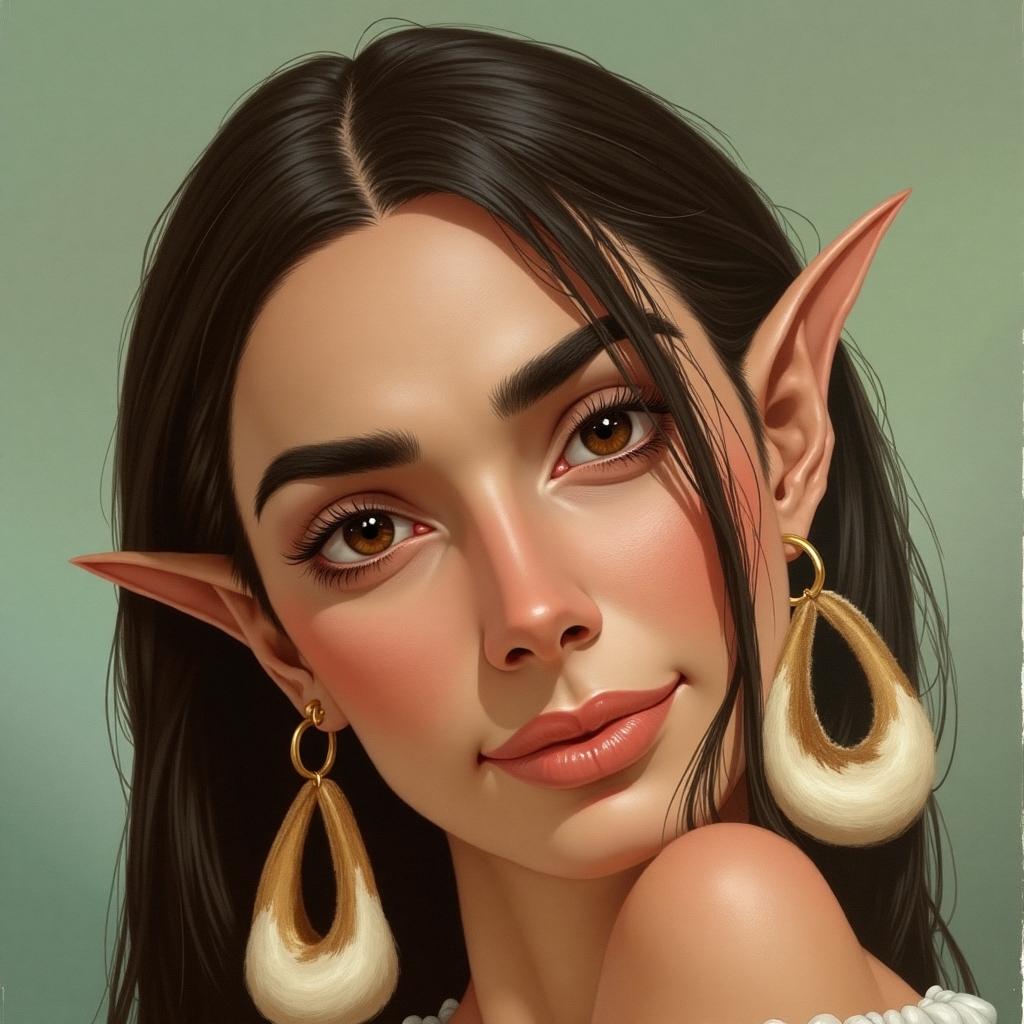} \\
        \multicolumn{5}{c}{``'' $\longrightarrow$ ``A portrait of a vampire'', RF-Inversion (Flux)} \\
        \includegraphics[width=0.195\linewidth]{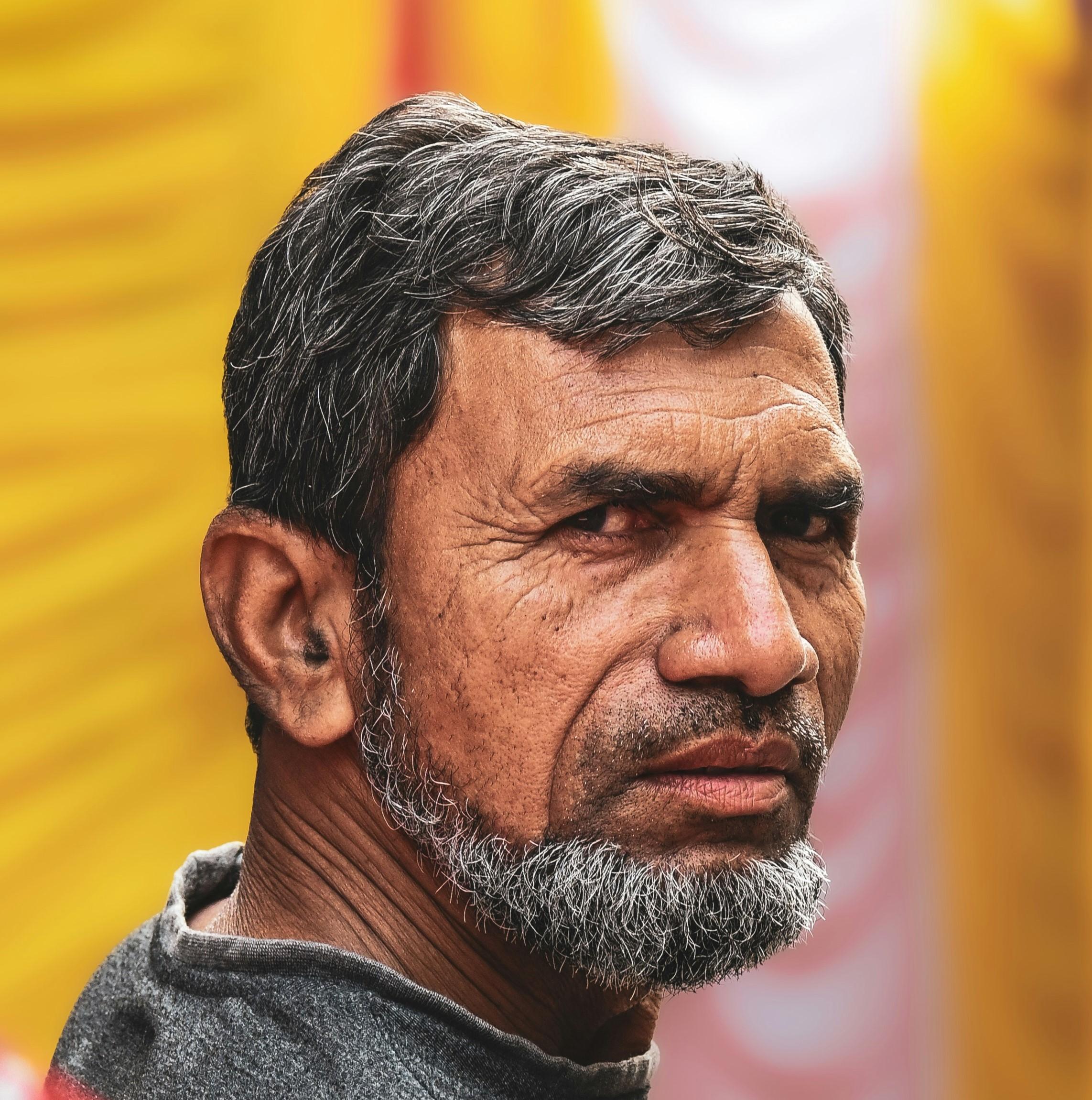}
        &
        \includegraphics[width=0.195\linewidth]{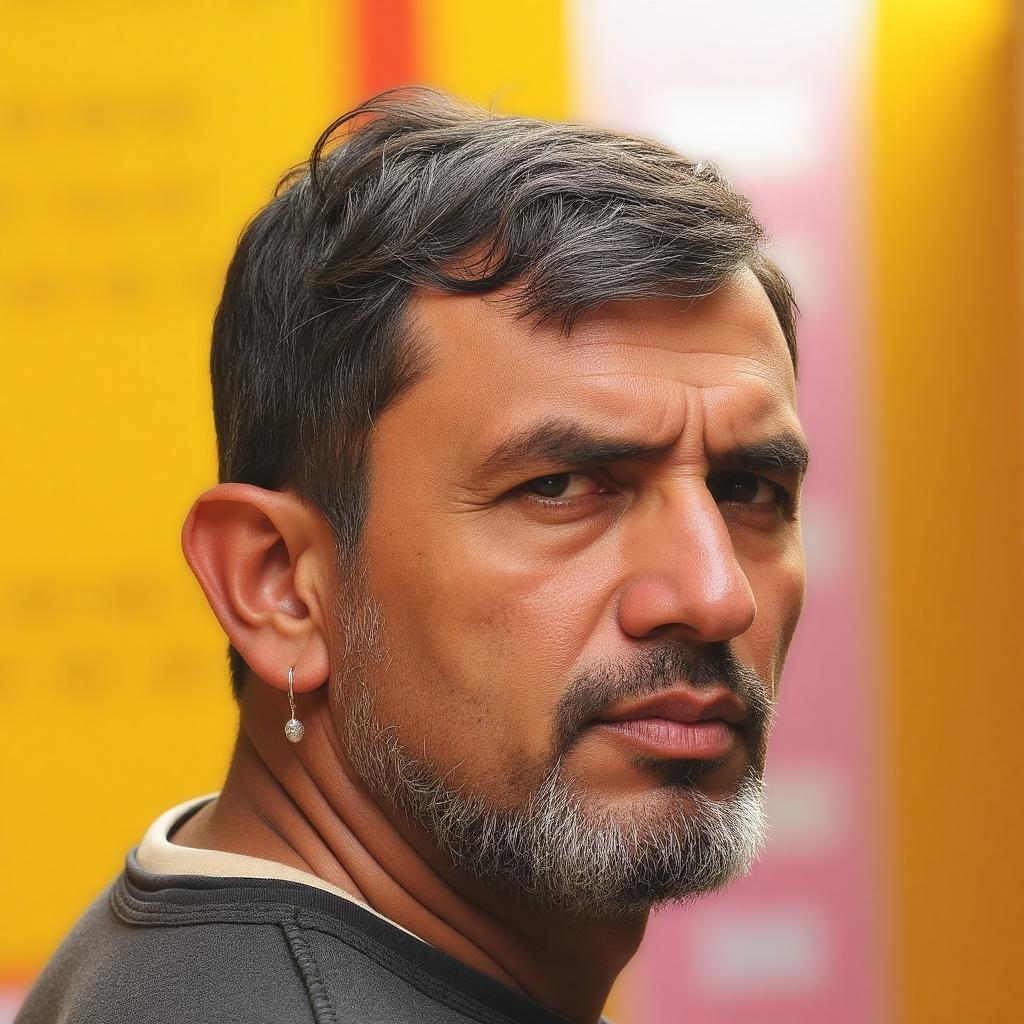} &
        \includegraphics[width=0.195\linewidth]{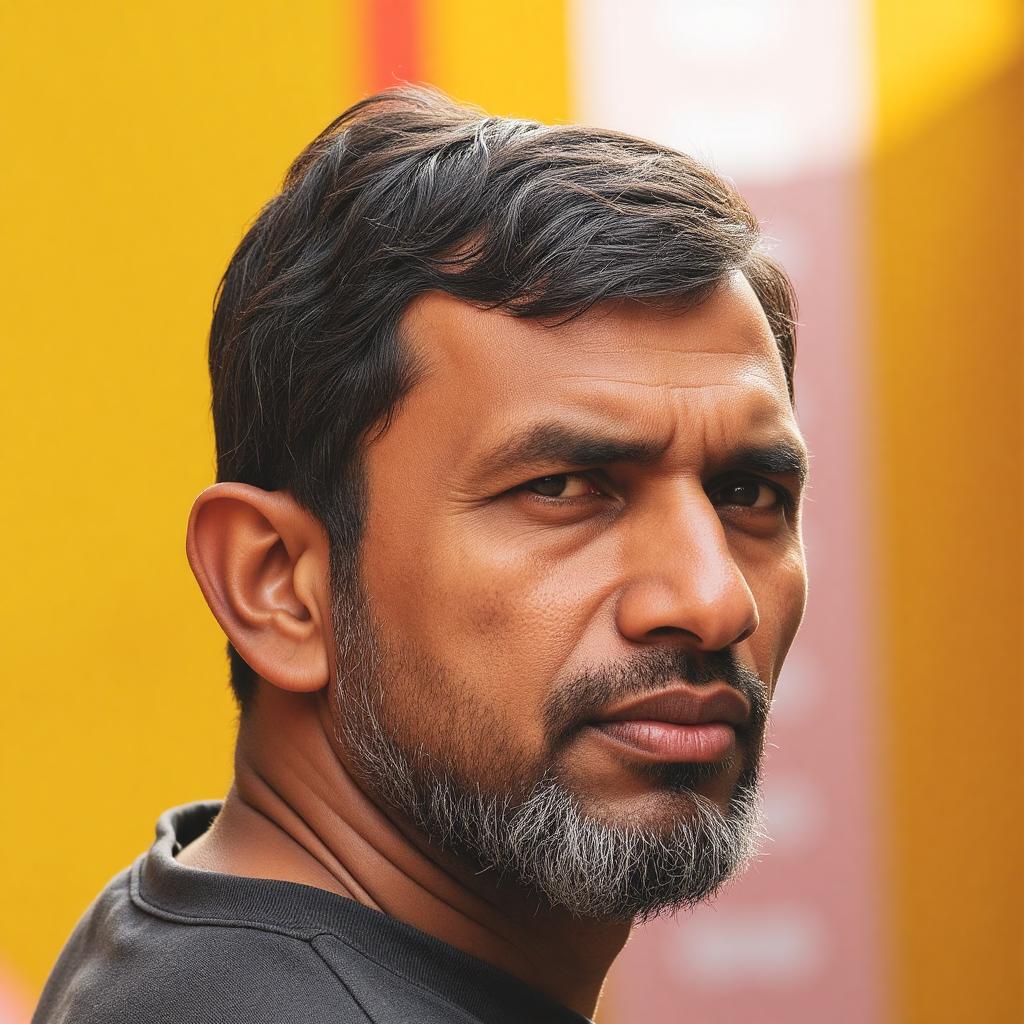} &
        \includegraphics[width=0.195\linewidth]{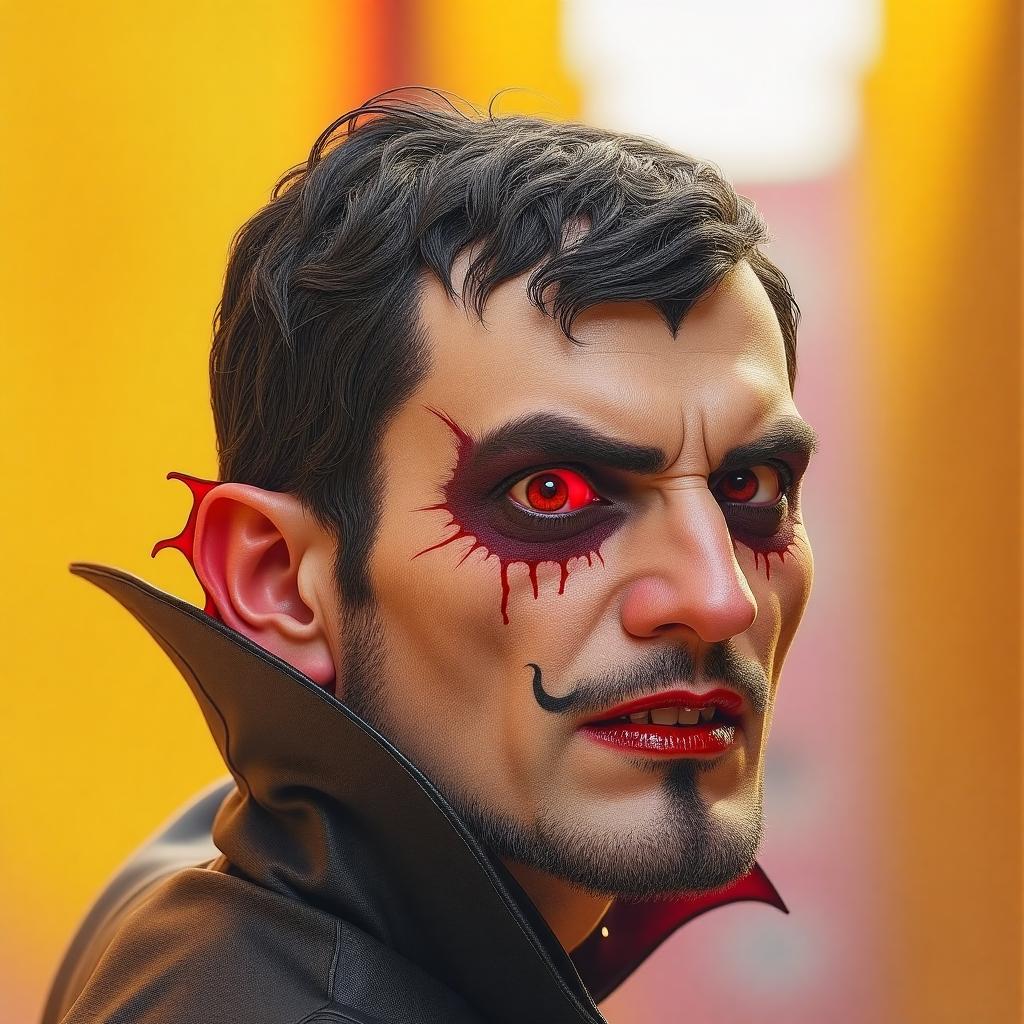} &
        \includegraphics[width=0.195\linewidth]{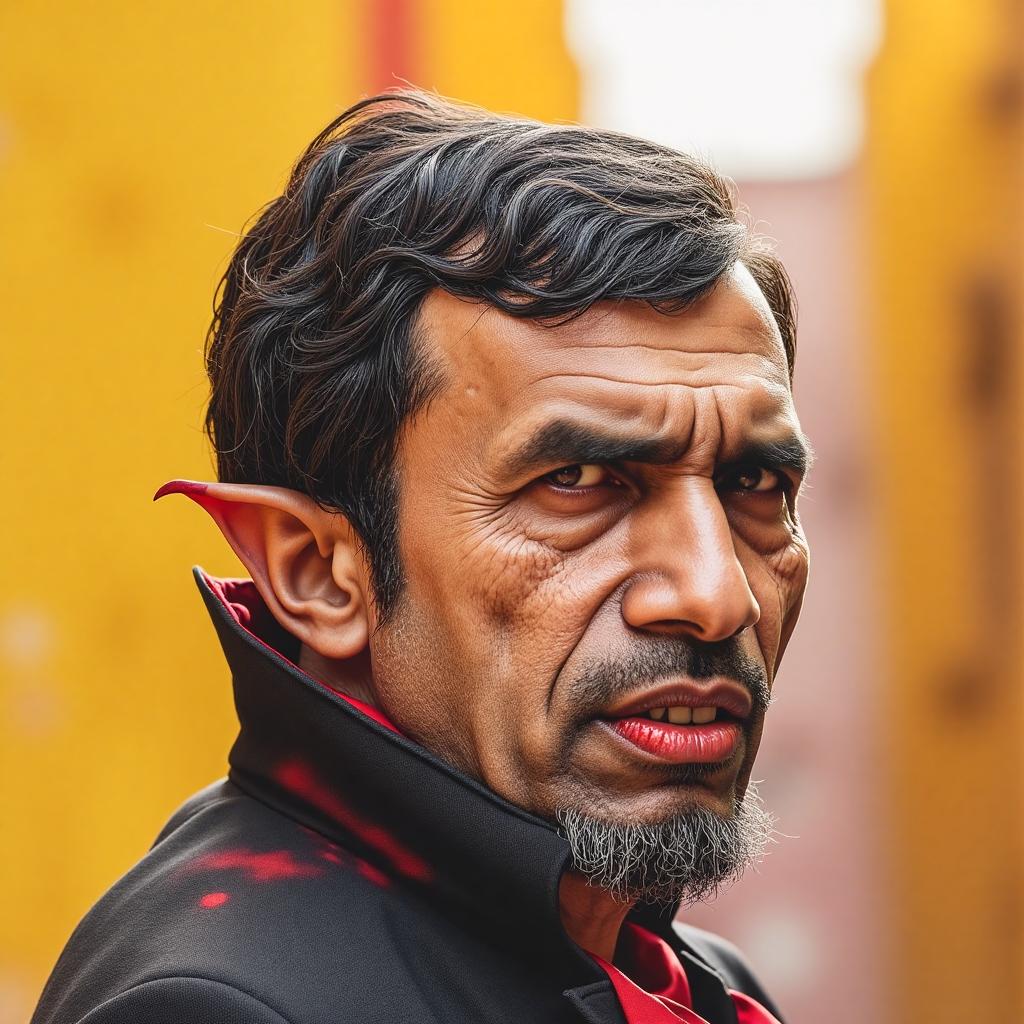} \\
        \multicolumn{5}{c}{``'' $\longrightarrow$ ``A bearded man wearing a hat'', RF-Inversion (Flux)} \\
        \includegraphics[width=0.195\linewidth]{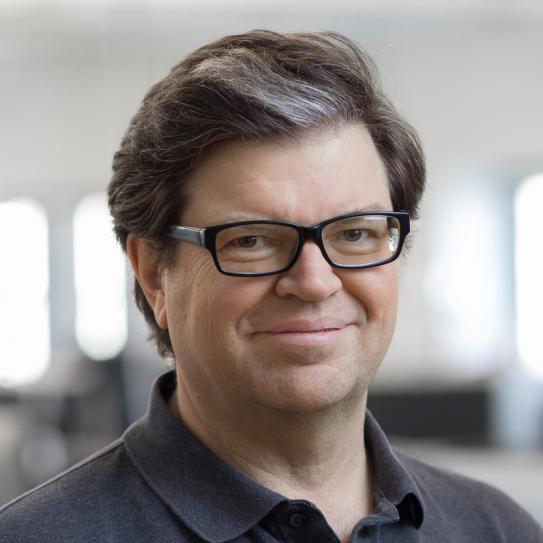} &
        \includegraphics[width=0.195\linewidth]{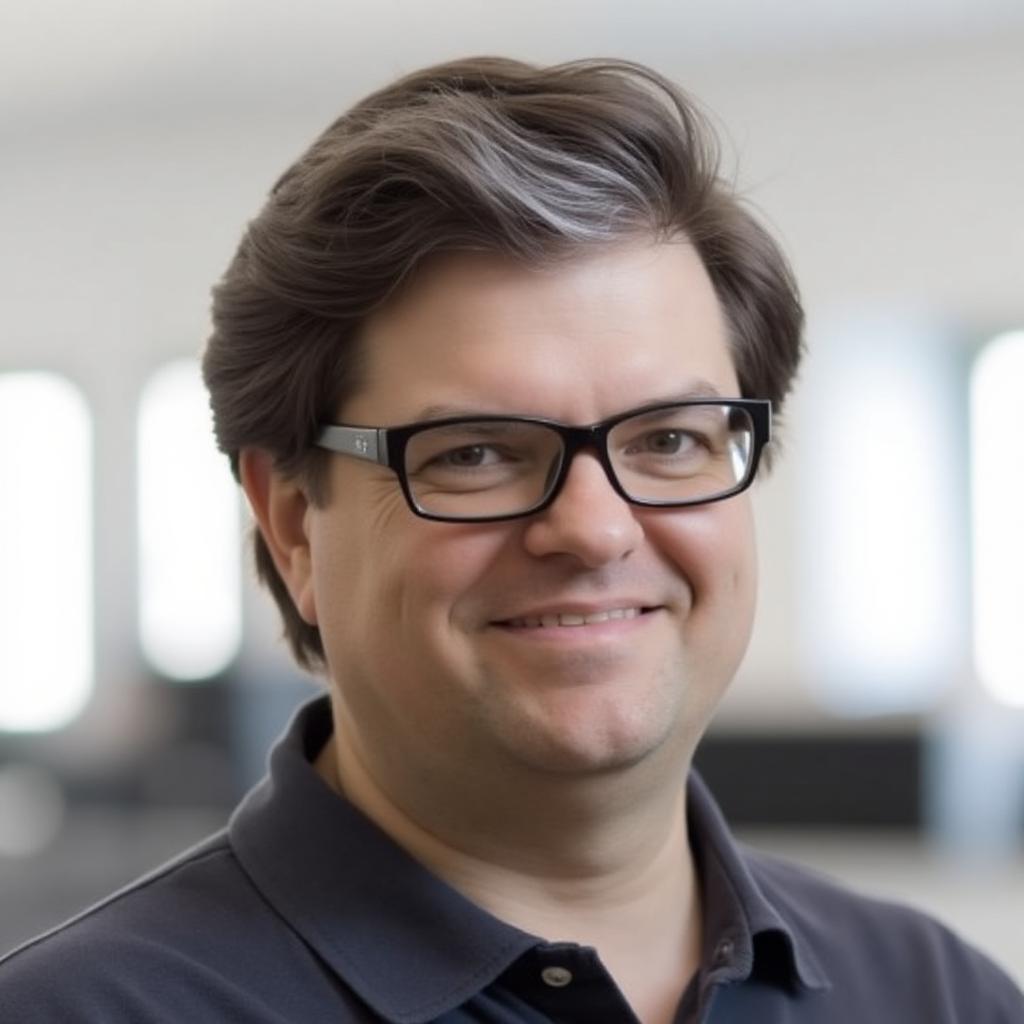} &
        \includegraphics[width=0.195\linewidth]{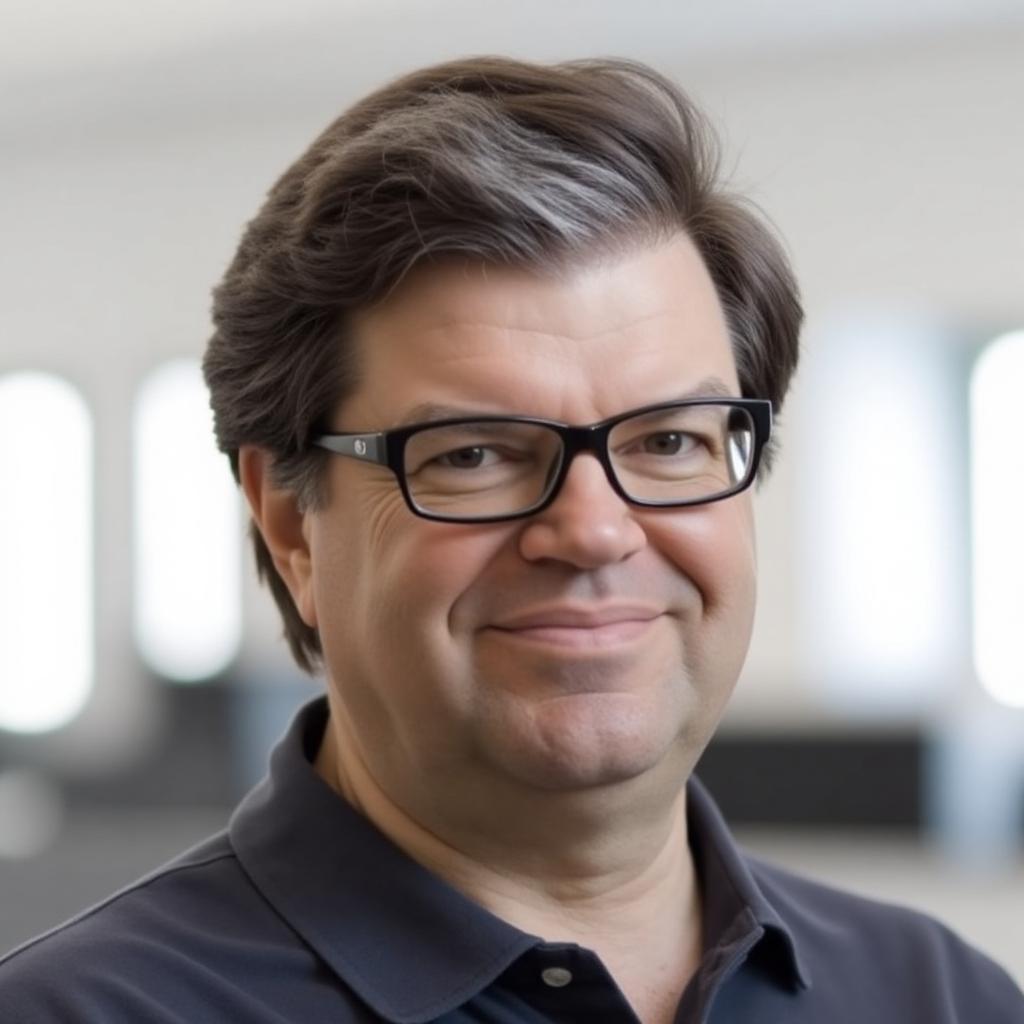} &
        \includegraphics[width=0.195\linewidth]{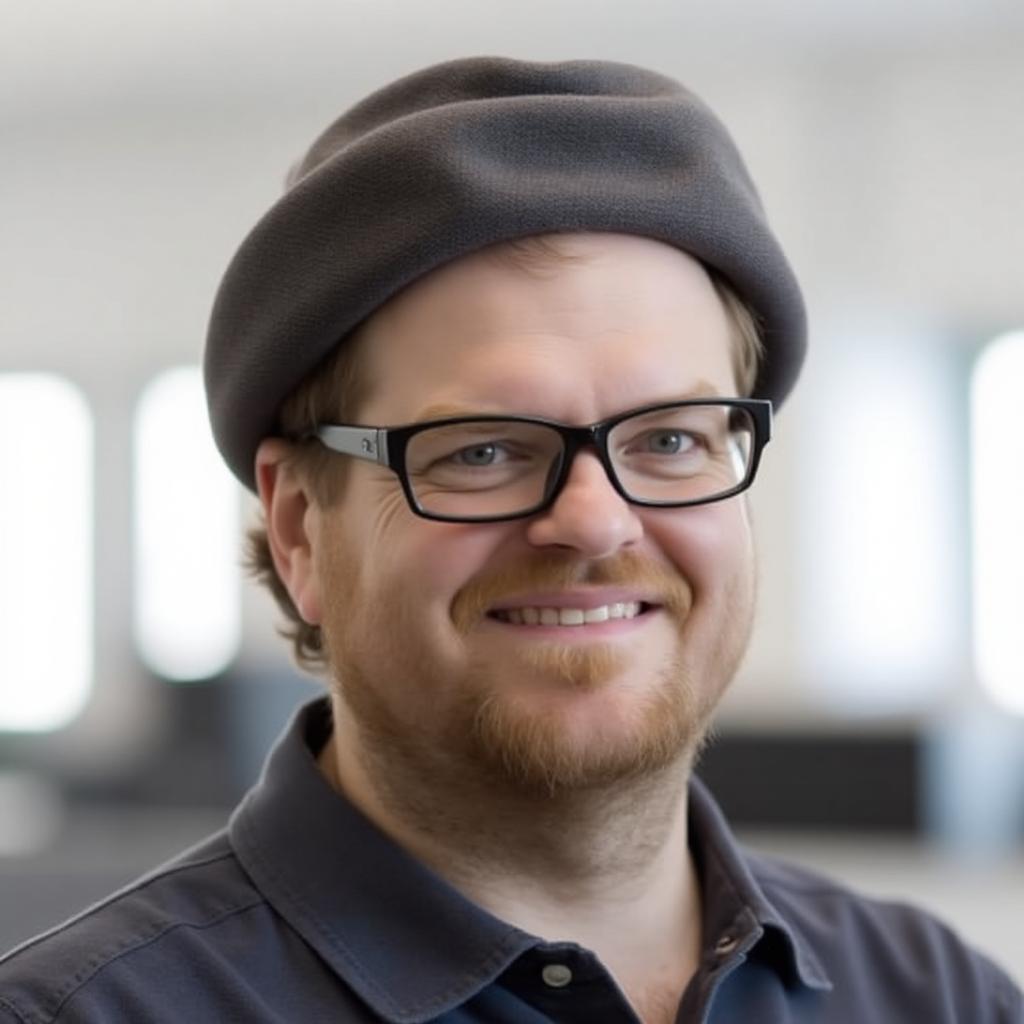} &
        \includegraphics[width=0.195\linewidth]{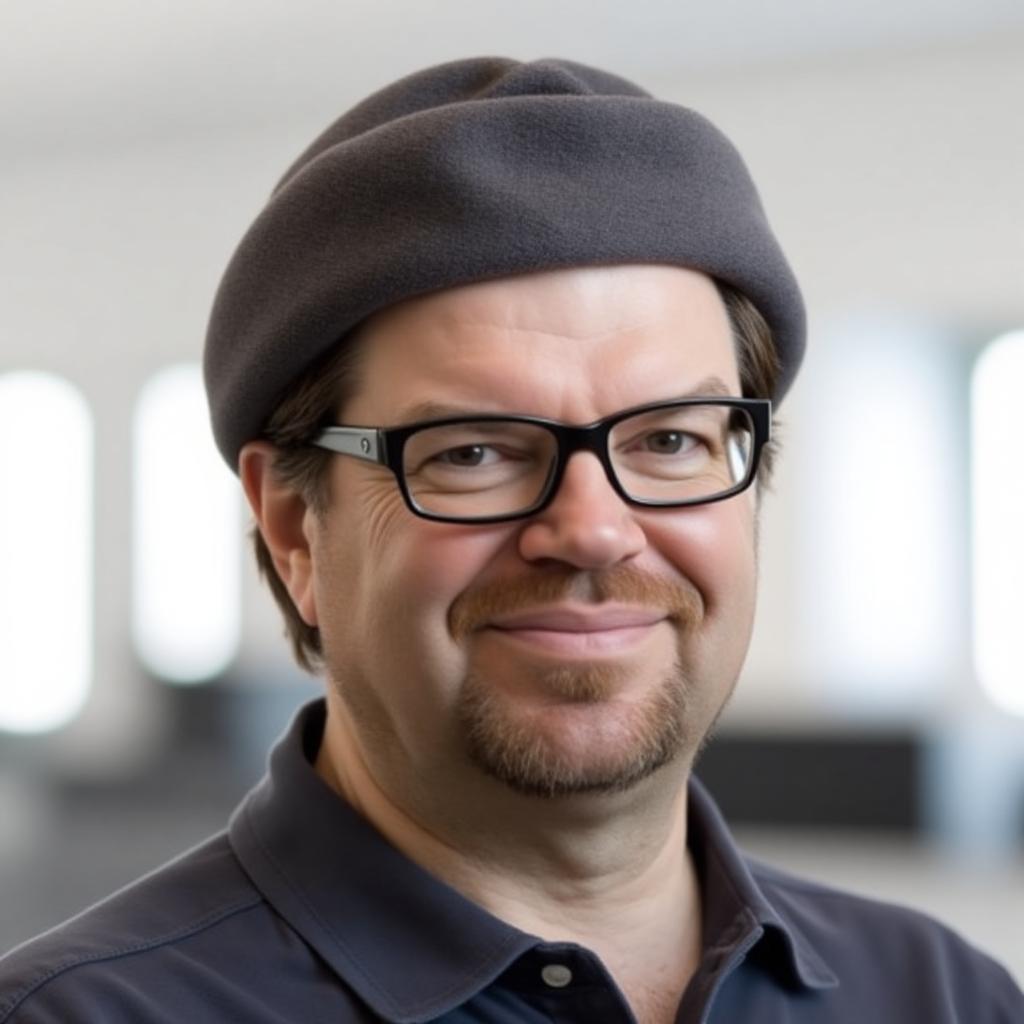} \\
        Input & Rec. w/o Tight & Rec. w/ Tight & Edit. w/o Tight & Edit. w/ Tight \\
    \end{tabular}
    }
    \caption{Combining Tight Inversion with various editing methods improves editability even in cases where the gap in reconstruction is negligible. Our method improves baseline methods for various editing types such as object addition, semantic modification, and pose modification.}
    \label{fig:edit-qualitative-comp}
\end{figure}

%% file: figures/challenging_edits_turbo.tex
\begin{figure}
    \centering
    \setlength{\tabcolsep}{1pt}
    \scriptsize{
    \begin{tabular}{cccccc}
        \multicolumn{5}{c}{``A picture of tea cups'' $\longrightarrow$ ``... white tea cups''} \\
        \includegraphics[width=0.195\linewidth]{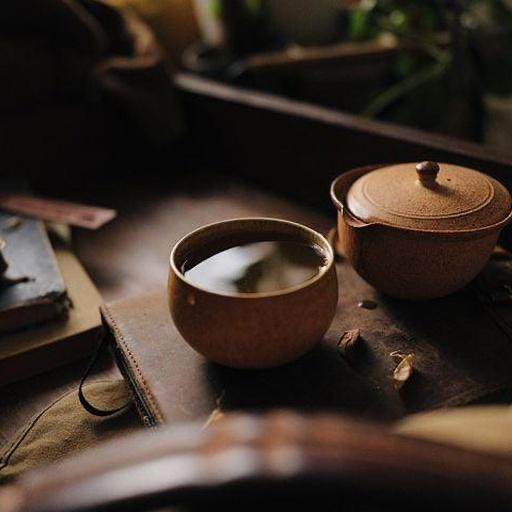} &
        \includegraphics[width=0.195\linewidth]{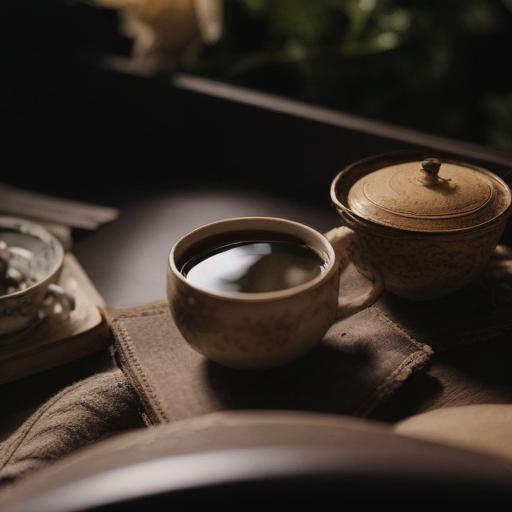} &
        \includegraphics[width=0.195\linewidth]{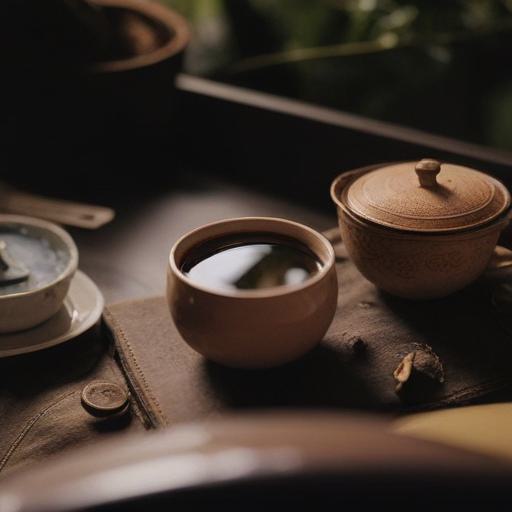} &
        \includegraphics[width=0.195\linewidth]{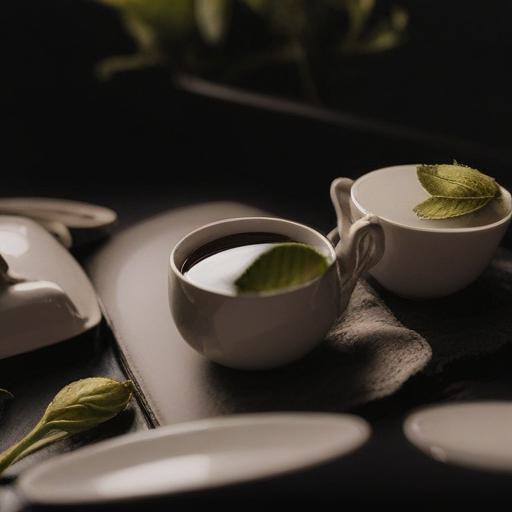} &
        \includegraphics[width=0.195\linewidth]{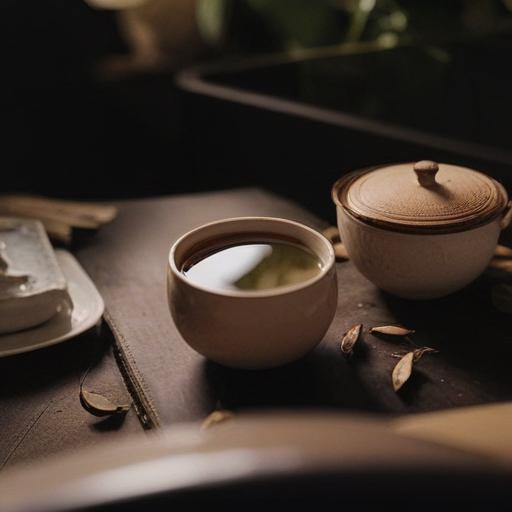} \\
        \multicolumn{5}{c}{``A woman is pulling a red suitcase down the sidewalk'' $\longrightarrow$} \\
        \multicolumn{5}{c}{ ``A woman is walking her dog and...''} \\
        \includegraphics[width=0.195\linewidth]{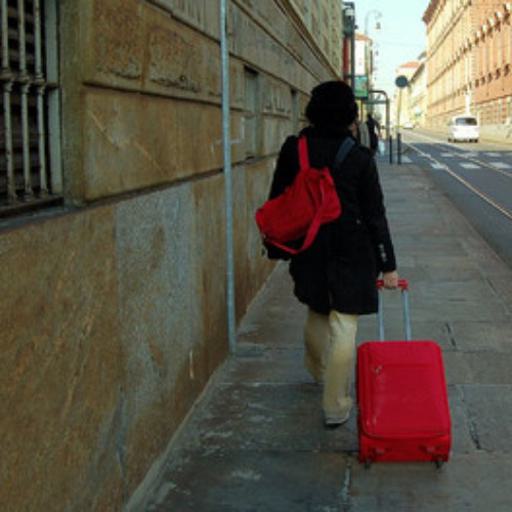} &
        \includegraphics[width=0.195\linewidth]{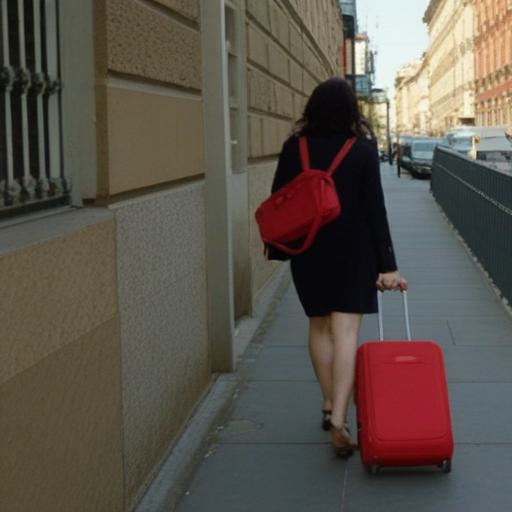} &
        \includegraphics[width=0.195\linewidth]{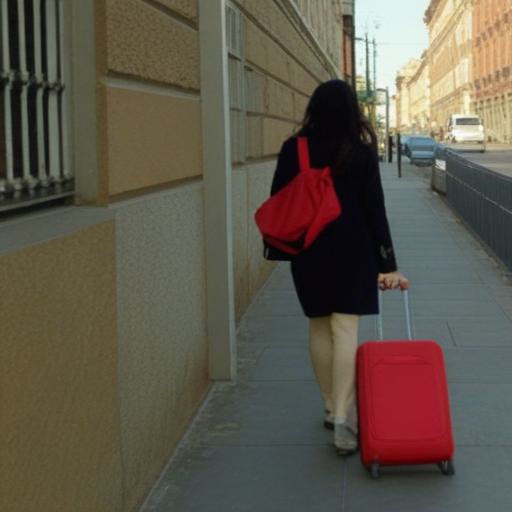} &
        \includegraphics[width=0.195\linewidth]{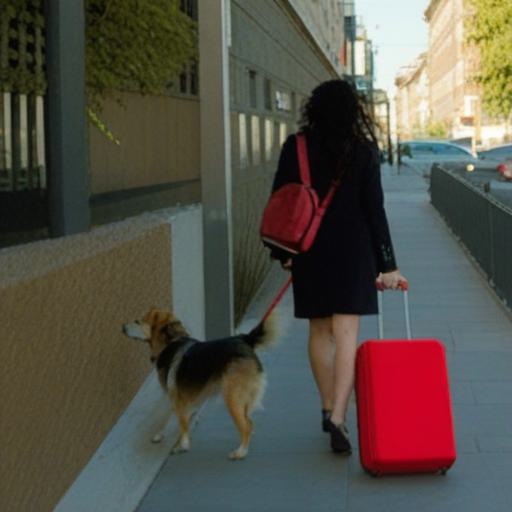} &
        \includegraphics[width=0.195\linewidth]{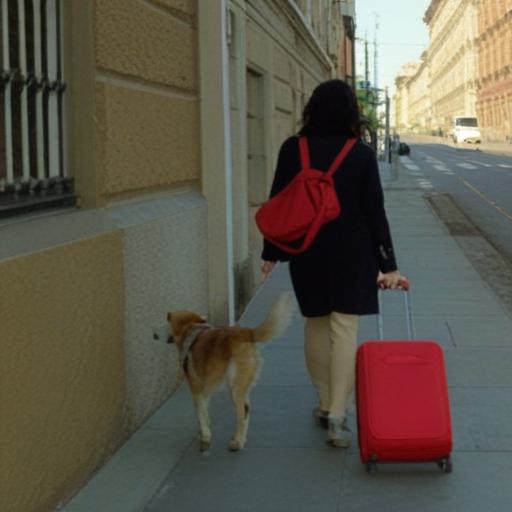} \\
        Input & Recon. w/o TiTi & Recon. w/ TiTi & Edit. w/o TiTi & Edit. w/ TiTi \\
    \end{tabular}
    }
    \caption{Combining Tight Inversion with ReNoise, where using SDXL-Turbo with 4 steps of denoising. The edit is applied by using a modified prompt in the denoising process.}
    \label{fig:turbo_results}
\end{figure}

%% file: sec/5_conclusion.tex
\vspace{6pt}
\section{Conclusions}

In this work, we explored the role of tight conditioning in addressing the challenges of the inversion task for diffusion-based image editing.
While significant progress has been made in image editing with diffusion models, these models continue to struggle with complex, real-world images that fall outside their training distribution—precisely the type of images users often wish to edit. This challenge motivated our focus on improving performance in such demanding scenarios.

We demonstrated the power of using an image as a conditioning input, reaffirming the adage that ``a picture is worth a thousand words''. Conditioning on an image significantly enhances inversion quality compared to relying solely on text prompts, offering a more robust solution for real-world cases.
Our method provides a plug-and-play enhancement that is compatible with any inversion technique. Experimental results show that Tight Inversion improves both reconstruction fidelity and editing quality, without imposing significant computational or runtime overhead.

However, our approach is not without limitations. It is constrained by the inherent tradeoff between reconstruction accuracy and editability, as excessively strong conditioning can reduce the flexibility required for effective editing.

In this work, we employed IP-Adapter and PuLID to condition the model on the source image. However, our method is versatile and can be integrated with other image conditioning mechanisms. As future work, we aim to develop novel image conditioning techniques specifically tailored to further enhance the inversion task.

%% file: sec/6_figures_only.tex
\input{figures/additional_edits}

%% file: figures/additional_edits.tex
\begin{figure*}[t]
    \centering
    \begin{minipage}[t]{0.48\textwidth}
        \centering
        \setlength{\tabcolsep}{1pt}
        \scriptsize{
        \begin{tabular}{ccc}
        \multicolumn{3}{c}{``A plate contains beef with a side of broccoli'' $\longrightarrow$ ``... + fries'', DDIM Inversion} \\
        \includegraphics[width=0.33\linewidth]{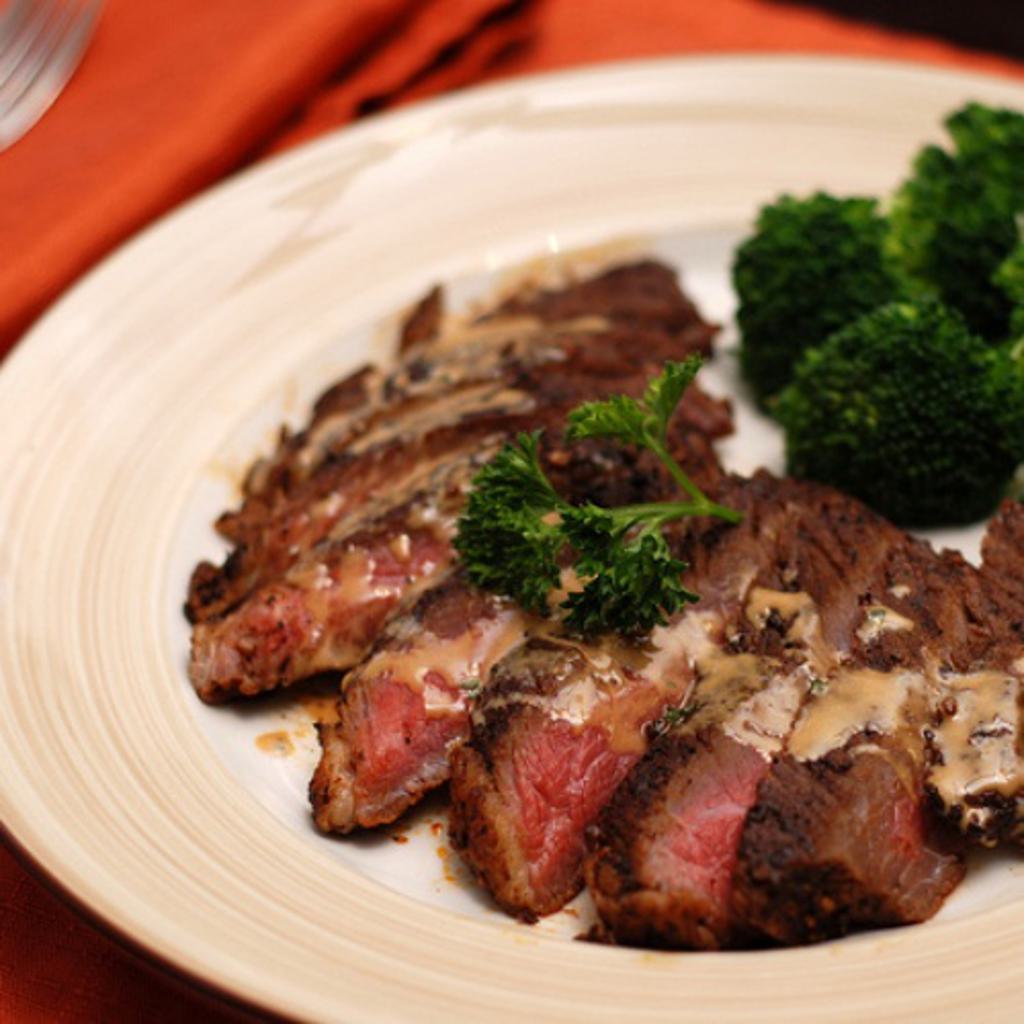} &
        \includegraphics[width=0.33\linewidth]{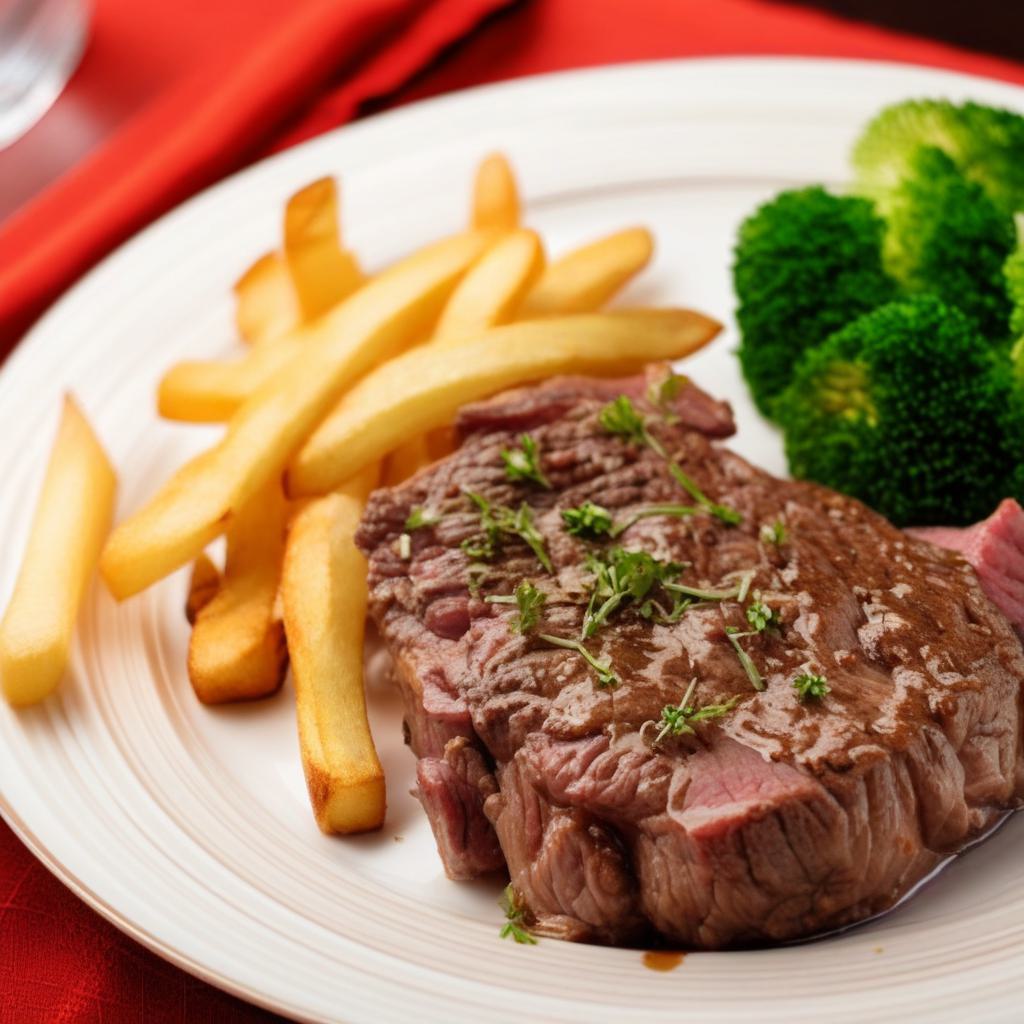} &
        \includegraphics[width=0.33\linewidth]{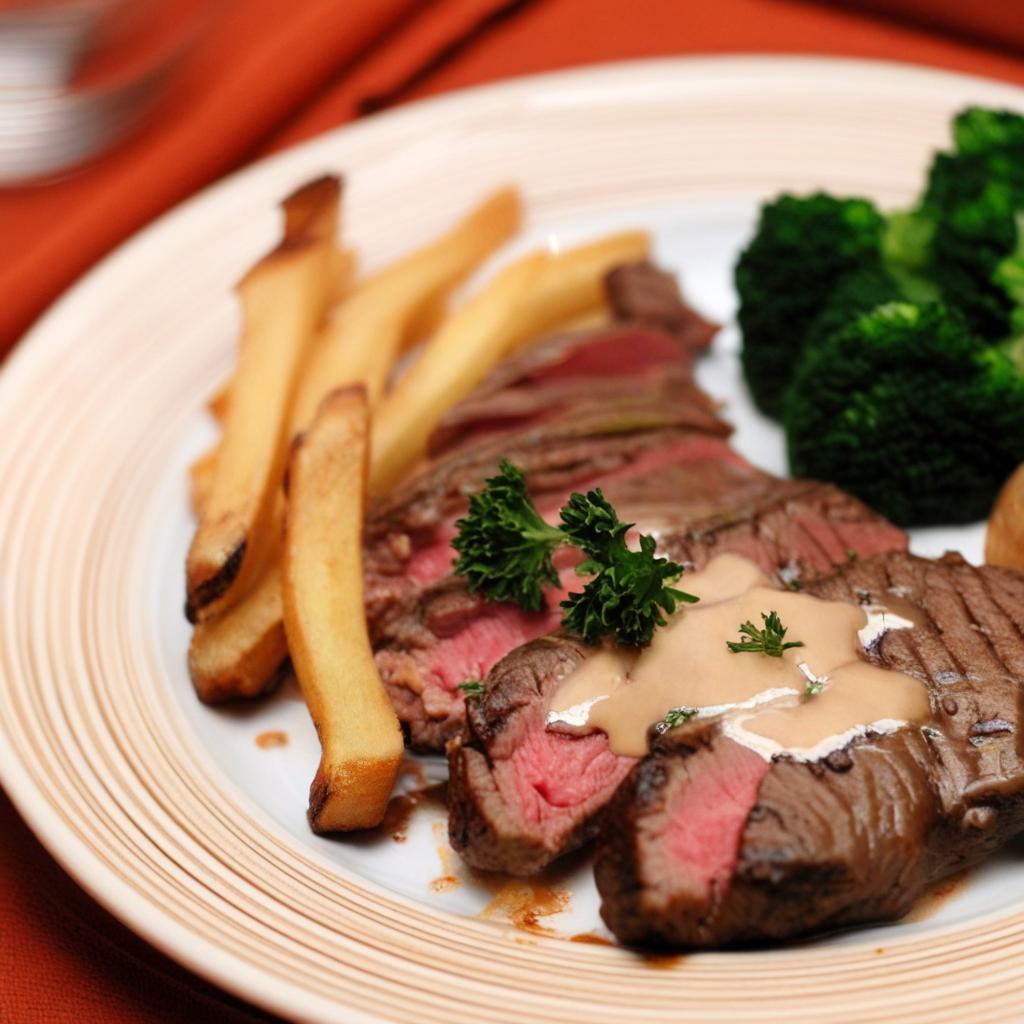}          
        \\
        \multicolumn{3}{c}{``A dog running in the forest'' $\longrightarrow$ ``A forest with no one around'', DDIM Inversion} \\
        \includegraphics[width=0.33\linewidth]{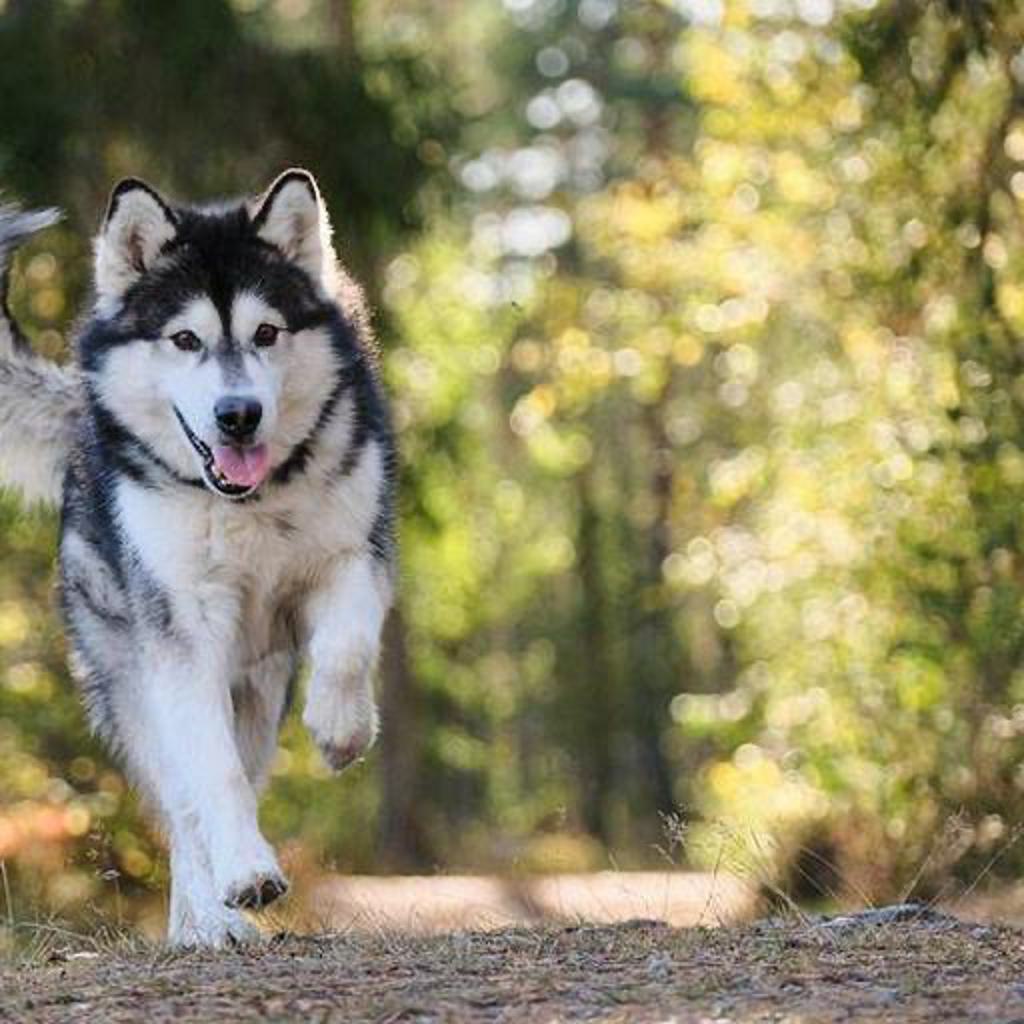} &
        \includegraphics[width=0.33\linewidth]{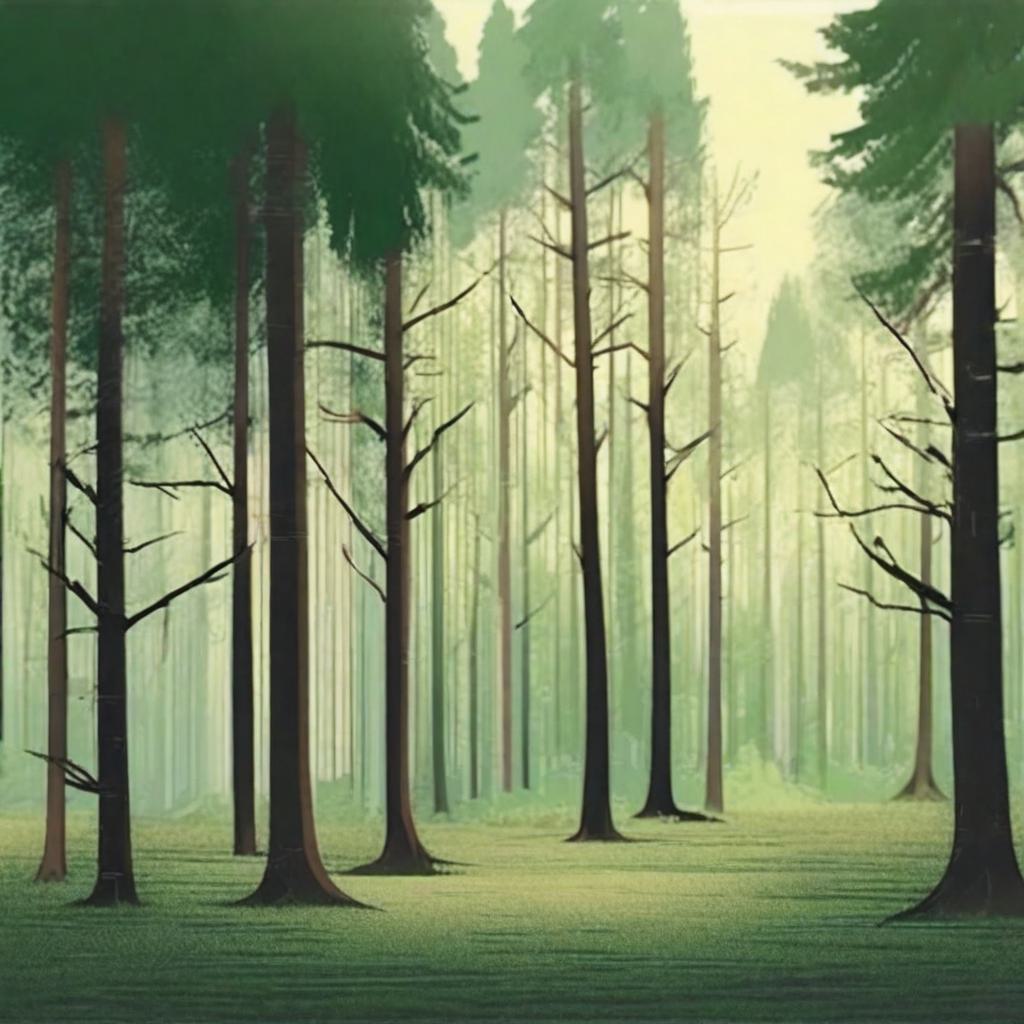} &
        \includegraphics[width=0.33\linewidth]{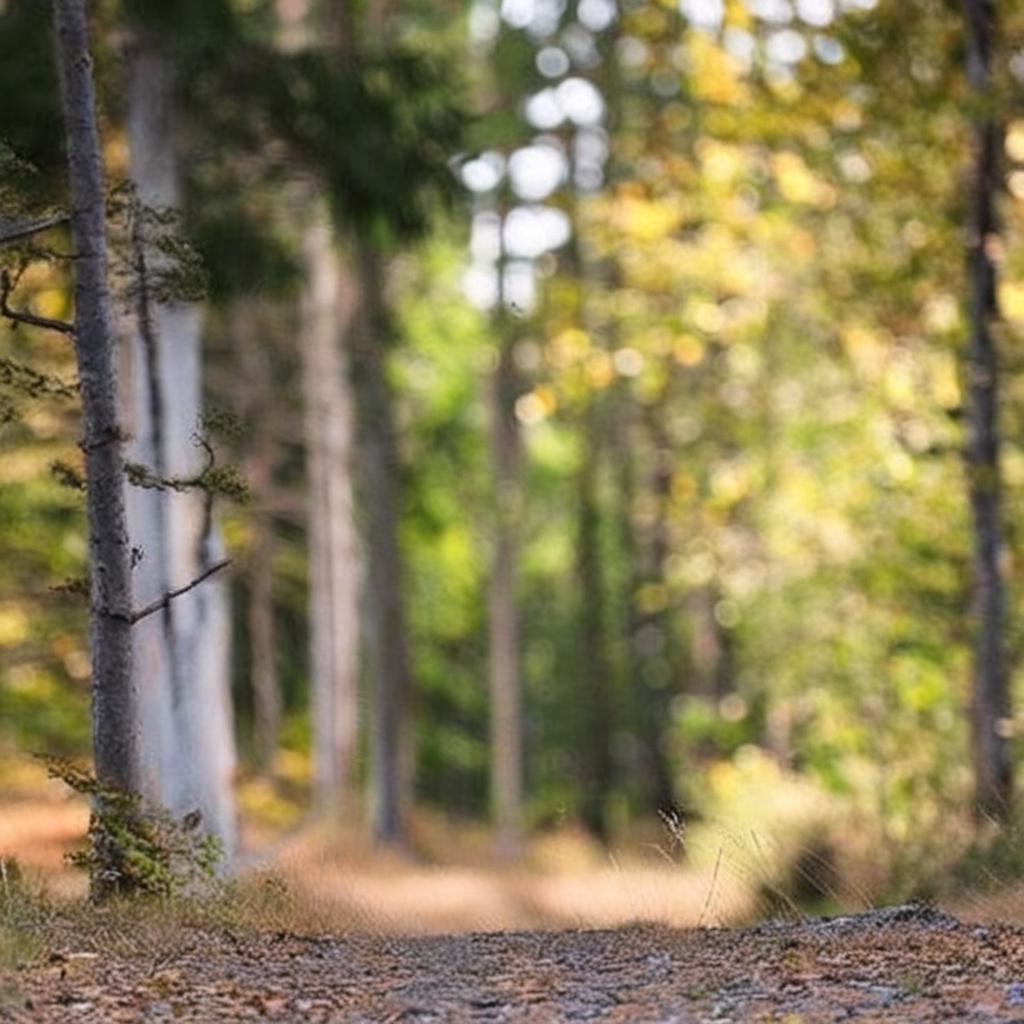}        
        \\
        \multicolumn{3}{c}{``A lion in the field'' $\longrightarrow$ ``... made of lego'', Negative Prompt Inversion} \\
        \includegraphics[width=0.33\linewidth]{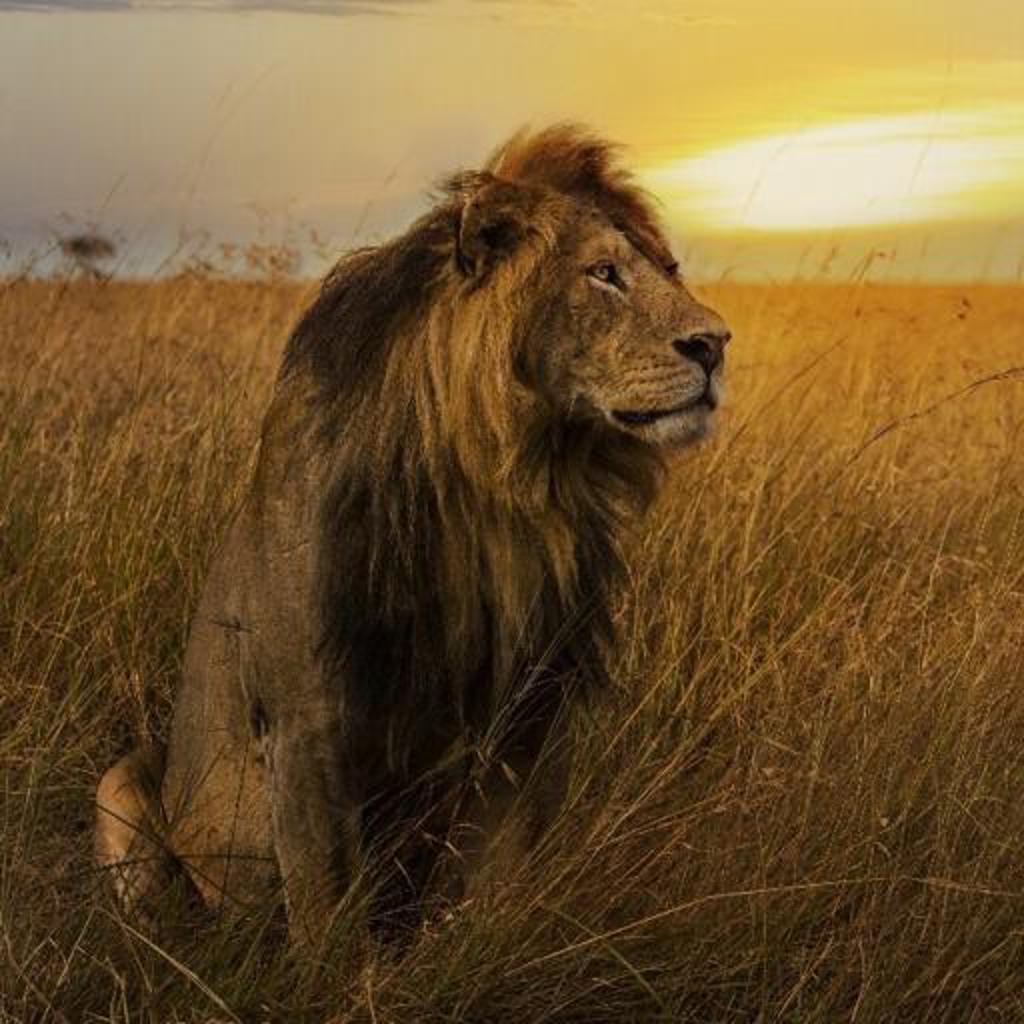} &
        \includegraphics[width=0.33\linewidth]{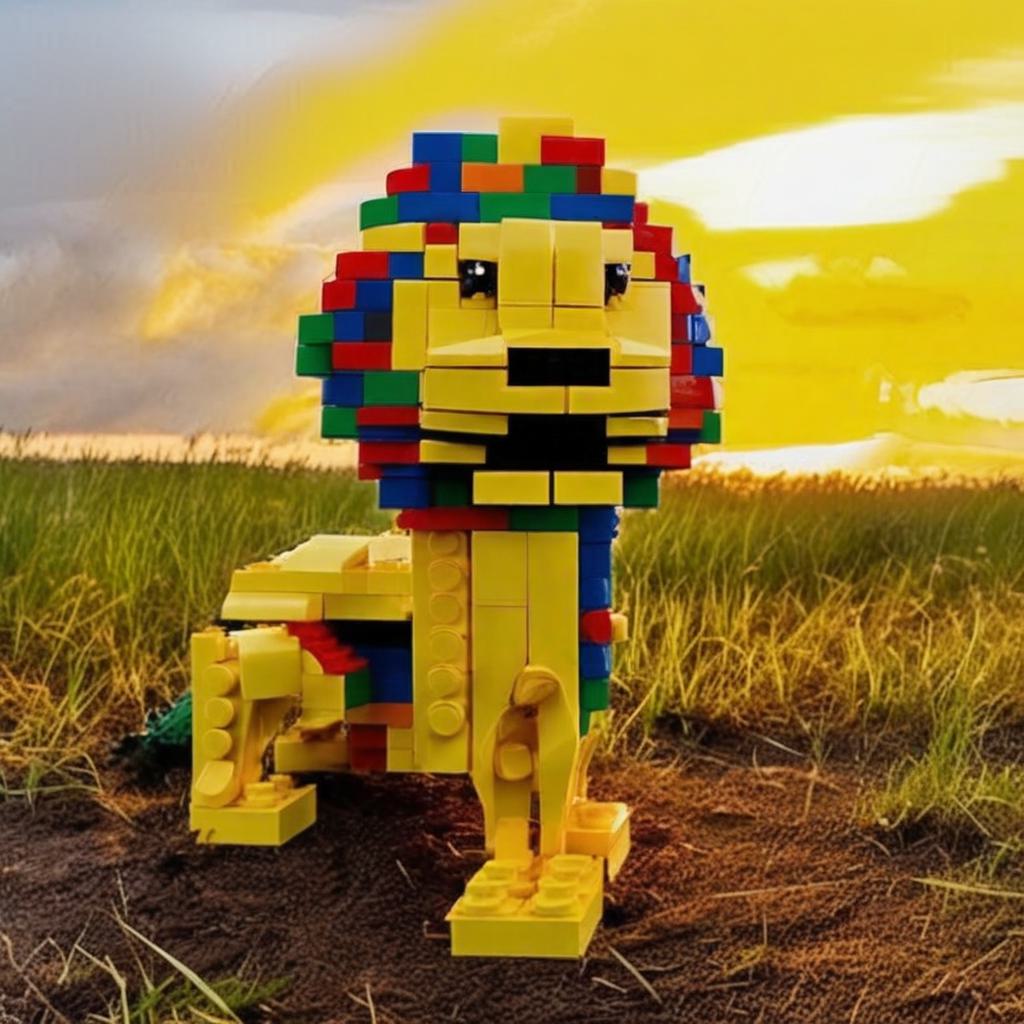} &
        \includegraphics[width=0.33\linewidth]{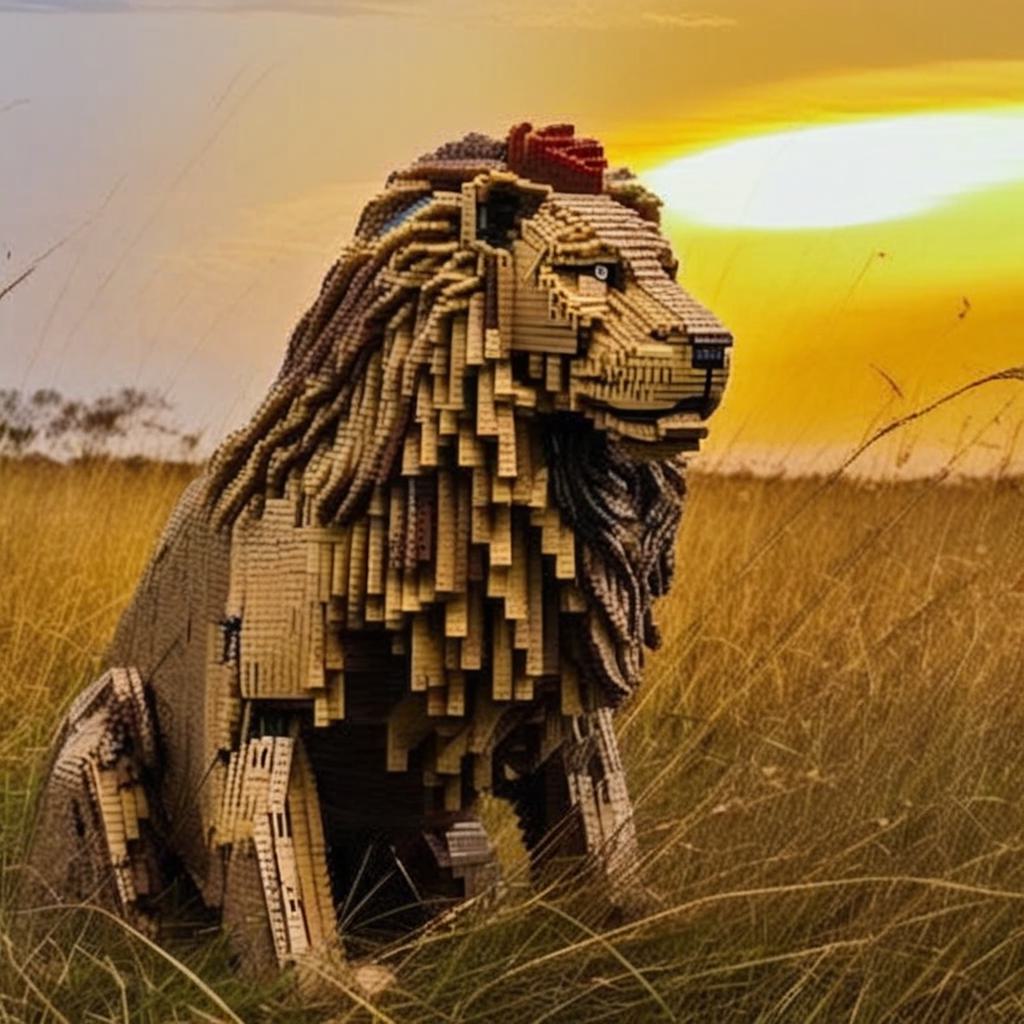}          
        \\
        \multicolumn{3}{c}{``Fresh fruits and vegetables displayed for sale'' $\longrightarrow$ ``Sliced watermelon and ...'', } \\
        \multicolumn{3}{c}{DDIM Inversion} \\
        \includegraphics[width=0.33\linewidth]{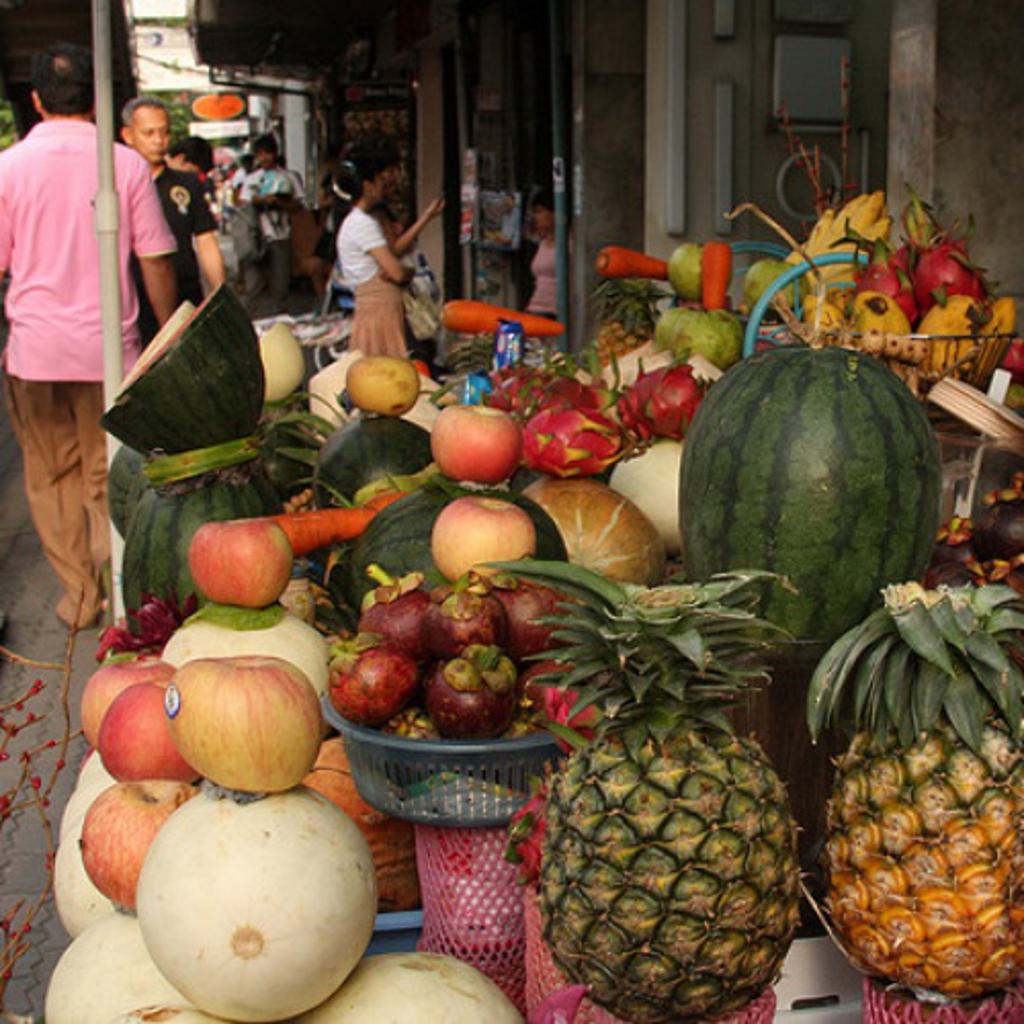} &
        \includegraphics[width=0.33\linewidth]{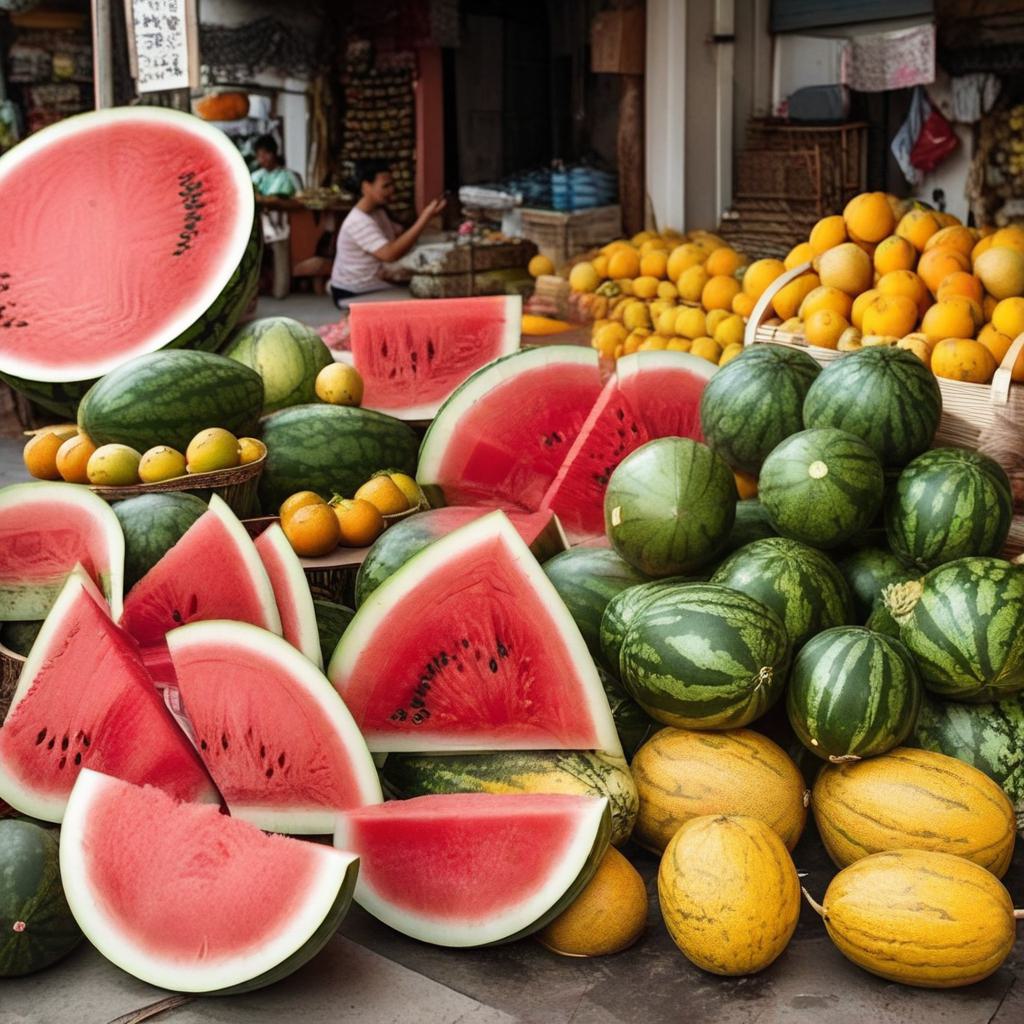} &
        \includegraphics[width=0.33\linewidth]{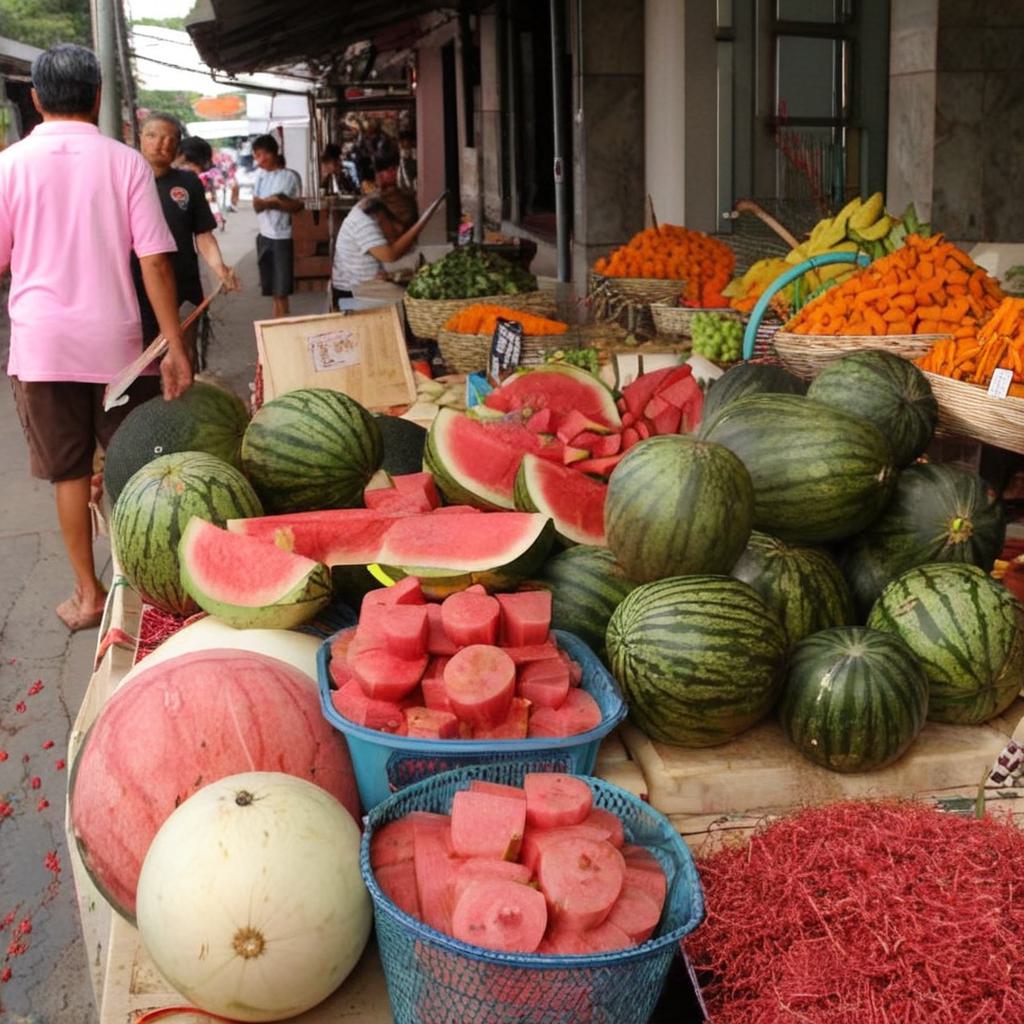}  
        \\
        \multicolumn{3}{c}{``A combination Angry Birds birthday and graduation cake'' $\longrightarrow$} \\
        \multicolumn{3}{c}{A combination flower themed ...'', LEDITS++} \\
        \includegraphics[width=0.33\linewidth]{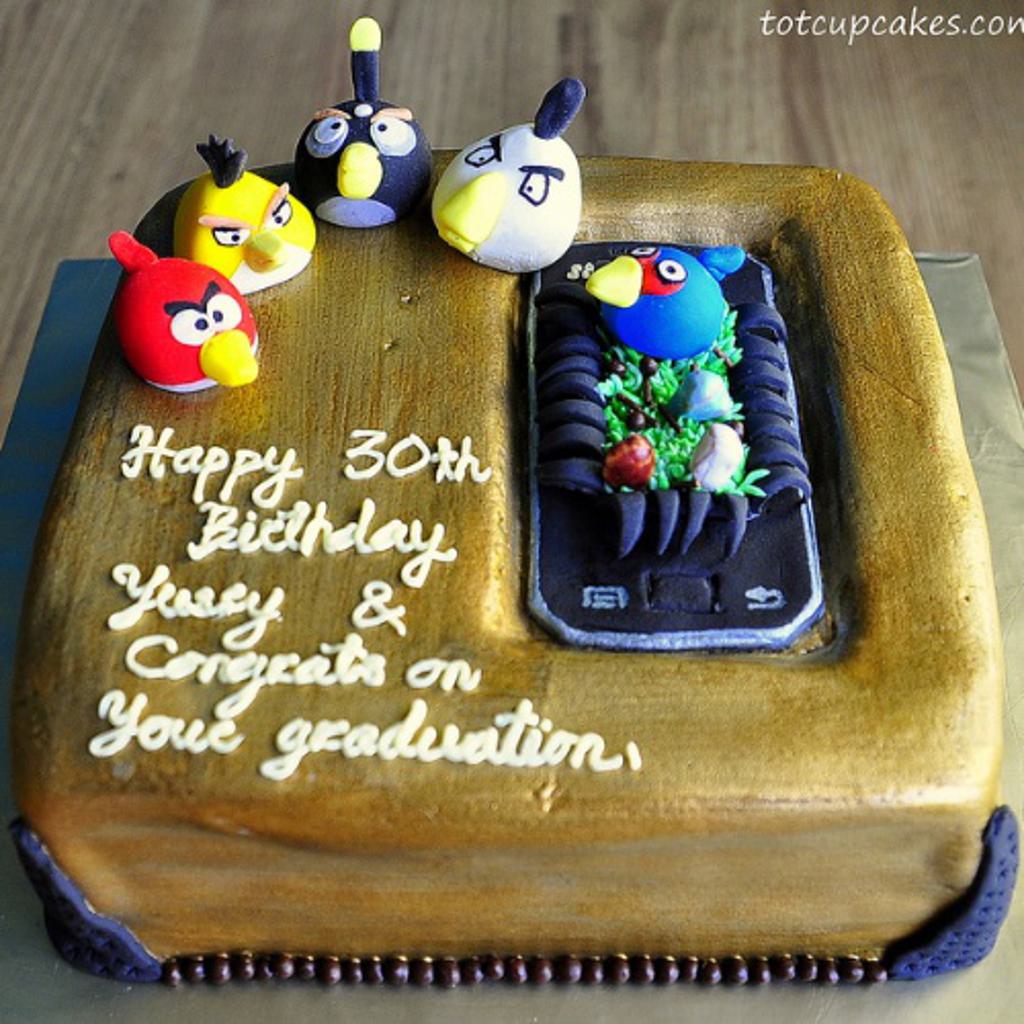} &
        \includegraphics[width=0.33\linewidth]{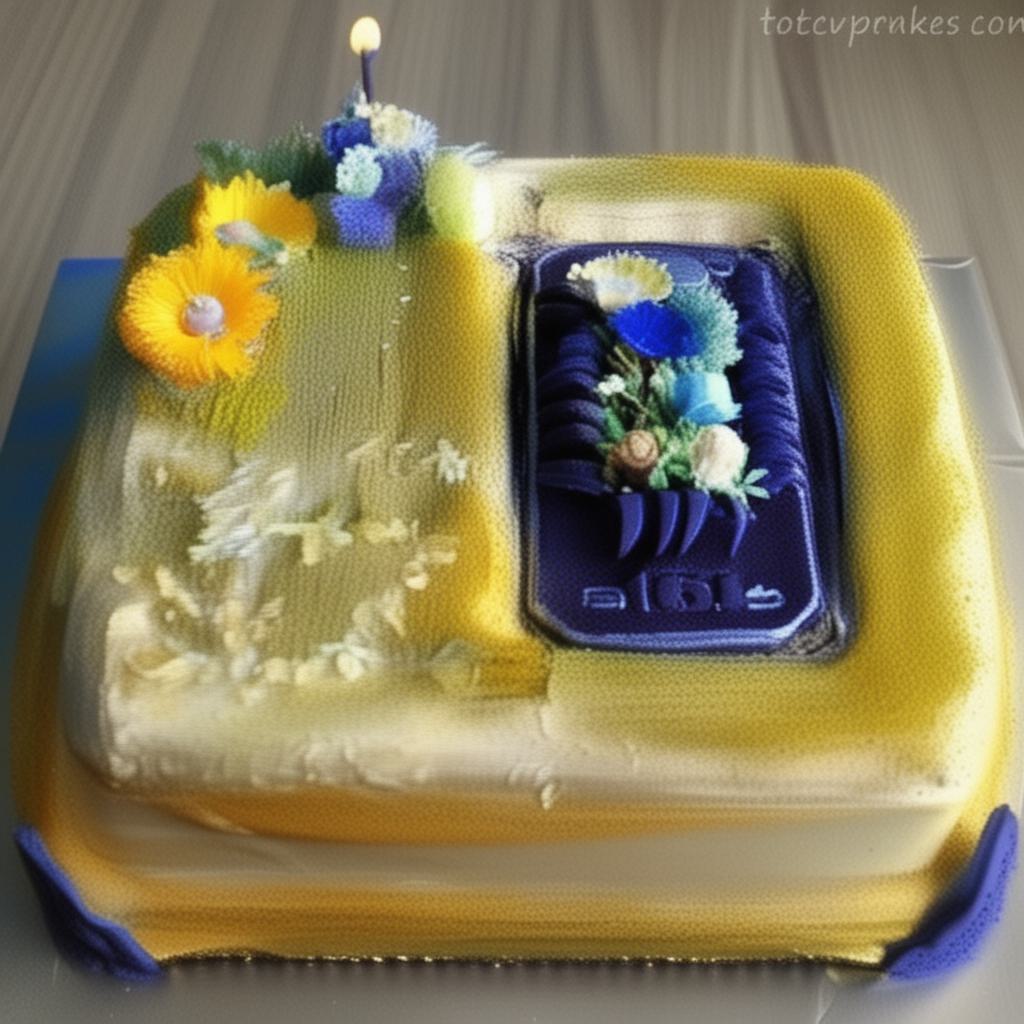} &
        \includegraphics[width=0.33\linewidth]{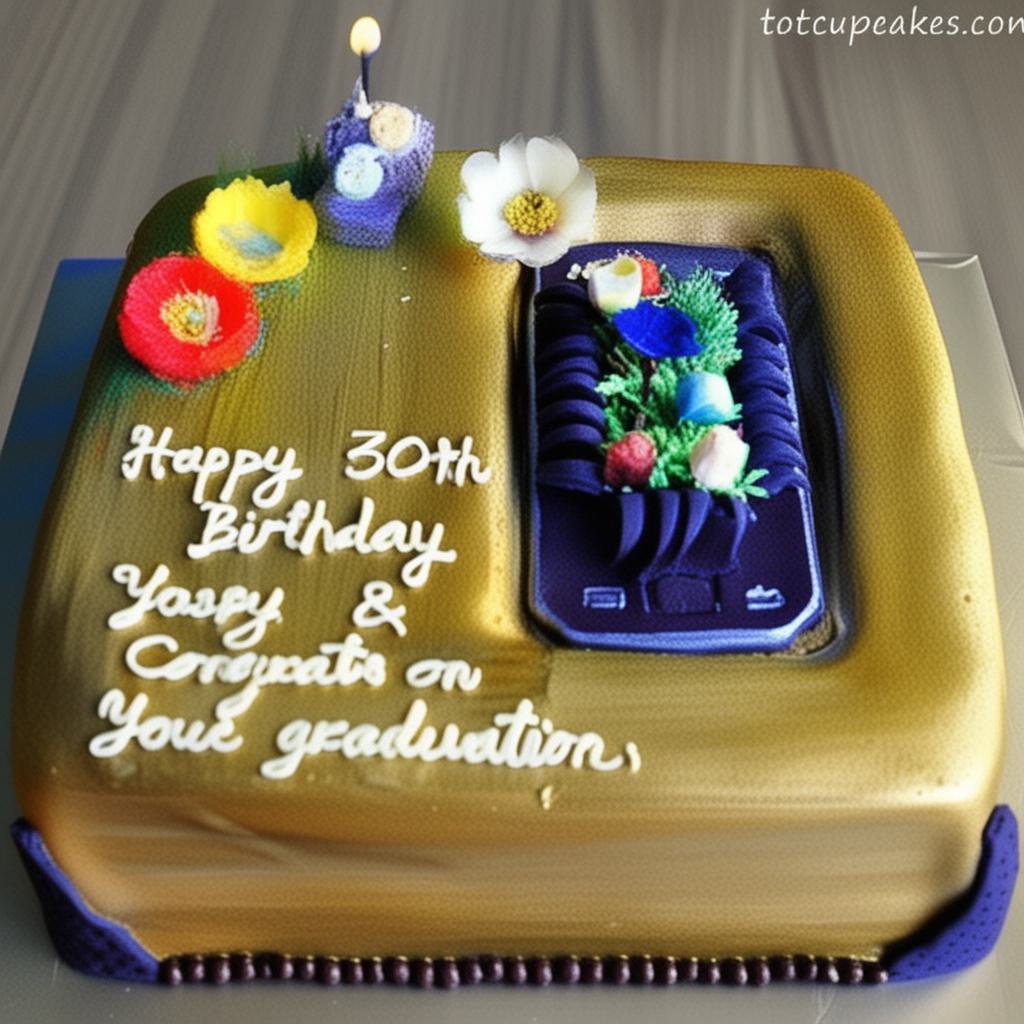} 
        \\
        \multicolumn{3}{c}{``A Teddy Bear cake on a wooden table for a 30th birthday celebration'' $\longrightarrow$} \\
        \multicolumn{3}{c}{``... 4th birthday celebration'', LEDITS++} \\
        \includegraphics[width=0.33\linewidth]{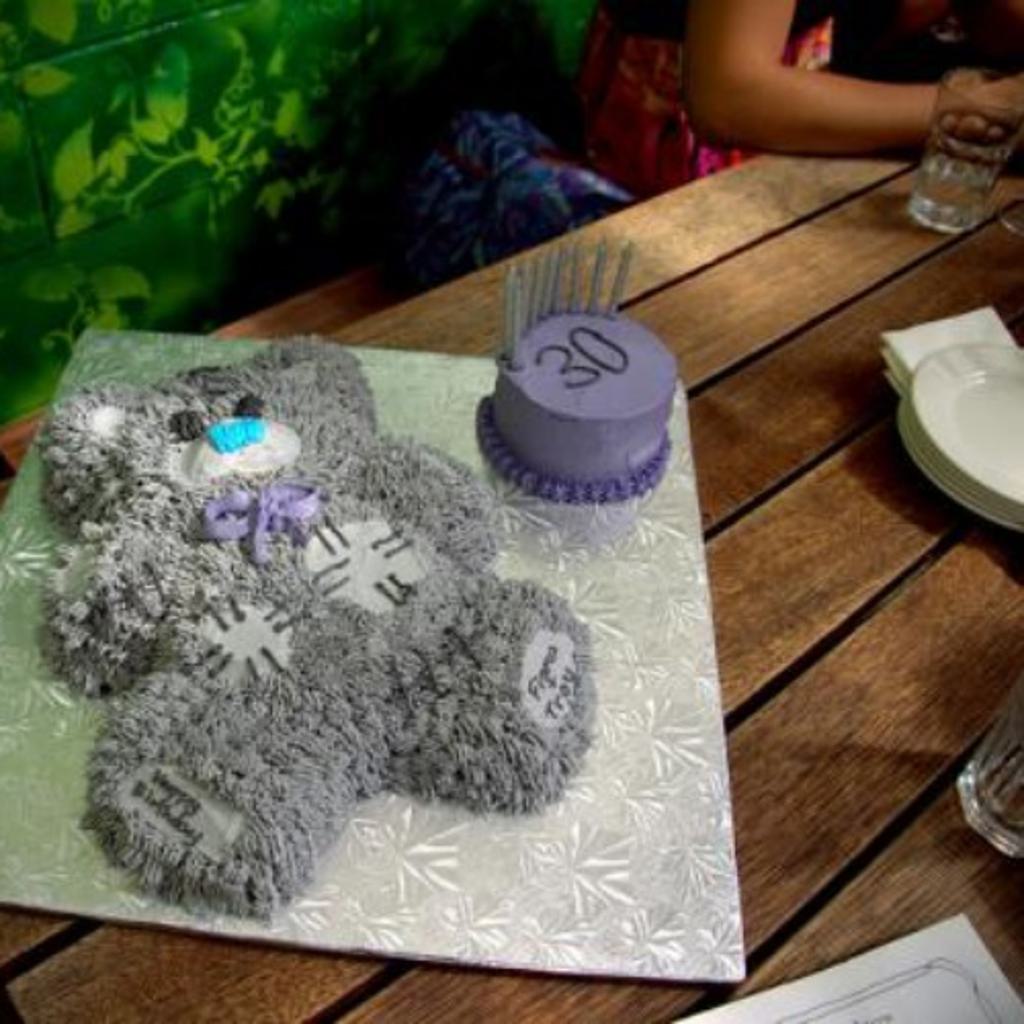} &
        \includegraphics[width=0.33\linewidth]{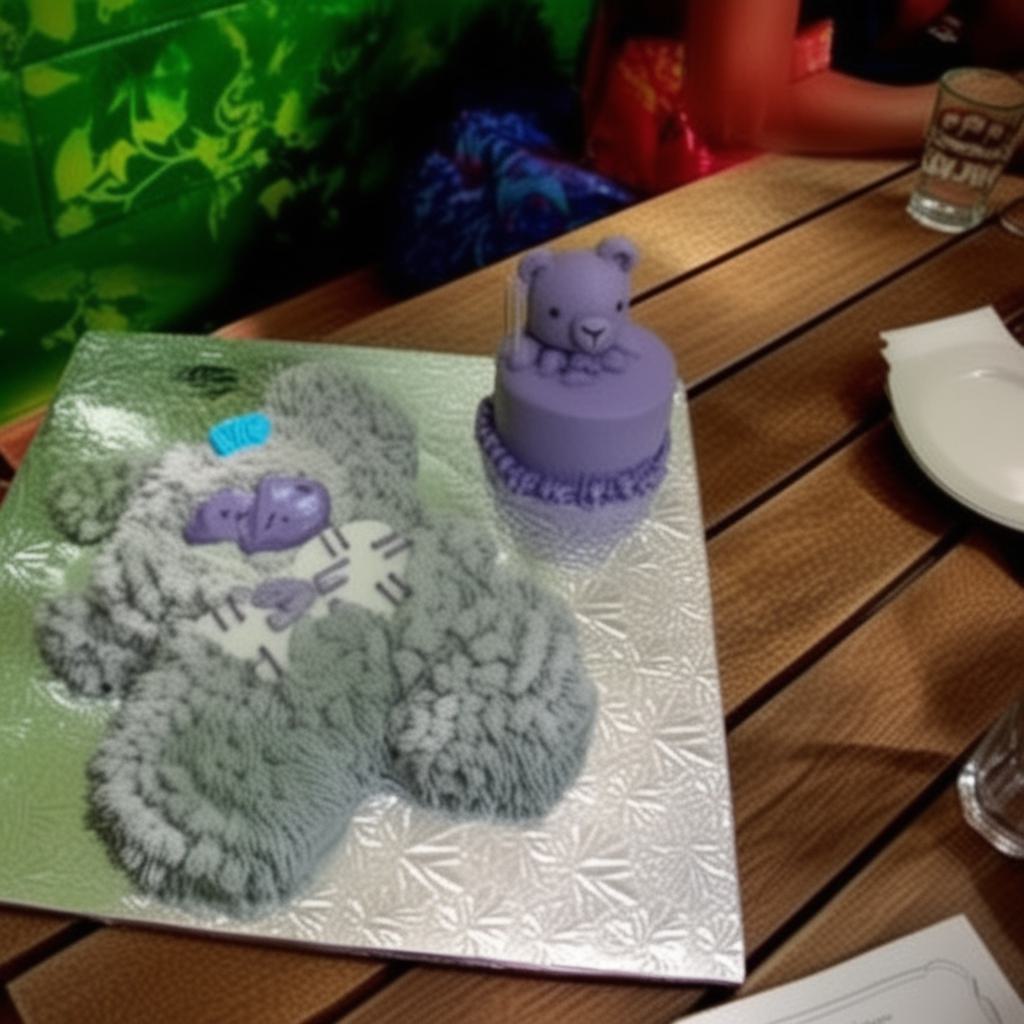} &
        \includegraphics[width=0.33\linewidth]{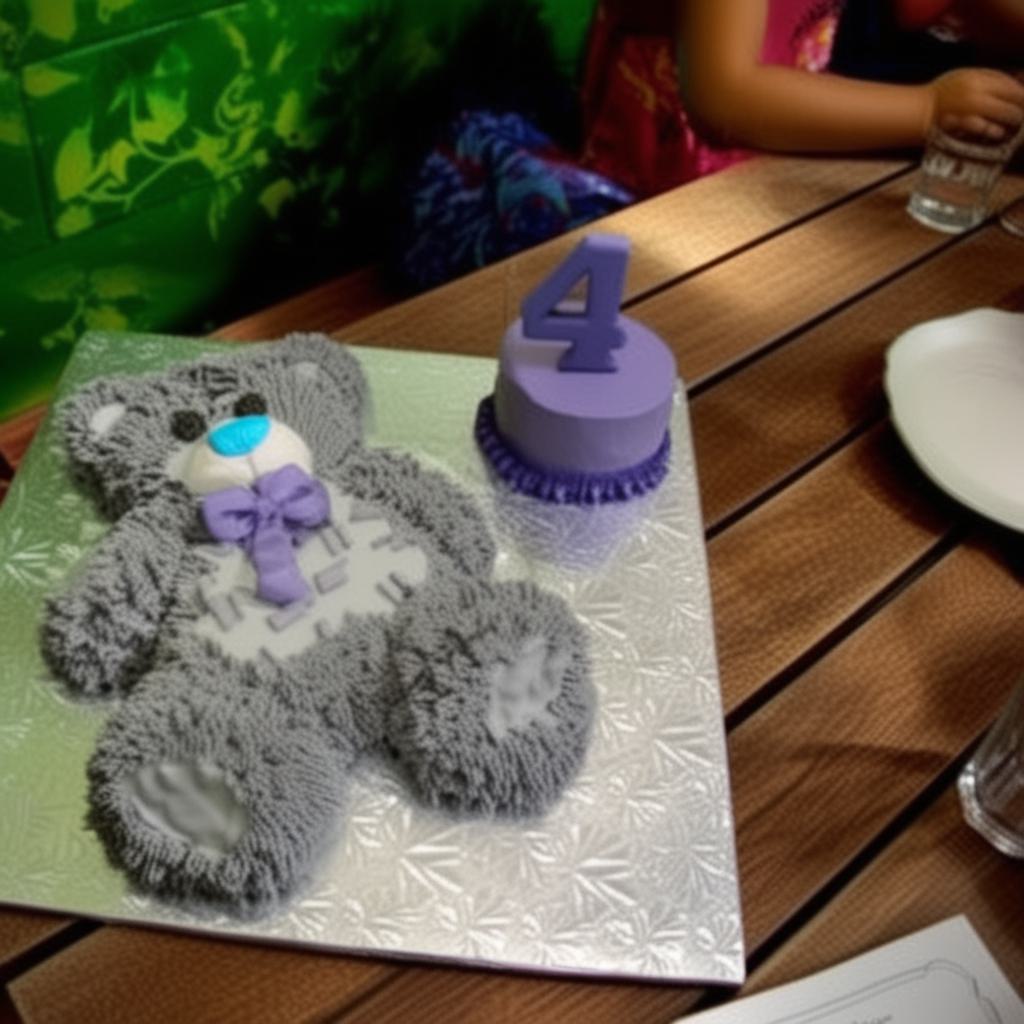} 
        \\
        Input & Edit. w/o Tight & Edit. w/ Tight 
        \end{tabular}
        }
    \end{minipage}%
    \hfill
    \begin{minipage}[t]{0.48\textwidth}
        \centering
        \setlength{\tabcolsep}{1pt}
        \scriptsize{
        \begin{tabular}{ccc}
            \multicolumn{3}{c}{``'' $\longrightarrow$ ``a portrait of a pirate'', RF-Inversion (Flux)} \\
            \includegraphics[width=0.33\linewidth]{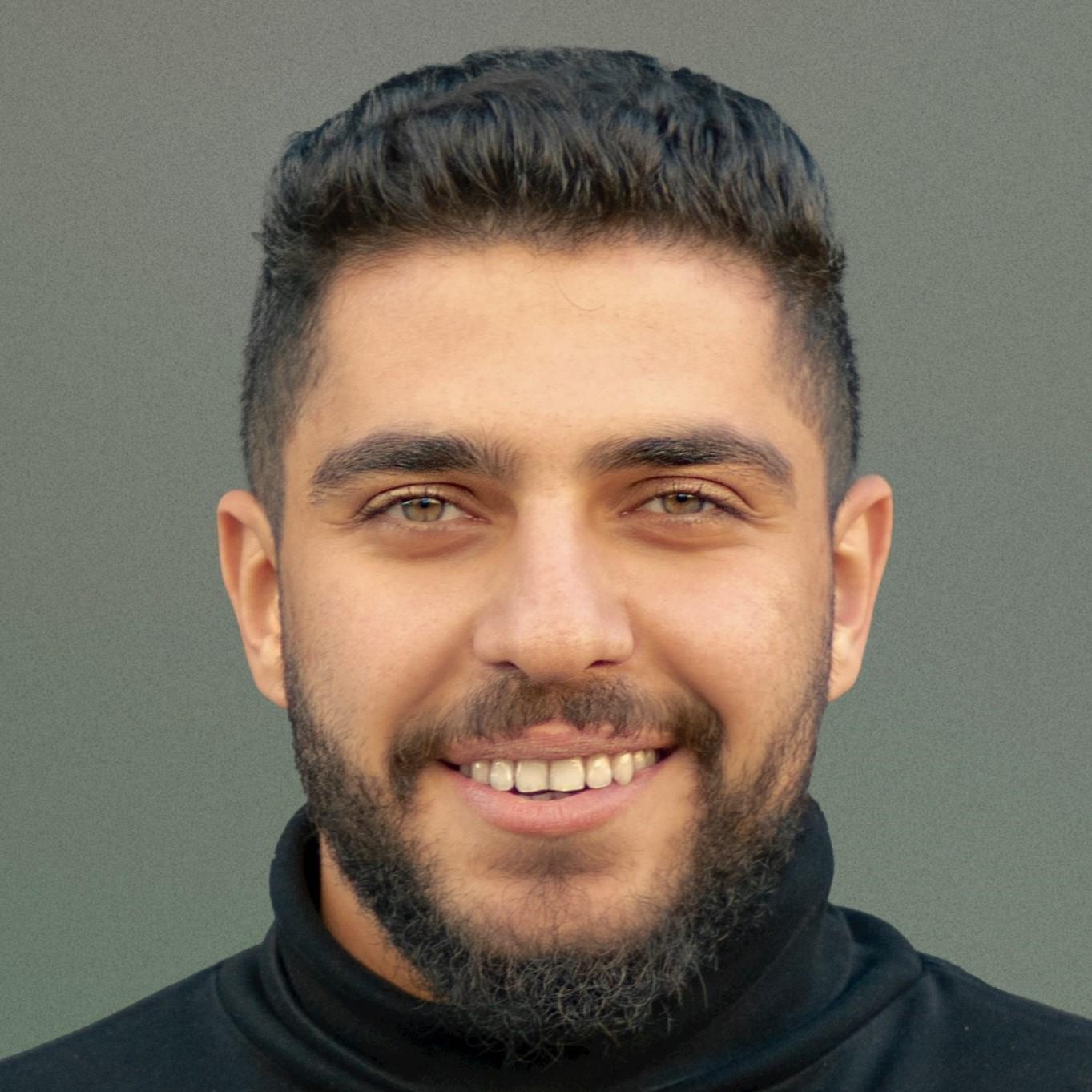} &
            \includegraphics[width=0.33\linewidth]{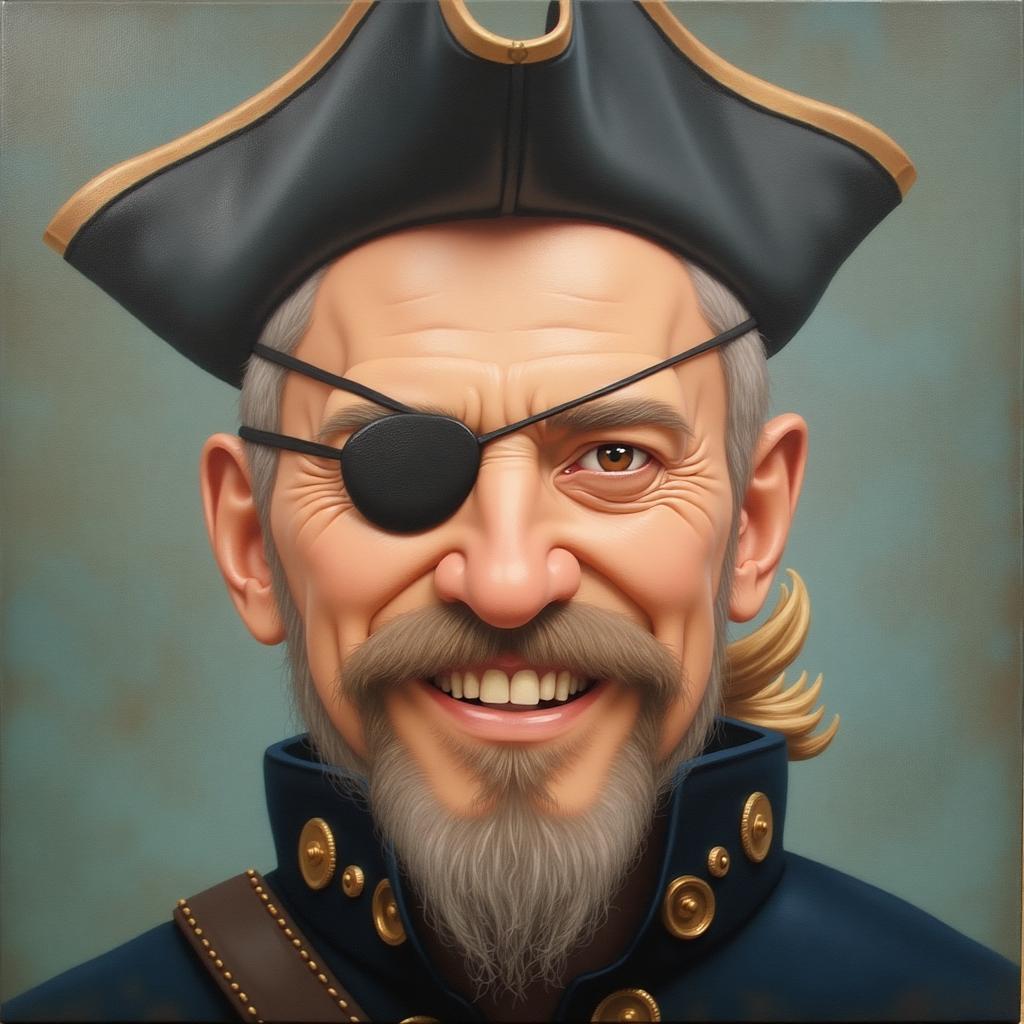} &
            \includegraphics[width=0.33\linewidth]{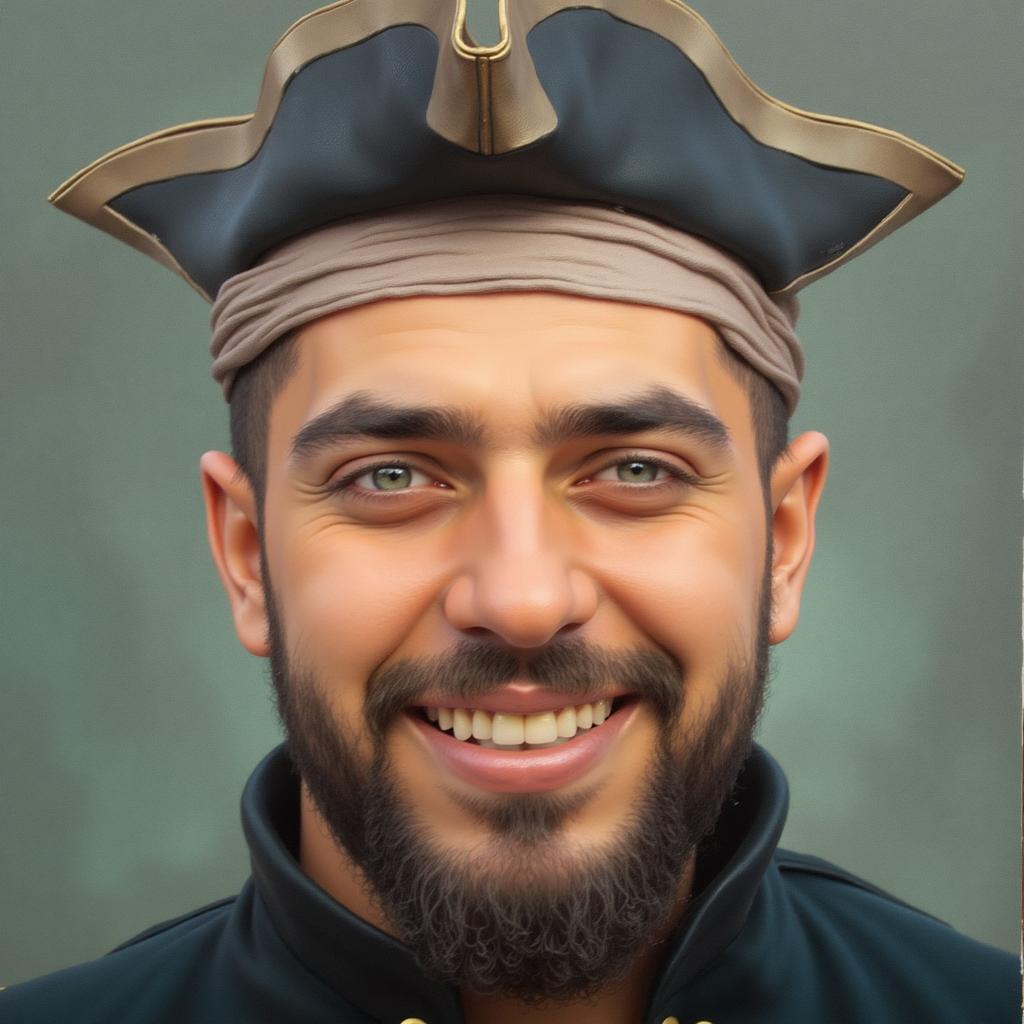} \\
            \multicolumn{3}{c}{``'' $\longrightarrow$ ``a portrait of a clown'', RF-Inversion (Flux)} \\
            \includegraphics[width=0.33\linewidth]{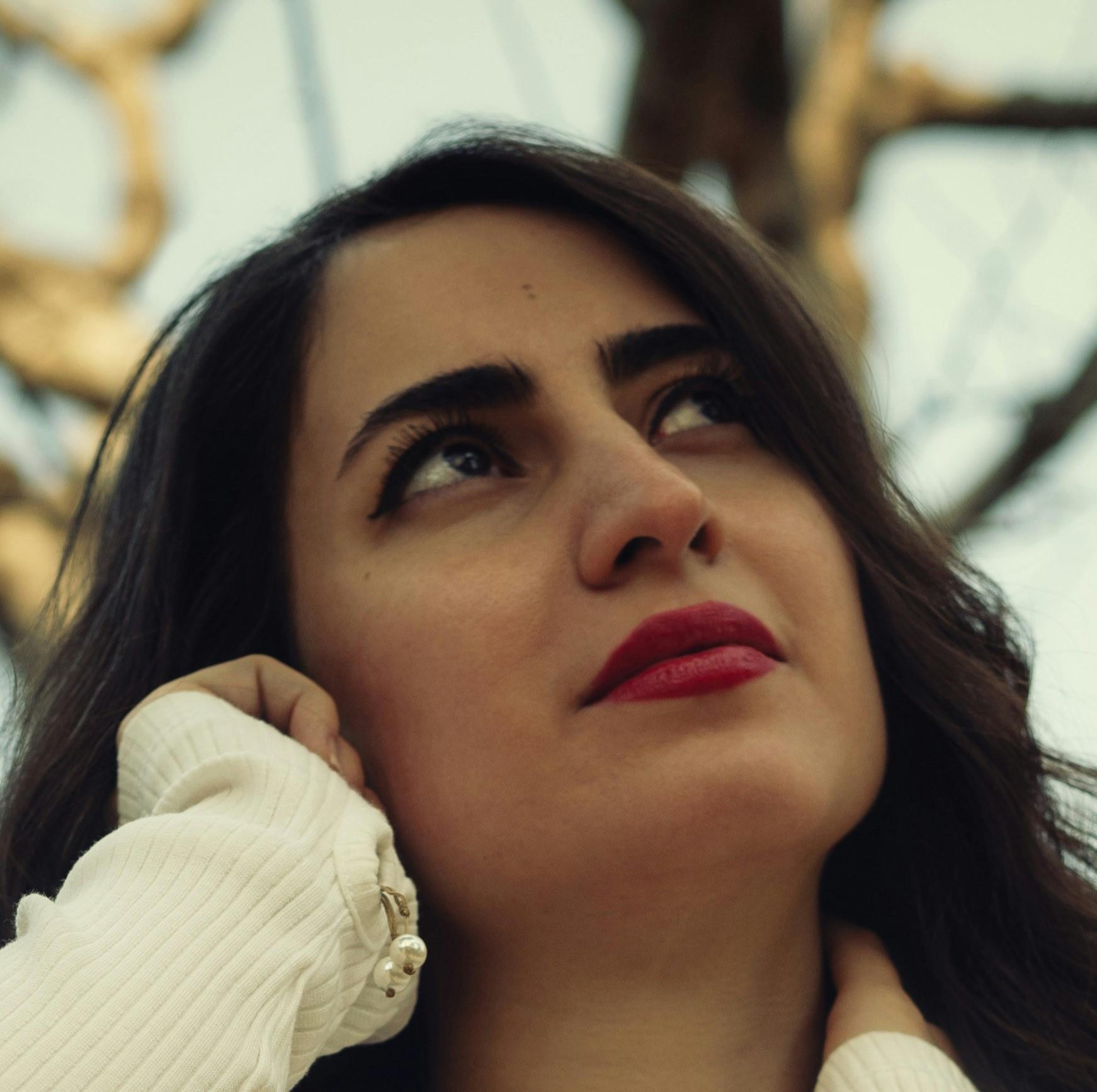} &
            \includegraphics[width=0.33\linewidth]{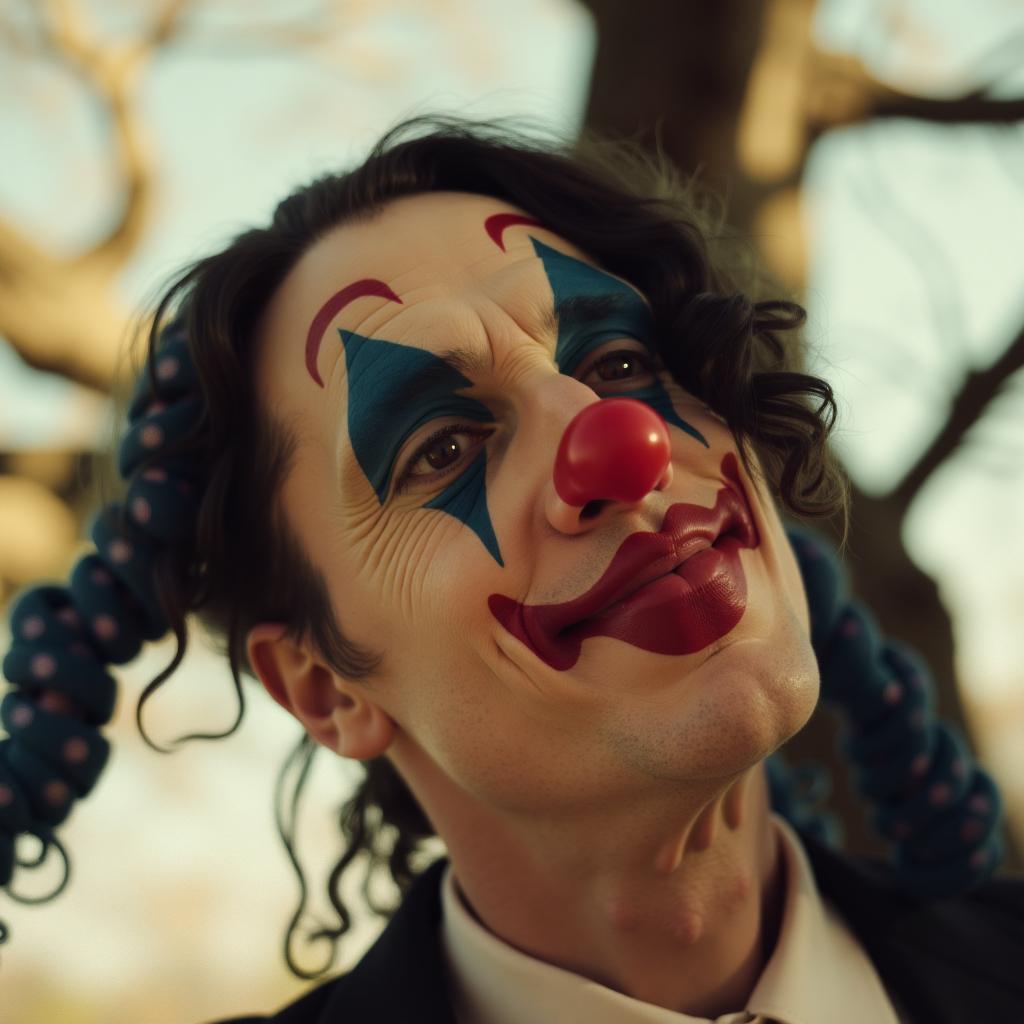} &
            \includegraphics[width=0.33\linewidth]{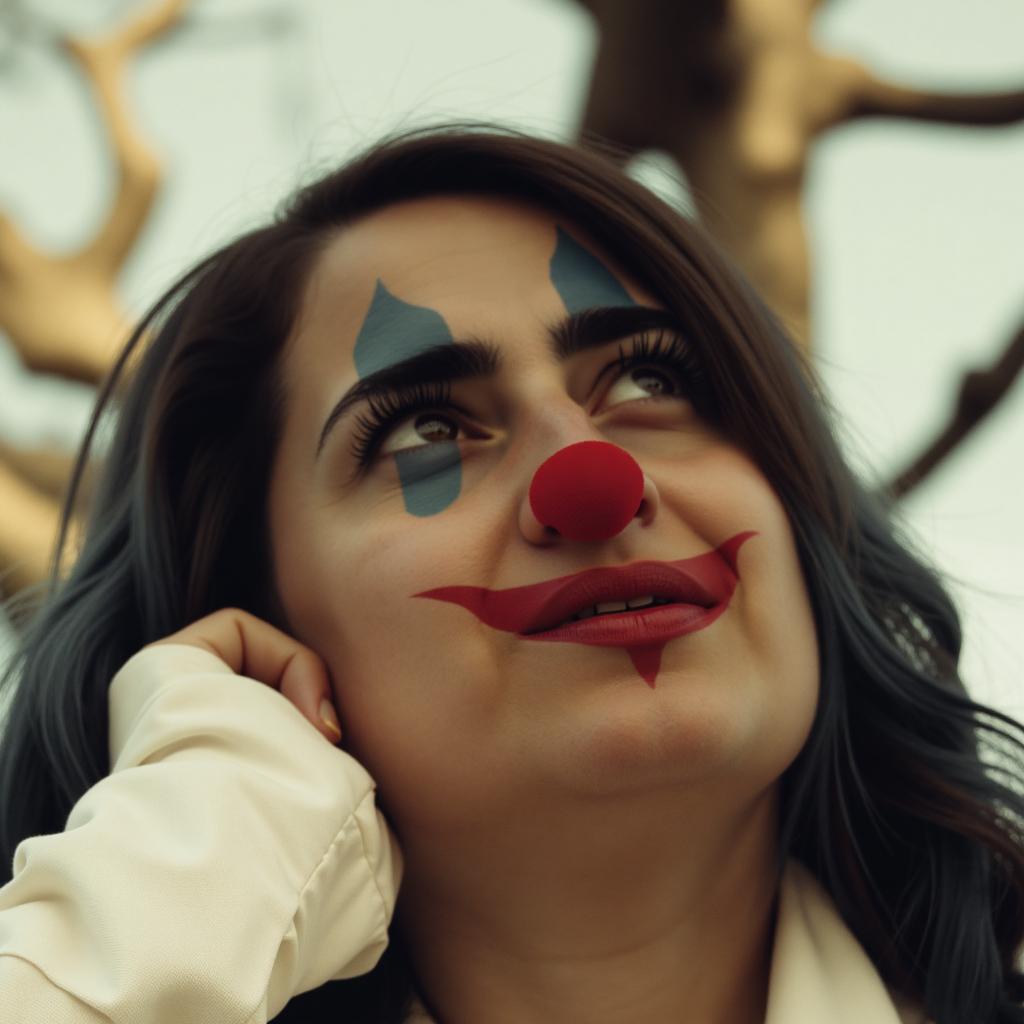} \\
            \multicolumn{3}{c}{``'' $\longrightarrow$ ``a portrait of a wizard'', RF-Inversion (Flux)} \\
            \includegraphics[width=0.33\linewidth]{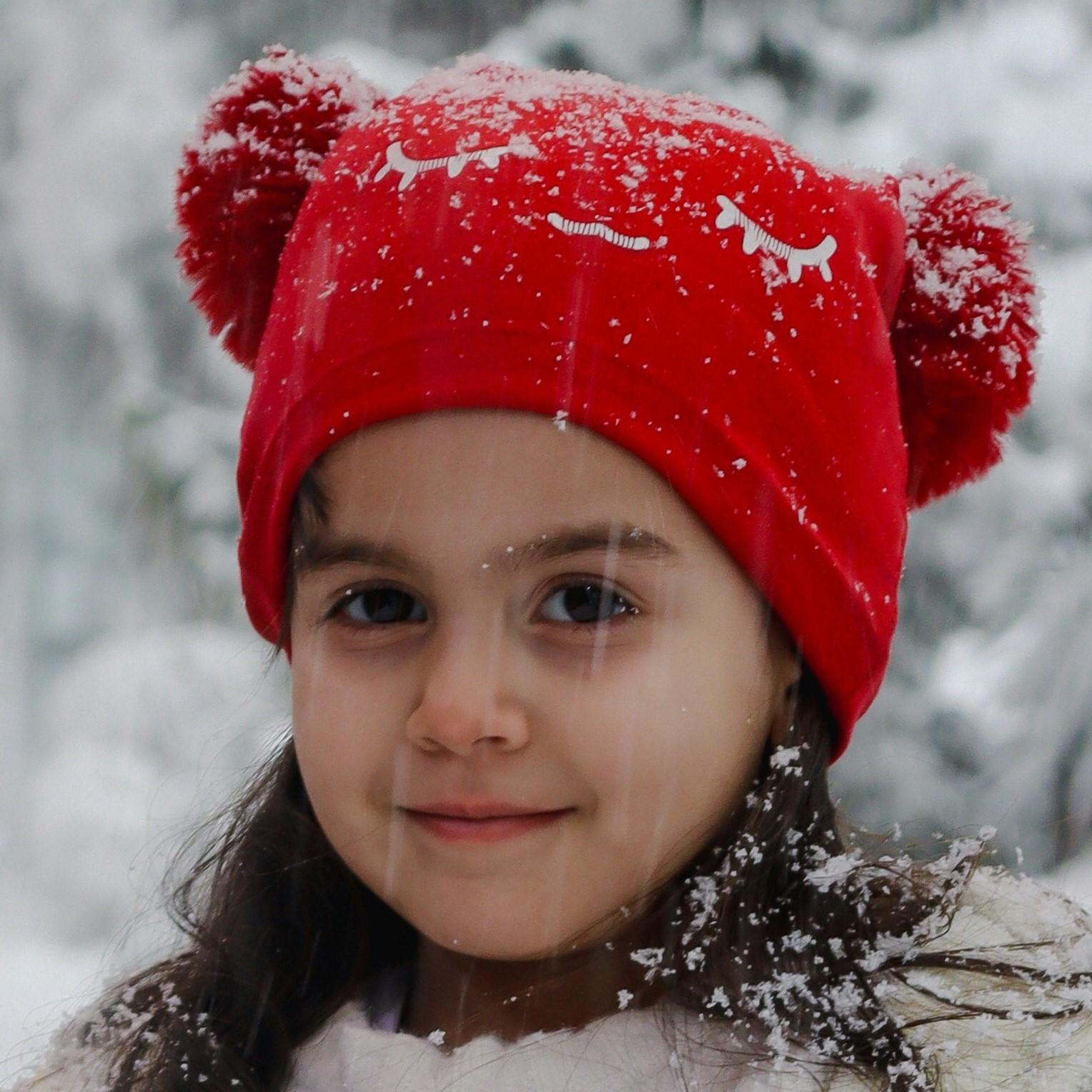} &
            \includegraphics[width=0.33\linewidth]{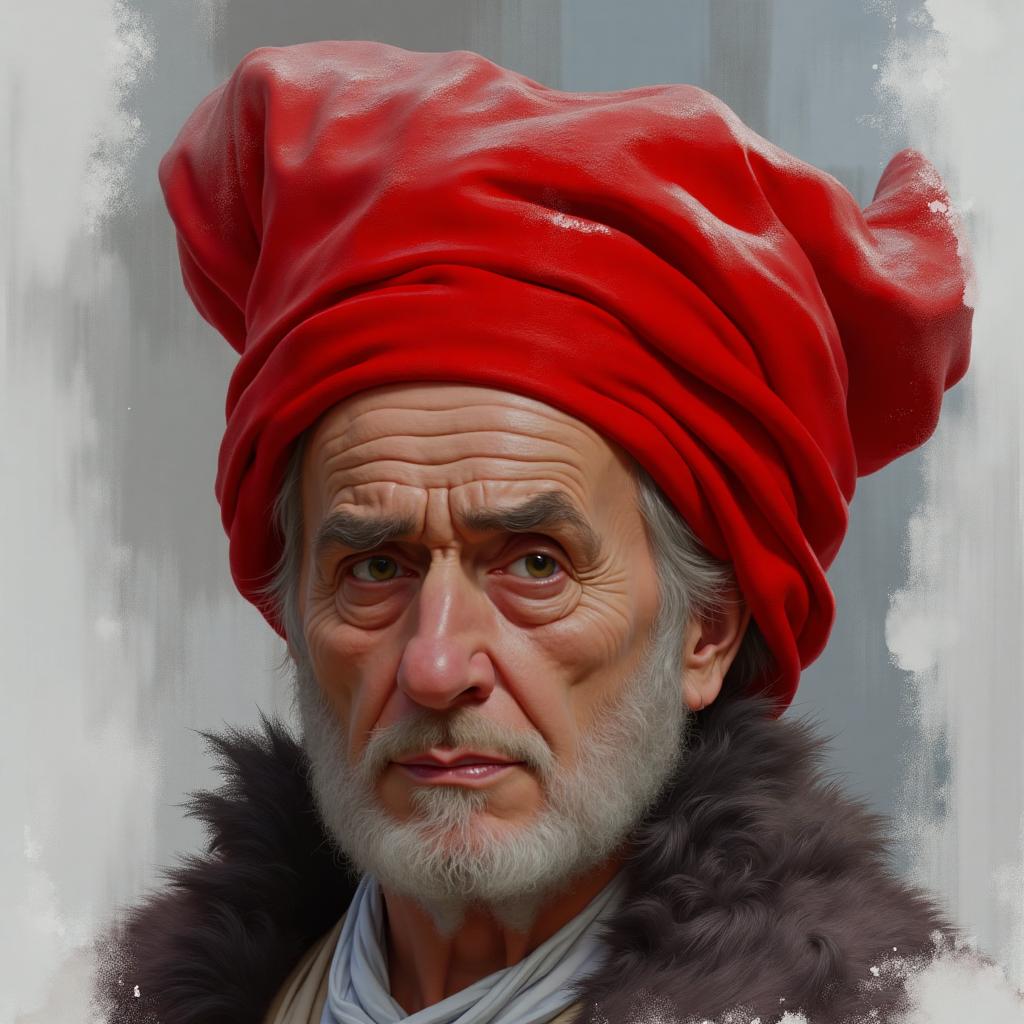} &
            \includegraphics[width=0.33\linewidth]{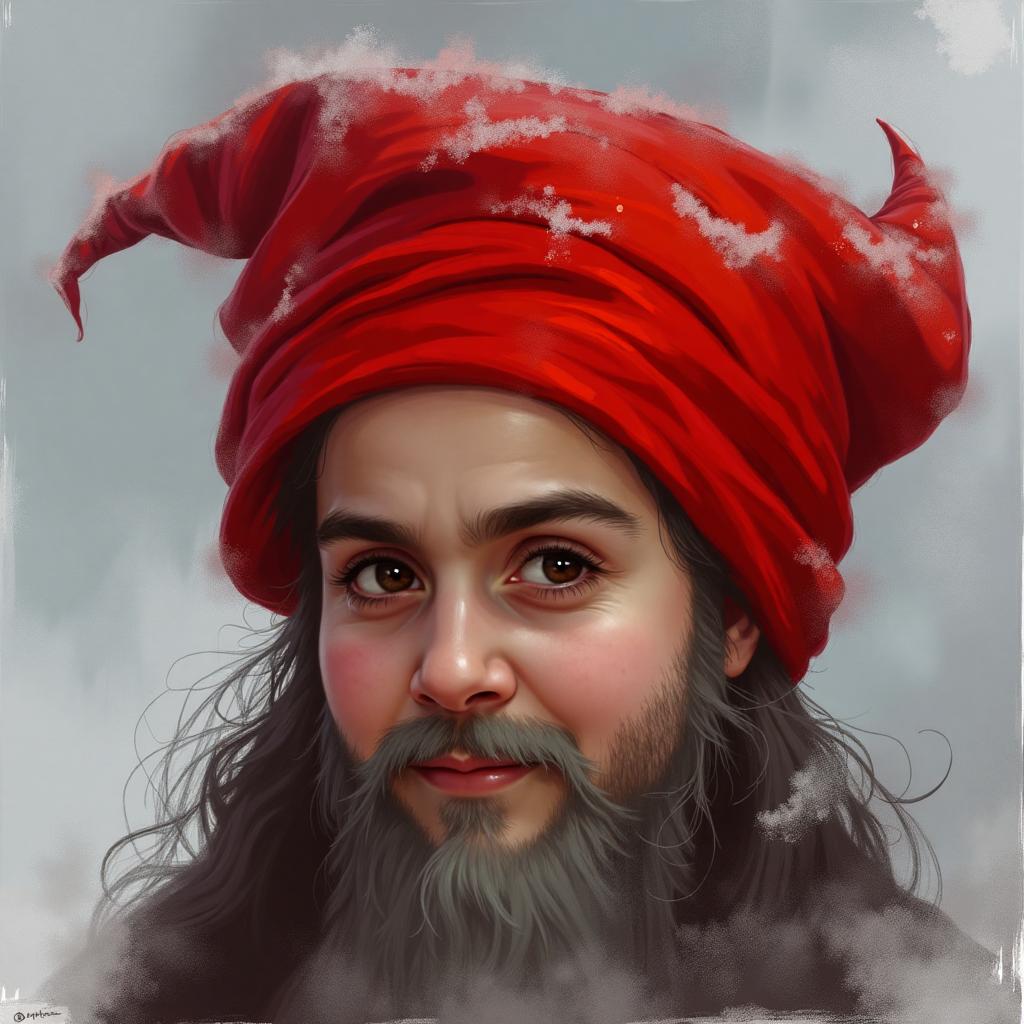} \\
            \multicolumn{3}{c}{``'' $\longrightarrow$ ``a portrait of an elf'', RF-Inversion (Flux)} \\
            \includegraphics[width=0.33\linewidth]{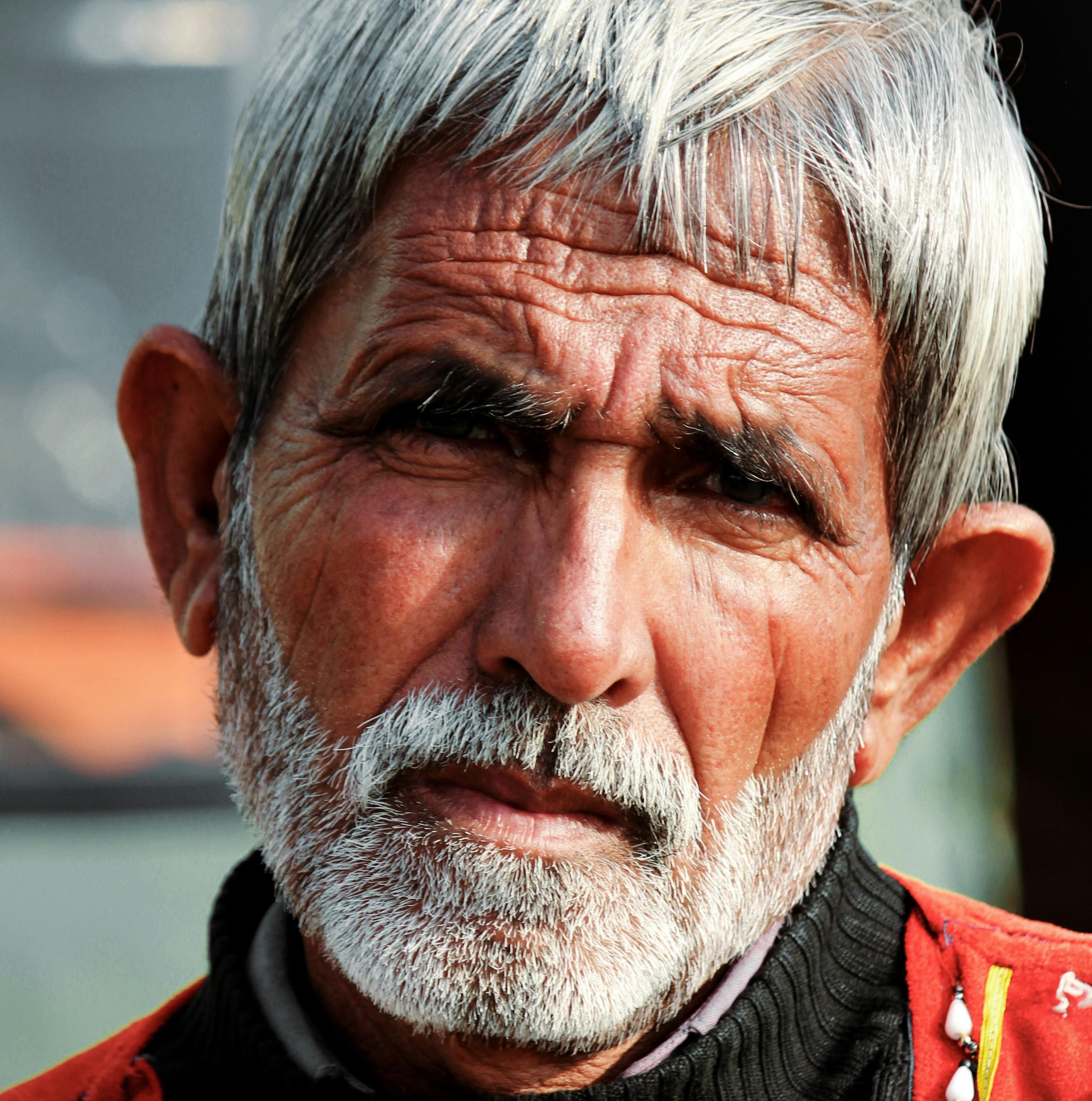} &
            \includegraphics[width=0.33\linewidth]{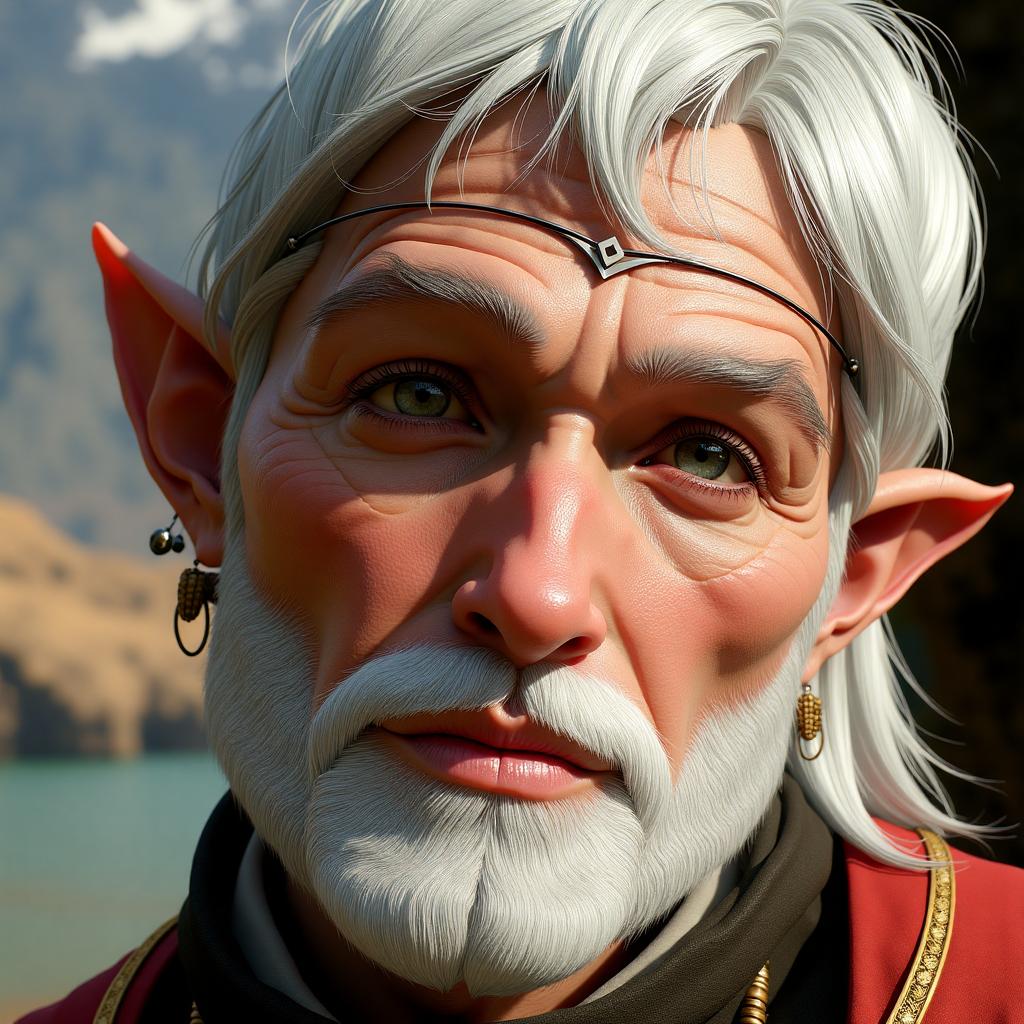} &
            \includegraphics[width=0.33\linewidth]{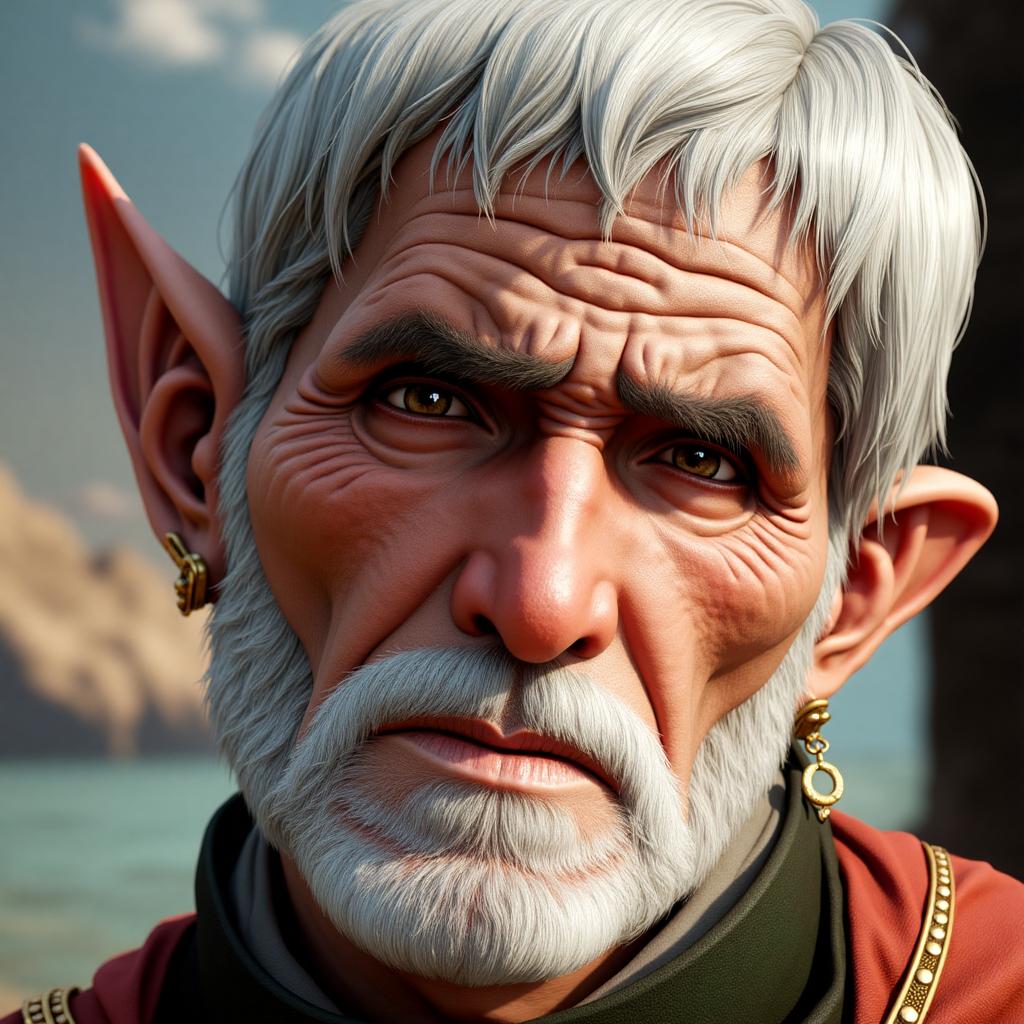} \\
            \multicolumn{3}{c}{``'' $\longrightarrow$ ``a portrait of a clown'', RF-Inversion (Flux)} \\
            \includegraphics[width=0.33\linewidth]{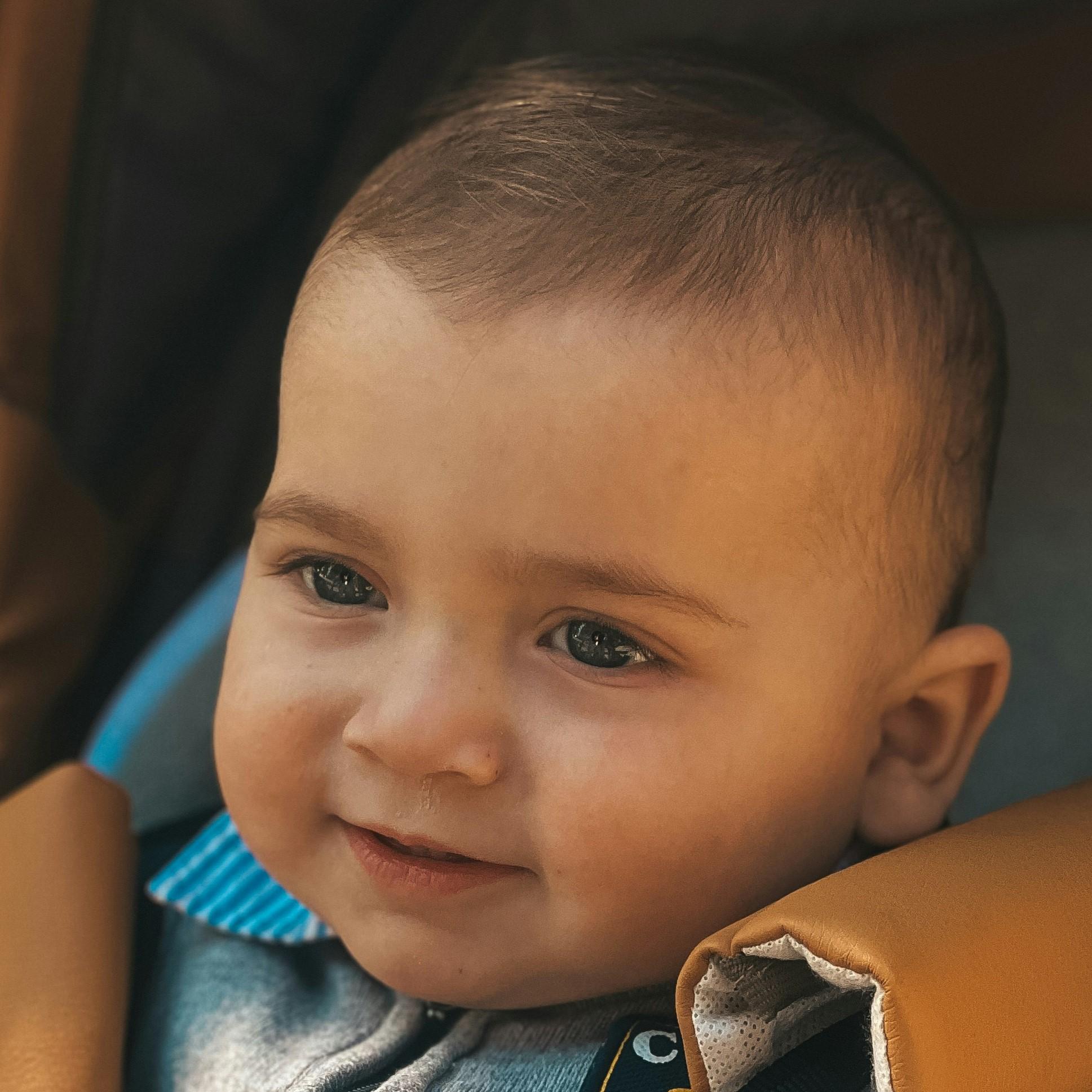} &
            \includegraphics[width=0.33\linewidth]{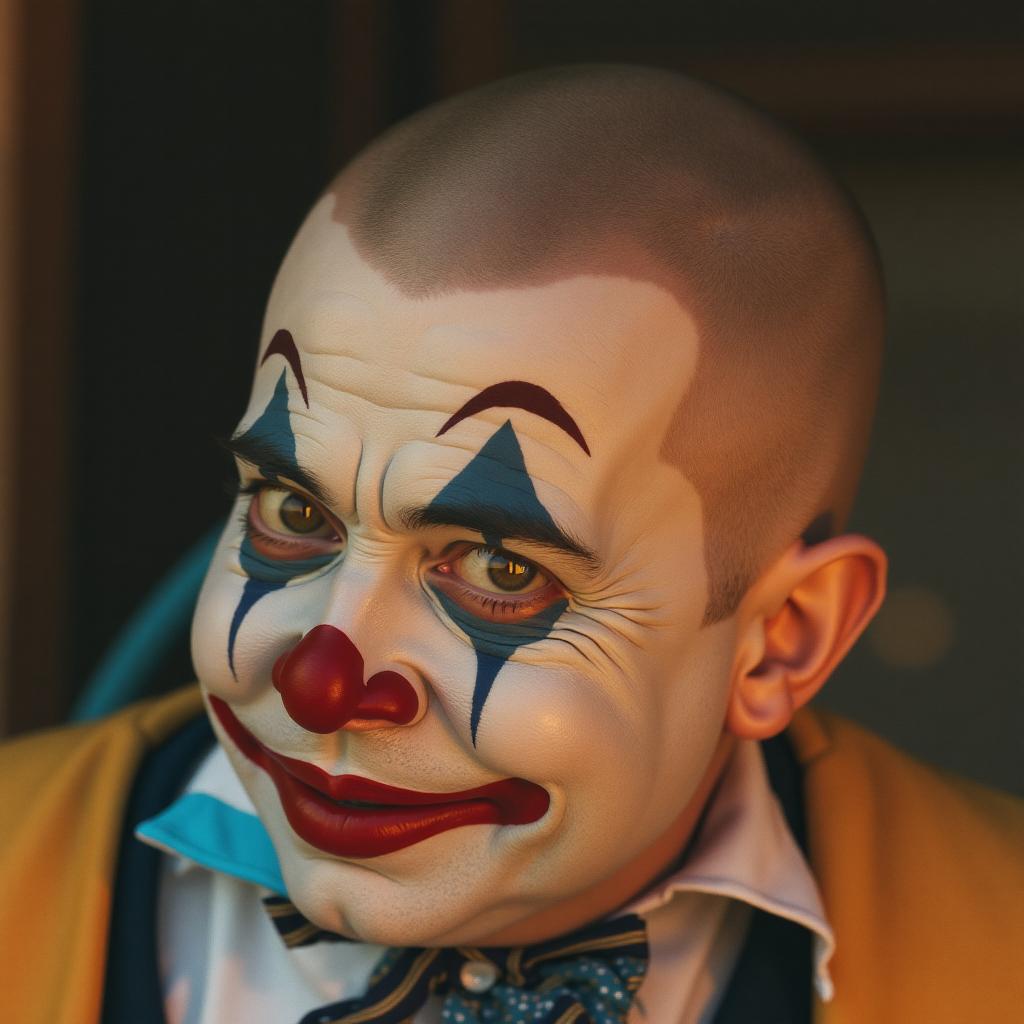} &
            \includegraphics[width=0.33\linewidth]{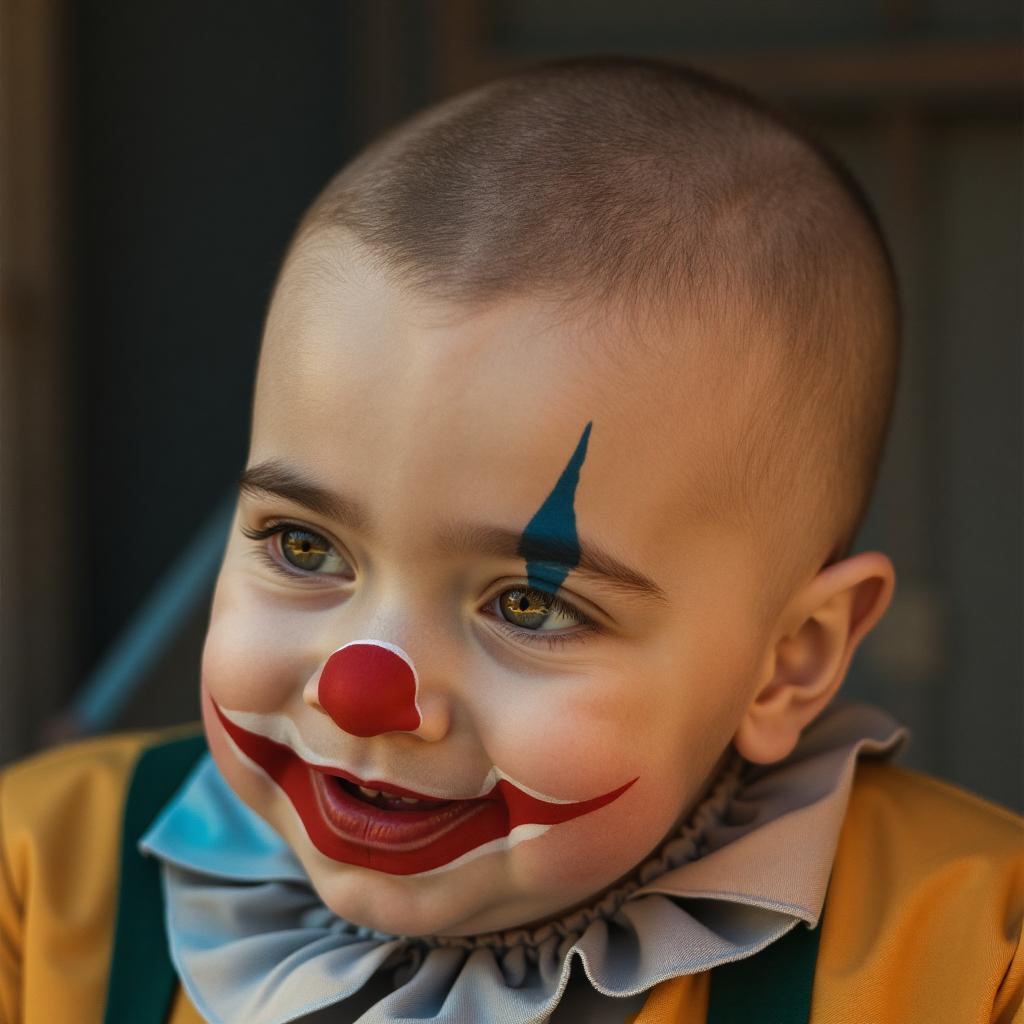} \\
            \multicolumn{3}{c}{``'' $\longrightarrow$ ``a portrait of an alien'', RF-Inversion (Flux)} \\
            \includegraphics[width=0.33\linewidth]{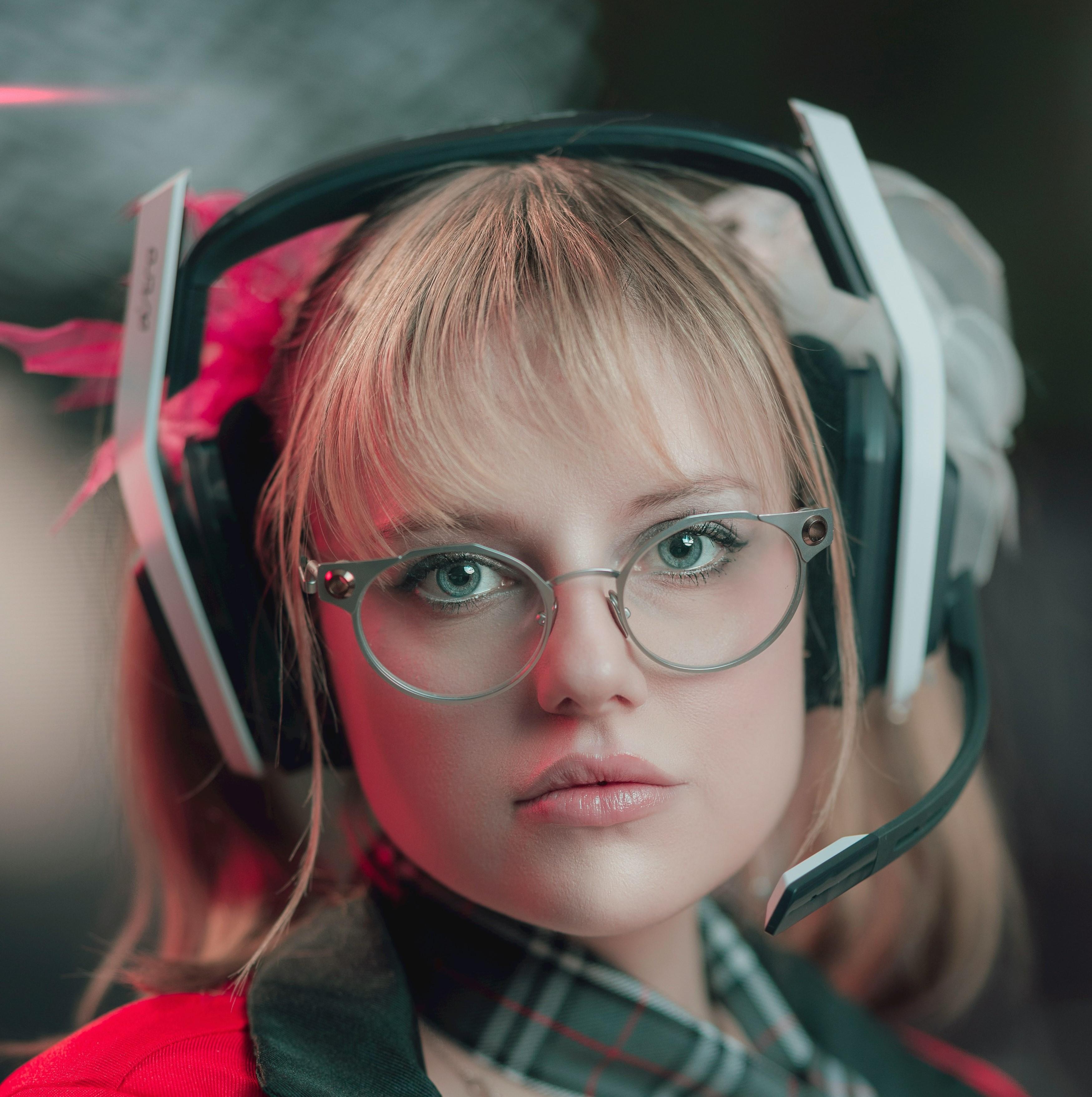} &
            \includegraphics[width=0.33\linewidth]{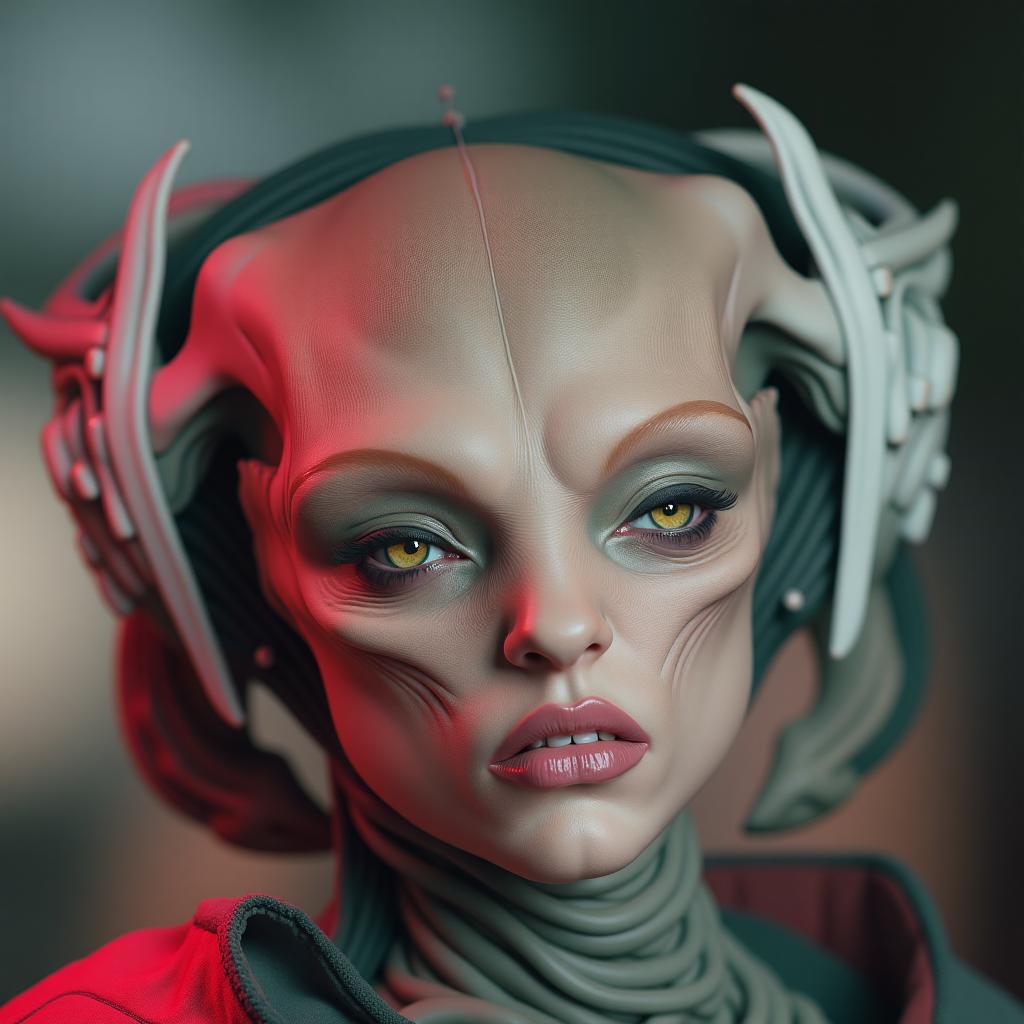} &
            \includegraphics[width=0.33\linewidth]{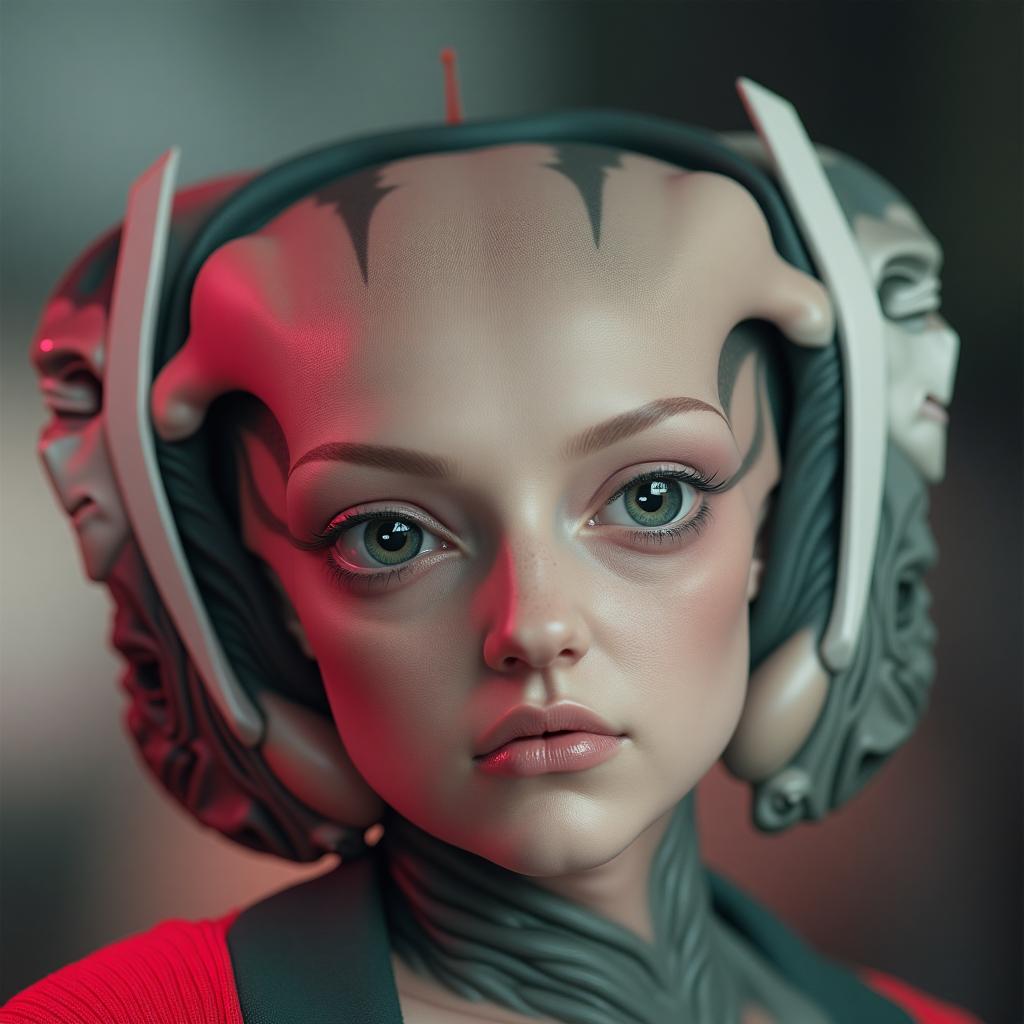} \\
            Input & Edit w/o Tight & Edit w/ Tight \\
        \end{tabular}
        }
    \end{minipage}

    \caption{Additional Editing Results.}
    \label{fig:more-results1}
\end{figure*}

\begin{figure*}[t]
    \centering
    \begin{minipage}[t]{0.48\textwidth}
        \centering
        \setlength{\tabcolsep}{1pt}
        \scriptsize{
        \begin{tabular}{ccc}
        \multicolumn{3}{c}{``A person riding a horse'' $\longrightarrow$ ``... with a cowboy hat'', DDIM Inversion} \\
        \includegraphics[width=0.33\linewidth]{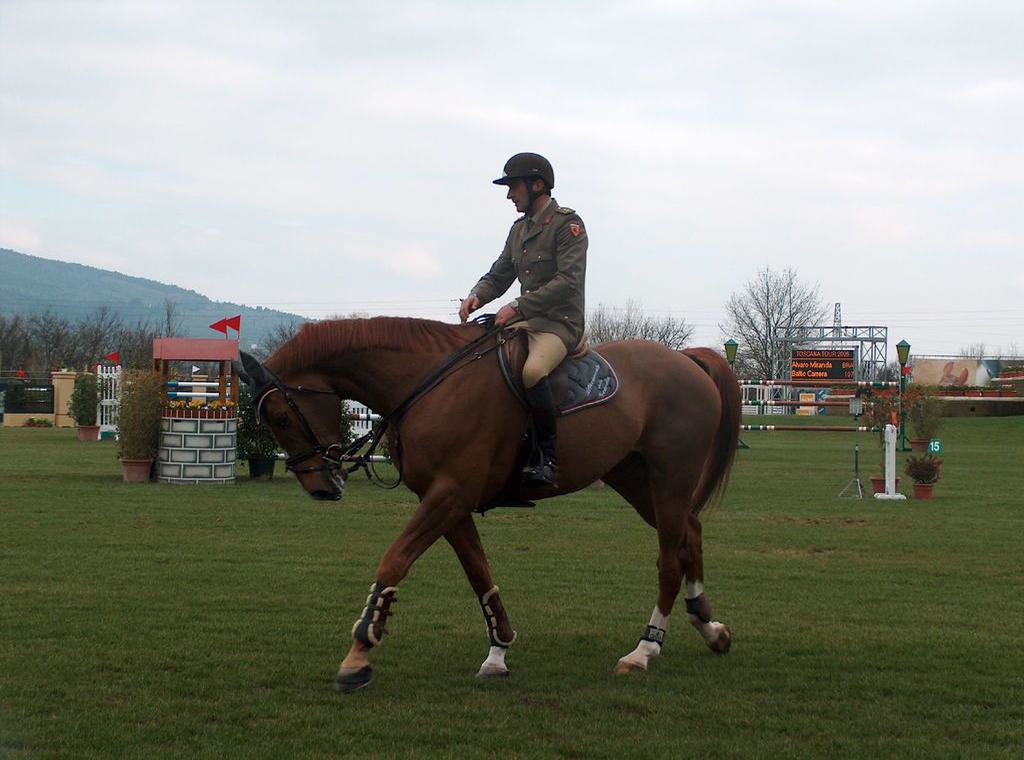} &
        \includegraphics[width=0.33\linewidth]{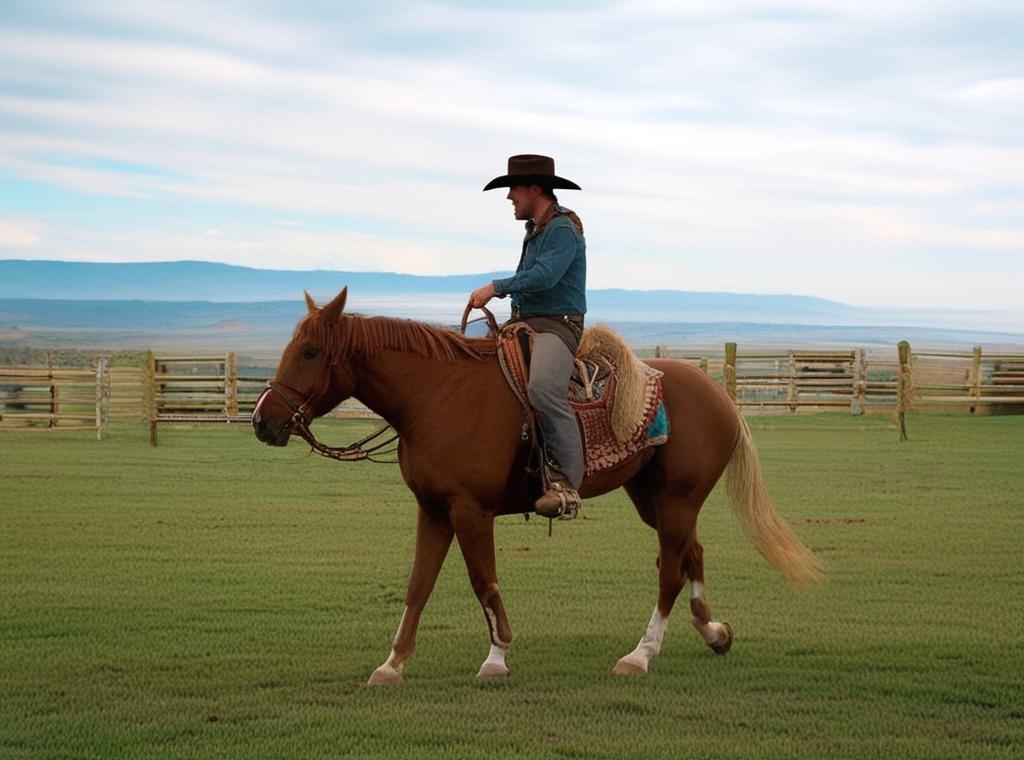} &
        \includegraphics[width=0.33\linewidth]{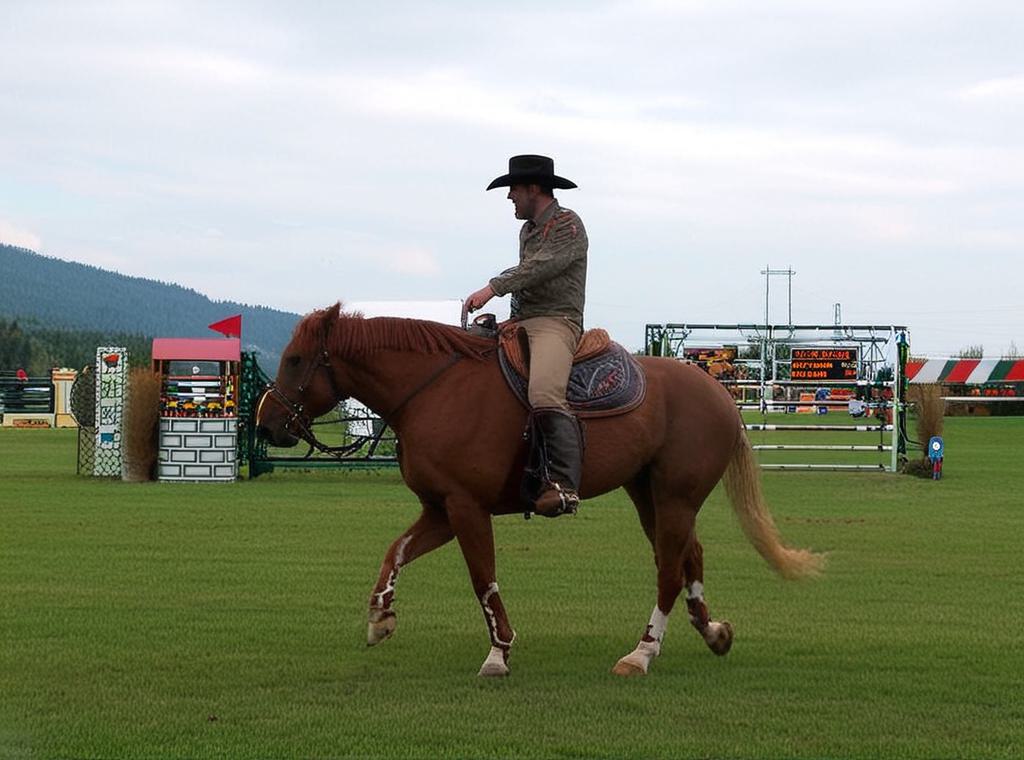}          
        \\
        \multicolumn{3}{c}{``A black car driving on the beach'' $\longrightarrow$ ``... red car...'', LEDITS++} \\
        \includegraphics[width=0.33\linewidth]{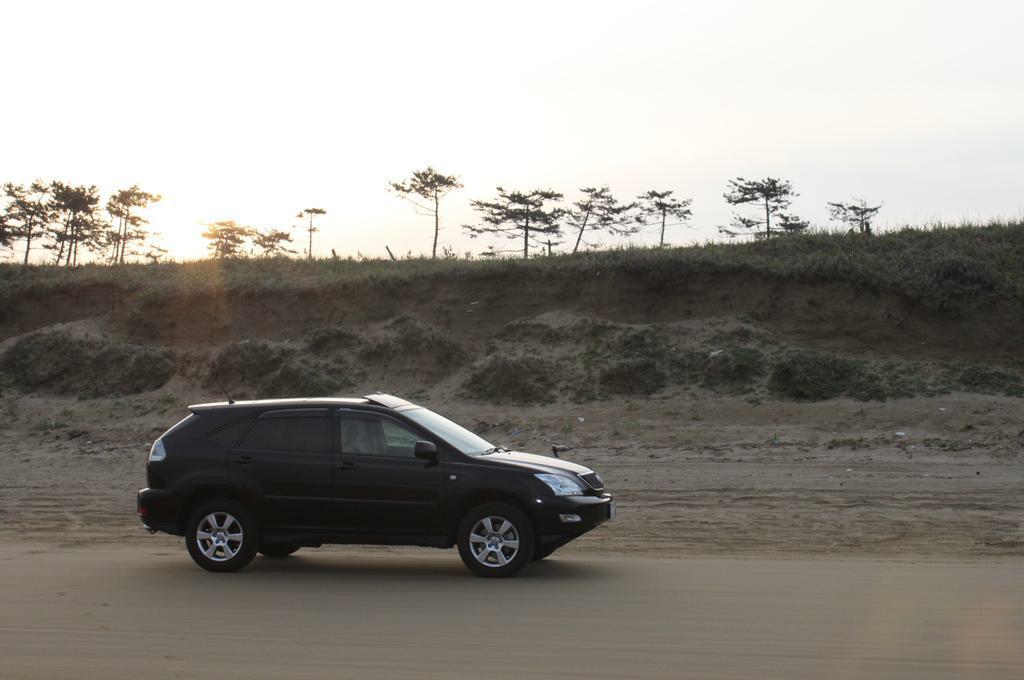} &
        \includegraphics[width=0.33\linewidth]{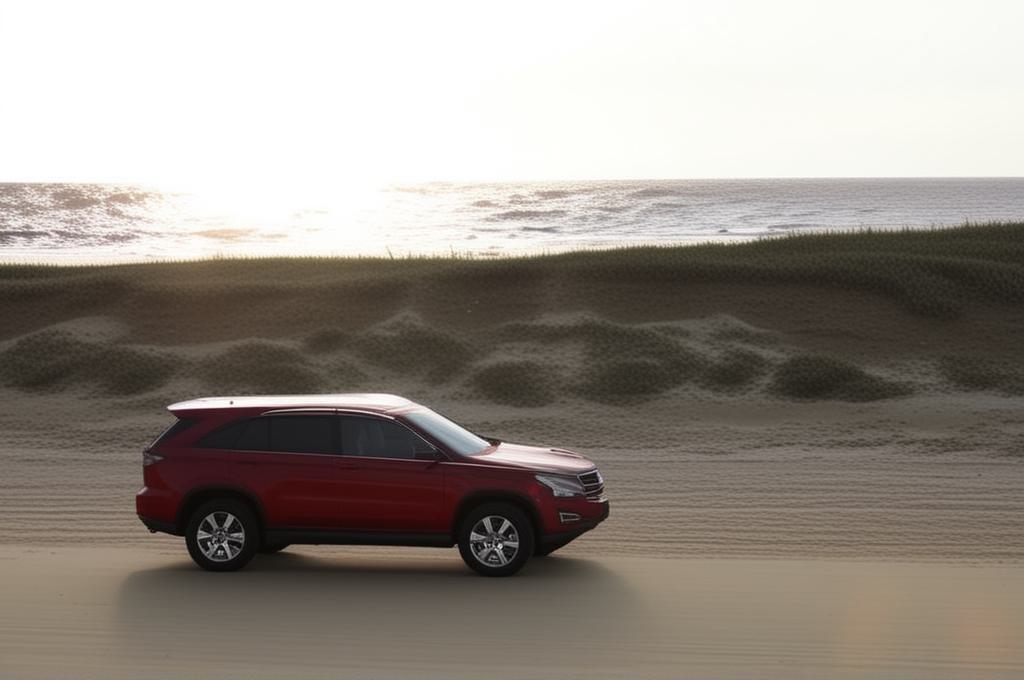} &
        \includegraphics[width=0.33\linewidth]{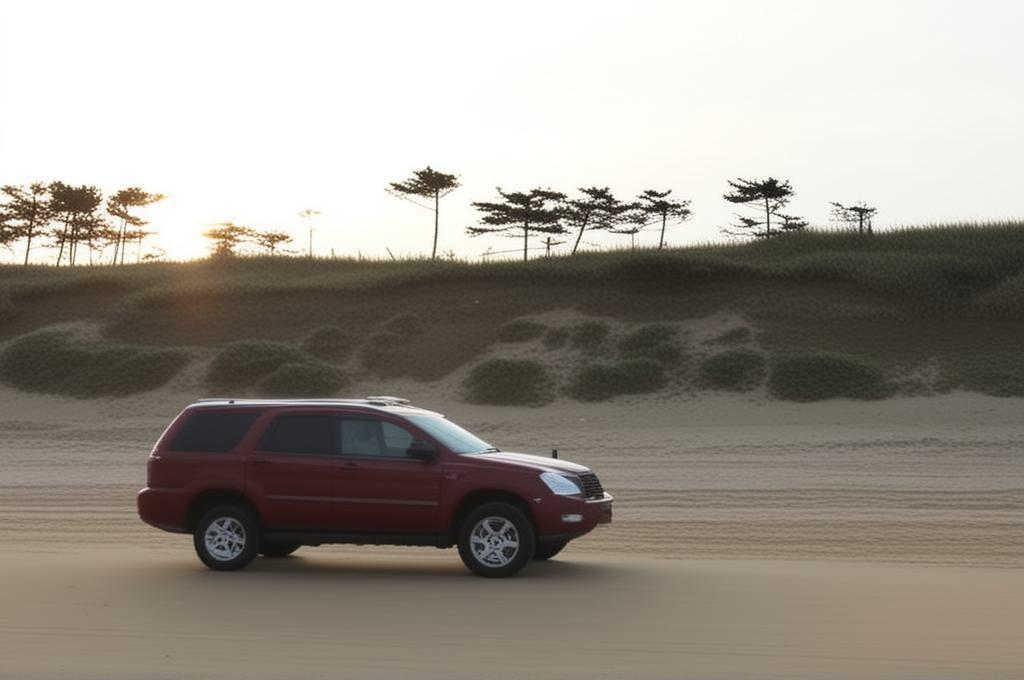}        
        \\
        \multicolumn{3}{c}{``A person in a ninja costume'' $\longrightarrow$ ``... with a mask'', LEDITS++} \\
        \includegraphics[width=0.33\linewidth]{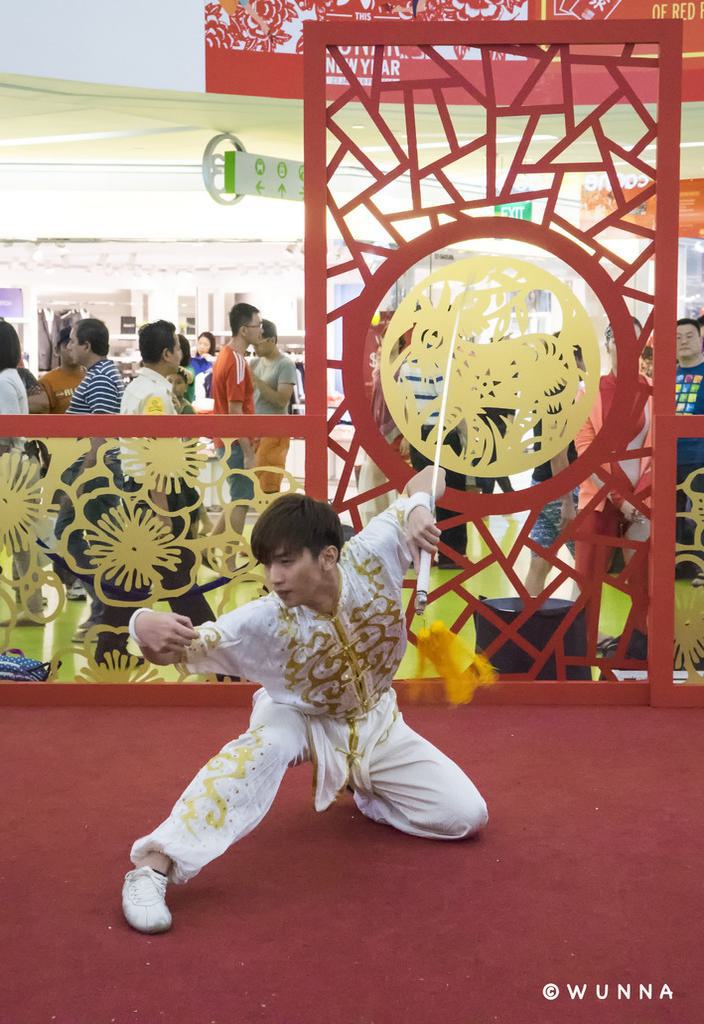} &
        \includegraphics[width=0.33\linewidth]{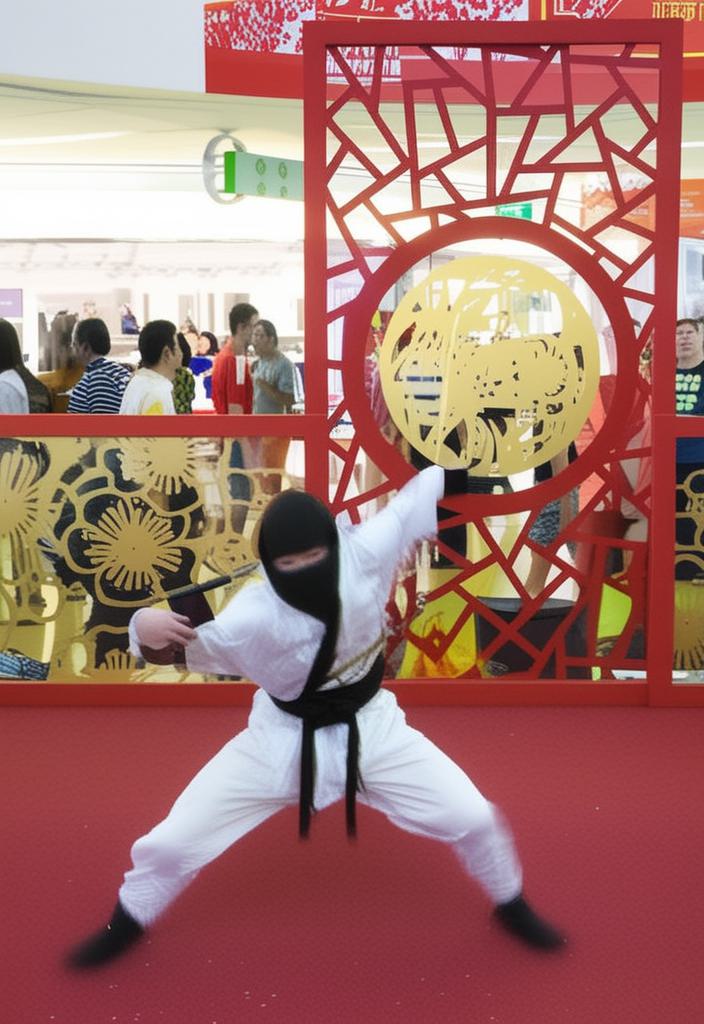} &
        \includegraphics[width=0.33\linewidth]{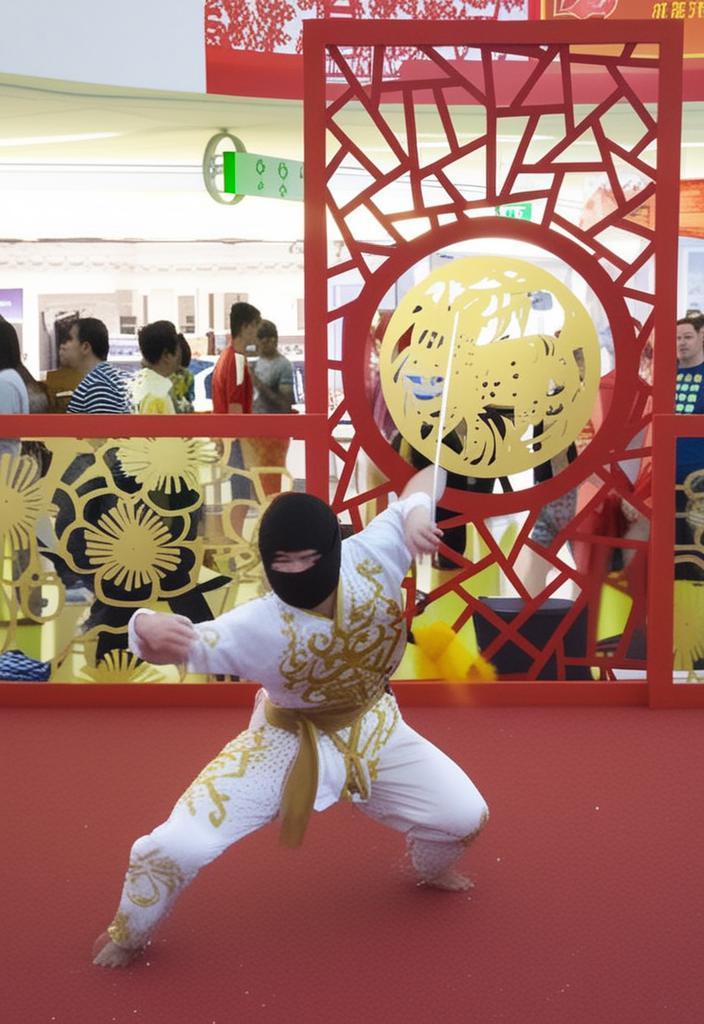}          
        \\
        \multicolumn{3}{c}{``A cat'' $\longrightarrow$ ``A cat wearing a bow tie'', ReNoise} \\
        \includegraphics[width=0.33\linewidth]{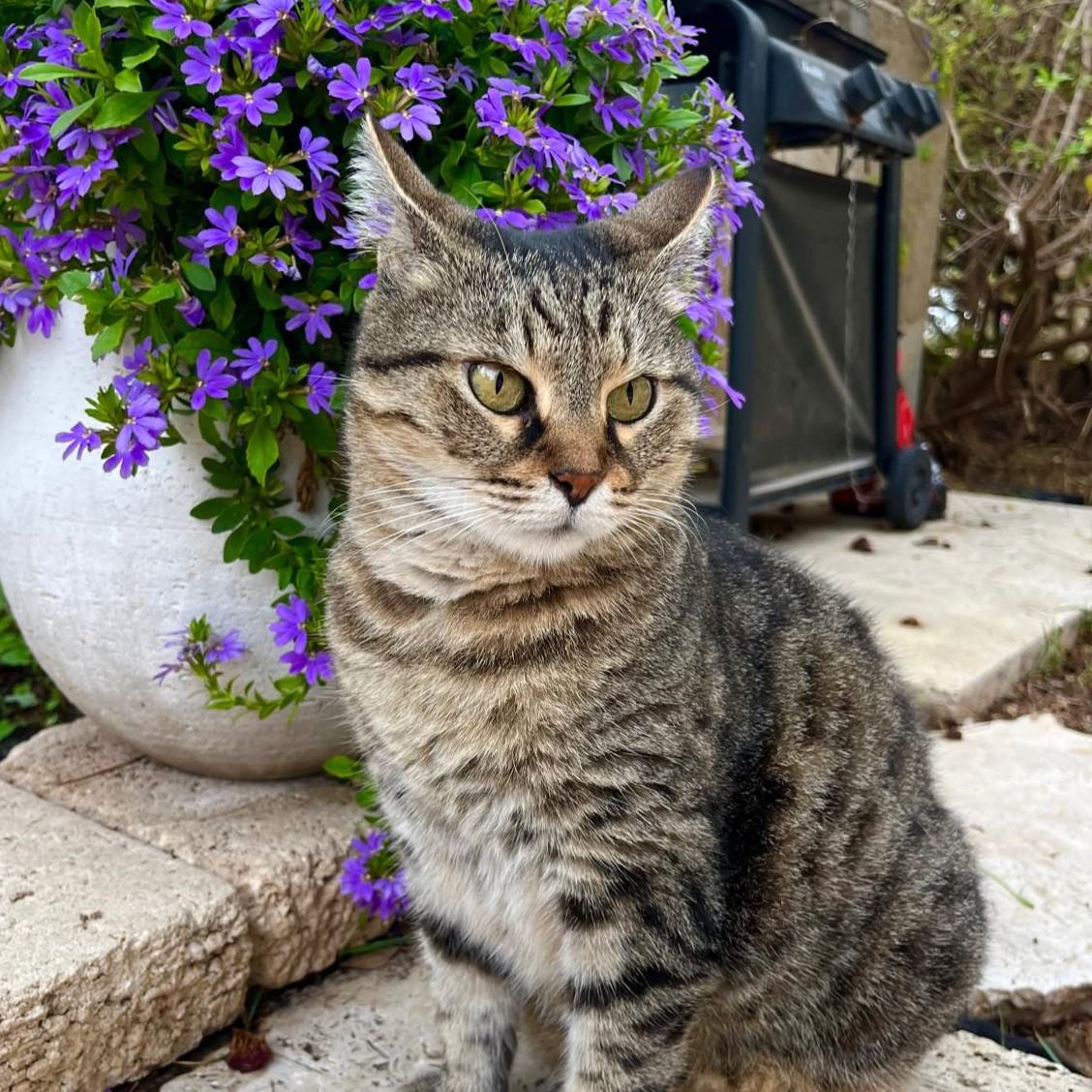} &
        \includegraphics[width=0.33\linewidth]{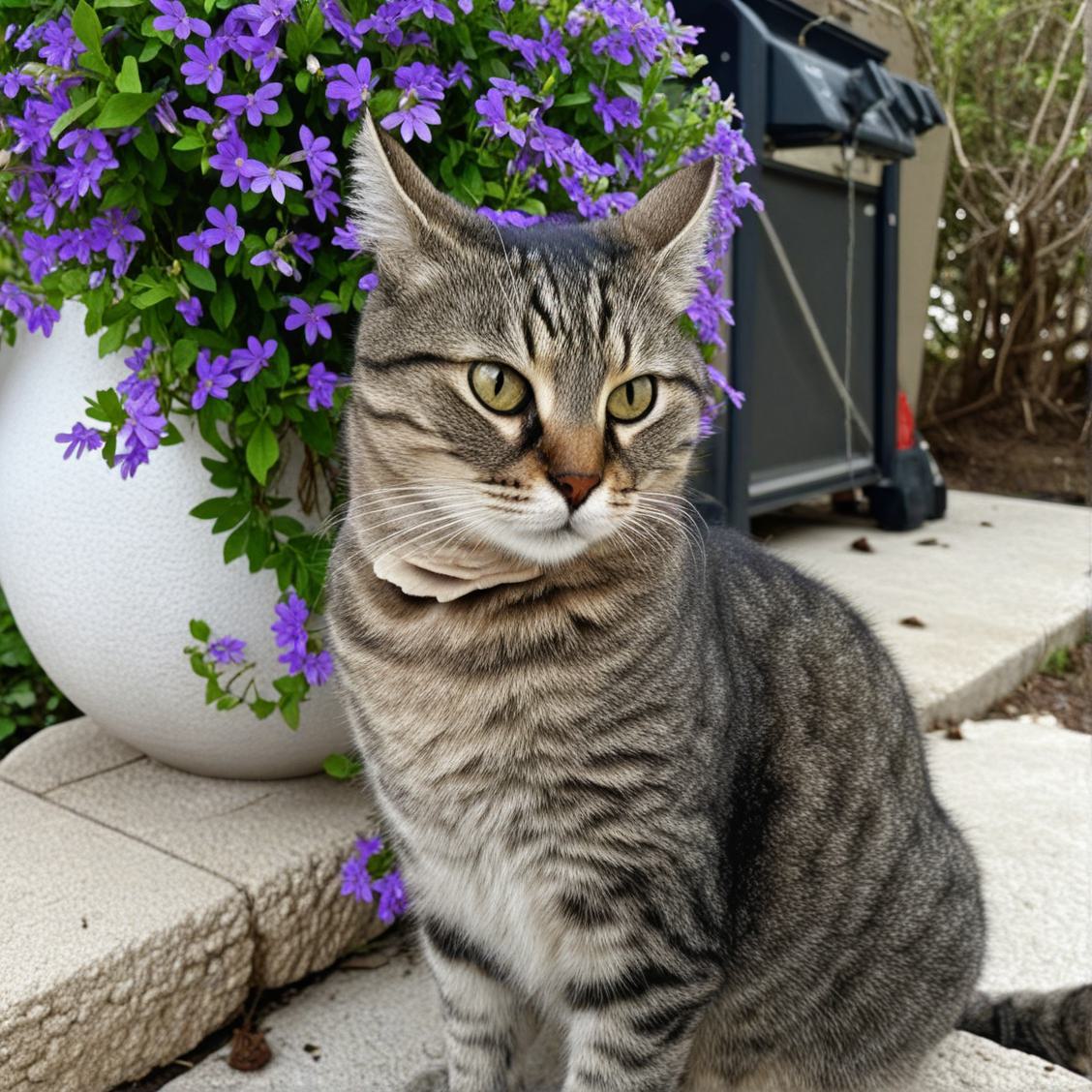} &
        \includegraphics[width=0.33\linewidth]{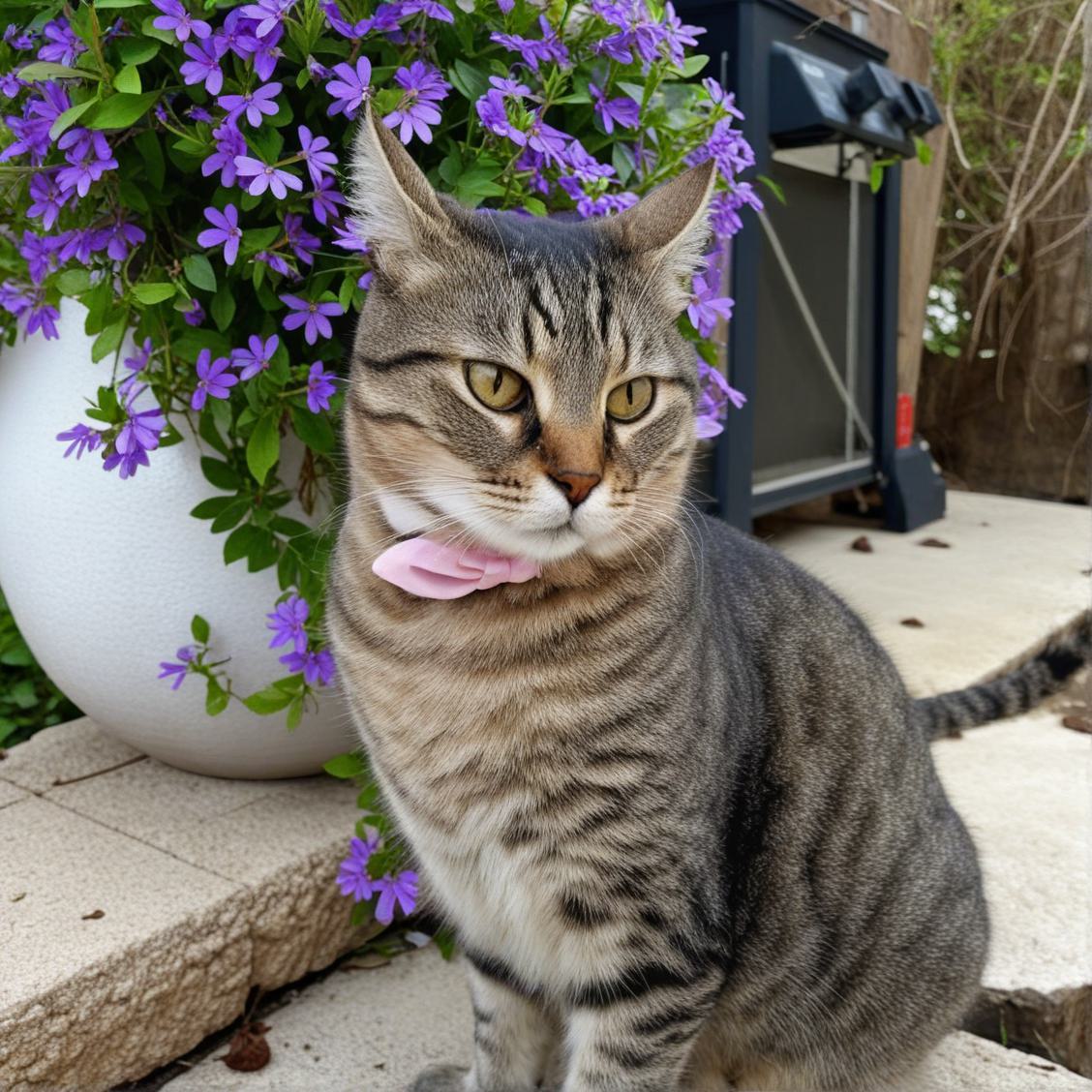}  
        \\
        \multicolumn{3}{c}{``A white metal bench next to a patch of grass'' $\longrightarrow$} \\
        \multicolumn{3}{c}{flower vase + dog resting underneath'', Negative Prompt Inversion} \\
        \includegraphics[width=0.33\linewidth]{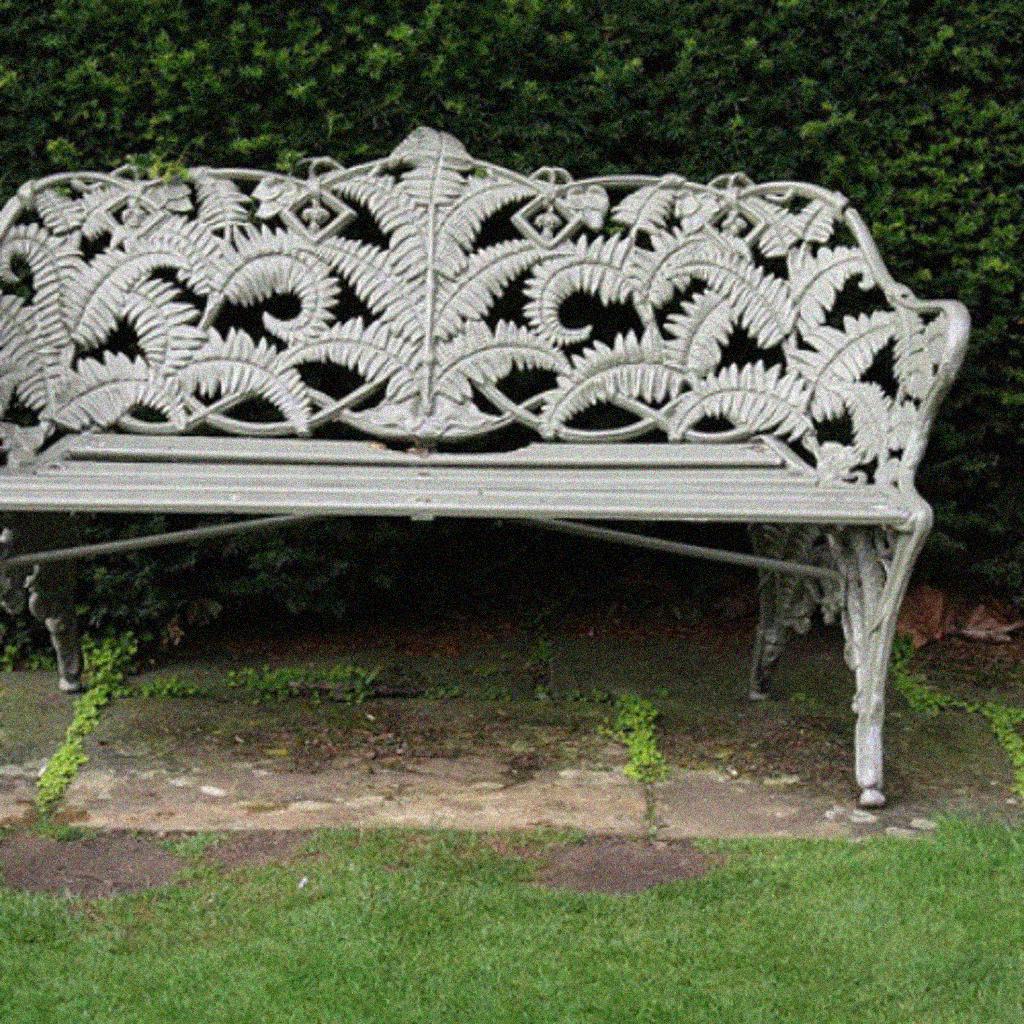} &
        \includegraphics[width=0.33\linewidth]{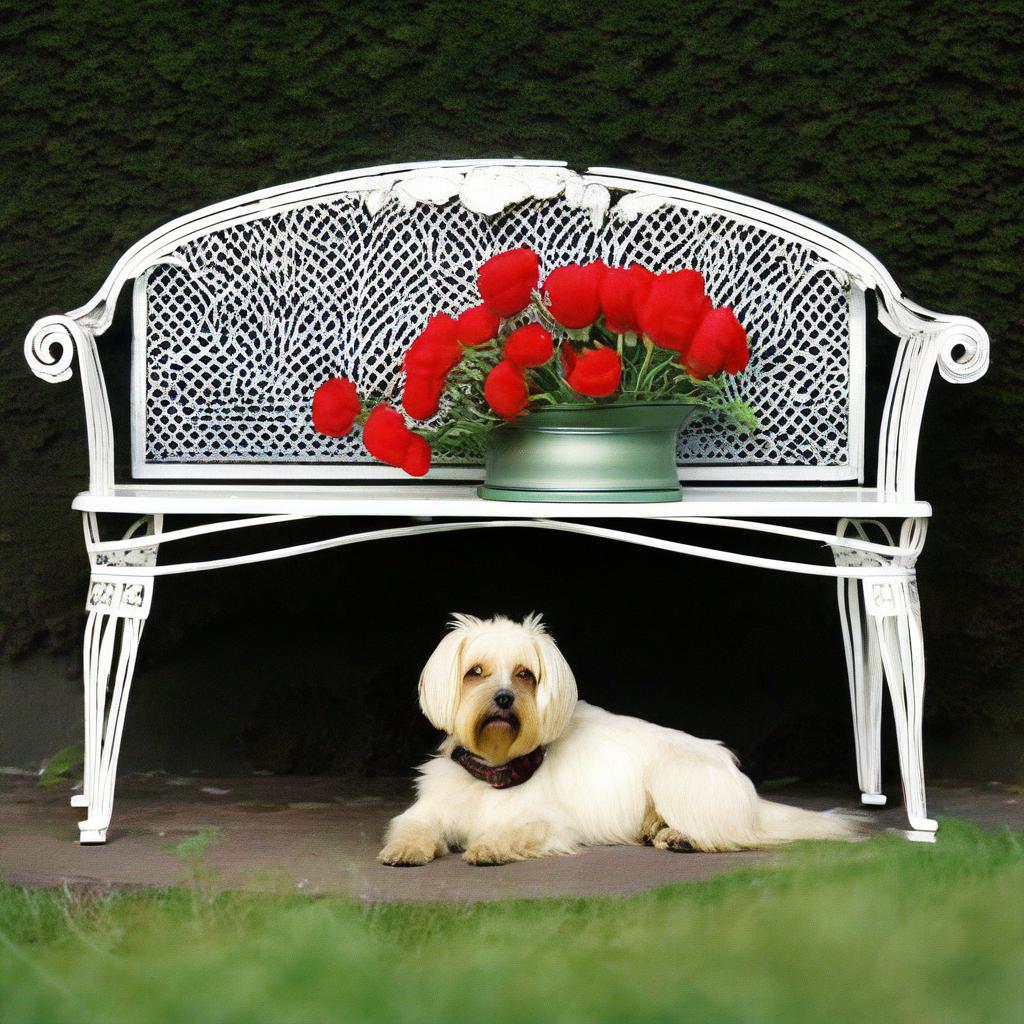} &
        \includegraphics[width=0.33\linewidth]{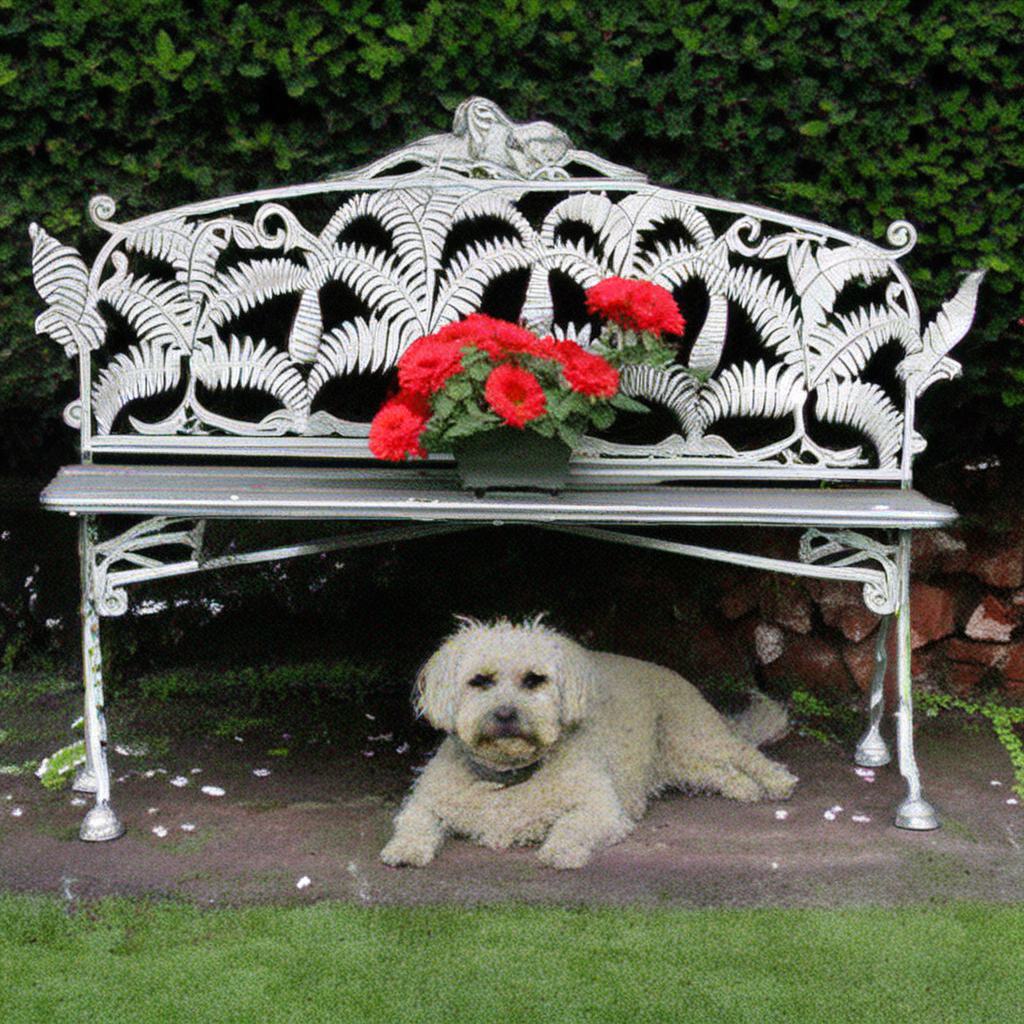} 
        \\
        \multicolumn{3}{c}{``A group of people riding on a boat across a lake'' $\longrightarrow$} \\
        \multicolumn{3}{c}{``... on a galley ... + skyscrapers + dolphin'', DDIM Inversion} \\
        \includegraphics[width=0.33\linewidth]{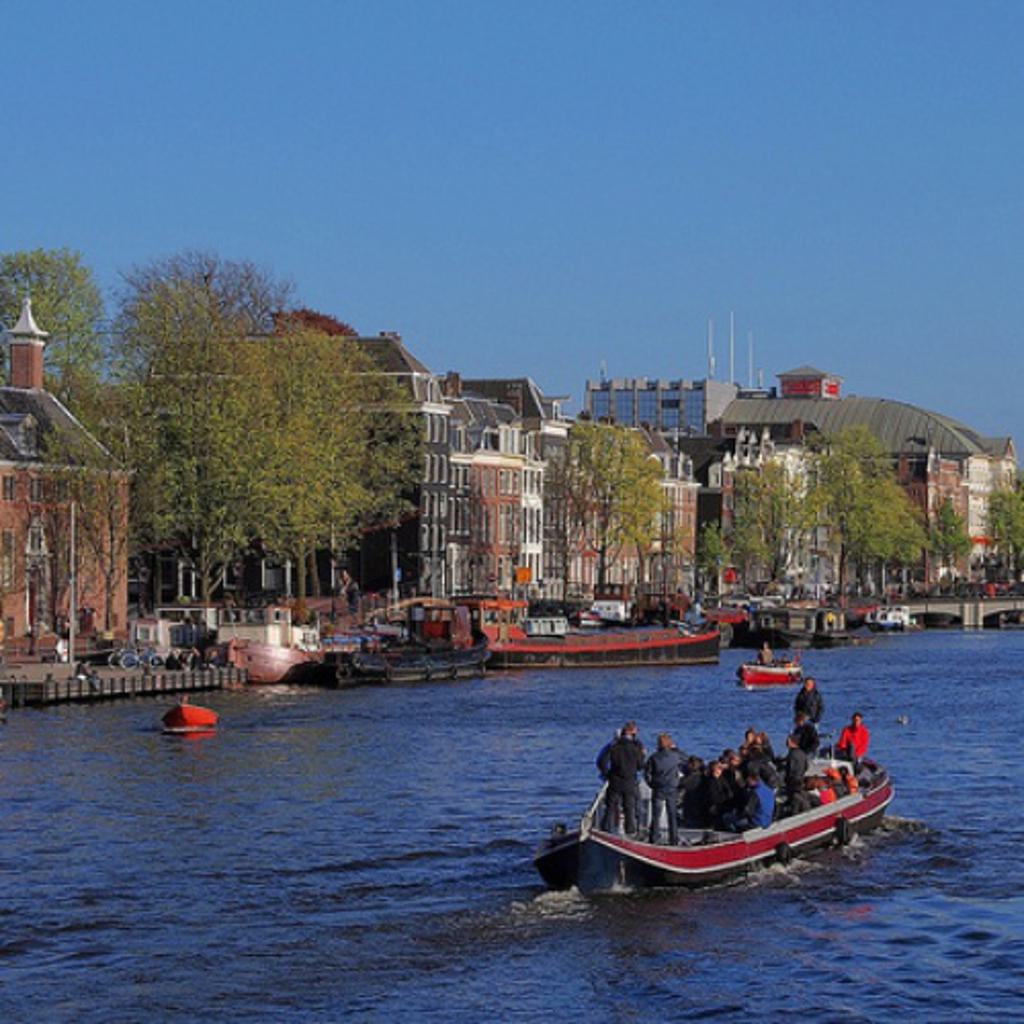} &
        \includegraphics[width=0.33\linewidth]{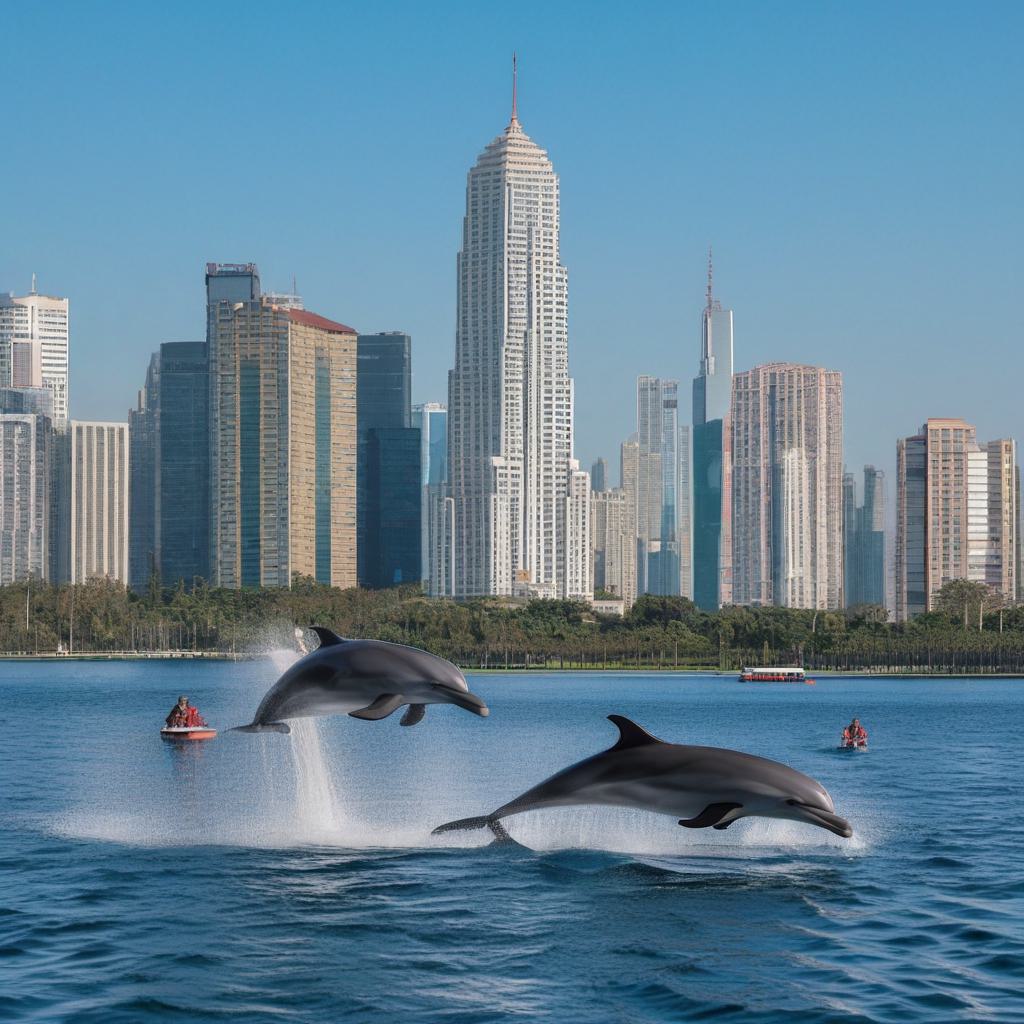} &
        \includegraphics[width=0.33\linewidth]{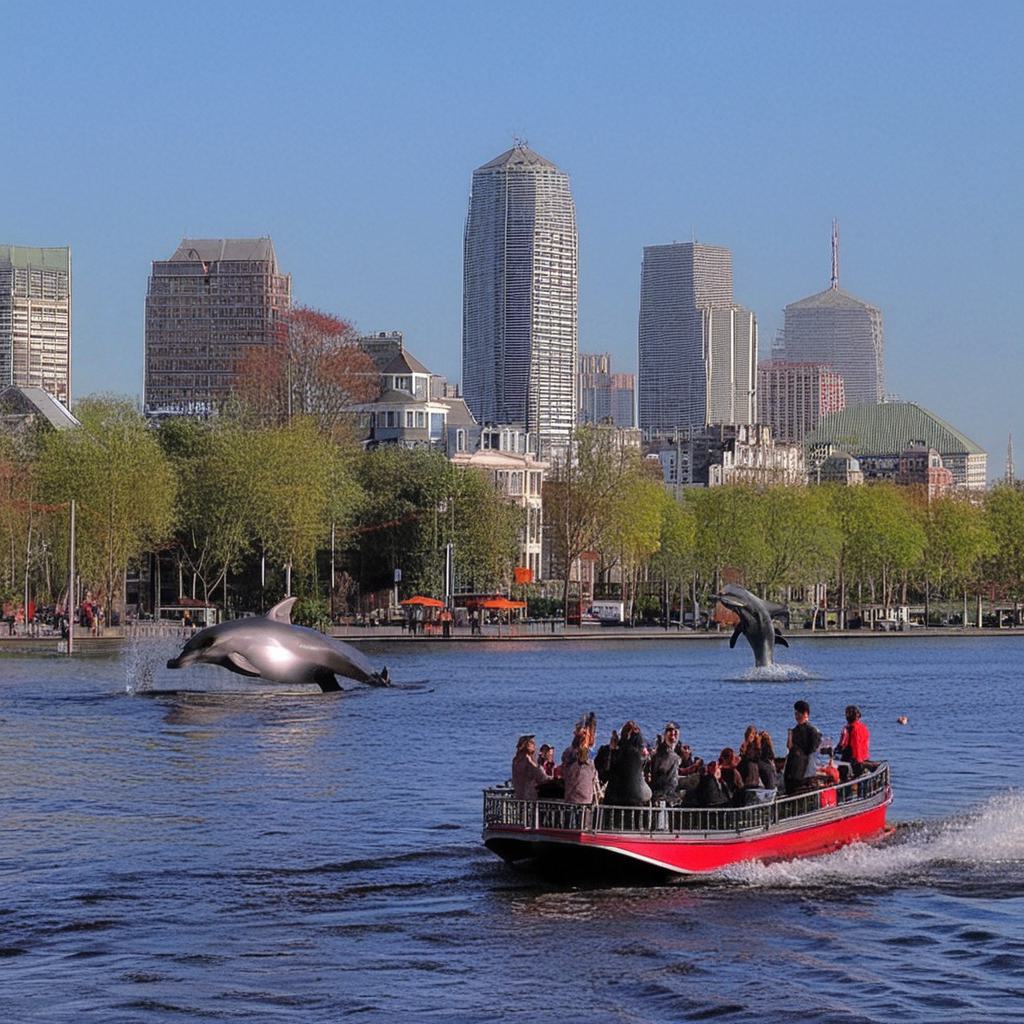} 
        \\
        Input & Edit. w/o Tight & Edit. w/ Tight 
        \end{tabular}
        }
    \end{minipage}%
    \hfill
    \begin{minipage}[t]{0.48\textwidth}
        \centering
        \setlength{\tabcolsep}{1pt}
        \scriptsize{
        \begin{tabular}{ccc}
            \multicolumn{3}{c}{``roasted coffee beans'' $\longrightarrow$ ``colorful candies'', DDIM Inversion} \\
            \includegraphics[width=0.33\linewidth]{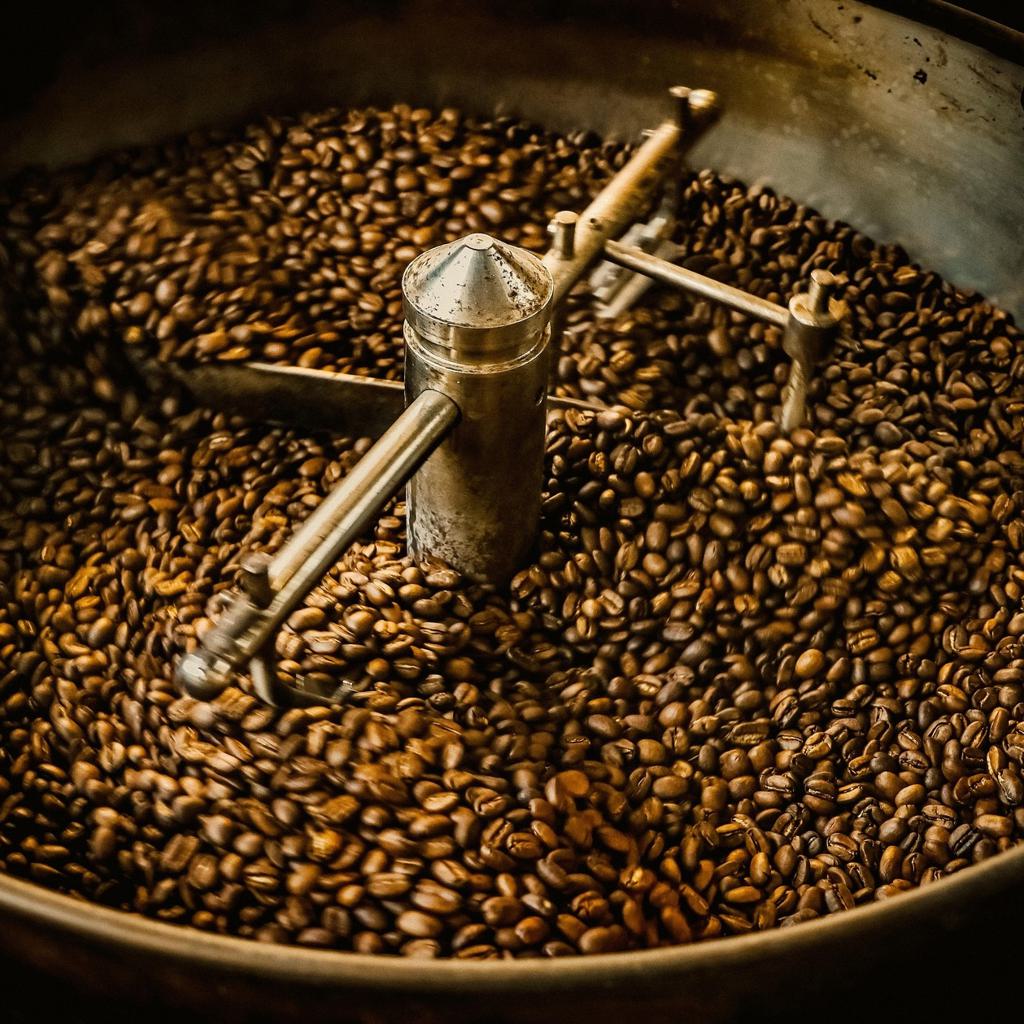} &
            \includegraphics[width=0.33\linewidth]{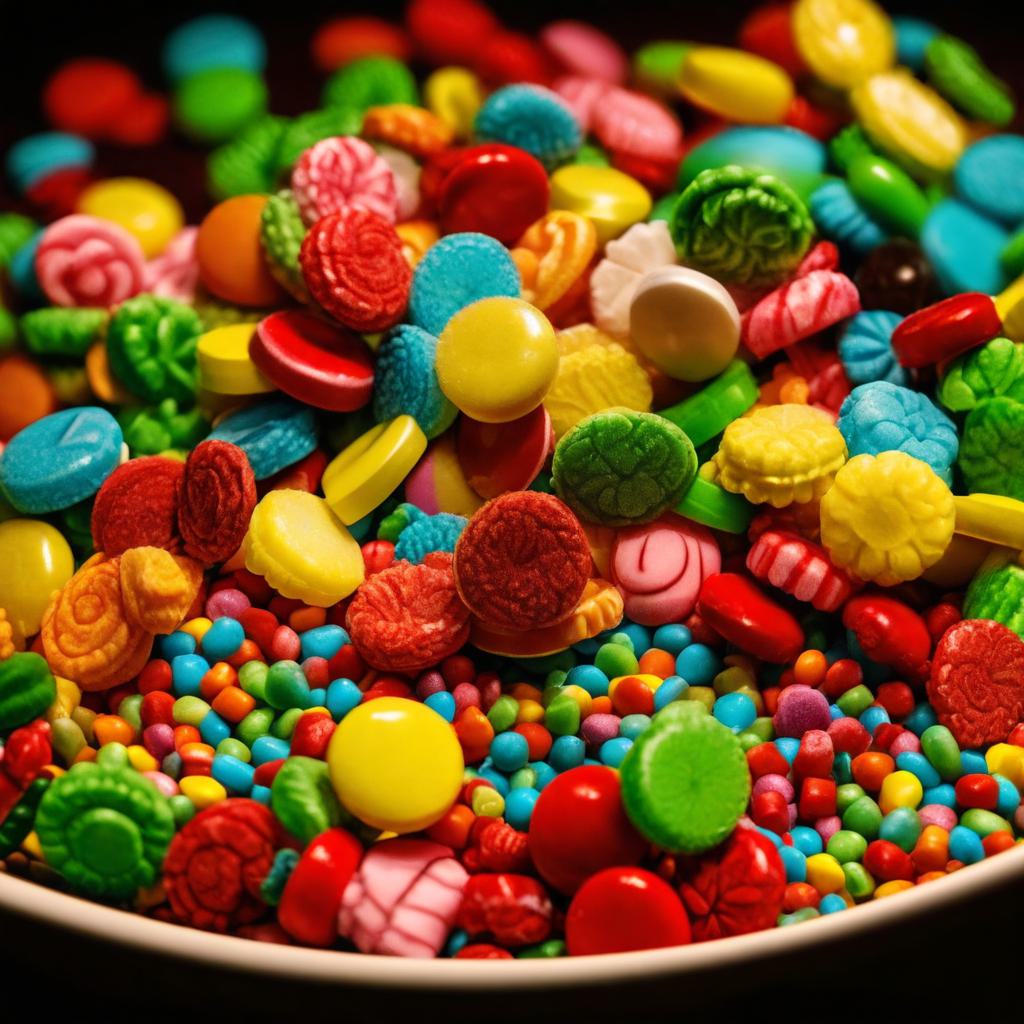} &
            \includegraphics[width=0.33\linewidth]{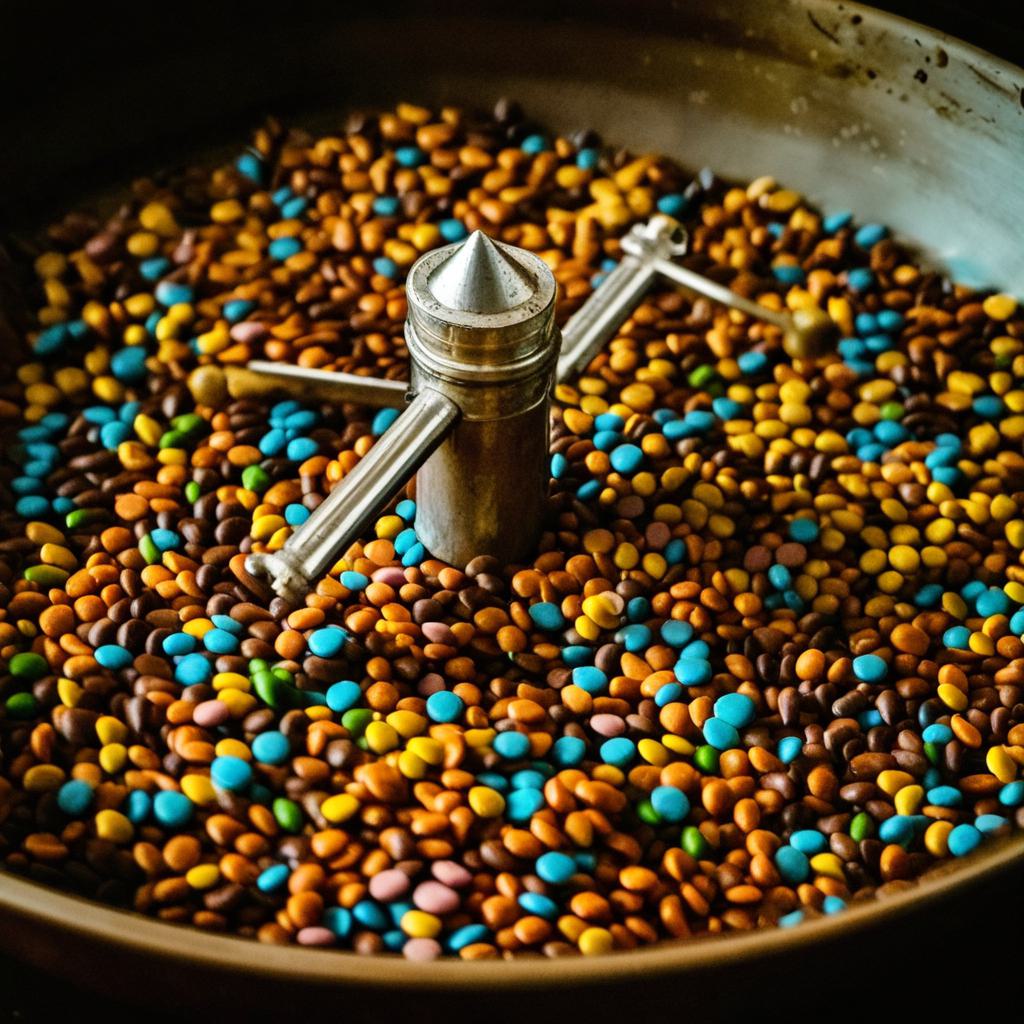} \\
            \multicolumn{3}{c}{``a man is riding a white horse in the sea'' $\longrightarrow$ ``... white lion ...'', Edit Friendly DDPM} \\
            \includegraphics[width=0.33\linewidth]{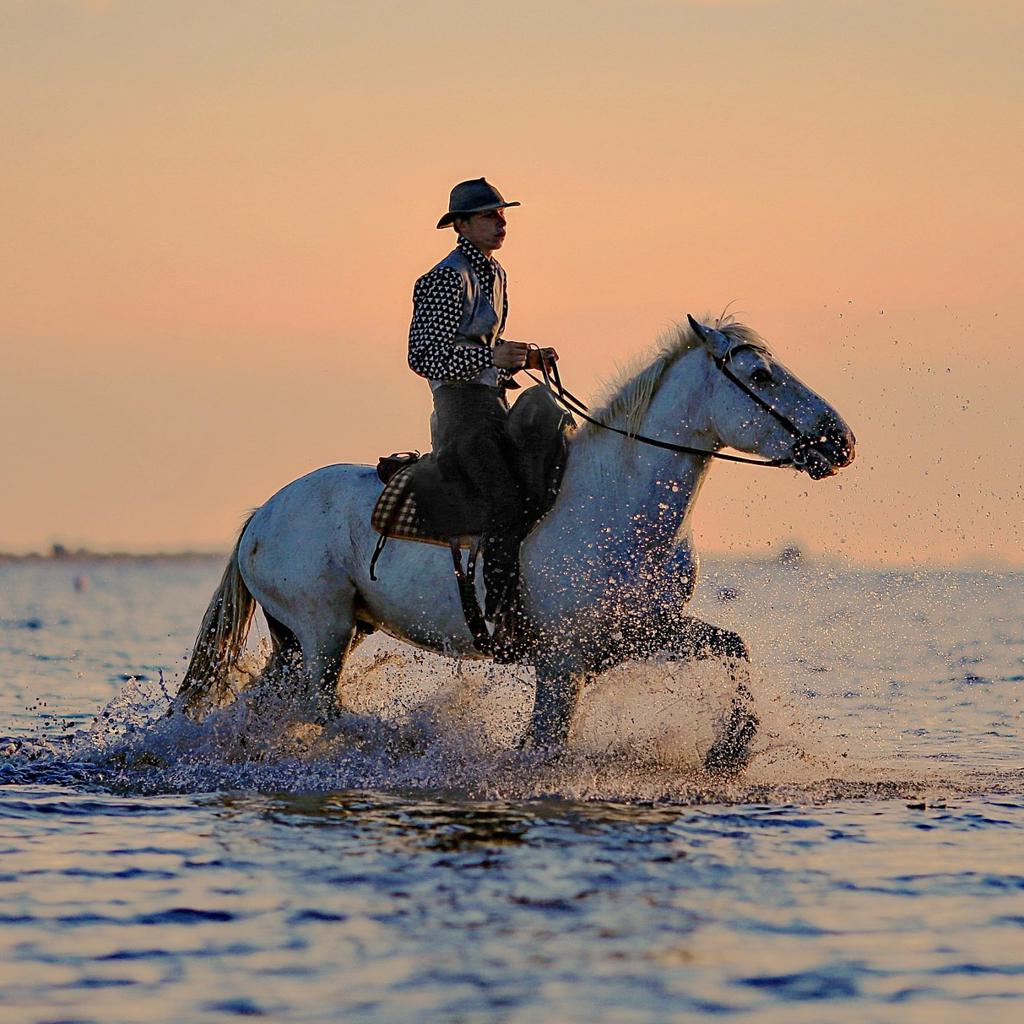} &
            \includegraphics[width=0.33\linewidth]{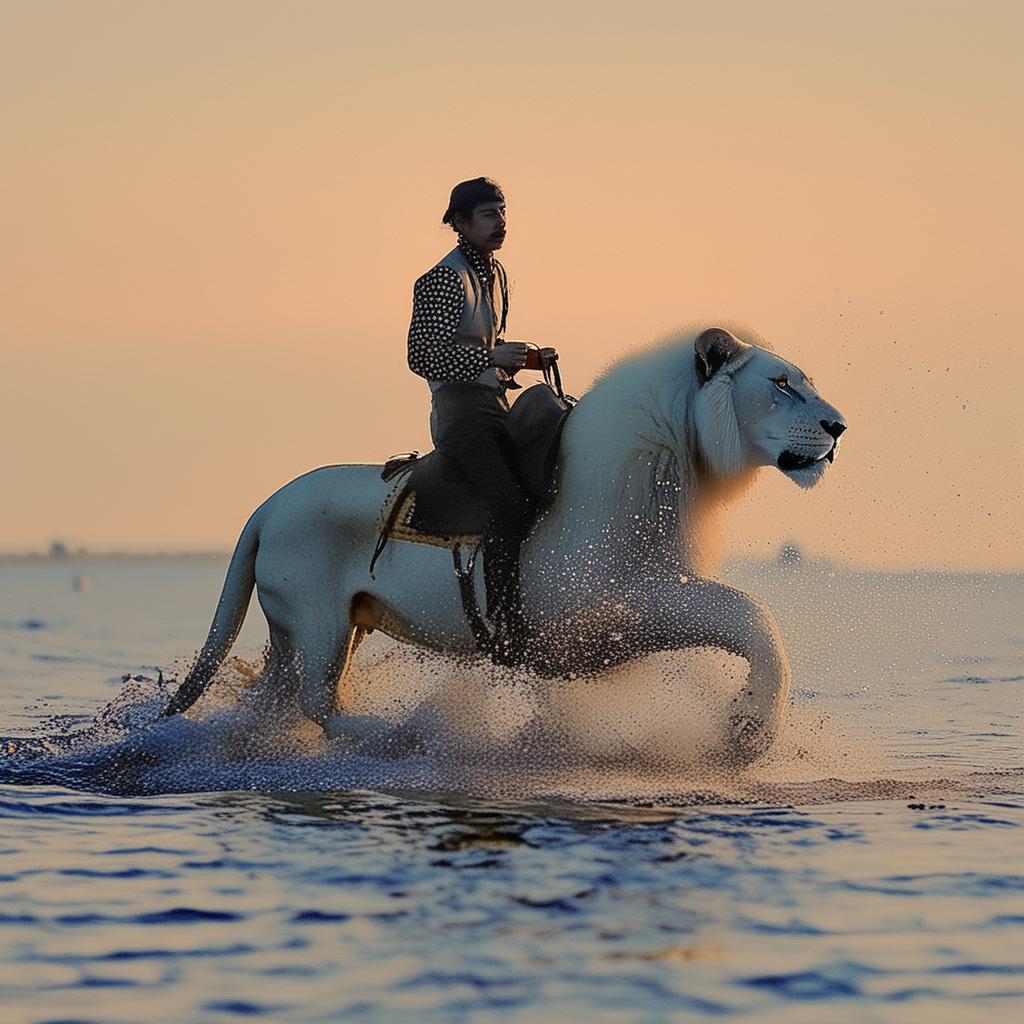} &
            \includegraphics[width=0.33\linewidth]{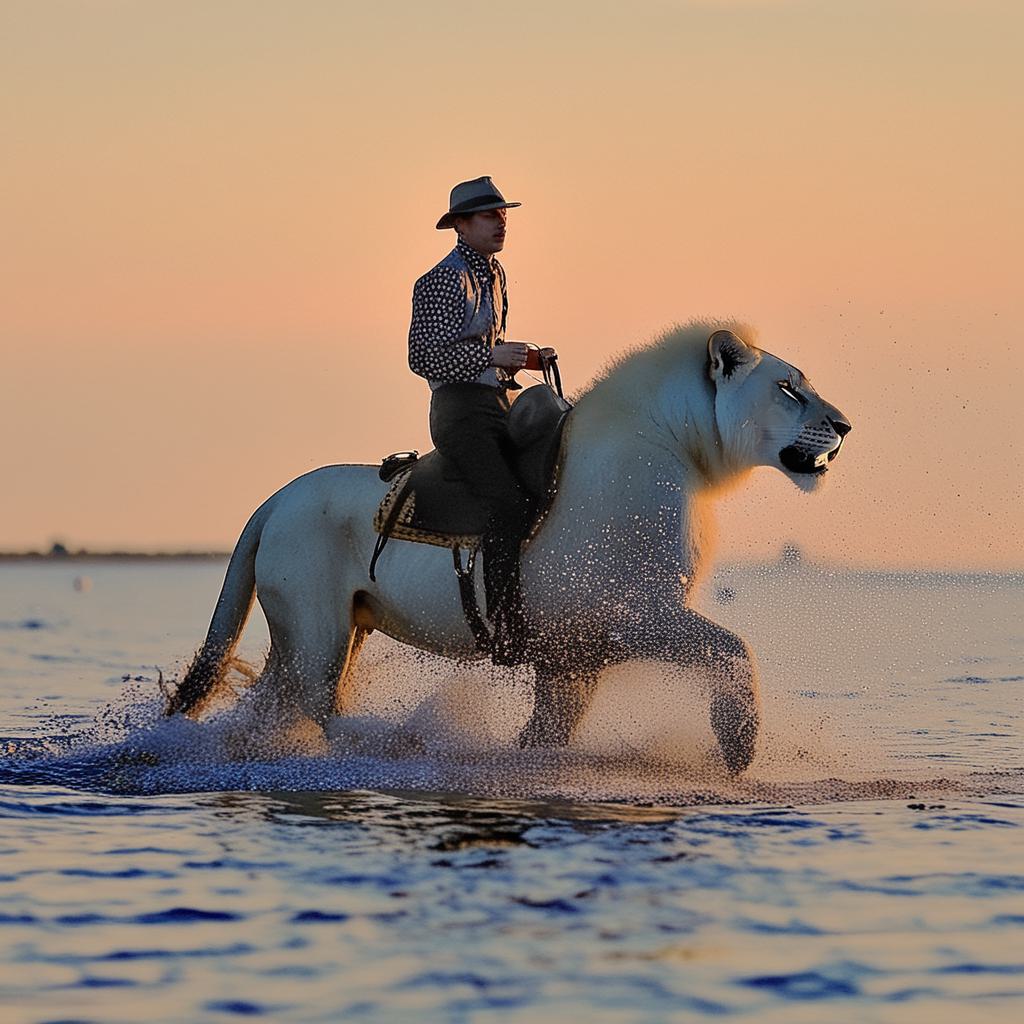} \\
            \multicolumn{3}{c}{``a marble statue'' $\longrightarrow$ ``... with sunglasses'', LEDITS++} \\
            \includegraphics[width=0.33\linewidth]{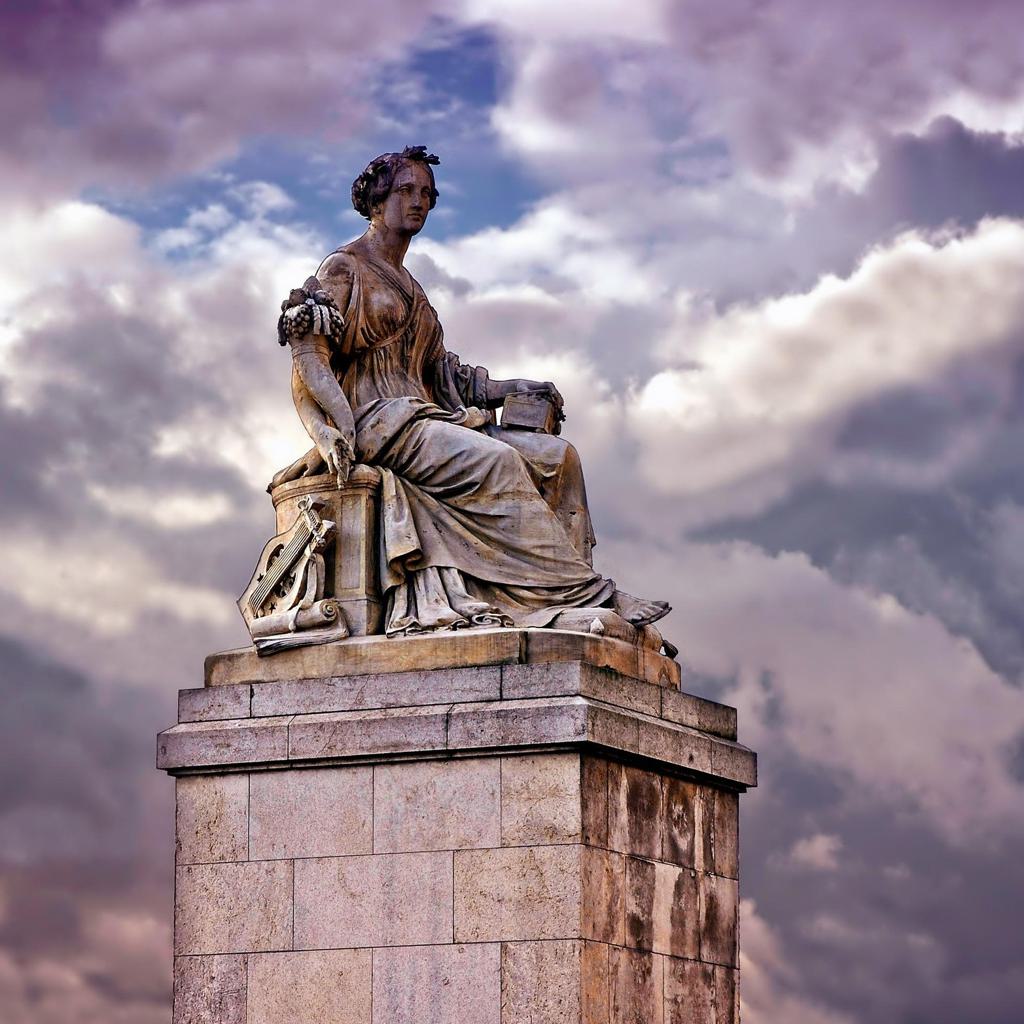} &
            \includegraphics[width=0.33\linewidth]{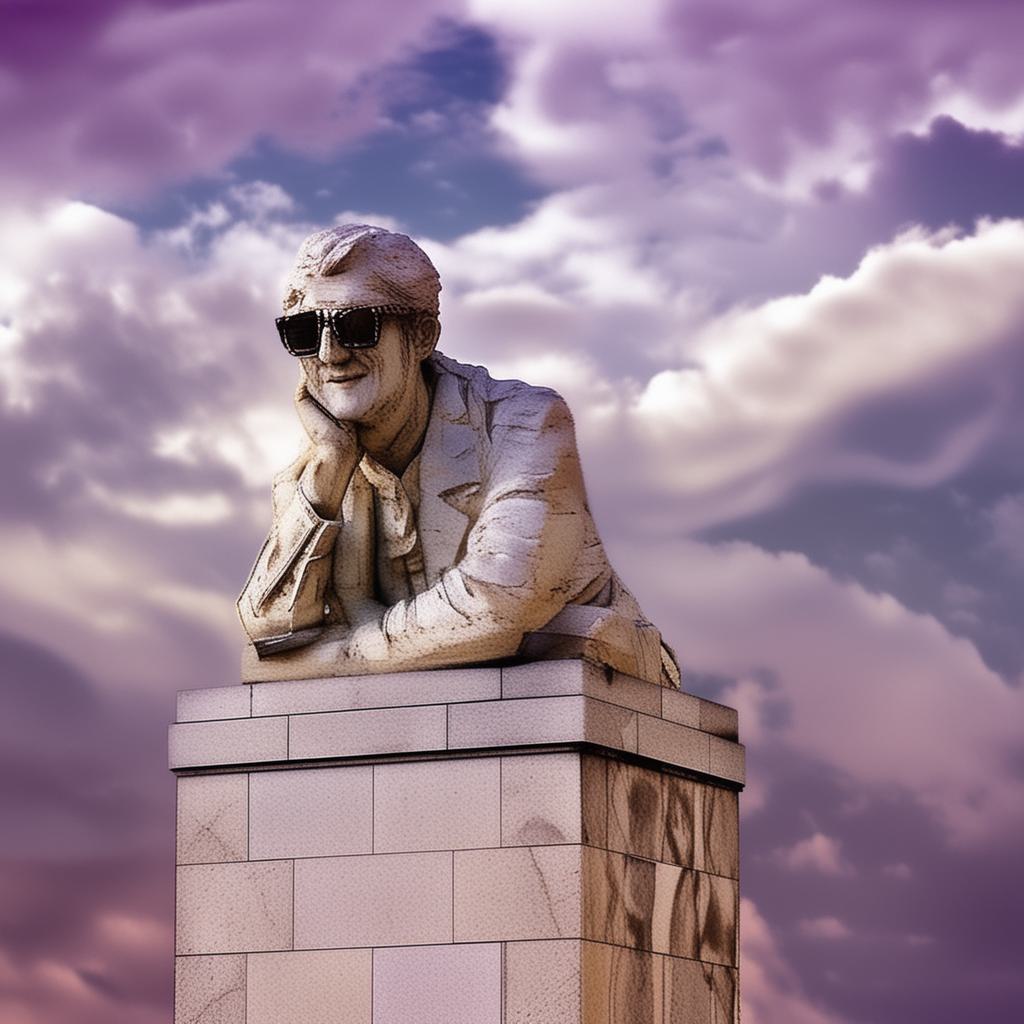} &
            \includegraphics[width=0.33\linewidth]{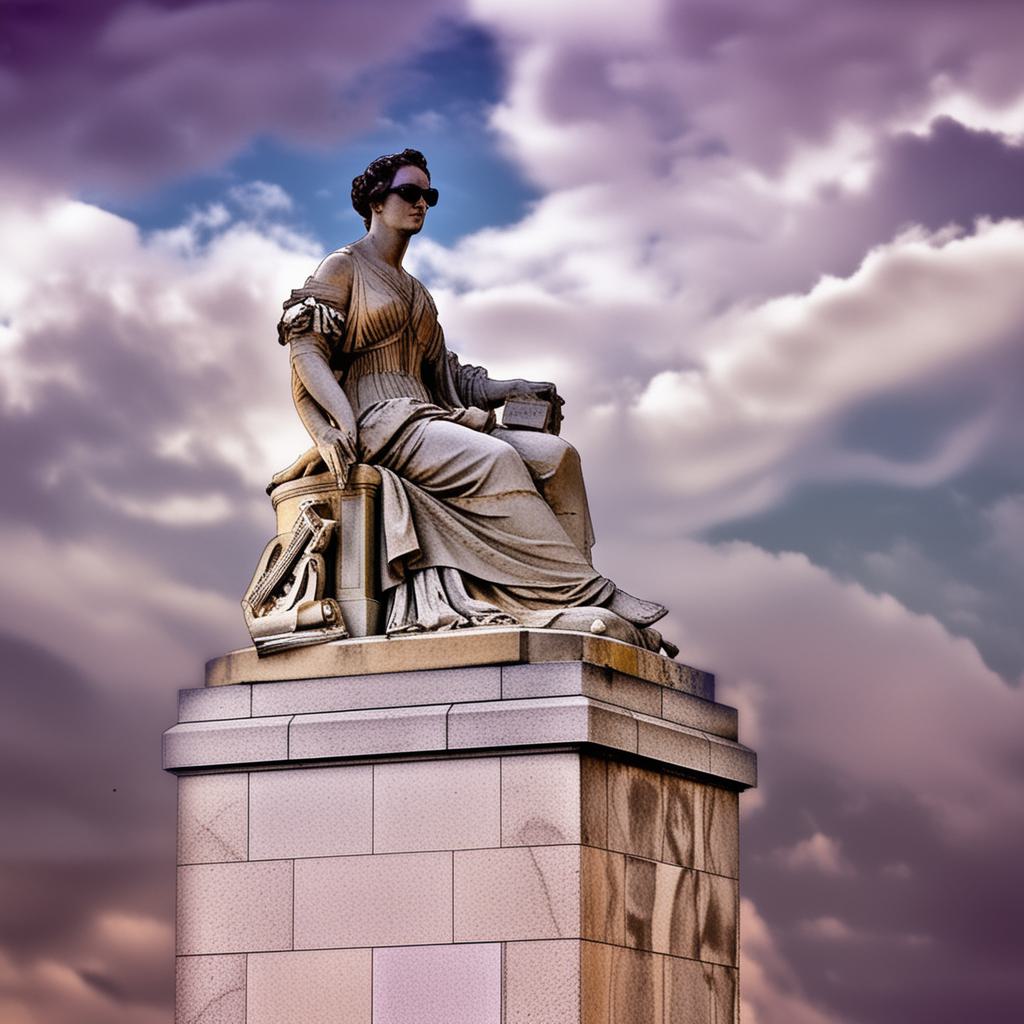} \\
            \multicolumn{3}{c}{``'' $\longrightarrow$ ``a person with a large thick beard'', RF-Inversion (Flux)} \\
            \includegraphics[width=0.33\linewidth]{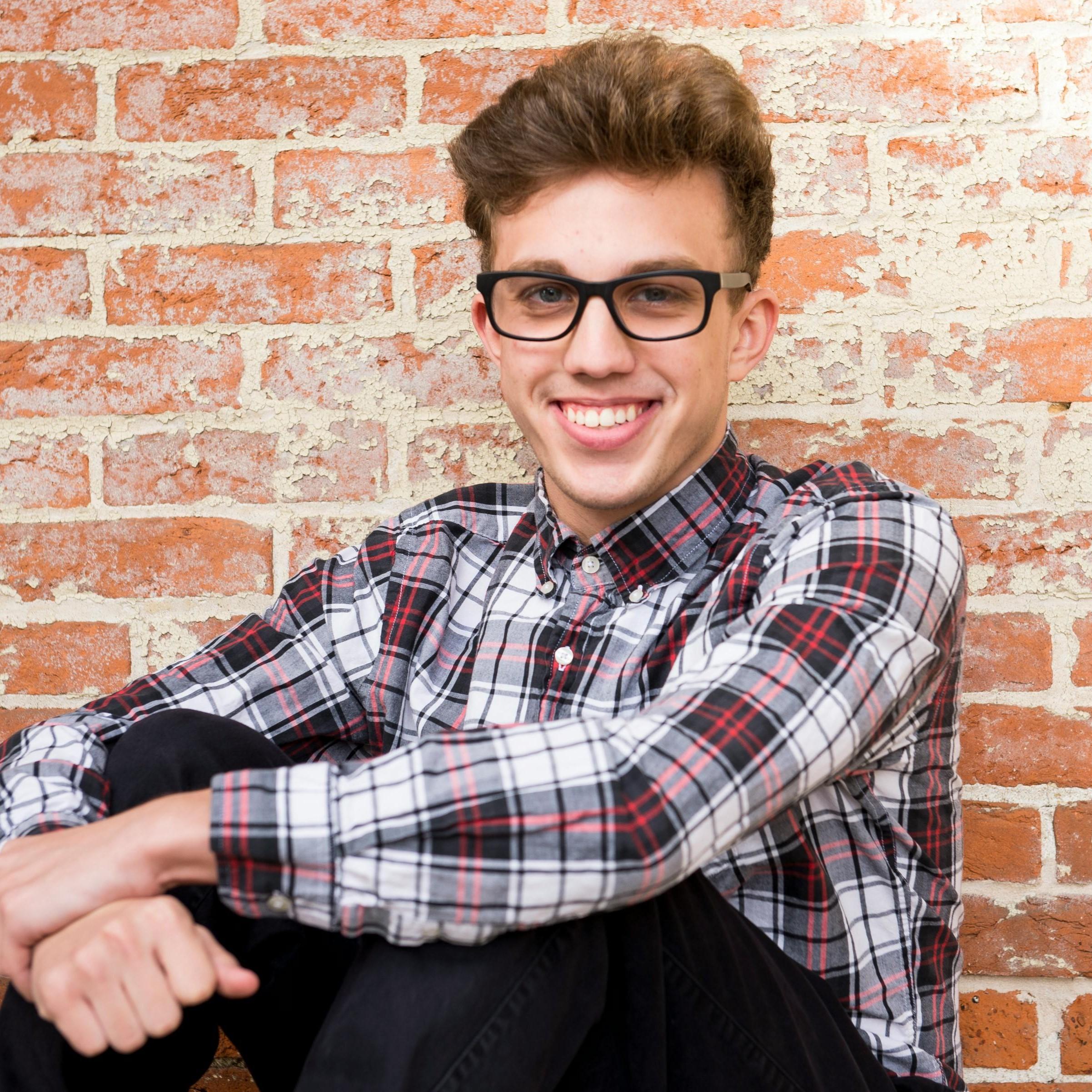} &
            \includegraphics[width=0.33\linewidth]{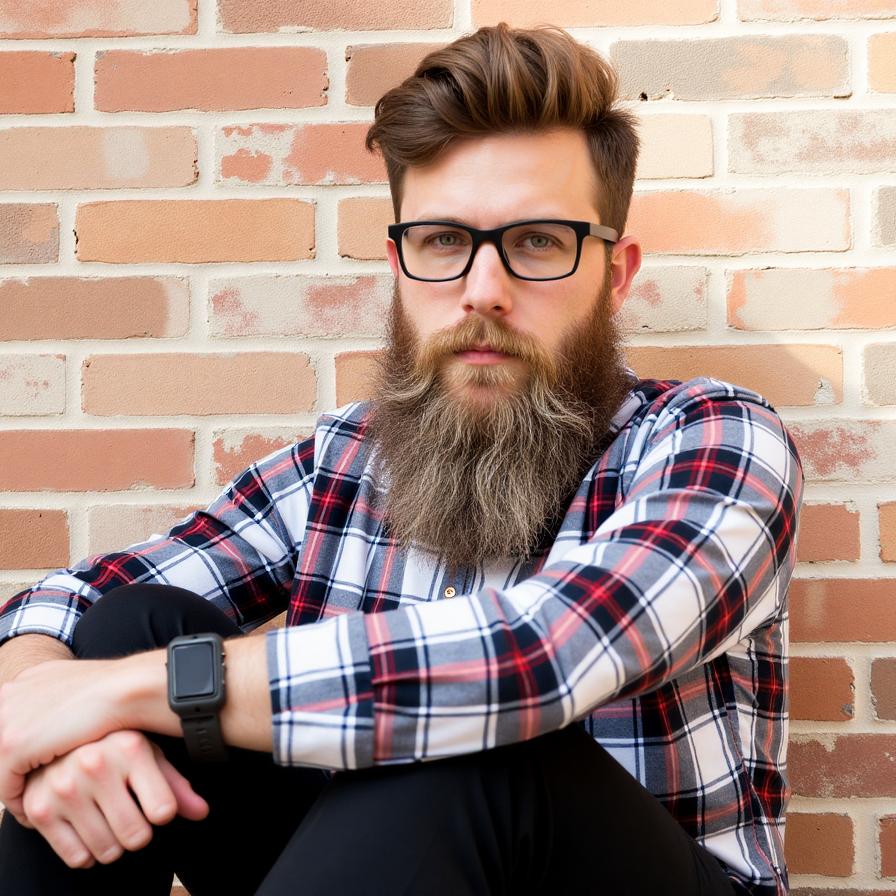} &
            \includegraphics[width=0.33\linewidth]{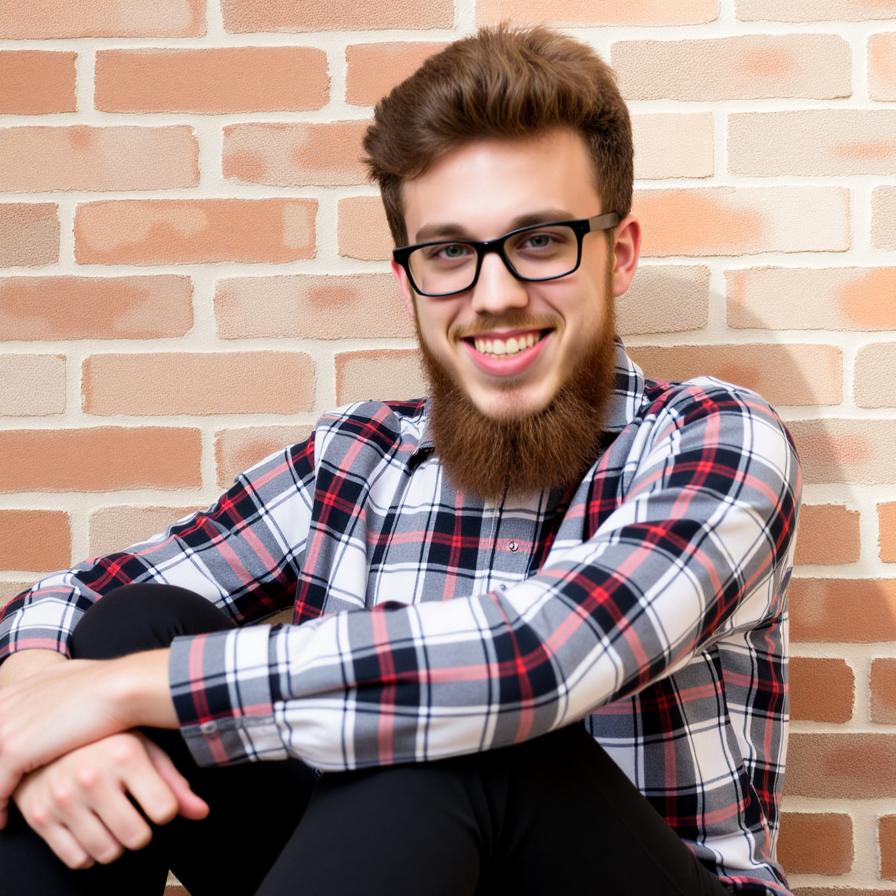} \\
            \multicolumn{3}{c}{``'' $\longrightarrow$ ``a person wearing a red hat with mountains in the background'', RF-Inversion (Flux)} \\
            \includegraphics[width=0.33\linewidth]{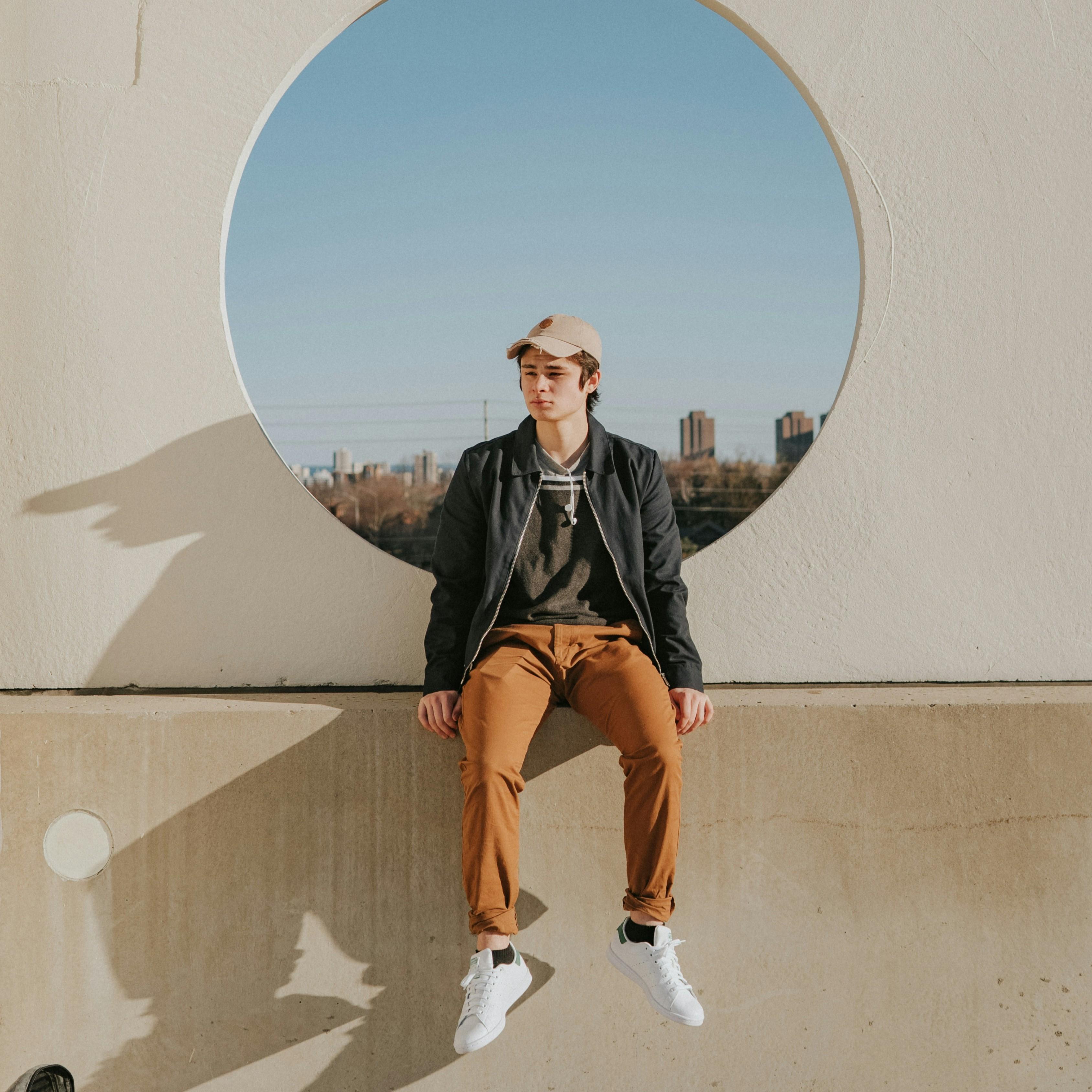} &
            \includegraphics[width=0.33\linewidth]{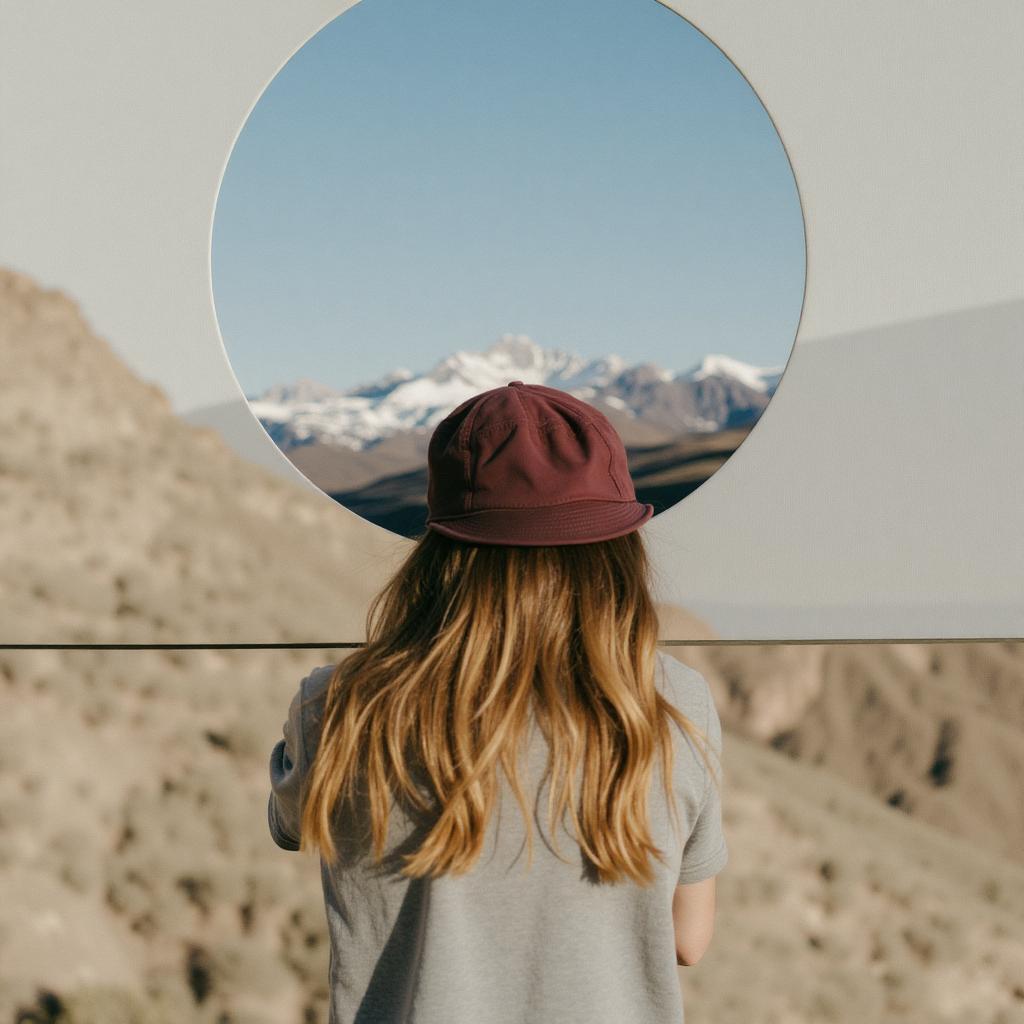} &
            \includegraphics[width=0.33\linewidth]{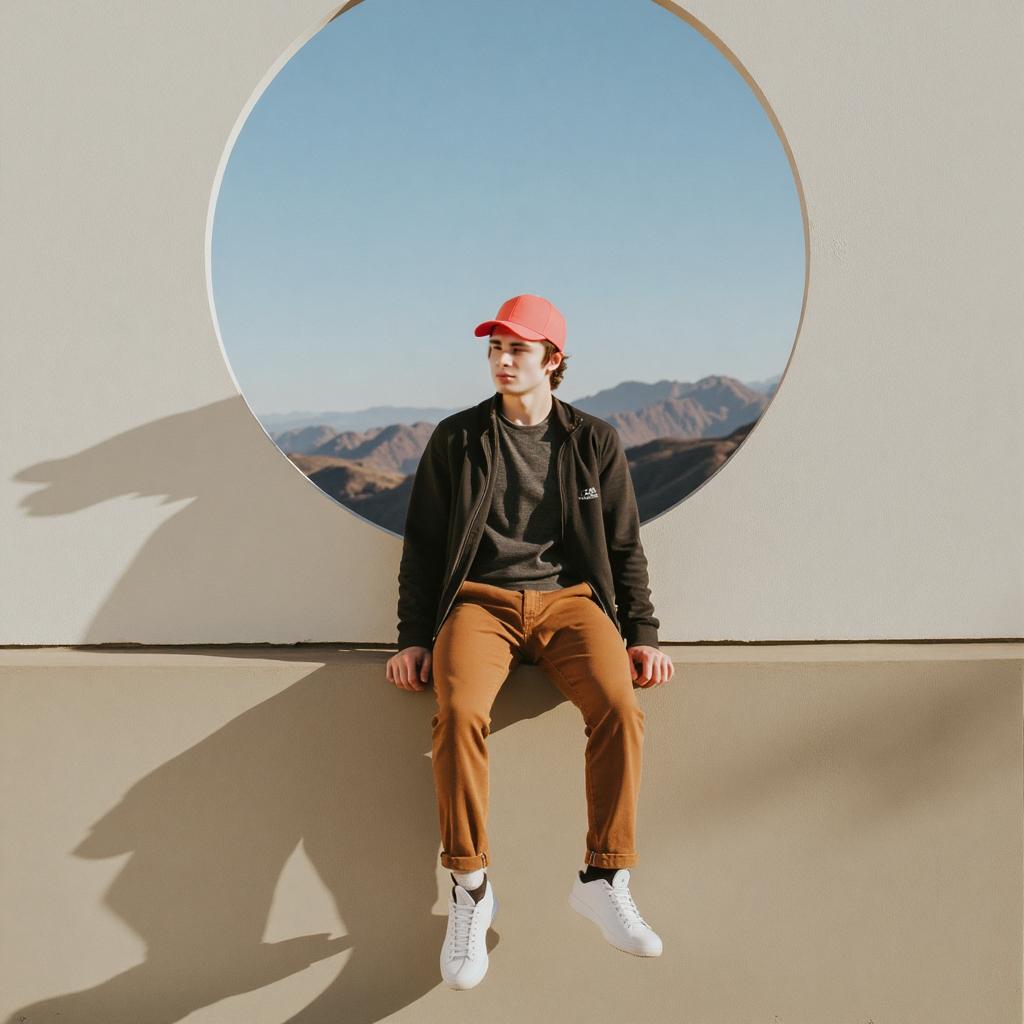} \\
            \multicolumn{3}{c}{``'' $\longrightarrow$ ``a person dressed in a formal suit and tie'', RF-Inversion (Flux)} \\
            \includegraphics[width=0.33\linewidth]{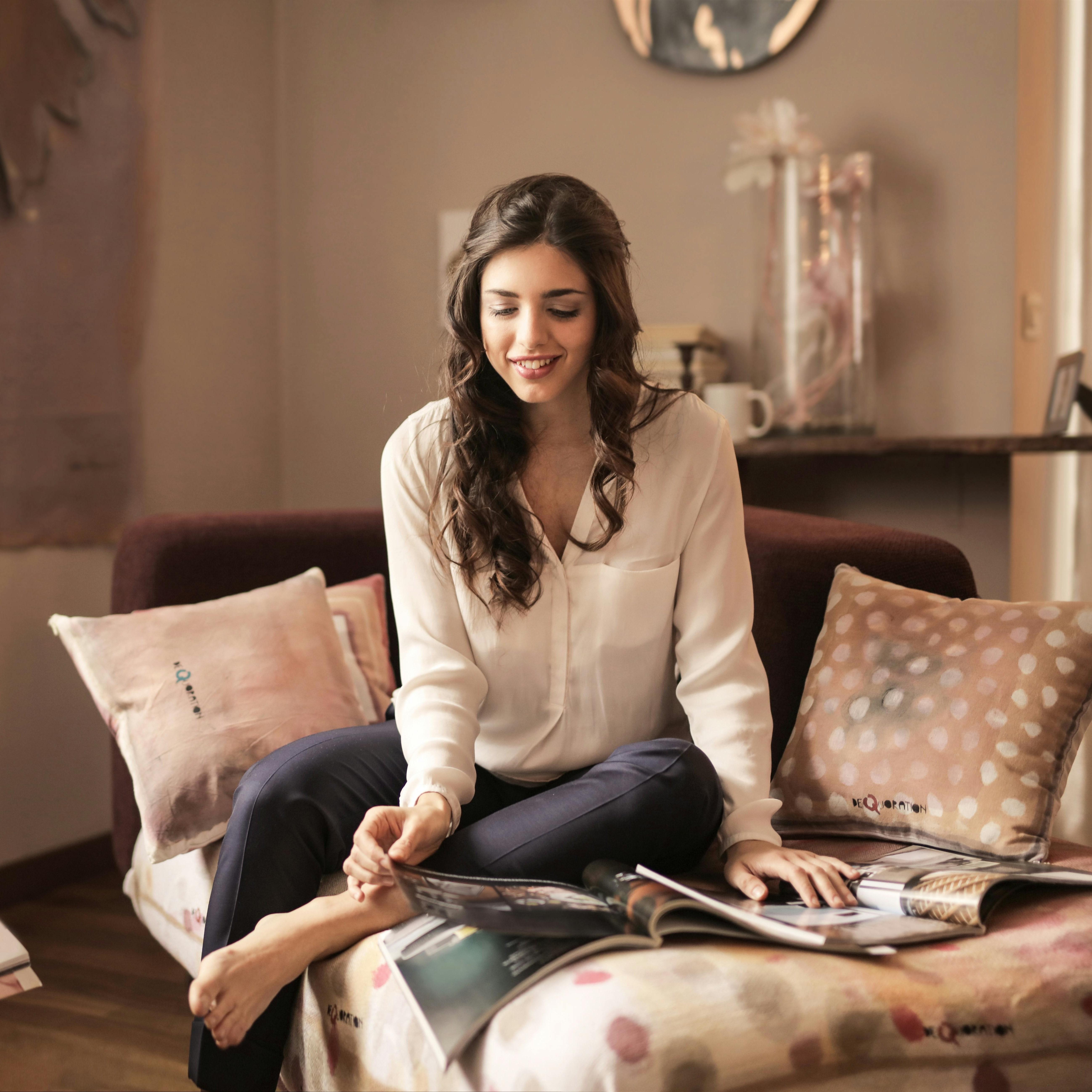} &
            \includegraphics[width=0.33\linewidth]{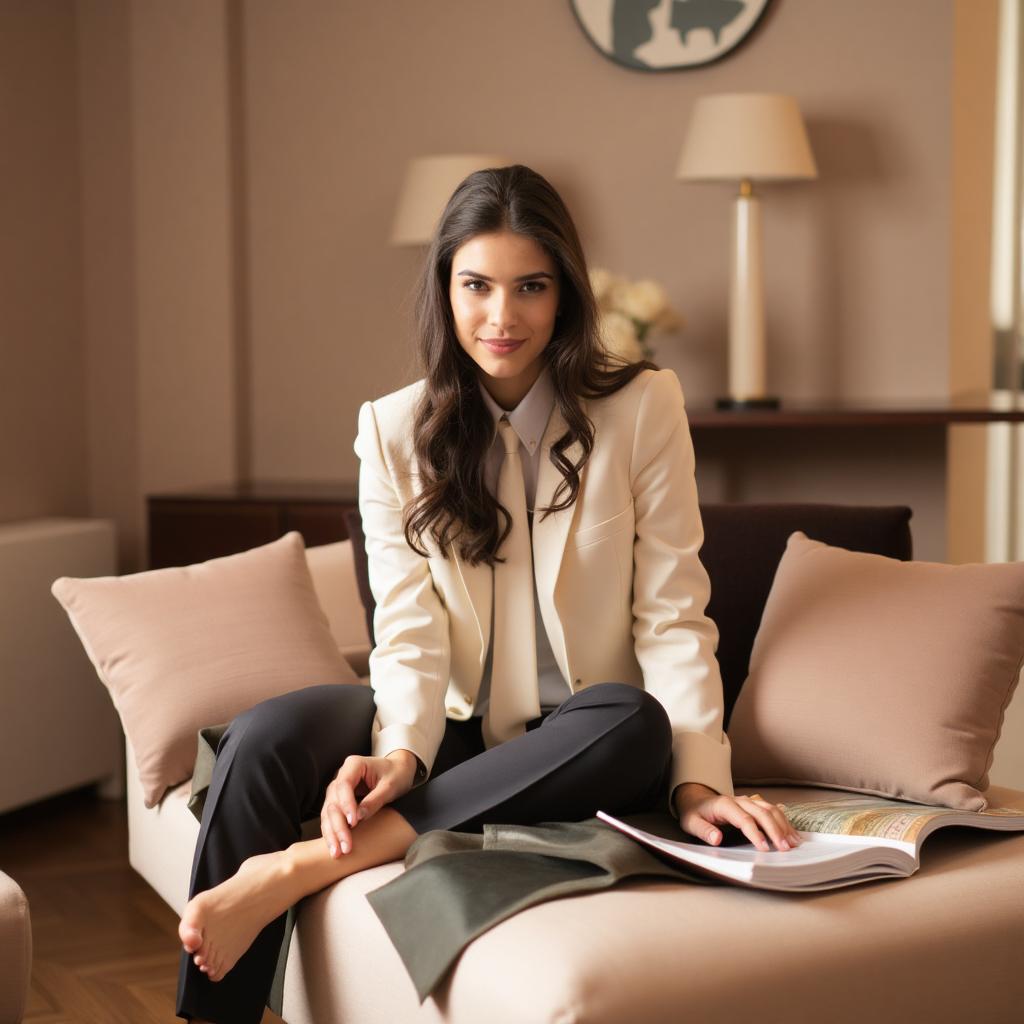} &
            \includegraphics[width=0.33\linewidth]{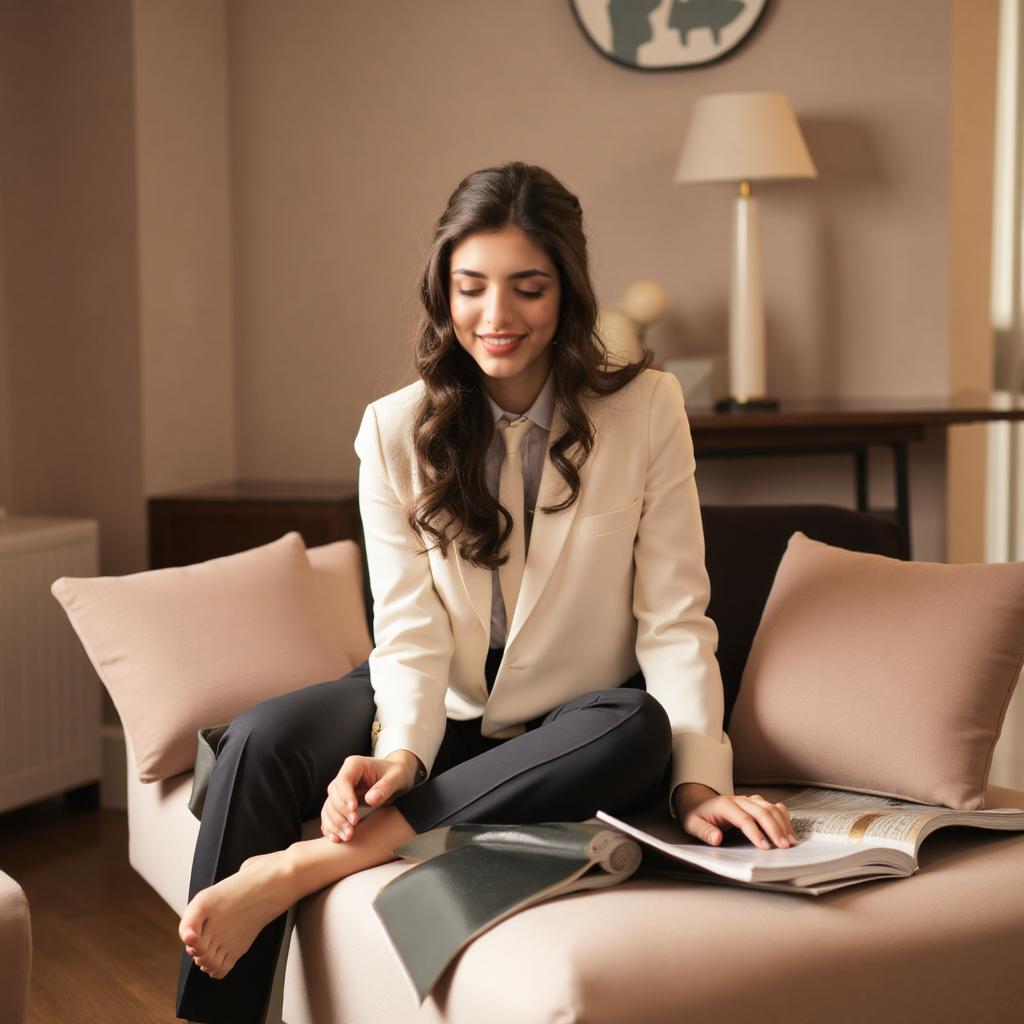} \\
            Input & Edit w/o Tight & Edit w/ Tight \\
        \end{tabular}
        }
    \end{minipage}

    \caption{Additional Editing Results.}
    \label{fig:more-results2}
\end{figure*}